\documentclass{aa}

\usepackage[varg]{txfonts}

\usepackage{graphicx,color}
\usepackage{natbib}
\usepackage[flushleft]{threeparttable}
\usepackage{rotating}
\usepackage{longtable, float}
\usepackage{pdflscape,lscape}
\usepackage{aas_macros}
\usepackage{subcaption} 
\usepackage{multicol,lipsum}
\usepackage{multirow}
\usepackage{dcolumn,booktabs}
\usepackage{siunitx}

\usepackage{url}

\newcommand{\mic}{$\mathrm{\mu m}$}

\newcommand{\msolyr}{{M$_{\odot}$}\,yr$^{-1}$}

\begin{document}

\title{PACS and SPIRE range spectroscopy of cool, evolved stars\thanks{The reduced spectra and the line subtracted spectra as well as 
Table~\ref{Table:LineF} are 
available in electronic form at the CDS via anonymous ftp to cdsarc.u-strasbg.fr (130.79.128.5) or via 
http://cdsweb.u-strasbg.fr/cgi-bin/qcat?J/A+A/.}\fnmsep 
%  \thanks{This is a second footnote} resulting in asymptotically faster convergence for the same amount of work per iteration} 
}

%\subtitle{II. An example text with infinitesimal scientific value\\ 
 % whose title and subtitle may also be split} 
 
\author{D. Nicolaes\inst{1,2}
  \and M.~A.~T. Groenewegen\inst{1} 
 % \thanks{\emph{Present address:} 
  % Department of Computer Science, Purdue University, West Lafayette, IN 47907, USA} 
     \and P. Royer\inst{2} 
     \and R. Lombaert\inst{3}
     \and T. Danilovich\inst{2} 
     \and L. Decin\inst{2}
     }

\institute{Koninklijke Sterrenwacht van Belgi\"e, 
  Ringlaan 3, 1180 Brussel, Belgium \\ \email{martin.groenewegen@oma.be}
  \and 
  Instituut voor Sterrenkunde, KU Leuven, 
  Celestijnenlaan 200D, 3001 Leuven, Belgium
  \and 
  Onsala Space Observatory, Department of Earth and Space Sciences, Chalmers University of Technology, 
  439 92, Onsala, Sweden
}

\date{Received * March 2018 / Accepted 20 July 2018}

%%%%%%%%%%%%%%%%%%%%%%%%%%%%%%%%%%%%%%%%%%%%%%%%%%%%%%%%%%%%%%%%%%%%%%%%%%%%%%%%%%%%%%%%%%%%%%%%%%%%%%%%%%%
%                      ABSTRACT
%%%%%%%%%%%%%%%%%%%%%%%%%%%%%%%%%%%%%%%%%%%%%%%%%%%%%%%%%%%%%%%%%%%%%%%%%%%%%%%%%%%%%%%%%%%%%%%%%%%%%%%%%%%
\abstract
%context
{At the end of their lives AGB stars are prolific producers of dust and gas. The details of this mass-loss process are
still not understood very well. {\it Herschel} PACS and SPIRE spectra which cover the wavelength range from 
$\sim$~55 to 670 $\mu$m almost continuously, offer a unique way of investigating properties of AGB stars in general 
and the mass-loss process in particular as this is the wavelength region where dust emission is prominent and 
molecules have many emission lines.
}
%aims
{We present the community with a catalogue of AGB stars and red supergiants (RSGs) 
with PACS and/or SPIRE spectra reduced according to the current state of the art.
}
%Methods 
{The {\it Herschel Interactive Processing Environment} (HIPE) software with the latest calibration is used to process
the available PACS and SPIRE spectra of 40 evolved stars. The SPIRE spectra of some objects close to the 
Galactic plane require special treatment because of the weaker fluxes in combination with the strong and 
complex background emission at those wavelengths. 
The spectra are convolved with the response curves of the PACS and SPIRE bolometers and compared to
the fluxes measured in imaging data of these sources.
Custom software is used to identify lines in the spectra, and to determine the central wavelengths and 
line intensities. 
Standard molecular line databases are used to associate the observed lines. Because of the limited spectral resolution 
of the PACS and SPIRE spectrometers ($\sim$~1500), several known lines are typically potential counterparts 
to any observed line. To help identifications in follow-up studies the 
relative contributions in line intensity of the potential counterpart lines are listed for three 
characteristic temperatures based on  local thermodynamic equilibrium (LTE) calculations and assuming optically thin emission.
}
%Results
{The following data products are released: the reduced spectra, 
the lines that are measured in the spectra with wavelength, intensity, potential identifications, 
and the continuum spectra, i.e. the full spectra with all identified lines removed.
As simple examples of how this data can be used in future studies we have fitted the continuum spectra
with three power laws (two wavelength regimes covering PACS, and one covering SPIRE) and find that the
few OH/IR stars seem to have significantly steeper slopes than the other oxygen- and carbon-rich 
objects in the sample, possibly related to a recent increase in mass-loss rate.
As another example we constructed rotational diagrams for CO (and HCN for the carbon stars) and fitted
a two-component model to derive rotational temperatures.
}
%Conclusion
{}


\maketitle

%%%%%%%%%%%%%%%%%%%%%%%%%%%%%%%%%%%%%%%%%%%%%%%%%%%%%%%%%%%%%%%%%%%%%%%%%%%%%%%%%%%%%%%%%%%%%%%%%%%%%%%%%%
%                    INTRODUCTION
%%%%%%%%%%%%%%%%%%%%%%%%%%%%%%%%%%%%%%%%%%%%%%%%%%%%%%%%%%%%%%%%%%%%%%%%%%%%%%%%%%%%%%%%%%%%%%%%%%%%%%%%%%

\section{Introduction} 
\label{intro}

The initial mass of a star determines its evolution and therefore also the final stages of its life. After leaving the main sequence, 
stars with an initial mass between $\mathrm{\sim 0.8~M_{\odot}}$ and $\mathrm{\sim 8~M_{\odot}}$ will climb the red giant and 
asymptotic giant branches (RGB and AGB), while more massive stars will go through a supergiant phase. 
During the AGB and supergiant phases, mass loss dominates the evolution and a star will expel a significant part of its 
initial mass via a stellar wind. 
%These winds are slow (typically $\mathrm{5 - 25\ km s^{-1}}$), while mass is lost at rates that vary 
%from $\mathrm{\sim 10^{-8} M_{\odot}yr^{-1}\ to\ 10^{-4} M_{\odot}yr^{-1}} $. 
The ejection of stellar material creates a cool and extended circumstellar envelope (CSE) containing dust grains and 
molecular gas-phase species. 
In this way, AGB and supergiant stars contribute significantly 
to the return of gas and dust to the interstellar medium (ISM) from which new generations of stars are born.

From a qualitative point of view it is known that the mass-loss processes are closely related to the intrinsic characteristics 
of the star, like mass, luminosity, variability and chemical composition \citep{Habing_and_Olofsson_2003, HO2018}. 
Despite extensive research efforts, stellar evolution models are not yet able to quantitatively predict the mass-loss history of 
AGB or supergiant stars from first principles. 
The details of the physical processes that govern the mass-loss dynamics and its variation in time remain unclear. 
A fulfilling description of the different key chemical processes that determine the wind's chemical composition is also lacking. 
Observationally characterising the full dynamical and chemical structure of the CSE from the stellar atmosphere up to the most 
outer parts of the wind will be helpful in clarifying the underlying mass-loss mechanism by providing models with as many 
constraints as possible. 

The Herschel Space observatory (hereafter \textit{Herschel}; \citealt{Pilbratt_etal_2010}) plays a key role in these analyses. 
\textit{Herschel} collected data at far-infrared and submillimetre wavelengths which cover a large part of the wavelength 
region where the gas and dust in the extended CSE emit most of their continuum and line radiation. In this way, \textit{Herschel} bridges 
the gap between ground-based instruments which are only able to obtain data in selected atmospheric windows at 
shorter (near- and mid-infrared) and those that can obtain data at longer (millimetre and radio) wavelengths. Due to the good spatial and 
spectral resolution of the instruments on board, \textit{Herschel} revealed new insights in the structure and chemistry of CSEs.  

This paper presents consistent and carefully reduced data of the PACS and SPIRE instruments on board \textit{Herschel} of all 
AGB and supergiants stars that were observed by the PACS and SPIRE spectrometers. 

A large fraction of the data presented here were obtained within the framework of the Mass loss of Evolved StarS (MESS) 
Guaranteed Time Key Programme \citep{Groenewegen_etal_2011} and were published in part in earlier publications, using the best available 
data reduction at that time.
%, sometimes in combination with additional ground-based data and with the aim of modelling individual objects. Examples are
\citet{Royer_etal_2010} presented PACS and SPIRE data on VY CMa and an initial model for the CSE, while a more elaborate analysis 
was presented by \cite{Matsuura2014}, using radiative transfer models to fit the $^{12}$CO, $^{13}$CO, SiO and water lines in these spectra 
and to derive mass-loss rates (MLRs) and the gas temperature profile in the CSE.
%\citet{Wesson2010} present SPIRE data on three C-rich post-AGB objects         science demonstration phase
\citet{Decin_etal_2010_Nature} presented the detection of high-excitation lines of H$_{\rm 2}$O in the carbon star CW Leo (IRC +10 216) 
and suggested that interstellar UV photons could penetrate deep into the clumpy CSE, an alternative scenario to the one 
proposed by \citet{Melnick01} of the vaporisation of a collection of orbiting icy bodies based on the detection of a single line with {\it SWAS}.
This analysis was later extended by \citet{Lombaert_etal_2016} who studied the water lines in 18 C stars (including six targets from the MESS program).
\citet{Danilovich2017} studied the water isotopologues in four M-type stars (R Dor, IK Tau, R Cas, and W Hya) including data from MESS.
Other studies analysed molecular line data (based partly on MESS data) for CW Leo \citep{DeBeck2012}, OH~127.8\,+0.0 \citep{Lombaert_etal_2013}, 
W Hya \citep{Khouri2014a, Khouri2014b}, W Aql \citep{Danilovich2017}, and R Dor \citep{VandeSande2018}, typically using radiative
transfer models to derive properties of the CSE, such as abundances or abundance profiles.

The present paper also discusses PACS and SPIRE imaging data, but only for the targets which have spectroscopic data.
Initial PACS and SPIRE photometry was presented in \citet{Groenewegen_etal_2011}, but not all observations had been completed at that time.
An overview of the PACS imaging of all 78 MESS targets can be found in \cite{Cox_etal_2012},  showing and discussing, amongst other things,
 four different classes of wind-ISM interaction observed in $\sim 40\%$ of the sample.

MESS imaging data have been discussed in more detail for individual objects as well.
\citet{Ladjal2010} discussed the bow shock around CW Leo seen in SPIRE data (discovered a few months earlier in GALEX UV data 
by \citealt{Sahai2010}) while \citet{Decin2011} presented the discovery of multiple shells around this object.
\cite{Decin2012} discussed the detection of the bow shock around Betelgeuse, while the interesting class of 
C-rich objects with detached shells have been discussed in \citet{Kerschbaum2010} (AQ And, U Ant, and TT Cyg) and \citet{Mecina2014} (S Sct and RT Cap).
The CSE of stars showing binary interaction have been discussed by \citet{Mayer2013} (R Aqr and W Aql) and \citet{Mayer2014} ($\pi$ Gru).

The paper is structured as follows. 
Section~\ref{S-Data} presents the data sample and describes the adopted data reduction and processing strategy. 
In Section~\ref{SectQual} the flux level of the PACS and SPIRE spectra is compared to that measured independently by the PACS and 
SPIRE bolometer arrays in order to have an estimate of the flux level consistency and to identify possible problematic stars or wavelength regions.
n Section~\ref{LineMeasurement} the strategy to extract the molecular lines is outlined, while 
Sect.~\ref{S_Cont} describes the determination of the dust continua.
Section~\ref{S_DA} discusses the identification of the molecular lines, the construction of rotational diagrams, and the derivation of 
rotational temperatures for CO (and HCN for the carbon stars), and the slopes of the dust continua.
Section~\ref{sect:discussion} summarises this paper.

When this paper was submitted we became aware of the article by \citet{THROES} that presents PACS range spectroscopy of 114 evolved stars.
The sample they consider also includes planetary nebula and post-AGB stars and is therefor larger than ours.
For the reader it is important to know that our effort and theirs were were carried out independently of each other.

%%%%%%%%%%%%%%%%%%%%%%%%%%%%%%%%%%%%%%%%%%%%%%%%%%%%%%%%%%%%%%%%%%%%%%%%%%%%%%%%%%%%%%%%%%%%%%%%%%%%%%%%%%%
%    DATA: sample, observations, data processing, line measuring, continuum extraction
%%%%%%%%%%%%%%%%%%%%%%%%%%%%%%%%%%%%%%%%%%%%%%%%%%%%%%%%%%%%%%%%%%%%%%%%%%%%%%%%%%%%%%%%%%%%%%%%%%%%%%%%%%%

\section{Data}
\label{S-Data}

\subsection{Target sample and observations} 
\label{sect:sample}

The sample consists of 37 AGB and 3 RSG stars observed by the 
PACS \citep{Poglitsch_etal_2010}\footnote{Also see the PACS observers manual at \url{http://herschel.esac.esa.int/Docs/PACS/pdf/pacs_om.pdf} 
or \url{http://herschel.esac.esa.int/Docs/PACS/html/pacs_om.html}.} and 
SPIRE \citep{Griffin_etal_2010}\footnote{Also see the SPIRE handbook at \url{herschel.esac.esa.int/Docs/SPIRE/spire_handbook.pdf} 
or \url{http://herschel.esac.esa.int/Docs/SPIRE/html/spire_om.html}.} instruments on board \textit{Herschel}. 
As this paper focuses on spectroscopy some of the main properties of the spectrometers are recalled below. 
The PACS spectrometer field of view consists of $5 \times 5$ spatial pixels (spaxels) of $9.4 \times 9.4$\arcsec\ each on the sky. 
Background subtraction is done using a classical chop-nod technique, with chopper throws of 1.5, 3.0 or 6.0\arcmin.

The SPIRE Fourier Transform Spectrometer (FTS) consists of two hexagonally close-packed arrays with 37 detectors 
in the short-wavelength array (SSW) and 19 in the long-wavelength array (SLW). 
The full instrument field of view is 2.6\arcmin\ in diameter.
The size of the beam and the resolution (represented as the width of an unresolved line) depend on wavelength and are given in Table~\ref{Table:Inst}.

\begin{table}
 \caption{ Basic properties of the PACS and SPIRE spectrometers. }
\label{Table:Inst}						
\centering	
\begin{tabular}{lllllll}
\hline\hline
band/range 	&  FWHM of an                     & PSF           \\
       	        &  unresolved line   ($\mu$m)      & (FWHM in \arcsec)  \\
\hline	

 B2A-B3A (55--72  $\mu$m) & 0.021--0.013  & $\sim 9.0$ \\
 B2B (72--105 $\mu$m) & 0.039--0.028  & $9.0-9.3$  \\
 R  (105--210 $\mu$m) & 0.10--0.13    & $9.3-14$  \\
SSW (191--318 $\mu$m) & 0.15--0.40 & 16.5-20.5 \\  % 1568 - 944 GHz
SLW (294--671 $\mu$m) & 0.35--1.8  & 31.0-42.8 \\  % 1018 - 447 
\hline	

\end{tabular}	
%\tablefoot{} 	
\end{table}

The main characteristics of the 40 targets are listed in Table~\ref{Table:sample}. 
Distances and MLRs are representative values taken from the 
literature and are not explicitly used in this paper. Together with the pulsation type and chemical type these parameters illustrate
the diversity of the sample. The expansion velocities are used in the paper (see Sect.~\ref{Sect:Identification}) to correct 
the central wavelength of the observed molecular lines to rest wavelengths.

More precisely, the Herschel Science Archive (HSA) was searched and all AGB and RSG 
targets with a PACS spectrum observed in spectral energy distribution (SED) mode were selected. In this mode, 
the full PACS wavelength range is covered by combining at least two astronomical observation requests (AORs). The three different 
SED-AOR options are\footnote{Nomenclature and wavelength ranges following the PACS observers manual Sect.~6.2.7.1. 
Short and long R1 are sometimes also designated R1A and R1B, respectively.}: 
B2A + short R1 ([51--73] $\mathrm{\mu m}$ + [102--146] $\mathrm{\mu m}$),  
B2B + long R1 ([70--105] $\mathrm{\mu m}$ + [140--220] $\mathrm{\mu m}$), and 
B3A + long R1  ([47--73] $\mathrm{\mu m}$ + [140--219] $\mathrm{\mu m}$). 
The bluest and reddest parts of the spectrum cannot be reliably calibrated (see Sect.~\ref{sectPACS}), 
and flux calibrated data is available from 55 to 95 and from 102 to 190~\mic.

The PACS spectra were complemented with single pointed 
%and high resolution 
SPIRE FTS data, when available. 
A complete SPIRE spectrum spans a wavelength range from 190 to 670 \mic\ covered by two bands: the SSW band (191 -- 317 \mic) and 
the SLW band (294 -- 670 \mic), which are simultaneously observed within 1 AOR.

Additionally, the HSA was searched for PACS and SPIRE photometric maps, which will be used as reference data. 
PACS data are simultaneously obtained
in the blue  ($\mathrm{\lambda_{ref}} =  70$ \mic) and red ($\mathrm{\lambda_{ref}}  = 160$ \mic), or simultaneously 
in the green ($\mathrm{\lambda_{ref}} = 100$ \mic) and red bands, 
while SPIRE maps are obtained in all three bands at once: the PSW ($\mathrm{\lambda_{ref}}  = 250$ \mic), 
the PMW ($\mathrm{\lambda_{ref}} = 350$ \mic) and the PLW ($\mathrm{\lambda_{ref}} = 500$ \mic) band. 
When imaging is performed in PACS-SPIRE parallel mode the PACS blue and red band, and all SPIRE bands are simultaneously covered. 

Details about the observations can be found in Table~\ref{Table:ObsPreview} in Appendix~\ref{Appen:Obs}.
The Mass loss of Evolved StarS (MESS) Guaranteed Time (GT) key programme \citep{Groenewegen_KP, Groenewegen_etal_2011} is 
the main contributor to the sample. 
The aim of the programme was to study the mass-loss processes in evolved stars and the structure of their circumstellar environments. 
The stars in the MESS sample were specifically chosen to be representative of the various types of objects in terms of 
spectral type (covering the oxygen-rich stars, S-stars, carbon stars), variability type (L, SR, Mira), and MLR 
(from $10^{-7}$ to $\sim~3 \cdot 10^{-4}$ \msolyr). The programme provided PACS spectra for 24 of our 40 targets, most of them obtained in the standard 
SED mode and combining the (B2A + short R1) and (B2B + long R1) bands.
% {\bf and using a chopper throw of 3\arcmin}. 
The PACS spectroscopic data of CW Leo were obtained in a 3$\times$1 raster, however, only the on-source pointed observation 
was selected for our study. Furthermore, a non-standard version of the PACS-SED observing mode was adopted to CW Leo and VY CMa, 
as described in \citet{Royer_etal_2010} and \citet{Decin_etal_2010_Nature}. For IK Tau, CW Leo and VY CMa a third-order B3A observation 
was also available. The B3A band covers approximately the same wavelength range as the B2A band, but surpasses 
the latter regarding spectral resolution. The second-order B2A band, however, possesses better continuum sensitivity. 
Therefore, the B2A and B3A bands are both added to the final data sample. The MESS programme also observed SPIRE FTS 
for 9 of the 24 selected targets and also for AGB target R Scl, which was not observed by the PACS spectrometer. 
The GT programme of PI Barlow \citep{Barlow_2011_HerschelProposal} provided SPIRE FTS observations for 
another 8 of these 24 targets as this programme was specifically proposed to obtain the complementary SPIRE data for some remaining MESS targets. 

No HSA PACS-SED spectroscopy products were found for OH 127.8\,+0.0. However, \citet{Lombaert_etal_2013} presented 
a complete PACS spectrum of this target, which was observed during calibration time and needed alternative data reduction. 
Their final data product was added to the data sample. 

Complementary SPIRE FTS data for OH 127.8\,+0.0 was obtained by the Open Time (OT) programme of PI Justtanont \citep{Justtanont_2011_HerschelProposal} 
as well as standard mode PACS and SPIRE spectroscopic data for five more targets. 
Their programme concentrated on studying the chemistry, cooling, and geometry of the circumstellar environment of M-type stars 
with very strong MLRs. Some other AGB targets were included in the \textit{Herschel} filler programme (OBSHerchel1) which 
observed six complete PACS spectra (B2A + short R1 and B2B + long R1) and additional SPIRE FTS for two of these targets. 
Finally, incomplete PACS spectra (only the B2A + short R1 band) were obtained for RR Aql during \mbox{OBSHerchel1}, and 
for T Cep and R Aql within the OT programme of \citet{Cami_2011_HerschelProposal}. 

Spectroscopic data could also be retrieved for the AGB stars OH\,30.7\,+0.4 \citep{Justtanont_2011_HerschelProposal}, ST Her, G Her, 
V438 Oph \citep{Cami_2011_HerschelProposal}
and RT Cap and AQ Sgr (OBSHerchel1). However, these targets were not included in the final data selection 
because of an insufficient signal to noise ratio.

PACS and SPIRE photometry could additionally be obtained for most of the sample targets. 
The MESS programme contains corresponding PACS (blue + red) scan maps for all MESS and OBSHerchel1 spectra, except for 
IRAS 09425$-$6040, AFGL 5379, AFGL 2513 and GY Aql, and it also has SPIRE large maps for half of these targets. 
Supplementary PACS green band scan maps are found for TX Cam, LL Peg, R Cas, $\alpha$ Ori \citep{Royer_2011_HerschelProposal} 
and  for EP Aqr \citep{Cox_2011_HerschelProposal}.
Photometry taken in PACS-SPIRE parallel mode within the \textit{Hi-GAL} programme \citep{Molinari_KP, Molinari_OT, Molinari_etal_2010, Molinari2016} 
produced large-area scan maps of the galactic plane region, containing another eight of our sample targets. 
Photometry for T Cep was obtained by \citet{Andre_2007_HerschelProposal}, also in PACS-SPIRE parallel mode.

\begin{table*}
 \caption{Basic properties of the sample stars. The sources are listed by the IRAS identifier.}				
\label{Table:sample}						
\centering	
\begin{tabular}{lllllll}
\hline\hline
IRAS name	&  Identifier	        &	Chem.	&	Puls. 	&	Distance (Ref.) & ${v}_\mathrm{LSR}$ (Ref.)          & $\dot{M}$ (Ref.)	            \\
	        &	        	&	type	&	type	&	(kpc)	 &	$\mathrm{\left(km\ s^{-1}\right)}$  & $\mathrm{\left(M_{\odot}\ yr^{-1}\right)}$ \\
\hline	
\multicolumn{2}{l}{\textit{AGB-stars}} \\  
\hline																	
01037+1219	& WX Psc  	  &	M		&	Mira 	&	0.74 (1)	&       $9.0$ (13)              &	$4.0 \times 10^{-5}$ (20)	\\
01246$-$3248	& R Scl	          &	C		&	SRb 	&	0.27 (2)	&	$-19.0$ (14)		&	$1.0 \times 10^{-6}$ (21)	\\
01304+6211	& OH\,127.8\,+0.0 &	M		&	Mira	&	2.10 (3)	&	$-55.0$ (13)		&	$5.0 \times 10^{-5}$ (3) \\ %V669 Cas
02168$-$0312	& $o$ Cet   	  &	M		&	Mira	&	0.09 (2)	&	$46.5$ (15)		&	$2.5 \times 10^{-7}$ (13)	\\
03507+1115	& IK Tau, NML Tau &	M		&	Mira	&	0.26 (4)	&	$34.0$ (16)		&	$1.0 \times 10^{-5}$ (20)	\\	
04361$-$6210	& R Dor	          &	M		&	SRb	&	0.05 (2)	&	$7.0$ (16)		&	$1.3 \times 10^{-7}$ (20)	\\
04566+5606	& TX Cam	  &	M		&	Mira	&	0.39 (1)	&	$11.4$ (16)		&	$1.0 \times 10^{-5}$ (20) \\
09425$-$6040	& $\ldots$	  &	C		&	Mira	&	1.30 (5)	&	$15.0$ (5)		&	$2.0 \times 10^{-6}$ (5)	\\
09452+1330	& CW Leo	&	C		&	Mira 	&	0.12 (6)	&	$-26.0$ (15)		&	$1.5 \times 10^{-5}$ (20)	\\
10131+3049	& RW LMi, CIT6  &	C		&	SRa 	&	0.32 (7)	&	$-1.8$ (13)		&	$6.0 \times 10^{-6}$ (20)	\\
10491$-$2059	& V Hya	        &	C		&	SRa 	&	0.60 (7)	&	$-16.0$ (17)		&	$8.3 \times 10^{-6}$ (7)	\\
13462$-$2807	& W Hya	        &	M		&	Mira    &	0.10 (2)	&	$40.5$ (16)		&	$1.3 \times 10^{-7}$ (22)	\\
15194$-$5115	& II Lup	&	C		&	Mira 	&	0.59 (7)	&	$-5.5$ (14)		&	$1.7 \times 10^{-5}$ (14)	\\
16011+4722	& X Her	        &	M		&	SRb	&	0.14 (2)	&	$-73.0$ (15)		&	$1.5 \times 10^{-7}$ (23)	\\
17411$-$3154	& AFGL 5379, OH 357.3$-$1.3 &	M	&	Mira	&	0.99 (1)	&	$-21.2$	 (13)	        &	$2.0 \times 10^{-4}$ (8)	\\
18257$-$1000	& OH\,21.5\,+0.5 &	M		&	Mira	&	2.50 (8)	&	$115.0$ (13)	        &	$2.6 \times 10^{-4}$ (8)	\\
18348$-$0526	& OH\,26.5\,+0.6 &	M		&	Mira 	&	1.30 (9)	&	$29.0$ (15)		&	$2.6 \times 10^{-4}$ (8)	\\
18460$-$0254	& OH\,30.1\,$-$0.7 &	M		&	Mira	&	1.75 (8)        &	$100.0$ (13)	        &	$2.2 \times 10^{-4}$ (8)	\\
18488$-$0107	& OH 32.0$-$0.5 &	M &	Mira	&	3.90 (8)	&	$75.0$ (13)		&	$3.6 \times 10^{-4}$ (8)	\\
18498$-$0017	& OH\,32.8\,$-$0.3 &	M		&	Mira	&	4.30 (8)	&	$60.0$ (13)		&	$3.1 \times 10^{-4}$ (8)	\\
19039+0809	& R Aql	        &	M		&	Mira	&	0.42 (2)        &	$46.0$ (18)		&	$3.5 \times 10^{-6}$ (24)	\\
19067+0811	& OH 42.3$-$0.1  &	M &	Mira	&	3.80 (8)	&	$60.0$ (13)		&	$2.7 \times 10^{-4}$ (12)	\\
19126$-$0708	& W Aql	        &	S		&	Mira 	&	0.68 (10)	&	$-27.5$ (15)		&	$2.2 \times 10^{-6}$ (12)	\\
19474$-$0744	& GY Aql	&	M		&	Mira	&	0.22 (2)	&	$33.0$ (18)		&	$6.0 \times 10^{-6}$ (11)	\\
19486+3247	& $\chi$ Cyg	&	S		&	Mira 	&	0.18 (2)	&	$10.0$ (15)		&	$3.8 \times 10^{-7}$ (20)	\\
19550$-$0201	& RR Aql	&	M		&	Mira	&	0.52 (2)	&	$28.0$ (14)		&	$2.4 \times 10^{-6}$ (14)	\\
20038$-$2722	& V1943 Sgr	&	M		&	SRb 	&	0.20 (2)	&	$-15.0$ (14)		&	$9.9 \times 10^{-8}$ (14)	\\
20072+3116	& AFGL 2513	&	C		&	Mira	&	1.76 (7)	&	$17.8$ (7)		&	$2.1 \times 10^{-5}$ (7)	\\
20077$-$0625	& IRC $-$10 529, V1300 Aql &	M       &	Mira 	&	0.66 (11)	&	$-18.0$ (14)		&	$3.0 \times 10^{-5}$ (20	\\
20248$-$2825	& T Mic	        &	M		&	SRb 	&	0.21 (2)	&	$25.0$ (19) 		&	$8.0 \times 10^{-8}$ (24)	\\
20396+4757	& V Cyg	        &	C		&	Mira 	&	0.35 (2)	&	$15.0$ (13)		&	$9.0 \times 10^{-7}$ (20	\\
21088+6817	& T Cep	        &	M		&	Mira	&	0.19 (2)        &	$-2.0$ (14)		&	$9.1 \times 10^{-8}$ (14	\\
21439$-$0226	& EP Aqr	&	M		&	SRb 	&	0.11 (2)	&	$-34.0$ (14)		&	$3.1 \times 10^{-7}$ (13)	\\
22196$-$4612	& $\pi$ Gru	&	S		&	SRb	&	0.16 (2)	&	$-12.0$ (14)		&	$8.5 \times 10^{-7}$ (13)	\\
23166+1655	& LL Peg, AFGL 3068 &	C	        &	Mira 	&	1.00 (7)	&	$-31.0$ (13)		&	$1.0 \times 10^{-5}$ (20	\\
23320+4316	& LP And, AFGL 3116 &	C	        &	Mira 	&	0.78 (7)	&	$-17.0$ (13)		&	$1.5 \times 10^{-5}$ (20)	\\
23558+5106	& R Cas	          &	M		&	Mira 	&	0.13 (2)	&	$25.0$ (16)		&	$5.0 \times 10^{-7}$ (20)	\\
\hline 
\multicolumn{2}{l}{\textit{Red Super Giants}} \\  
\hline											
05524+0723	& $\alpha$ Ori	        &	M		&	SRc 	&	0.15 (2)	&	$3.5$ (13)		&	$2.0 \times 10^{-6}$ (25)	\\
07209$-$2540	& VY CMa		&	M		&	Lc	&	1.17 (12)	&	$17.0$ (15)		&	$3.0 \times 10^{-4}$ (25)	\\
$\ldots$        & NML Cyg		&	M		&	$\ldots$	&	1.61 (12)	&	$-1.0$ (15)		&	$8.7 \times 10^{-5}$ (13)	\\
\hline						
\end{tabular}	
\tablefoot{Column~1 lists the IRAS name, when available, and Col.~2 some common names.
Column~3 gives the chemical type (M, S, or C), and Col.~4 the variability type.
Column~5, 6, and 7 give the distance, velocity of the object (on the local standard of rest scale), and MLR, with the reference between parenthesis.
} 	
\tablebib{(1) \citet{Olivier2001}, (2) \citet{vanLeeuwen2007}, (3) \citet{Lombaert_etal_2013},  (4) \citet{Richards2012}, (5) \citet{Molster2001}, 
(6) \citet{GroenewegenCWL}, (7) \citet{Groenewegen2002}, (8) \citet{Justtanont2006}, (9) \citet{vanLangevelde1990}, (10) \citet{Groenewegen_DeJong_1998}, 
(11) \citet{Loup1993}, (12) \citet{Reid2014}, (13) \citet{DeBeck_etal_2010}, (14) \citet{Danilovich2015}, (15) \citet{deVincente_etal_2016}, 
(16) \citet{Danilovich2016}, (17) \citet{Sahai2009}, (18) \citet{Desmurs2014}, (19) \citet{GonzalezD2003}, (20) \citet{Schoier2013}, 
(21) \citet{Maercker2016a}, (22) \citet{Khouri2014a}, (23) \citet{Olofsson2002}, (24) \citet{Knapp1985},  (25) \citet{Smith2009}.
}
\end{table*}

\subsection{Data processing}
\label{sect:processing}

\subsubsection{PACS spectroscopy}
\label{sectPACS}

All data was reduced with the standard interactive pipeline in HIPE \citep{Ott_etal_2010}, version 14.1 in combination with 
version 78 of the calibration tree. The flux calibration (which includes spatial flat-fielding) is performed via 
normalisation to the telescope background. The final 
re-binning is performed with an oversampling of two, thus with Nyquist sampling with respect to the instrumental resolution. 
The spectra were extracted assuming a point source approximation and a point source correction was applied for every target.
This point source correction is wavelength dependent so that the spectral fluxes are to be understood as measured in an infinite aperture.

AGB stars with especially heavy mass loss are not perfect point sources. However they never fill the beam as assumed in the extended source calibration.
Only CW Leo is extended beyond the 3$\times$3 central spaxels that are used to derive the total flux (see below).

For most of the targets, the spectrum was extracted from the central spaxel (hereafter c1) and subsequently scaled by the flux level of 
the summed spectrum of the 3$\times$3 central spaxels (c9), resulting in a 
%"c129" 
spectrum which accounts for the flux lost from the central spaxel.  
An exception is TX Cam which shows high flux values %($>200\%$) 
in the surrounding spaxels compared to c1. In this case 
% the c9-spectrum was selected, i.e. 
the summed spectrum in the 3$\times$3 central spaxel box
was taken, with the c9-to-total point source correction applied. 
In case of OH\,30.1\,$-$0.7 and OH\,32.8\,$-$0.3 the c1-spectrum without subsequent scaling was preferred because inhomogeneous 
background emission affects the scaling to c9. The choice of c1 is also justified by the fact that very little flux ($<10\%$) is 
found in the surrounding spaxel with respect to the central one, hence rendering the flux correction too uncertain. 

The observations for AFGL 5379 and OH\,21.5\,+0.5 suffered from mispointing by about 18\arcsec. 
In these cases, the spectrum was extracted from the spaxel containing the actual source,
and applying the proper point source correction for a single spaxel.
These sources and the others discussed below which showed some problems or needed some alternative treatment are flagged 
in Table~\ref{Table:ObsPreview} in Appendix~\ref{Appen:Obs}.

The standard pipeline eliminates the noisiest parts of the spectra and the regions affected by light leaks, and the wavelength ranges 
of the spectral bands are confined to: B2A segments between 55 and 72~\mic, B3A segments between 55 and 70~\mic, 
B2B segments between 70 and 95~\mic\ and R1 segments between 102 and 190~\mic.

The spectra for all targets and all individual spectral bands are available in ASCII format at the CDS.
These files contain wavelength, flux, and flux uncertainty. The flux uncertainty results from the production of the final spectrum. 
The wavelength bins of the final spectrum are composed of multiple data points which are obtained in the different detectors and 
during different positions of the grating. These single spectra are averaged and the corresponding standard deviation of the data 
is a good measure for the scatter in the flux level.
According to the PACS observers manual the relative flux calibration accuracy within a PACS band is 10\% (20\% in R1), 
while the absolute flux calibration error is $\sim 12\%$ across the entire wavelength range\footnote{See Sect.~4.10 in the PACS observers manual.}.

\subsubsection{SPIRE spectroscopy}
The HIPE pipeline version 14.1 and the latest set of calibration files of version 14.3 were used to reprocess the SPIRE FTS data. 
Assuming that all targets were point sources or only slightly extended sources at SPIRE spatial resolution, 
only the central detector feed horns of the SSW and the SLW bolometer arrays were reduced. 

The standard pipeline subtracts the telescope background emission, which generally approximates the real background spectrum to within 1\%. 
Strong far-IR emission from the Galactic plane, however, heavily affected the calibration of OH 127.8\,+0.0, TX Cam, II Lup, 
AFGL 5379, OH\,21.5\,+0.5, OH\,26.5\,+0.6, OH\,30.1\,$-$0.7, IRAS 18488$-$0107, OH\,32.8\,$-$0.3, and IRAS 19067+0811. 
The extra background contamination lead to a discontinuity between the SSW and SLW band in the reprocessed spectrum. 
To correct for this, we performed an additional background subtraction, following the method described in 
the \textit{SPIRE Data Reduction Guide}\footnote{\url{http://herschel.esac.esa.int/hcss-doc-13.0/load/spire_drg/html/spire_drg.html}}. 
In this method, the averaged off-axis detectors are subtracted from the central detector. 
The low-frequency information was extracted from the off-axis spectra by smoothing them with a wide kernel. 
An illustration of this kind of contamination and a detailed description of the corresponding background subtraction 
is given in Appendix~\ref{Appen:Background} for OH\,30.1\,$-$0.7. 

All spectra are kept unapodised to preserve the original, sinc-function line profiles with a spectral resolution of 1.4 GHz (FWHM).
The SSW and SLW spectra are available in ASCII format at the CDS.
These files contain wavelength, flux, and flux uncertainty, which is the standard deviation of the averaging 
process of the data points from the different detectors and observation scans. 
The SPIRE handbook states that there is a possible continuum offset error of 0.4 Jy for SLW and 0.3 Jy for SSW, and a calibration error of 6\%.

\begin{figure*}
\centering
\includegraphics[width=0.95\textwidth]{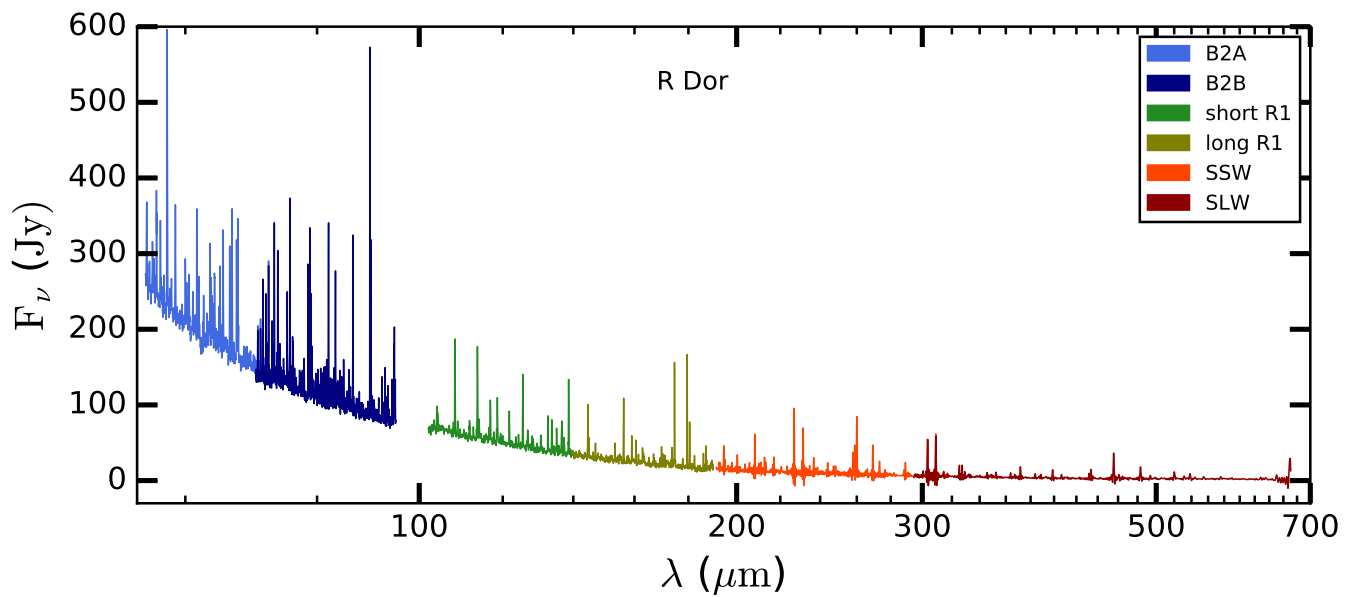}
\caption{Full-range PACS and SPIRE spectra for R Dor. 
The PACS spectrum consists of bands B2A (55--72~\mic), B2B (70--95~\mic), short R1 (102--146~\mic) and long R1 (140--190~\mic). 
The SPIRE spectrum is made up from the SSW (191--317~\mic) and SLW band (294--670~\mic).
The spectra of all stars in the sample are available at the CDS.
}
\label{Fig:FullSpec}
\end{figure*}

An example of a combined full-range PACS and SPIRE spectrum is shown in Fig.~\ref{Fig:FullSpec} for the M-type AGB-star R Dor.

\subsubsection{PACS and SPIRE imaging}
The latest PACS level 2.5 scan maps were downloaded from the HSA. The data are the final legacy products generated in an automated 
fashion by the \textit{Herschel Standard Product Generator} pipelines, using HIPE version 14.2 of the software. We used the maps 
generated by the JScanam pipeline, obtained by combining the orthogonal scan and cross-scan AORs. 

The SPIRE large maps are the best science quality level 2 products created with version 14.1 of HIPE that were obtained from the HSA 
for all bands and for both the point source and the extended source calibration. 

Scan maps in PACS-SPIRE parallel mode of level 2 and 2.5 were also downloaded from the HSA. 
The maps were obtained by two orthogonal scans, adopting the default fast scan velocity of 60 \arcsec/s and they were 
created by the HIPE 14.2 and 14.1 software version for PACS and SPIRE wavelengths, respectively.

%%%%%%%%%%%%%%%%%%%%%%%%%%%%%%%%%%%%%%%%%%%%%%%%%%%%%%%%%%%%%%%%%%%%%%%%%%%%%%%%%%%%%%%%%%%%%%%%%%%%%%%%%%%
%                 QUALITY CHECK OF FLUX CALIBRATION
%%%%%%%%%%%%%%%%%%%%%%%%%%%%%%%%%%%%%%%%%%%%%%%%%%%%%%%%%%%%%%%%%%%%%%%%%%%%%%%%%%%%%%%%%%%%%%%%%%%%%%%%%%%

\section{Quality check of the flux calibration}
\label{SectQual}

In order to check the calibration quality of the spectra, the flux level is compared to the source flux measured
in the photometer maps. 
As the spectra were extracted by assuming point source 
%like 
targets (see Section \ref{sect:processing}), the imaged photometry was derived following a point source assumption for consistency. 
The details on the methods used to obtain the source photometry fluxes are given in Sect.~\ref{sect:SourcePhot}. 

The flux measured in the imaged maps depends on the spectral shape of the source spectrum and the response of the 
overall instrumental setup, while on the other hand, the spectral fluxes can be considered to be monochromatic. 
To compare the flux measured by the bolometers to that of the spectra, the \textit{synthetic photometry} was calculated by convolving 
the flux spectra with 
the response functions of the photometric system, as explained in Sect.~\ref{sect:SynthPhot}. 
The obtained photometric source fluxes and the derived synthetic fluxes can then directly be compared to each other. 
The results of this comparison are given in Sect.~\ref{sect:Compare}.

\subsection{Imaged source photometry} 
\label{sect:SourcePhot}

PACS source photometry was carried out by performing aperture photometry using the \textit{annularSkyAperturePhotometry} task that comes with HIPE. 
The recommended\footnote{\url{https://nhscsci.ipac.caltech.edu/workshop/Workshop_Oct2014/Photometry/PACS/PACS_phot_Oct2014_photometry.pdf}}  
source and background apertures for point sources were adopted (Table~\ref{Table:Apertures}) and the \textit{daophot} option was chosen to estimate 
the sky background. Subsequently, a point source correction was performed by the HIPE \textit{photApertureCorrectionPointSource} task. 
The method gave unreliable results for OH\,21.5\,+0.5, OH\,30.1\,$-$0.7, IRAS 18488$-$0107, OH\,32.8\,$-$0.3, and IRAS 19067$-$0811 
due to their weak brightness and the strong galactic background emission. For these sources a source aperture which visually contained 
all source flux was applied and a representative surrounding annulus was chosen as background aperture.

The recommended method to obtain SPIRE point source photometry is to perform fitting of the timeline data. This method fits source and 
background simultaneously in the level 1 timeline data products and was adopted for most of our sources. 
The sources OH\,21.5\,+0.5, IRAS 18488-0107, OH\,32.8\,$-$0.3, and IRAS 19067$-$0811 were too weak at SPIRE wavelengths to derive 
a reliable photometric flux in any of the SPIRE bands, while OH30.1\,$-$0.7 was only sufficiently bright in the 250 \mic\ band. 
The timeline fitting also failed to produce reliable results for R Scl and R Aql as the best-fit PSF exceeded the confidence limits. 
In these two cases, aperture photometry (HIPE DaoPhot task) was instead carried out on the maps calibrated for extended emission 
for all SPIRE bands of \mbox{R Scl} and for the 250 \mic\ and 350 \mic\ bands of R Aql.

The results for the source photometry can be found in Table~\ref{Table:QualityCheck}.
We repeat that all photometry is derived by 
using apertures optimised for point sources, and then applying a point source correction.
For targets showing extended circumstellar emission, the 
flux values can therefore be lower than those derived with source apertures that include this extended emission. Likewise, they can also 
significantly differ from lower spatial resolution literature photometry that does not resolve the extended emission. The sample targets with a 
significant contribution from extended emission are: CW Leo, $\alpha$ Ori, $o$ Cet, W Hya, X Her, R Cas, \mbox{R Scl} and V1943 Sgr. 

No colour correction was applied to the derived values. 
The reader can find values for the colour corrections in the instruments manuals, and they are also available in the HIPE calibration files.
They are a few percent at both PACS \citep{Poglitsch_etal_2010} and SPIRE wavelengths (SPIRE handbook) for energy distributions 
typical of late-type stars.

\begin{table}
\centering
 \caption{Source apertures and sky background annuli used to obtain PACS source photometry.}	
\label{Table:Apertures}		
\begin{tabular}{llcc}  
\hline	
Name			        & band 	&	source 	&	sky  ($\mathrm{R_{inner}, R_{outer}}$) \\
                                &       &   annulus     &      annulus               \\
				&	&	(\arcsec)	& (\arcsec , \arcsec) \\
\hline									
\textit{recommended}            & blue	&	12		&	(35 , 45)	\\	
				& green	&	12		&	(35 , 45)	\\	
				& red	&	22		&	(35 , 45)	\\	
OH\,21.5\,+0.5		        & blue	&	40		&	(40 , 60)	\\	
				& red	&	15		&	(40 , 60)	\\			
OH\,30.1\,$-$0.7		& blue	&	40		&	(40 , 60)	\\
				& red	&	25		&	(40 , 60)	\\			
IRAS 18488$-$0107	        & blue	&	40		&	(40 , 60)	\\
				& red	&	15		&	(40 , 60)	\\	
OH\,32.8\,$-$0.3		& blue	&	25		&	(25 , 30)	\\	
				& red	&	25		&	(25 , 40)	\\		
IRAS 19067+0811	                &  blue	&	30		&	(40 , 60)	\\	
				& red	&	10		&	(20 , 25)	\\							
\hline						
\end{tabular}					
\tablefoot{Source apertures and sky background annuli used for the HIPE \textit{annularSkyAperturePhotometry} task to obtain PACS source photometry. 
The recommended apertures are adopted for all sources, except for the ones that are explicitly listed. }

\end{table}

\subsection{Synthetic photometry from the spectra} 
\label{sect:SynthPhot}

SPIRE synthetic photometry was obtained by use of the built-in \textit{spireSynthPhotometry} task in HIPE. This task uses 
the SPIRE photometer relative spectral response functions (RSRFs) available in the calibration files to calculate 
the synthetic photometry at the overlapping photometer wavelength bands by weighting the spectrum with the RSRFs of the different  bands.

A similar method was carried out for PACS.
The RSRFs of the PACS photometer bands 
are closely approximated by combining the different filter transmission curves and the PACS bolometer response curve which 
are both available in the HIPE calibration files. To fully cover the green band, the gap in the 
spectrum between the B2B and R1A segment was bridged by a power law function $F_{\lambda} = a \, \lambda^{-b}$ which best fitted the spectrum 
between 80 and 120~\mic. Also, the PACS spectrum was extended to longer wavelengths as the red photometric band exceeds 
the 190~\mic\ wavelength limit in the spectrometer. When a SPIRE spectrum is available, both PACS and SPIRE spectra are combined to 
cover the full wavelength range of the red band.
When no SPIRE spectrum is available, 
the long R1 segment of the spectrum is extrapolated to longer wavelengths (320 \mic) using a power law 
function $F_{\lambda} = a\lambda^{-b}$ which best fitted the spectrum between 130 and 180 \mic.
As a second step, the integrated fluxes are divided by the width of the bands to obtain a monochromatic flux $F_{\lambda_{\rm eff}}$ at 
the effective wavelength of each band: $F_{\lambda_{\rm eff}} = \frac{\int{F_{\lambda} \; d\lambda}}{\Delta_{\lambda}}$. 
The reasoning behind this and the values for the band widths and the effective wavelengths are listed in the 
PACS Technical note PICC-CR-TN-044, issue 1.1\footnote{\url{http://herschel.esac.esa.int/twiki/pub/Public/PacsCalibrationWeb/PICC-CR-TN-044.pdf}}.

\subsection{Imaged vs. synthetic photometry} \label{sect:Compare}

The results of the comparison between imaged and synthetic photometry are summarised in Table~\ref{Table:QualityCheck}. 
No imaged PACS or SPIRE photometry is available for IRAS 09425$-$6040, AFGL 2513, RR Aql and GY Aql. On the other hand, no synthetic 
photometry could be derived for T Cep, R Aql and RR Aql as these spectra lack the B2B and long R1 segments.
For most of the targets, synthetic and imaged photometry are in agreement within the mutual 15\% uncertainty level. 
The targets for which a significant difference is found between imaged and synthetic fluxes in at least one colour are flagged 
with a $\dagger$ symbol and the corresponding flux values are indicated in bold. 
These sources are also flagged in Table~\ref{Table:ObsPreview}.
For these targets it is recommended that the PACS and SPIRE spectra, as well as the imaging, be treated with care. 

Except for the difference between SLW and PSW fluxes for R Scl (and where the flux is very low anyway) none of the larger
differences is in a source where extended emission might play a role in the derivation and comparison of fluxes derived from
the spectra and the imaging. 

Essentially all of our targets are variable, and many belong to the Mira class showing the largest variations.
As the imaging and spectra are not taken at the same time, differences in flux levels between the two could be due to variability. 
However, there is not enough data available to be more specific. Only CW Leo has been observed over its pulsation cycle \citep{GroenewegenCWL} 
indicating peak-to-peak variations of about 25\%, 23\%, and 21\% in the PSW, PMW and PLW filters, respectively, or about 0.2 mag.
At PACS wavelengths it has not been measured, but we expect the variations to be larger. 
For comparison, the peak-to-peak amplitude in the $K$-band for CW Leo is 2.0 mag \citep{LeBertre1992}.

\begin{sidewaystable*}

\caption{Synthetic photometry from the spectra is compared to imaged photometry. 
}	
\label{Table:QualityCheck}	
\centering
%\begin{tabular}{l||ll|ll|ll|ll|ll|ll|} 
\begin{tabular}{lllllllllllll} 
\hline        
Identifier & PACS sp. & PACS im. & PACS sp. & PACS im. & PACS sp. & PACS im. & SPIRE sp. & SPIRE im. & SPIRE sp. & SPIRE im. & SPIRE sp. & SPIRE im. \\
 & $F_{70}$ (Jy) & $F_{70}$ (Jy) & $F_{100}$ (Jy) & $F_{100}$ (Jy) & $F_{160}$ (Jy) & $F_{160}$ (Jy) & $F_{250}$ (Jy) & $F_{250}$ (Jy) & $F_{350}$ (Jy) & $F_{350}$ (Jy) & $F_{500}$ (Jy) & $F_{500}$ (Jy) \\
\hline                 
\multicolumn{2}{l}{\textit{AGB-stars}} \\ 
\hline 
WX Psc $\dagger$ & \textbf{150.} & \textbf{94.6} & 64.3 &  $\ldots$ & 18.3 & 15.2 & 3.5 & $\ldots$  & 0.84 & $\ldots$ & 0.52 & $\ldots$  \\
R Scl $\dagger$  & $\ldots$    & 17.5 & $\ldots$ & $\ldots$  &  $\ldots$ & 5.7 & 0.98 & 1.2 & 0.87 & 0.68 & \textbf{0.62} & \textbf{0.30} \\
OH\,127.8\,+0.0  & 87.5 & 86.9 & 33.9 & $\ldots$ & 8.9 & 10.3 & 2.21 & 2.0 & 0.66 & 0.52 & 0.17 & $\ldots$  \\
$o$ Cet          & 143. & 158. & 62.6 & $\ldots$ & 20.2 & 23.0 & 7.5 & 7.04 & 3.6 & 3.2 & 1.9 & 1.6 \\
IK Tau $\dagger$ & 169  & 167. & 70.0 & 60.5 & 19.2 & 21.6 & \textbf{3.8} & \textbf{5.1} & 1.9 & 2.0 & 0.83 & 0.95 \\
R Dor            & 159  & 154 & 76.6 & $\ldots$ & 27.1 & 29.3 & 11.5 & 11.6 & 5.6 & 5.6 & 3.1 & 2.8 \\
TX Cam           & 63.3 & 61.1 & 27.6 & 24.8 & 9.2 & 8.8 & 1.65 & $\ldots$ & 1.1 & $\ldots$ & 0.51 & $\ldots$  \\
09425$-$6040     & 14.1 & $\ldots$ & 6.6 & $\ldots$ & 2.1 &  $\ldots$ &  $\ldots$ &  $\ldots$ &  $\ldots$ &  $\ldots$ & $\ldots$ & $\ldots$  \\
CW Leo           & 2461. & 2367. & 985. & 850. & 280. & 334. & 74.5 & 64.9 & 41.2 & 33. & 17.7 & 16. \\
RW Lmi          & 157. & 142. & 71.3 &  $\ldots$ & 23.3 & 24.6 & 7.4 & 6.6 & 4.0 & 3.5 & 2.1 & 1.6 \\
V Hya           & 60.6 & 58.0 & 28.2 &  $\ldots$ & 9.4 & 9.9 & 2.8 & 2.8 & 1.6 & 1.6 & 0.79 & 0.69 \\
W Hya           & 107. & 119. & 53.1 &  $\ldots$ & 22.3 & 23.0 & 8.4 & 8.5 & 3.8 & 4.0 & 2.1 & 2.0 \\
II Lup $\dagger$ & \textbf{66.5} & \textbf{89.5} & 31.9 & $\ldots$ & \textbf{11.3} & \textbf{15.9} & 3.8 & $\ldots$ & 1.8 & $\ldots$ & 0.96 &$\ldots$   \\
X Her            & 16.6 & 17.1 & 8.1 &  $\ldots$ & 2.9 & 3.1 &  $\ldots$ & $\ldots$  &  $\ldots$ &  $\ldots$ & $\ldots$  & $\ldots$  \\
AFGL 5379        & 661. & 633. & 271. &  $\ldots$ & 64.1 & 79.3 & 20.5 & $\ldots$ & 5.8 & $\ldots$  & 1.7 & $\ldots$  \\
OH\,21.5\,+0.5 $\dagger$ & \textbf{123.} & \textbf{36.3} & 45.3 & $\ldots$ & \textbf{9.7} & \textbf{6.0} & 1.3 & $\ldots$ & 0.06 & $\ldots$ & $-$0.3 & $\ldots$  \\
OH\,26.5\,+0.6     & 373. & 323. & 143 & $\ldots$ & 35.9 & 36.7 & 7.4 & 8.4 & \textbf{1.1} & \textbf{2.0} & \textbf{0.02} & \textbf{0.64} \\
OH30.1\,$-$0.7      & 133. & 148. & 59.3 &  $\ldots$ & 16.0 & 15.5 & 4.0 & 4.0 & $-$0.12 & $\ldots$ & $-$0.35 & $\ldots$  \\
18488$-$0107 $\dagger$ & \textbf{33.8} & \textbf{17.8} & 15.2 & $\ldots$ & \textbf{5.6} & \textbf{3.0} & 2.2 & $\ldots$ & 0.9 & $\ldots$ & 0.19 & $\ldots$  \\
OH\,32.8\,$-$0.3 $\dagger$ & \textbf{116.} & \textbf{56.3} & 43.4 & $\ldots$ & \textbf{11.6} & \textbf{7.0} & 3.8 & $\ldots$ & 0.05 & $\ldots$ & $-$0.20 & $\ldots$  \\
R Aql                 &  $\ldots$ & 23.7 &   $\ldots$&  $\ldots$ &  $\ldots$ & 4.8 & $\ldots$ & 1.58 & $\ldots$ & 1.0 & $\ldots$ & $\ldots$ \\
19067+0811         & 21.6 & 24.1 & 8.6 & $\ldots$ & 1.8 & 2.1 & 0.86 & $\ldots$ & $-$0.31 & $\ldots$ & $-$0.47 & $\ldots$  \\
W Aql             & 41.2 & 46.7 & 19.1 & $\ldots$ & 6.0 & 7.8 & $\ldots$ & $\ldots$ & $\ldots$ & $\ldots$ &  $\ldots$ &  $\ldots$ \\
GY Aql            & 31.2 & $\ldots$ & 13.3 & $\ldots$ & 4.3 & $\ldots$ & $\ldots$  & $\ldots$ & $\ldots$ & $\ldots$ & $\ldots$  & $\ldots$  \\
$\chi$ Cyg        & 42.6 & 42.7 & 19.4 & $\ldots$ & 6.4 & 7.6 & $\ldots$ & $\ldots$  & $\ldots$ & $\ldots$ & $\ldots$ & $\ldots$  \\
%RR Aql & $\ldots$ &  & $\ldots$ &  & $\ldots$ &  & $\ldots$ &  & $\ldots$ &  & $\ldots$ &  \\
V1943 Sgr         & 13.1 & 12.8 & 6.0 & $\ldots$ & 2.2 & 2.4 & 1.1 & $\ldots$ & 0.37 & $\ldots$ & 0.11 & $\ldots$  \\
AFGL 2513         & 15.5 & $\ldots$ & 7.6 & $\ldots$ & 2.9 & $\ldots$ & $\ldots$ & $\ldots$ & $\ldots$ & $\ldots$ & $\ldots$ & $\ldots$ \\
IRC $-10$ 529 $\dagger$ & \textbf{109.} & \textbf{72.4} & 47.0 & $\ldots$ & 13.2 & 11.1 & 2.9 & $\ldots$ & 1.2 & $\ldots$ & 0.42 & $\ldots$ \\
T Mic             & 15.7 & 15.8 & 6.63 & $\ldots$ & 2.6 & 3.0 & $\ldots$ &  $\ldots$ & $\ldots$ & $\ldots$ & $\ldots$ & $\ldots$ \\
V Cyg             & 25.4 & 23.9 & 12.0 & $\ldots$ & 4.2 & 5.0 & $\ldots$ & $\ldots$  & $\ldots$ & $\ldots$ & $\ldots$ & $\ldots$ \\
T Cep             & $\ldots$ & 19.8 & $\ldots$ & $\ldots$  & $\ldots$ & 4.0 & $\ldots$ &$\ldots$  & $\ldots$ & $\ldots$ & $\ldots$ & $\ldots$ \\
EP Aqr            & 23.4 & 22.3 & 10.5 & 10.0 & 3.8 & 3.7 & $\ldots$ & $\ldots$ & $\ldots$ & $\ldots$  & $\ldots$ & $\ldots$ \\
$\pi$ Gru            & 31.0 & 33.1 & 13.8 & $\ldots$ & 5.0 & 5.4 & $\ldots$ & 1.6 & $\ldots$ & 0.76 & $\ldots$ & 0.39 \\
LL Peg            & 174. & 173. & 82.4 & 79.9 & 27.0 & 27.3 & 6.0 & 7.7 & 3.1 & 3.3 & 1.4 & 1.3 \\
LP And            & 82.0 & 63.6 & 36.8 & $\ldots$ & 11.7 & 10.9 & $\ldots$ & $\ldots$ & $\ldots$ & $\ldots$ & $\ldots$ & $\ldots$ \\
R Cas            & 52.0 & 55.5 & 24.6 & 22.3 & 8.4 & 10.0 & 3.3 & 3.2 & 1.6 & 1.5 & 1.0 & 0.76 \\
\hline           
\multicolumn{2}{l}{\textit{Red Super Giants}} \\  
\hline           
$\alpha$ Ori $\dagger$ & 127. & 129. & 57.1 & 53.6 & 18.1 & 20.7 & \textbf{4.5} & \textbf{6.5} & 2.7 & 3.3 & 1.4 & 1.7 \\
VY CMa            & 1125. & 1083. & 454. & $\ldots$ & 132. & 148. & 41.6 & $\ldots$ & 16.7 & $\ldots$ & 7.1 & $\ldots$  \\
NML Cyg            & 707. & 726. & 339. & $\ldots$ & 108. & 119. & 37.1 & $\ldots$ & 18.4 & $\ldots$ & 8.2 & $\ldots$  \\
\hline 
\end{tabular} 
\tablefoot{Synthetic photometry from the spectra (PACS/SPIRE sp.) is compared to imaged photometry (PACS/SPIRE im.). 
Values not in agreement within the respective $15\%$ error levels (that is, differ by more than 30\%) are indicated 
in \textbf{bold} and the corresponding targets are flagged with a $\dagger$.
}  
\end{sidewaystable*}

\section{Data analysis}
\label{S_DA}

\subsection{Line detection and measurement}
\label{LineMeasurement}

The selection and measuring of the spectral lines was performed following a general and consistent algorithm for all targets. 
This algorithm differed slightly for PACS and SPIRE spectra, 
since the SPIRE instrument is a Fourier transform spectrometer, providing sinc-shaped spectral profiles.

In the case of PACS, all local maxima were first detected in the spectrum. From this pre-selection, only the maxima meeting the following 
two conditions made it to the final list of spectral lines: (1) they must be separated by at least one full width at half maximum (FWHM) 
to ensure that they can be distinguished from neighbouring spectral lines and (2) the flux density corresponding to the 
wavelength of the local maximum must be sufficiently high relative to the local noise level in order to minimise 
the selection of spurious lines. After extensive testing, 
$F_{\lambda_{\mathrm{max}}}\geq 4.25 \times \mathrm{MAD_{36 FWHM}}$ was adopted as a criterion, where the 
median absolute deviation (MAD) was calculated locally for a wavelength region spanning 36 FWHM around 
the local maximum in question. 
This corresponds to 0.015--0.021~\mic\ in B3A, 0.034--0.039~\mic\ in B2A and B2B, and 0.11--0.13~\mic\ in R1 around each peak.
The MAD was used instead of a traditional root mean square to measure the noise level as it is more stable to outliers.
The flux level criterion and the width of the wavelength region used in the procedure were determined by running 
the algorithm looking for `negative lines' and then minimising the detection of actual noise peaks as true lines.

\begin{figure*}
 \begin{subfigure}{0.49\textwidth}
 \centering
 \includegraphics[width=88mm]{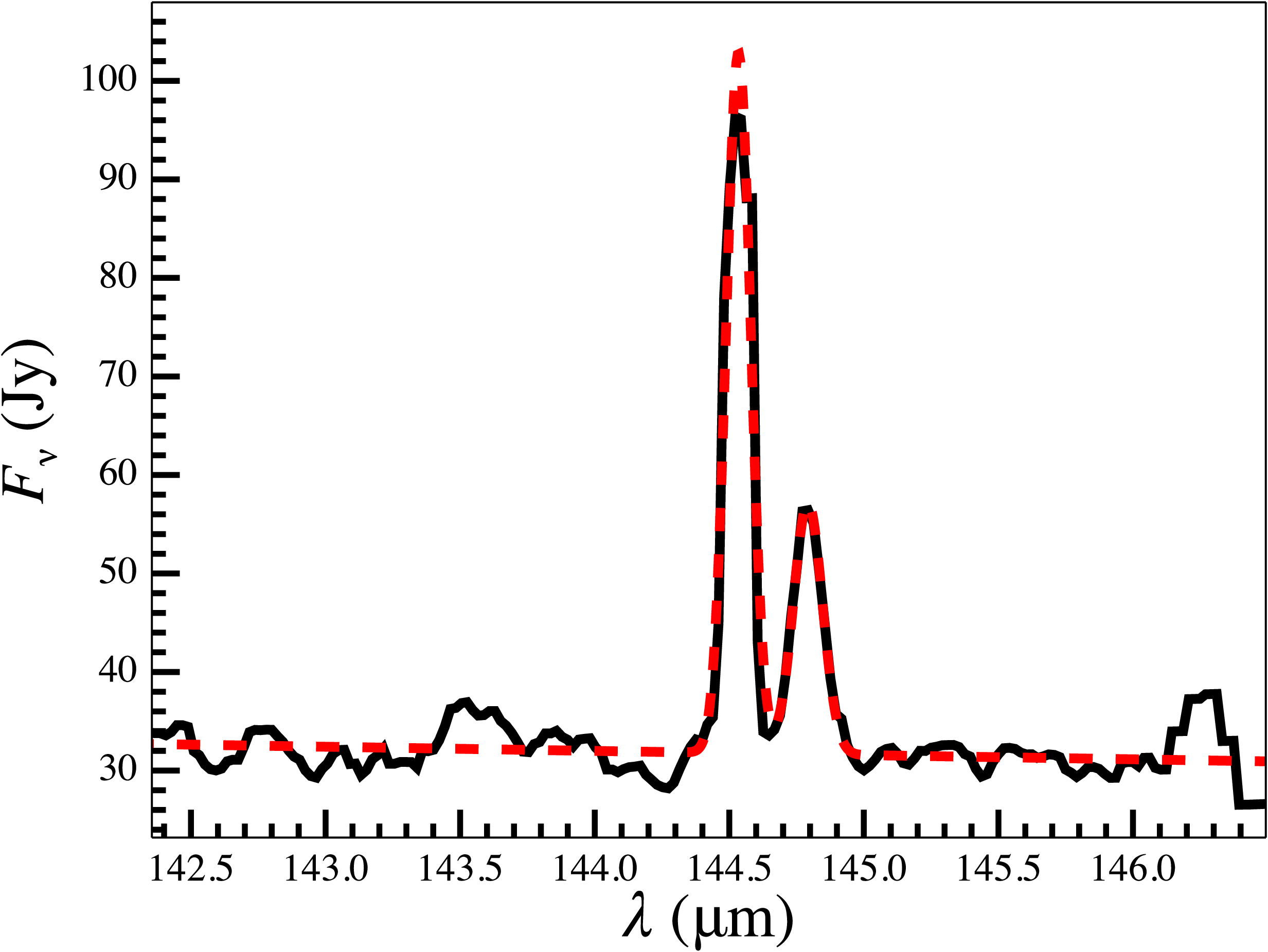}
% \caption{Example of PACS line measurement for R Dor.}
%\label{Fig:PACSLineFit}
 \end{subfigure}%
\hfill
 \begin{subfigure}{0.49\textwidth}
 \centering
 \includegraphics[width=88mm]{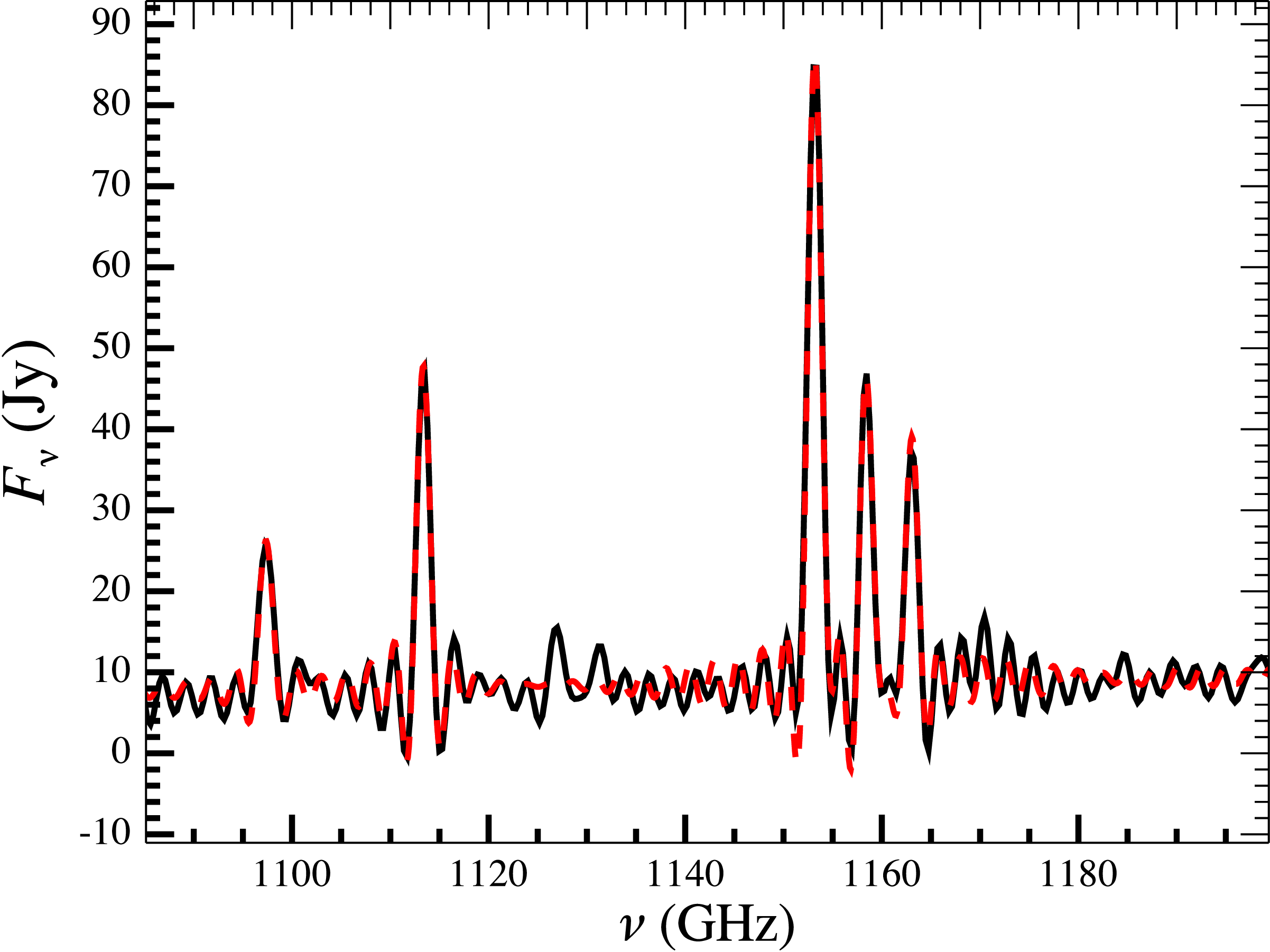}
 %\caption{Example of SPIRE line measurement for R Dor.}
%\label{Fig:SPIRELineFit}
 \end{subfigure}
\caption{Example of line fitting (red dotted line) for the PACS and SPIRE spectrum (black full line) of the M-type AGB-star R Dor. 
The fitting functions are Gaussians for PACS (left panel) and sync functions for SPIRE (right panel).
The local maxima at 143.5 and 146.2~\mic\ and 1130 GHz have not been selected as lines by the algorithm.
}
\label{Fig:LineFit}
\end{figure*}

Subsequent to the line selection, the integrated fluxes were measured by fitting a Gaussian to the line profile, 
while locally approximating the continuum by a first-order polynomial. Neighbouring lines were fitted together 
when this improved the goodness of the fit and to account for line blends. The quality of the fits was manually checked, 
resulting in a removal of less than 10\% of the lines which were judged unsatisfactory.
An illustration of such a fit with multiple lines in the PACS range of R Dor is given in the left panel of Fig.~\ref{Fig:LineFit}.

In case of SPIRE, spectral lines were selected and measured in an iterative way. During each iteration step, the strongest 
local maximum that satisfied the following two conditions, was selected as a spectral line: 
(1) they must be separated by at least one FWHM from spectral lines detected during previous iterations and 
(2) $F_{\lambda_{\mathrm{max}}}\geq 8.0 \times \mathrm{MAD_{36 FWHM}}$. The stricter criterion is justified as the Fourier transform nature 
and the lower resolution of the spectra make the line-noise distinction harder and was derived by minimising the detection 
of false lines when running the algorithm on negative noise peaks.

After a line is selected, the integrated flux is 
measured by simultaneously fitting the spectral line in question, together with the continuum and the lines selected during 
all previous iteration steps. The fitted model consisted of a fifth-order polynomial, which approximates the continuum, 
plus a sinc function for each line. The residual spectrum then served as input for the next iteration step. 
The algorithm continued until no more lines were found that satisfied the conditions as described above. 
The right panel of Fig.~\ref{Fig:LineFit} shows an example of line fitting in the SPIRE range of R Dor.

The detected lines and the derived line fluxes are reported for each target in Appendix~\ref{Appen:lines}. 
The reported total uncertainty represents the fitting uncertainty. 
Line identifications are discussed in Sect.~\ref{Sect:Identification} and are reported in Appendix~\ref{Appen:lines}.

The line sensitivity varies from source to source as the number of repetitions of a line scan (for PACS) or spectral map (for SPIRE) varies.
The best line sensitivities achieved in some sources are
8.4 $\cdot 10^{-18}$ W/m$^2$ (in the  65--72 $\mu$m range in B2A), 
6.3 $\cdot 10^{-18}$ W/m$^2$ (in the  75--94 $\mu$m range in B2B), 
3.6 $\cdot 10^{-18}$ W/m$^2$ (in the 120--140 $\mu$m range in R1 short),
2.9 $\cdot 10^{-18}$ W/m$^2$ (in the 140--190 $\mu$m range in R1 long),
1.9 $\cdot 10^{-18}$ W/m$^2$ (in the 270--300 $\mu$m range in SSW),
1.5 $\cdot 10^{-18}$ W/m$^2$ (in the 300--450 $\mu$m range in SLW), while
the typical sensitivities are a factor of 3 (SPIRE) and 4--6 (PACS) worse.

\subsection{Line identification}
\label{Sect:Identification}

The resulting line lists were cross-referenced to the molecular spectroscopic catalogues of 
the CDMS\footnote{The Cologne Database for Molecular Spectroscopy, available at \url{http://www.astro.uni-koeln.de/cdms}, 
see \citet{CDMS2001}.} 
and JPL\footnote{Available at \url{http://spec.jpl.nasa.gov/}, see \citet{JPL1998}.} databases in order to identify 
the corresponding chemical species. For each observed spectral line, molecular transitions are searched for within 
a wavelength region of half the FWHM around the observed central wavelength. The spectral lines were 
all corrected for the velocity of the local standard of rest ($v_\mathrm{LSR}$) which are listed in Table~\ref{Table:sample}.

As multiple transitions of different chemical species could belong to a specific spectral line, an abundance analysis assuming 
local thermodynamic equilibrium (LTE) was performed to indicate the most probable contributors. In LTE, the level 
populations $n_{i}$ 
%of a certain chemical species 
are given by 
$$ \frac{n_{i}}{n} = g_{i}\frac{e^{-E_{i}/k_{\rm b}T}}{Q(T)},$$
with $n$ the total abundance of the species, $g_{i}$ the degeneracy and $E_{i}$ the energy of state level $i$, 
$T$ the excitation temperature of the particles, and $Q(T)$ the partition function of the species. 
The level populations corresponding to all possible molecular 
transitions for each spectral line were calculated. In the case of optically thin emission the line intensity is proportional 
to the number density and the Einstein A coefficient of the transition.
The relative contribution of each transition of the different chemical species to the observed spectral line could thus be estimated. 

The level transitions and molecule-specific properties like statistical weights, transition energies and partition functions 
were taken directly from the CDMS and JPL catalogues. Additional partition functions 
for $\mathrm{H_{2}O}$ and $\mathrm{C_{2}H_{2}}$ were 
found in \citet{Chen2000} and \citet{Amyay2011}, respectively. 
The molecules that were considered for the identification are listed in Table~\ref{Table:Species}, together with their 
typical fractional abundances as found in the literature. The species SO, $\mathrm{SO_{2}}$, PO and PN were 
only considered for M- and S-type stars as they either have very low abundances or are undetected in C-type stars. 
In the same way, the carbon-rich species $\mathrm{C_{2}H_{2}}$, CCH, $\mathrm{C_{4}H}$, 
$\mathrm{HC_{3}N}$, $\mathrm{SiC_{2}}$, $\mathrm{\element[][29]{Si}C_{2}}$, and $\mathrm{\element[][30]{Si}C_{2}}$ 
are only used for the line identification in C-type stars. 

The \element[][13]{C} isotopologues of CO, HCN, CS and CN are also considered. 
Typical $\mathrm{\element[][12]{C} / \element[][13]{C}}$ ratios of 13, 25 and 34 for respectively M-type, S-type and C-type stars 
are adopted \citep{Ramstedt_Olofsson_2014}. 

The \element[][17]{O} and \element[][18]{O} isotopologues are only considered for CO and $\mathrm{H_{2}O}$, as these isotopologues are 
typically 2 to 3 orders of magnitude less abundant then the \element[][16]{O} ones, which would lead to undetectable amounts 
of  
isotopologues of other molecules.
Previous studies (recently e.g. \citealt{Hinkle_etal_2016} or \citealt{DeNutte2017} and references therein)     
also show a wide range in  \element[][16]{O}/\element[][17]{O} and \element[][16]{O}/\element[][18]{O} ratios. 
For simplicity, an isotopic ratio of 1000 for both \element[][16]{O}/\element[][17]{O} and \element[][16]{O}/\element[][18]{O} was chosen. 

Finally, the \element[][29]{Si} and \element[][30]{Si} isotopologues of SiO, SiS, and $\mathrm{SiC_{2}}$ were also considered. 
Solar isotopic ratios \citep{Asplund2009} of 20 and 30 for \element[][29]{Si}/\element[][28]{Si} and \element[][30]{Si}/\element[][28]{Si}, 
respectively are adopted. This is justified as studies of evolved low-mass stars find similar 
values (e.g. \cite{VelillaPrieto2017}, \citet{Decin2010, Schoier_etal_2011, Danilovich2014}).                

The LTE abundance analysis was performed for three different temperatures 75~K, 300~K, and 500~K. In doing so, a large part of the 
CSE from the warmer inner envelope to the colder outer regions is covered. Also, a biased identification due to a spurious choice 
of temperature is avoided. The choice of 500~K as a the maximum value is justified by the fact that chemical data and properties are 
often lacking for higher temperatures. 
For each target, the complete line list with possible identifications for the three temperatures is given in Appendix~\ref{Appen:lines}. 
The uncertainty on the identification is dependent on the validity of the LTE assumption, the assumption of optically thin line emission, 
the adopted abundances  and temperatures and on the uncertainties and incompleteness of the data provided in the CDMS and JPL catalogues. 
An extra uncertainty is added when transitions of isotopologues are involved, as their real isotopic ratios are often highly dependent 
on the  initial mass and the evolutionary status of the target, see e.g. \citet{Cristallo2015, Marigo2016, Karakas2016, DeNutte2017}.

Atomic transitions have not been explicitly searched for using the methodology described above. 
Atomic transitions have not been studied very extensively in AGB stars, 
although the C{\sc i} (1-0) line at 609~$\mu$m (492 GHz) has been detected in a few AGB stars using heterodyne techniques.
In Appendix~\ref{Appen-Atomic} we briefly discuss the presence of some well-known atomic lines.

\begin{table}
\setlength{\tabcolsep}{1.05mm}
\centering
 \caption{Input molecular species and abundances relative to $\mathrm{H_2}$ (in exponential notation) adopted 
for the identification of the spectral lines.}
\label{Table:Species}															
\begin{tabular}{llrlrlr}		
\hline	
\hline	
\multirow{2}{*}{Species}		&	\multicolumn{6}{c}{$n\left(\mathrm{X}\right)/n\left(\mathrm{H_2}\right)$}  	\\
\cline{2-7}
					&	    M-type	& Ref.	&	S-type 	& Ref.	&	C-type & Ref.	\\
\hline		
CO					&	3.0 (-4) & 1,2  &       6.0 (-4) & 3,4	&	1.0  (-3) & 2	\\
$\mathrm{\element[][13]{C}O}$		&	2.3 ({-5}) & 	&	2.4 ({-5}) &	&	3.0  ({-5}) &	\\
$\mathrm{C\element[][17]{O}}$		&	3.0 ({-7}) & 	&	6.0 ({-7}) &	&	1.0  ({-6}) &	\\
$\mathrm{C\element[][18]{O}}$		&	3.0 ({-7}) &	&	6.0 ({-7}) &	&	1.0  ({-6}) &	\\
$\mathrm{H_{2}O}$			&	1.0 (-4) & 5	&	1.0 (-5) & 2	&	1.0  (-5) & 6	\\
$\mathrm{H_{2}\element[][17]{O}}$	&	1.0 ({-7}) &	&	1.0 ({-8}) &	&	1.0  ({-8}) &	\\
$\mathrm{H_{2}\element[][18]{O}}$	&	1.0 ({-7}) &	&	1.0 ({-8}) &	&	1.0  ({-8}) &	\\
SiO					&	6.0 (-6) & 1	&	6.0 (-6) & 1	&	1.0  (-6) & 1	\\
$\mathrm{\element[][29]{Si}O}$		&	3.0 ({-7}) &	&	3.0 ({-7}) &	&	5.0  ({-8}) &	\\
$\mathrm{\element[][30]{Si}O}$		&	2.0 ({-7}) &	&	2.0 ({-7}) &	&	3.3  ({-8}) &	\\
SO					&	3.5 (-6) & 10,11,12	& 3.5 ({-6}) &	&	$\ldots$        &  \\
$\mathrm{SO_{2}}$			&	2.5 (-6) & 10,11,12	& 2.5 ({-6}) &	&	$\ldots$ 	&  \\
SiS					&	3.0 (-7) & 13	&	8.0 ({-7}) &	&	3.0  (-6) & 13	\\
$\mathrm{\element[][29]{Si}S}$		&	1.5 ({-8}) &	&	4.0 ({-8}) &	&	1.5  ({-7}) &	\\
$\mathrm{\element[][30]{Si}S}$		&	1.0 ({-8}) &	&	2.7 ({-8}) &	&	1.0  (-7) & 	\\
HCN					&	1.2 (-7) & 14	&	7.0 (-7) & 14	&	2.9  (-5) & 14 \\
$\mathrm{H\element[][13]{C}N}$	        &	9.2 (-9) &	&	2.8 ({-8}) &	&	8.5  (-7) &	\\
PO					&	1.0 (-7) & 15   &	1.0 ({-7}) &	&			$\ldots$  &		\\
PN					&	1.0 (-7) & 15	&	1.0 ({-7}) &	&			$\ldots$  &		\\
CS					&	1.0 (-7) & 10,12,16	& 5.0 ({-7}) &	&	1.0  (-6) & 17,18,19 	\\
$\mathrm{\element[][13]{C}S}$		&	7.7 ({-9}) &	&	2.0 ({-8}) &	&	3.0  (-8)	\\
CN					&	1.0 (-7) & 10,12,16  &	1.0 (-5) & 3,4	&	2.0  (-5)& 18,19\\
$\mathrm{\element[][13]{C}N}$		&	3.9 ({-9}) &	&	4.0 ({-7}) &	&	5.9  ({-7}) &	\\
$\mathrm{NH_{3}}$			&	7.5 (-7) & 20	&	7.5 ({-7}) &	&	7.5  ({-7}) &	\\
$\mathrm{H_{2}S}$			&	1.0 (-8) & 21	&	1.0 ({-8}) &	&	4.0  (-9) & 17	\\
$\mathrm{C_{2}H_{2}}$			&		$\ldots$ 	&	&	$\ldots$ 	& &	1.0  (-5) & 22	\\
CCH					&		$\ldots$ 	&	&	$\ldots$ 	& &	5.0  (-6) & 18	\\
$\mathrm{C_{4}H}$			&		$\ldots$ 	&	&	$\ldots$ 	& &	3.0  (-6) & 18,19	\\
$\mathrm{HC_{3}N}$			&		$\ldots$ 	&	&       $\ldots$ 	& &	1.0  (-6) & 18,19	\\
$\mathrm{SiC_{2}}$			&		$\ldots$ 	&	&	$\ldots$ 	& &	2.0  (-7) & 18,19,23	\\
$\mathrm{\element[][29]{Si}C_{2}}$	&		$\ldots$ 	&	&	$\ldots$ 	& &	1.0  ({-8}) &	\\
$\mathrm{\element[][30]{Si}C_{2}}$	&		$\ldots$ 	&	&	$\ldots$ 	& &	6.7  ({-9}) &	\\
										
\hline					
\end{tabular}
%\tablefoot{}		
\tablebib{(1) \citet{Ziurys2009}, (2) \citet{Danilovich2015}, (3) \citet{Danilovich2014}, (4) \citet{Schoier_etal_2011}, 
(5) \citet{Maercker_2016_2}, (6) \citet{Lombaert_etal_2016}, (7) \citet{GonzalezD2003}, (8) \citet{Ramstedt_etal_2009}, 
(9) \citet{Schoier2006}, (10) \citet{Decin2010}, (11) \citet{Danilovich2016}, (12) \citet{VelillaPrieto2017}, 
(13) \citet{Schoier2007}, (14) \citet{Schoier2013}, (15) \citet{DeBeck2013}, (16) \citet{Kim2010}, (17) \citet{Agundez2012}, 
(18) \citet{ZhangAndKwok2009}, (19) \citet{Woods2003}, (20) \citet{Wong2015}, (21) \citet{Gobrecht2016}, 
(22) \citet{Fonfria2008}, (23) \citet{Cernicharo2011b}. References for the isotopic ratios are listed in the main text. }
\end{table}

\subsection{Rotation diagrams}
\label{S-RD}

The rotation diagrams technique was used to estimate the typical excitation temperatures 
of the circumstellar CO gas for all targets and of HCN for the carbon stars in the sample. 
These two molecules are the only ones for which there is sufficient multi-transitional data available.
Assuming that the spectral lines are optically thin and that the level populations are in local thermal equilibrium (LTE) corresponding 
to a uniform rotational temperature ($T_{\mathrm{rot}}$) in the gas shell, the following relation holds:
$$\ln\frac{N_{\mathrm{u}}}{g_{\mathrm{u}}} = \ln\frac{N}{Q\left(T_{\mathrm{rot}}\right)} - \frac{E_{\mathrm{u}}}{k_{\mathrm{b}}T_{\mathrm{rot}}}.$$
Here $N_{\mathrm{u}}$, $g_{\mathrm{u}}$, and $E_{\mathrm{u}}$ are the population, degeneracy, and excitation energy of the upper level, $Q$ is the 
partition function of the species and $k_{\mathrm{b}}$ is the Boltzmann constant. If sufficient transitions are detected in the PACS and/or 
SPIRE ranges which cover different energy ranges, $T_{\mathrm{rot}}$ can be derived by a linear fit of 
$\ln ({N_{\mathrm{u}}}/{g_{\mathrm{u}}})$ 
versus $\frac{E_{\mathrm{u}}}{k_{\mathrm{b}}T_{\mathrm{rot}}}$ (see e.g. \citet{Goldsmith1999} for 
a didactic outline of this classical method).

Deviations from the above assumptions, like optically thick lines, gas temperatures gradients in the CSE and deviations from LTE will give rise to 
a non-linear behaviour in the rotation diagrams, meaning that in most cases a single temperature will not describe the data very well. 
The next simplest approach was adopted by dividing the data into two components when both a PACS and SPIRE spectrum were available. 
Rotational temperatures were derived by a linear least-squares fit for each of the different components. 
When at least one of the two components 
contained fewer than three data points a one-component linear least-squares fit is performed. 
The fitting is weighted by considering the absolute calibration error and the uncertainty in the line measurement 
as obtained in Sect.~\ref{LineMeasurement}.

\begin{figure}
 \begin{subfigure}{0.35\textwidth}
 \centering
 \includegraphics[width = \textwidth]{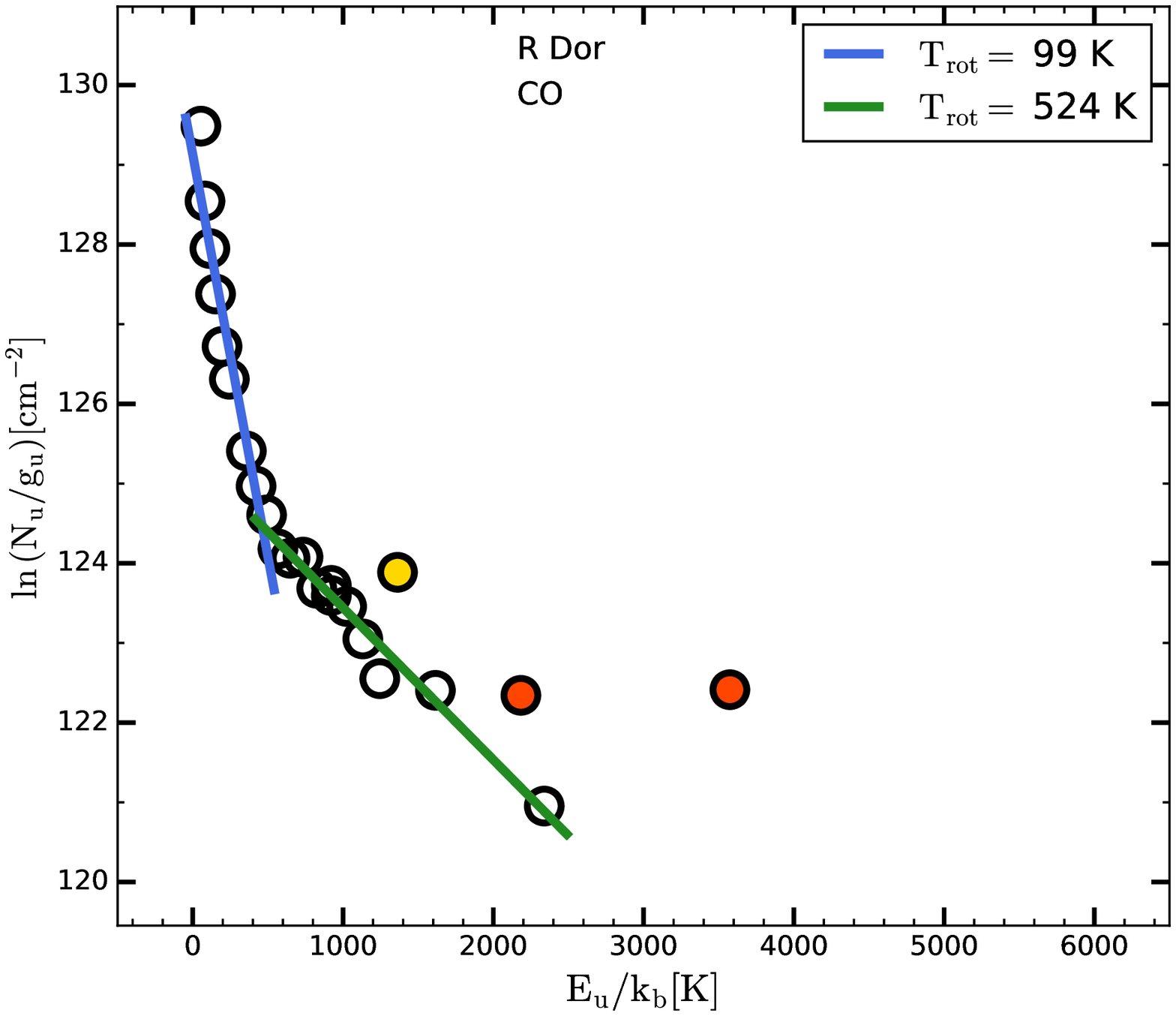}
% \label{Fig:RotRdorCO}
 %\caption{a}
 \end{subfigure}%
\hfill
 \begin{subfigure}{0.35\textwidth}
 \centering
 \includegraphics[width = \textwidth]{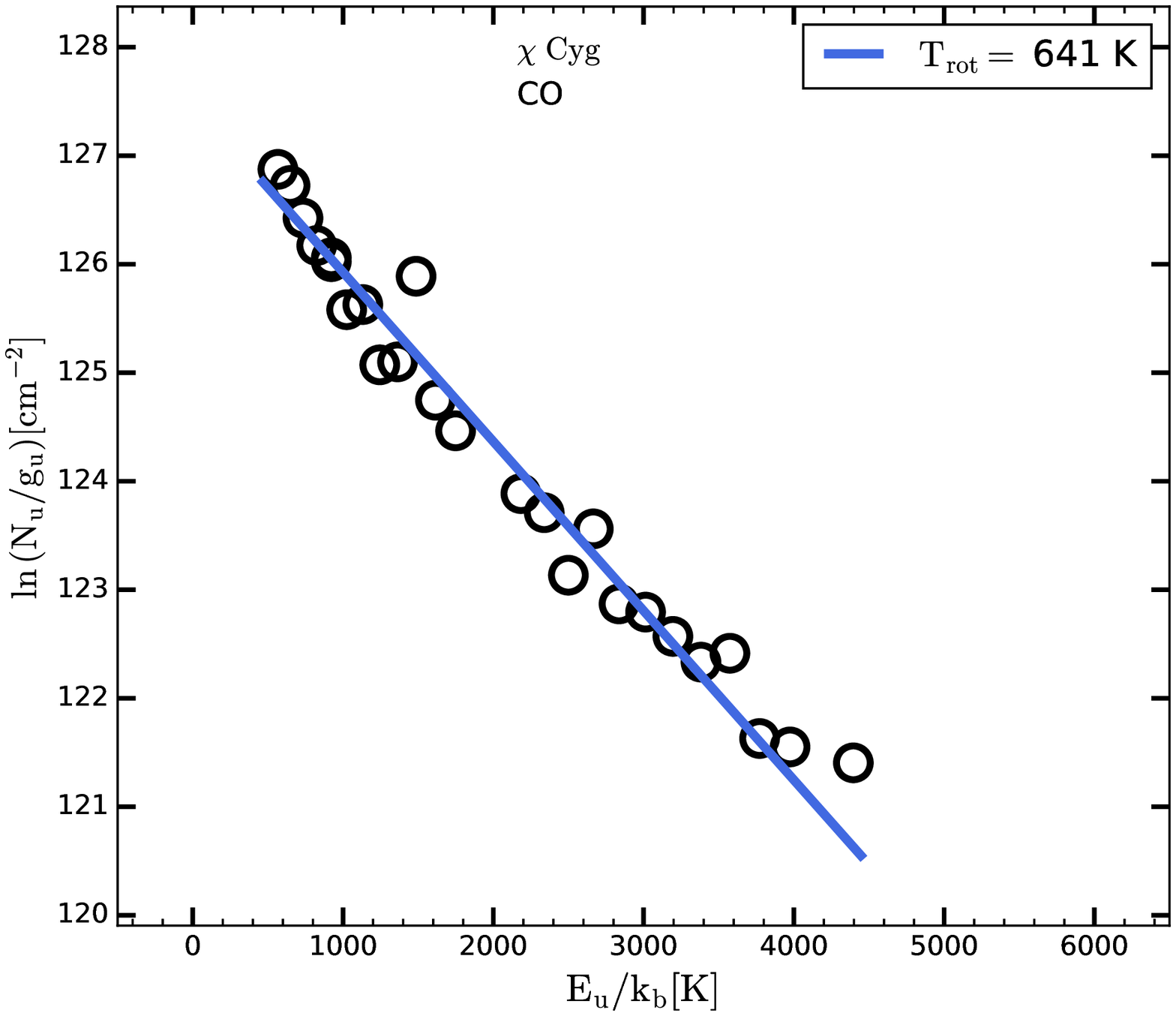}
 %\label{Fig:RotKhicygCO}
 %\caption{a}
 \end{subfigure}%
 \hfill
  \begin{subfigure}{0.35\textwidth}
 \centering
 \includegraphics[width = \textwidth]{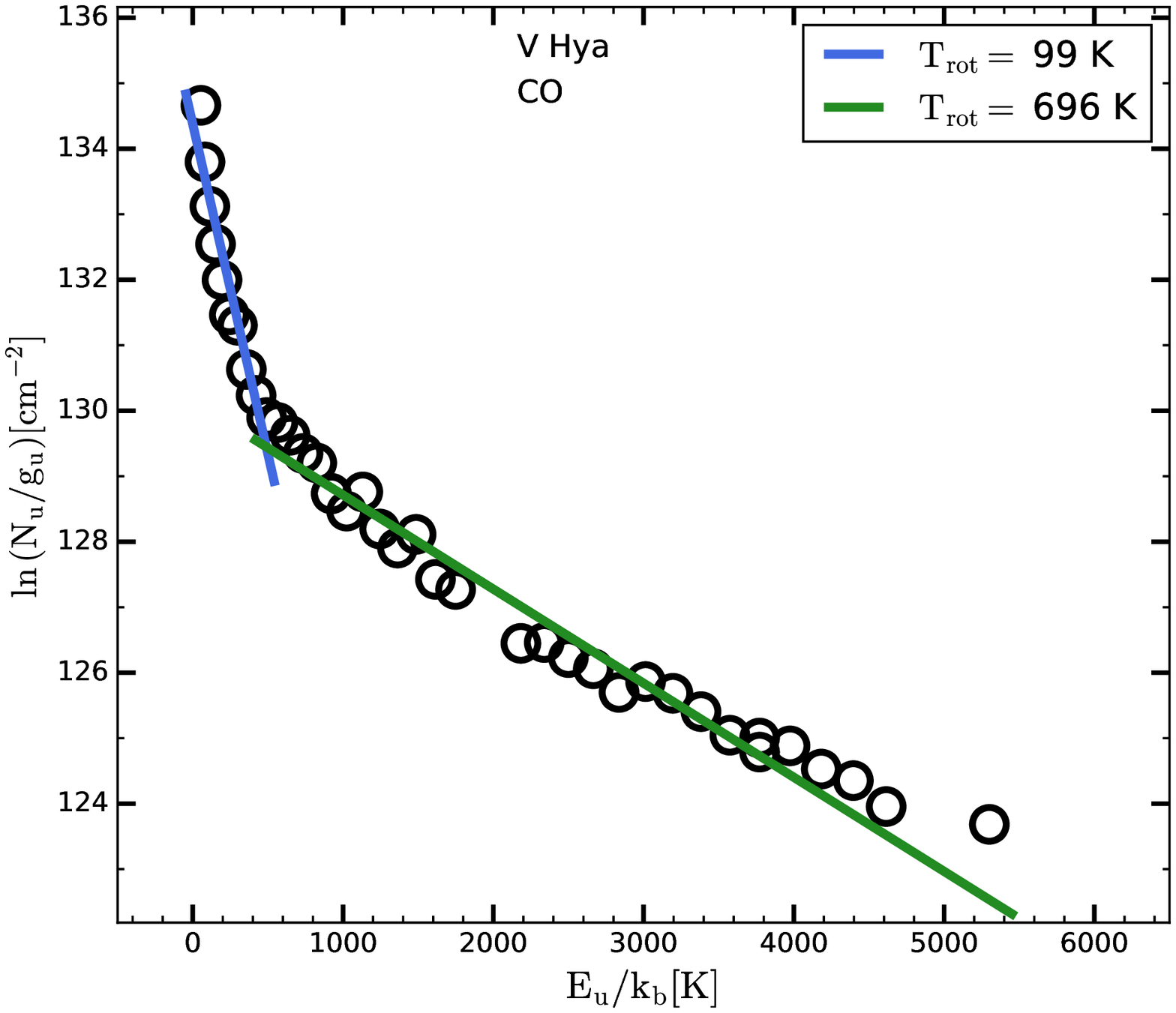}
 %\label{Fig:RotVhyaCO}
 %\caption{a}
 \end{subfigure}%
\hfill
 \begin{subfigure}{0.35\textwidth}
 \centering
 \includegraphics[width = \textwidth]{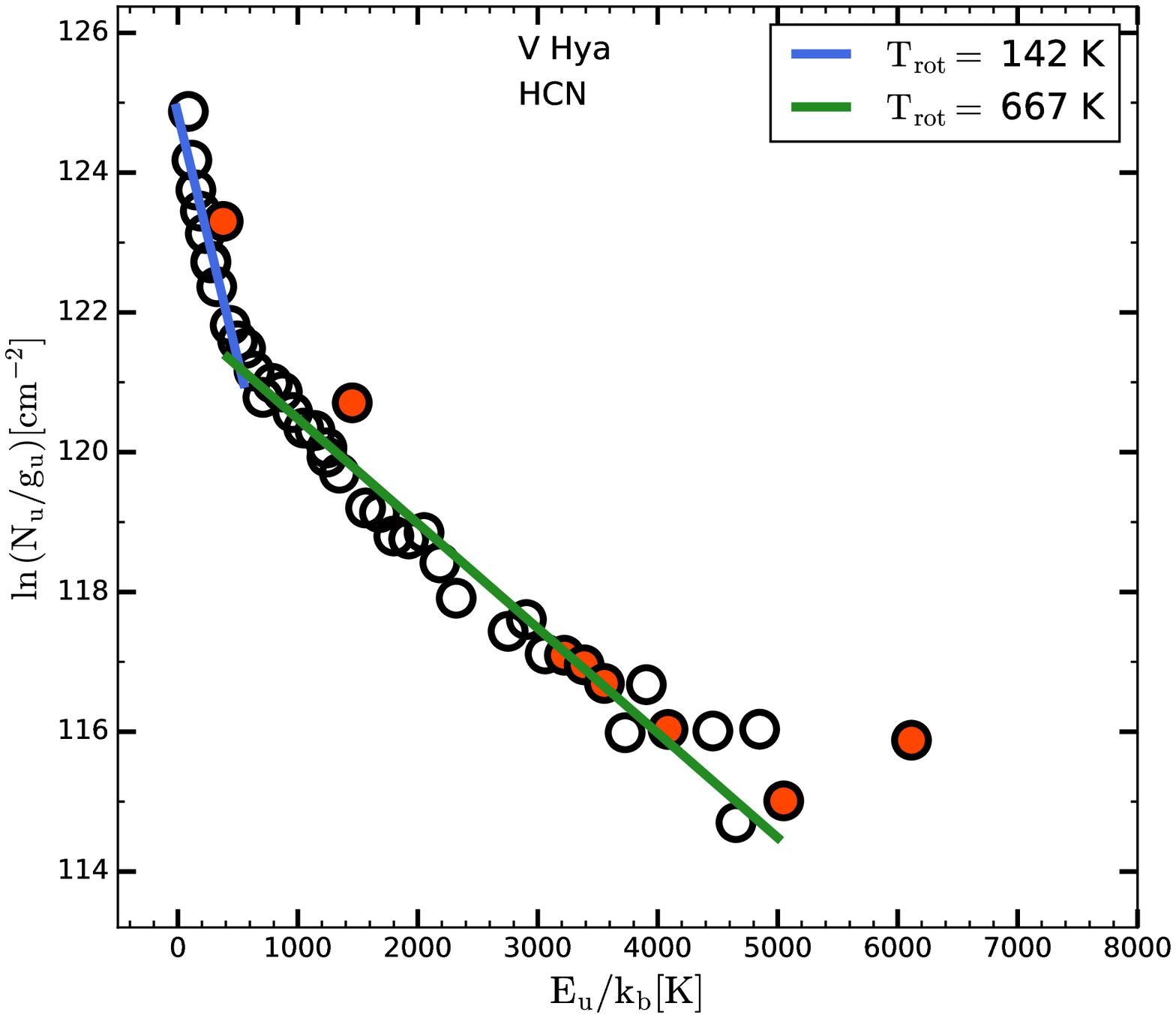}
 %\label{Fig:RotVhyaHCN}
 %\caption{a}
 \end{subfigure}%
 \caption{\label{Fig:RotDiagrams} Examples for rotational diagrams for CO in R Dor and $\chi$ Cyg, and CO and HCN in V Hya. 
The other diagrams are given in Appendix~\ref{Appen:RotDiag}. Yellow and red symbols are points excluded from the fit (see text for details).
%The error bar on a data point is smaller than the plot symbol.
}
\end{figure}

The procedure that was adopted included two iterations steps. Firstly, all lines with a possible contribution
from CO or HCN were included in the least-squares fits. This resulted in 
first estimates for $T_{\mathrm{rot}}$. 
Secondly, all rotation diagrams were re-fitted, while eliminating 
the outliers and possible erroneous data points.
Data points were traced automatically by considering the relative contribution of the molecule to 
the spectral line which was calculated for temperatures of 75~K, 300~K, and 500~K as explained in Sect.~\ref{Sect:Identification} and 
listed in the tables in Appendix~\ref{Appen:lines}. The contributions corresponding to the temperature that was closest to the 
derived $T_{\mathrm{rot}}$ from the first iteration step were considered. The CO data points were removed from the fit when their contribution 
was conservatively estimated to be less than 90\% (for HCN a 75\% criterion was adopted). In the plots these points were flagged in red. 
Any remaining obvious outliers were removed manually from the fit and are indicated in yellow in the final plots.
The final results are listed in Table~\ref{Table:pop-tempdiagram}, while Fig.~\ref{Fig:RotDiagrams} shows examples of 
CO rotation diagrams for R Dor, $\chi$ Cyg and V Hya, and also the HCN rotation diagram of the latter target. 
The rotation diagrams of the other targets can be found in Appendix~\ref{Appen:RotDiag}. 
Due to a lack of detected lines, no CO rotation temperature could be derived for OH 21.5 +0.5, IRAS~19067, R Aql, and RR Aql, 
while no result for HCN could be obtained for R Scl and IRAS 09525-6040.
The results are briefly discussed in Sect.~\ref{SS-RT}.

\begin{table*}	
 \caption{Rotational temperatures.} 

\label{Table:pop-tempdiagram}										
\centering
\begin{tabular}{lllll}	
\hline
\multirow{3}{*}{Name}	&	\multicolumn{2}{c}{CO}					&  		\multicolumn{2}{c}{HCN}	\\ 
\cline{2-5}		& $T_\mathrm{rot, cool}$ ($\sigma_{\rm T, c}$) & $T_\mathrm{rot, hot}$ ($\sigma_{\rm T, h}$)  & $T_\mathrm{rot, cool}$ ($\sigma_{\rm T, c}$) & $T_\mathrm{rot, hot}$ ($\sigma_{\rm T, h}$) \\ 
    			&	(K)		&	(K)	&	(K) & (K)		\\ 
\hline				
WX Psc			& 93 (13)  	& 481 (53) 	&$\ldots$ 		&$\ldots$ 					\\
R Scl			& 56 (12)       &$\ldots$ 	  	&  $\ldots$      	 	&$\ldots$ 	 	\\
OH\,127.8\,+0.0	        &$\ldots$ 		& 939 (33) 	&$\ldots$ 		&			\\
$o$ Cet			& 94 (8) 	& 520 (155)	&$\ldots$ 		&$\ldots$ 					\\
IK Tau			& 96 (11) 	& 463 (39)	&$\ldots$ 		&$\ldots$ 					\\
R Dor			& 99 (11) 	& 524 (46) 	&$\ldots$ 		&$\ldots$ 					\\
TX Cam			& 80 (7) 	& 396 (55)	&$\ldots$ 		&$\ldots$ 					\\
IRAS 09425$-$6040       &$\ldots$ 	 & 365 (202)     &$\ldots$ 		&  $\ldots$ 			\\
CW Leo			& 109 (10) 	& 692 (22) 	& 119 (12)	& 753 (34) 	\\
RW LMi			& 101 (9) 	& 642 (27) 	& 118 (13) 	& 662 (35) 		\\
V Hya			&  99 (10) 	& 696 (25) 	& 142 (13) 	& 667 (29)		\\
W Hya			& 112 (9) 	& 460 (111)	&$\ldots$ 	        &$\ldots$ 					\\
II Lup			&  99 (10) 	& 449 (18)	& 118 (13) 	& 452 (19) 		\\
X Her			&$\ldots$ 		& 1722 (314) 	&$\ldots$ 		&$\ldots$ 			\\
AFGL 5379	        & 94 (10) 	& 768 (511) 	&$\ldots$ 		&$\ldots$ 				\\
%OH\,21.5\,+0.5		&	/	&	 /	&	/	&	/	&		&		&		&			\\
OH\,26.5\,+0.6	        & 88 (16) 	& 582 (75) 	&$\ldots$		&$\ldots$					\\
OH\,30.1\,$-$0.7	& 58 (11) 	&$\ldots$ 		&$\ldots$		&$\ldots$						\\
IRAS 18488$-$0107       & 41 (6) 	& 785 (168) 	&$\ldots$		&	$\ldots$					\\
OH\,32.8\,$-$0.3	& 15 (1) 	&$\ldots$ 		&$\ldots$		&$\ldots$					\\
%R Aql			&	/	&	 /	&	/	&	/	&		&		&		&			\\
%IRAS 19067+0811	&	/	&	 /	&	/	&	/	&		&		&		&			\\
W Aql			&$\ldots$ 		& 527 (14) 	&$\ldots$		&$\ldots$		\\ % HCN 534 (251)
GY Aql			&$\ldots$ 		& 363 (36) 	&	$\ldots$	&$\ldots$					\\
$\chi$ Cyg	        &$\ldots$ 		& 641 (28) 	&	$\ldots$	&$\ldots$			\\ % HCN 607 (99)
%RR Aql			&	/	&	 /	&	/	&	/	&		&		&		&		\\
V1943 Sgr		& 72 (6) 	& 743 (105) 	&	$\ldots$	&$\ldots$					\\
AFGL 2513		&$\ldots$ 		& 608 (27) 	&	$\ldots$	& 668 (37)		\\
IRC $-$10 529		& 132 (10) 	& 471 (54) 	&$\ldots$		&$\ldots$					\\
T Mic			&$\ldots$ 	 	& 458 (53) 	&$\ldots$ 		&$\ldots$			\\
V Cyg			&$\ldots$ 		& 564 (24)      &$\ldots$ 		& 462 (38) 		\\
T Cep			&$\ldots$ 		& 568 (155) 	&$\ldots$ 		&$\ldots$				\\
EP Aqr			&$\ldots$ 		& 911 (235) 	&$\ldots$ 		&$\ldots$					\\
$\pi$ Gru		&$\ldots$ 		& 668 (27) 	&$\ldots$ 		&$\ldots$					\\
LL Peg			& 92 (8) 	& 459 (30) 	& 115 (7) 	& 484 (38) 		\\
LP And			& 109 (9) 	& 606 (21) 	& 125 (13) 	& 650 (30)		\\
R Cas			& 92 (9) 	& 455 (50) 	&$\ldots$ 		&$\ldots$				\\
$\alpha$ Ori	        & 96 (8) 	& 681 (41) 	&$\ldots$ 		&$\ldots$				\\
VY CMa			& 119 (7) 	& 471 (32) 	&$\ldots$ 		&$\ldots$					\\
NML Cyg			& 100 (10)	& 528 (36) 	&$\ldots$ 		&$\ldots$				\\
\hline									
\end{tabular}	
\tablefoot{
Rotational temperatures, either a single value or separated into a cool and hot component. Error bars are given between parenthesis.	
}				
\end{table*}

\subsection{Continuum extraction}
\label{S_Cont}

The determination of the dust thermal continuum from the complete PACS and SPIRE spectra was done by eliminating 
the contribution of the detected spectral lines.

The PACS continuum was obtained by replacing each spectral line by the best-fitting first-order polynomial which approximates 
the local continuum (see Sect.~\ref{LineMeasurement}). Doing this for a wavelength region spanning 3$\sigma$ around 
the central wavelength (i.e. $1.27 \cdot$ FWHM of the spectral line), results in a removal of about 99\% of the line's contribution.

The contribution of SPIRE spectral lines is contained in the corresponding best-fitting sinc functions as obtained from 
the algorithm explained in Sect.~\ref{LineMeasurement}. Following the algorithm, the SPIRE continuum was approximated 
by a fifth-order polynomial. To determine the final estimation of the real SPIRE continuum, this fifth-order polynomial 
is summed with the residual spectrum of the last iteration step. In this way, information about 
the spectral noise is preserved and possible quality restrictions of the polynomial fit are eliminated.

\begin{figure*}
\centering
\includegraphics[width = 18cm]{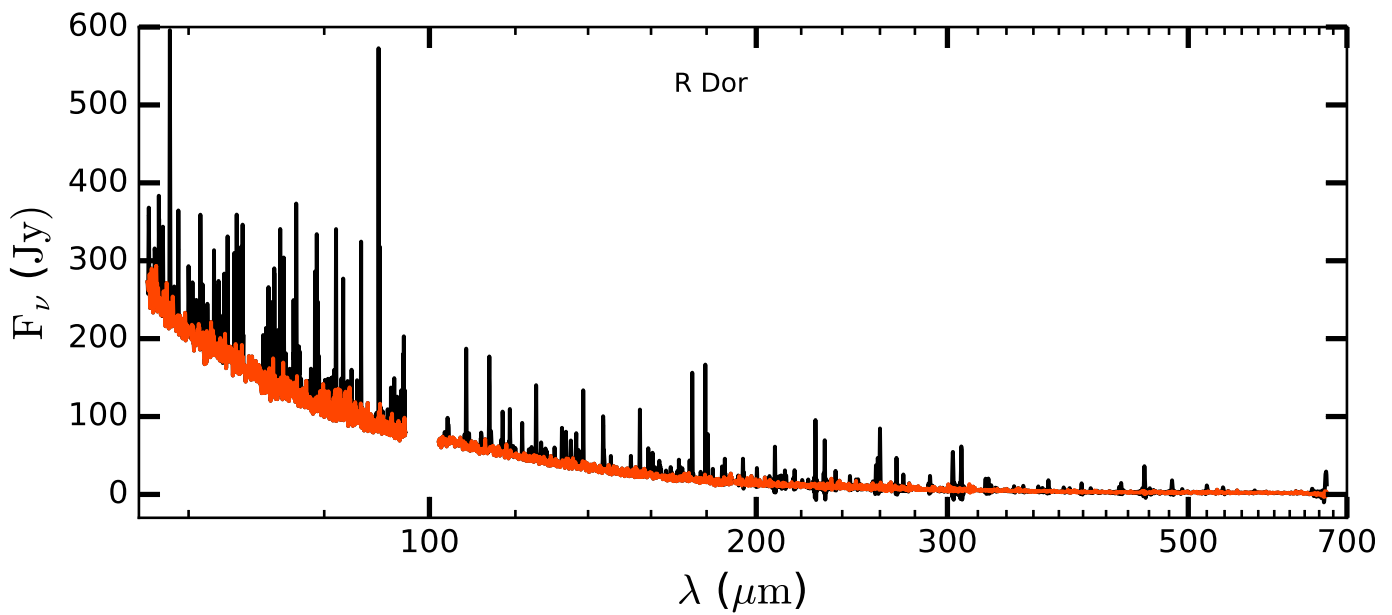}
\caption{Derived continuum spectrum (red) of R Dor, extracted from the complete spectrum (black) and 
removal of all identified spectral lines. 
In this best estimate of the dust continuum spectrum weaker lines and the effect of imperfect removal of the strongest lines 
remain visible.
The continuum spectra of all stars in the sample are available at the CDS.
}
\label{Fig:FullCont}
\end{figure*}

As an example, the PACS and SPIRE continuum of R Dor is plotted on top of the original spectrum in Fig.~\ref{Fig:FullCont}.
All continua spectra are also available in ASCII format at the CDS.

The spectrum looks noisy especially in the PACS blue range, where the spectral line density is highest.
An obvious limitation of the procedure to estimate the continuum is that only the detected spectral lines can be removed.
Lines of low intensity are still present in these spectra. The Gaussian approximation for the shape of the spectral lines and 
the imperfect removal of the strongest lines may also leave residuals that will be present in these `dust continua' spectra.

\subsection{Power law fitting of the dust continuum}
\label{S-PL}

\begin{figure*}
\centering
\includegraphics[width = 18cm]{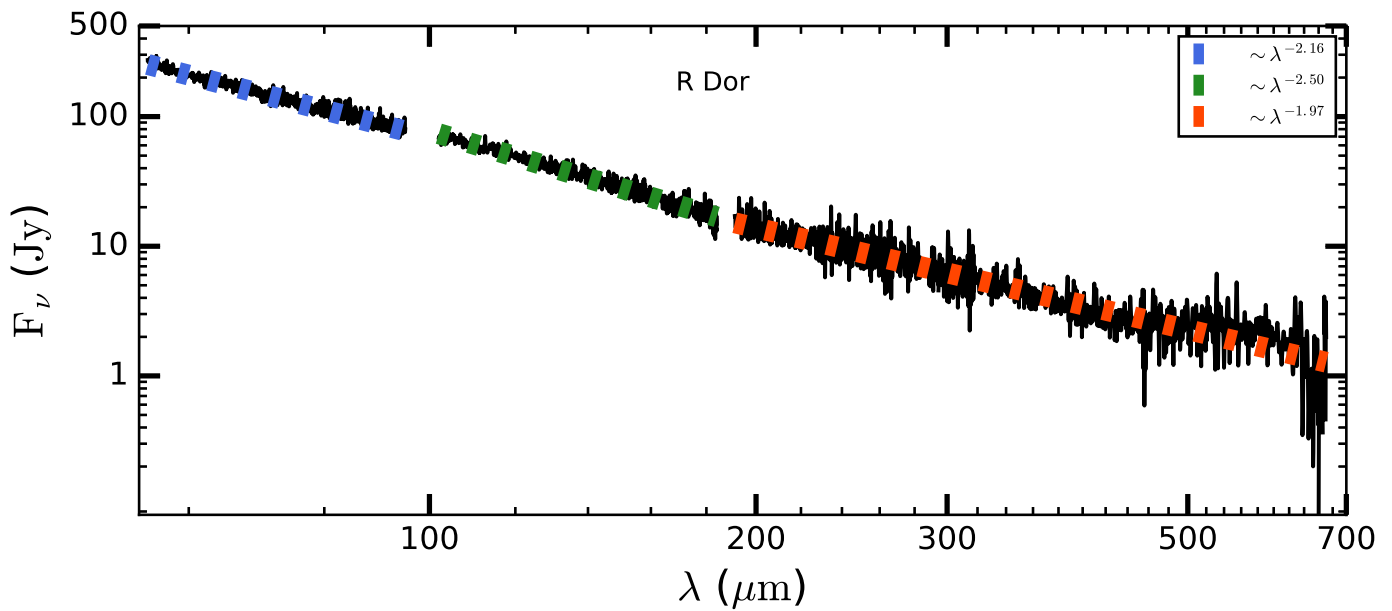}
\caption{Example of the power law fit to three segments of a full PACS and SPIRE line-clipped continuum spectrum. 
}
\label{Fig:Powerlaw}
\end{figure*}

Simple power law fitting of the PACS and SPIRE continua is performed to highlight possible differences in dust grain properties 
and/or in the density structure of the CSEs.
The far-infrared and submillimetre energy distribution of evolved stars undergoing mass loss is dominated by thermal 
emission from dust grains in their CSEs. The dust emission depends on the chemical and morphological properties of the dust grains.
Apart from some narrow wavelength bands (a few $\mathrm{\mu}$m wide) showing some characteristic dust features, these opacities 
are generally approximated by a power law: $\kappa_{\lambda} \sim \lambda^{-\beta}$ in the Rayleigh--Jeans regime. 
In this regime, thermal black-body emission is characterised by $F_{\nu} \sim \lambda^{-2}$. When all dust grains have the same temperature, 
the far infrared energy spectrum can be approximated by a power law: $F_{\nu} \sim \lambda^{-2-\beta} = \lambda^{-p}$. In reality, dust 
grain temperatures range from about 1000 K at the dust formation zone to a few tens of Kelvin in the outer regions. At the shortest PACS 
wavelengths, the emission of the coldest grains could therefore deviate from the Rayleigh--Jeans tail regime. The temperature gradient, 
in combination with optical depth effect and the morphological structure of the dusty CSE will also deflect the spectra from being a 
true power law. To account for this, multiple power law fits $\sim \lambda^{-p}$ were used to approximate the PACS and SPIRE spectra. 
A least-squares fit was performed for the $55--100\ \mathrm{\mu}$m and $100--190\ \mathrm{\mu}$m PACS segments separately. 
The SPIRE continuum as a whole was fitted individually. 
For some targets only the SSW segment of continuum was considered as the signal-to-noise ratio of the SLW segment was considered too low.
The fitting is weighted by a combination of the absolute calibration error, the standard deviation of the spectra 
and a possible continuum offset error of 0.4 Jy for SLW and 0.3 Jy for SSW (see Sect.~\ref{sect:processing}).
The results are listed in Table~\ref{Table:powerlawresults}, while Fig.~\ref{Fig:Powerlaw} shows an example 
of the fit for R Dor. The power law fits of the other targets can be found in Appendix~\ref{Appen:Powerlaw}.
In some cases there is a mismatch between the PACS and SPIRE spectra and these sources are flagged in Table~\ref{Table:ObsPreview}.

\begin{table}	
\setlength{\tabcolsep}{1.4mm}
 \caption{Power law fits to the different segments of the PACS and SPIRE spectra.}
\label{Table:powerlawresults}											
\centering	
\begin{tabular}{llll}		
\hline\hline 
%\multirow{3}{*}{Name}&	power ($\sigma_{\mathrm{power}}$)	&	power	($\sigma_{\mathrm{power}}$)	&	power ($\sigma_{\mathrm{power}}$)	\\
\multirow{3}{*}{Name} &	$p$ ($\sigma_{\mathrm{p}}$)	&	$p$	($\sigma_{\mathrm{p}}$)	&	$p$ ($\sigma_{\mathrm{p}}$)	\\
    				&							&			 \\
\cline{2-4} 
			& 55--100~\mic	&  100--190~\mic	&  200--670~\mic	\\
\hline	
WX Psc			& 2.41 (0.01) 	& 2.96 (0.01) 	& 3.44\tablefootmark{a} (0.03) 		\\
R Scl			&$\ldots$ 	  	&$\ldots$ 	 	& 0.71 (0.03) 	\\
OH\,127.8\,+0.0		& 2.38 (0.01)   & 3.12 (0.02) 	& 3.55\tablefootmark{a} (0.03)	\\
$o$ Cet			& 2.48 (0.01)	& 2.57 (0.01) 	& 2.15 (0.01) 	\\
IK Tau			& 2.49 (0.01) 	& 3.15 (0.01) 	& 2.19 (0.02) 	\\
R Dor			& 2.16 (0.01) 	& 2.50 (0.01) 	& 1.97 (0.01) 	\\
TX Cam			& 2.51 (0.01) 	& 2.61 (0.01) 	& 1.99 (0.03) 	\\
IRAS 09425$-$6040       & 2.02 (0.01) 	& 2.96 (0.01) 	&$\ldots$ 		\\
CW Leo			& 2.66 (0.01) 	& 2.91 (0.01) 	& 2.43 (0.01) 	\\
RW LMi			& 2.36 (0.01) 	& 2.81 (0.01) 	& 2.18 (0.01) 	\\
V Hya			& 2.32 (0.01) 	& 2.76 (0.01) 	& 2.09 (0.02) 	\\
W Hya			& 1.98 (0.01) 	& 1.83 (0.01) 	& 2.08 (0.01) 	\\
II Lup			& 2.17 (0.01) 	& 2.53 (0.01) 	& 2.19 (0.02) 	\\
X Her			& 2.07 (0.01) 	& 2.48 (0.02) 	&	$\ldots$ 		\\
AFGL 5379		& 2.37 (0.02) 	& 3.83 (0.02) 	& 3.65\tablefootmark{a} (0.01) 	\\
OH\,21.5\,+0.5		& 2.34 (0.01) 	& 3.79 (0.02) 	& 2.43\tablefootmark{a} (0.07) 	\\
OH\,26.5\,+0.6		& 2.70 (0.01) 	& 3.37 (0.01) 	& 3.62\tablefootmark{a} (0.01) 	\\
OH\,30.1\,$-$0.7	& 2.13 (0.01) 	& 3.07 (0.01) 	& 3.42\tablefootmark{a} (0.02) 	\\
IRAS 18488$-$0107       & 2.27 (0.01) 	& 2.20 (0.01) 	& 1.27\tablefootmark{a} (0.03) 	\\
OH\,32.8\,$-$0.3	& 2.69 (0.01) 	& 3.21 (0.01) 	& 3.24\tablefootmark{a} (0.02) 	\\
R Aql			& 1.97 (0.03) 	& 2.47 (0.03) 	&	$\ldots$ 			\\
IRAS 19067+0811		& 2.48 (0.01) 	& 4.41 (0.01) 	& 0.91\tablefootmark{a} (0.06) 	\\
W Aql			& 2.25 (0.01) 	& 3.00 (0.01) 	&	$\ldots$ 			\\
GY Aql			& 2.55 (0.01) 	& 2.76 (0.01) 	&	$\ldots$ 			\\
$\chi$ Cyg		& 2.37 (0.01) 	& 2.79 (0.01) 	&	$\ldots$ 			\\
RR Aql			& 2.28 (0.03) 	& 2.58 (0.03) 	&	$\ldots$ 			\\
V1943 Sgr		& 2.37 (0.01) 	& 2.44 (0.01) 	& 2.30\tablefootmark{a} (0.05) 	\\
AFGL 2513		& 2.20 (0.01) 	& 2.41 (0.01) 	&$\ldots$ 				\\
IRC $-$10 529		& 2.25 (0.01) 	& 2.79 (0.01) 	& 3.00 (0.02) 	\\
T Mic			& 2.32 (0.01) 	& 2.38 (0.01) 	&$\ldots$ 				\\
V Cyg			& 2.23 (0.01) 	& 2.74 (0.02) 	&$\ldots$ 				\\
T Cep			& 2.31 (0.03) 	& 2.45 (0.03) 	&$\ldots$ 			\\
EP Aqr			& 2.62 (0.01) 	& 2.57 (0.01) 	&$\ldots$ 				\\
$\pi$ Gru		& 2.48 (0.01) 	& 2.80 (0.01) 	&$\ldots$ 				\\
LL Peg			& 2.24 (0.01) 	& 2.60 (0.01)   & 2.18 (0.01) 	\\
LP And			& 2.39 (0.01) 	& 2.82 (0.01) 	& 2.13 (0.02) 	\\
R Cas			& 2.20 (0.01) 	& 2.67 (0.01) 	& 2.07 (0.02) 	\\
$\alpha$ Ori		& 2.46 (0.01) 	& 2.55 (0.01) 	& 1.68 (0.01) \\
VY CMa			& 2.56 (0.01) 	& 2.74 (0.01) 	& 2.68 (0.01) 	\\
NML Cyg			& 2.16 (0.01) 	& 2.66 (0.01) 	& 2.17 (0.01)  \\ 
\hline 
\end{tabular} 
\tablefoot{
Power law fits to the different segments of the PACS and SPIRE spectra. 
For each segment the best-fit power index and the 1-$\sigma$ fitting error (between parenthesis) is given.
\tablefoottext{a}{The fitting of the 3rd segment was limited to SSW as the signal-to-noise ratio of SLW segment was judged to be too low.}
}
\end{table}

%%%%%%%%%%%%%%%%%%%%%%%%%%%%%%%%%%%%%%%%%%%%%%%%%%%%%%%%%%%%%%%%%%%%%%%%%%%%%%%%%%%%%%%%%%%%%%%%%%%%%%%%%%
\section{Discussion} \label{sect:discussion}

\subsection{Continuum slopes}

\begin{figure*}
\centering
\includegraphics[width=0.7\hsize]{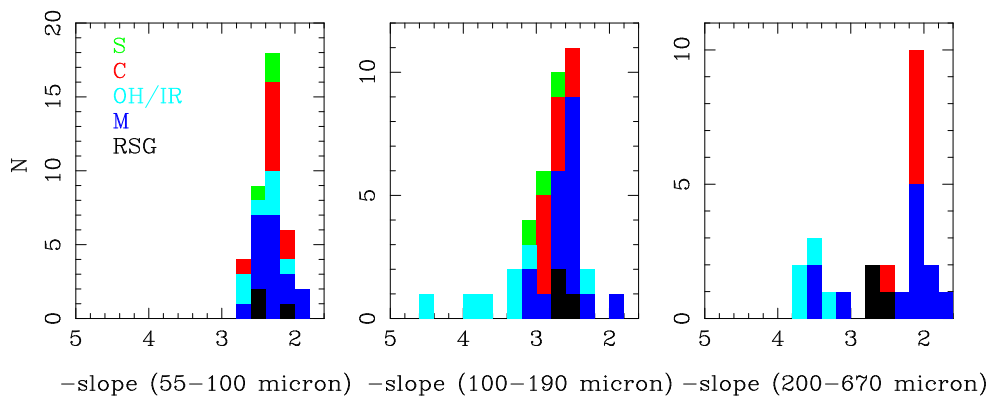}
\caption{Histogram of the continuum slopes for the three wavelength regions. 
}
\label{Fig:HistSlopes}
\end{figure*}

Figure~\ref{Fig:HistSlopes} shows histograms of the continuum slopes based on the data in Table~\ref{Table:powerlawresults}.
The seven stars with the highest MLRs (in excess of $10^{-4}$~\msolyr) are colour-coded explicitly as OH/IR stars 
(OH\,21.5, OH\,26.5, OH\,30.1, OH\,32.8, and AFGL~5379, IRAS~18448, IRAS~19067, which all three are also known OH maser sources).
The source OH~127.8 is not included in this subsample as its MLR is high but nevertheless a factor of two lower 
than the lowest MLR of the other seven stars.

In the 55--100~\mic\ region the continuum slopes for M-, S-, and C-stars; OH/IR stars; and RSGs are very similar, roughly 
between 1.8 and 2.8.
What is interesting is that the spread in the slopes is larger in the 100--190~\mic\ region, and that 
some of the OH/IR stars stand out as having very steep slopes ($>$3.5), while for the other classes the slope is similar to that 
at shorter wavelengths.
In the SPIRE regime there are fewer stars, and sometimes the slope is only based on the SSW part, but the trend is the same.

It is beyond the scope of the current paper to perform the detailed radiative transfer modelling that would be required to discuss this observation 
in more detail, but qualitatively a steeper slope at longer wavelengths could point to a lower MLR in the past, or, to
reverse the timeline, to a recent start of the so-called superwind phase that would indicate the beginning of the end of AGB evolution.
This scenario was invoked  a long time ago \citep{Heske1990} based on the relative weakness of the lower-J CO lines and has been
discussed in the literature since (e.g. \citealt{Justtanont1996,Justtanont2006,Decin2006,Decin2007,Justtanont2013,deVries2014}).

\subsection{Solid-state features}

The {\it Infrared Space Observatory} (ISO) 
%with its two spectrometers 
has revolutionised our knowledge of dust spectroscopy (see e.g. the review by \citealt{Henning2010}).
Based on that success one of the aims of the MESS {\it Herschel} Key Programme \citep{Groenewegen_KP} was to look for (new) dust features.
\citet{Posch2005} discuss solid-state features that are potentially observable in the PACS range, and that include some ices and silicate dust species.

%{Groenewegen_etal_2011}

Figures~\ref{Fig:DustOH}--\ref{Fig:DustS} show the continuum spectra divided by a power law for the M-, C- and S-type targets.
This is useful in order to assess the potential presence of solid-state features
Previously, \citet{Sylvester1999} discussed the spectra of OH/IR stars based on ISO-SWS (2.4--45~\mic) and LWS (45--197~\mic).
In the wavelength region covered by PACS they claim the detection of water ice in emission at 62~\mic\ in OH~127.8, OH~26.5, 
and AFGL 5379 (all in our sample), and detect forsterite silicate at 69~\mic\ in OH~127.8, OH~26.5, OH~32.8, and AFGL 5379 (all in our sample).

Unfortunately, the 61--63~\mic\ region is a difficult one for PACS as the relative spectral response function (RSRF) is 
not easy to calibrate due to the presence of a spectral feature in one of the filters in the light path, resulting 
in a spectral feature that depends on the spatial structure of the source and on the pointing error of the corresponding observation.
The variation due to the RSRF can be judged from the spectra of the C-stars where no water ice is expected to be present.
The typical shape of the spectra of the M-stars in that region is not markedly different.
Unfortunately, the problematic PACS data in that wavelength region do not allow us to confirm or deny the claim 
by \citet{Sylvester1999} regarding the presence of water ice in OH~127.8, OH~26.5, and AFGL 5379.

The location and shape of the forsterite feature near 69~\mic\ in PACS spectra of evolved stars has been extensively 
discussed in \citet{Blommaert2014} and \citet{deVries2014}. No new sources showing forsterite have been found, 
nor any new possible solid-state features, at least down to a level of $\sim$~5\% of the continuum.

\begin{figure*}
\centering
\includegraphics[width=0.9\hsize]{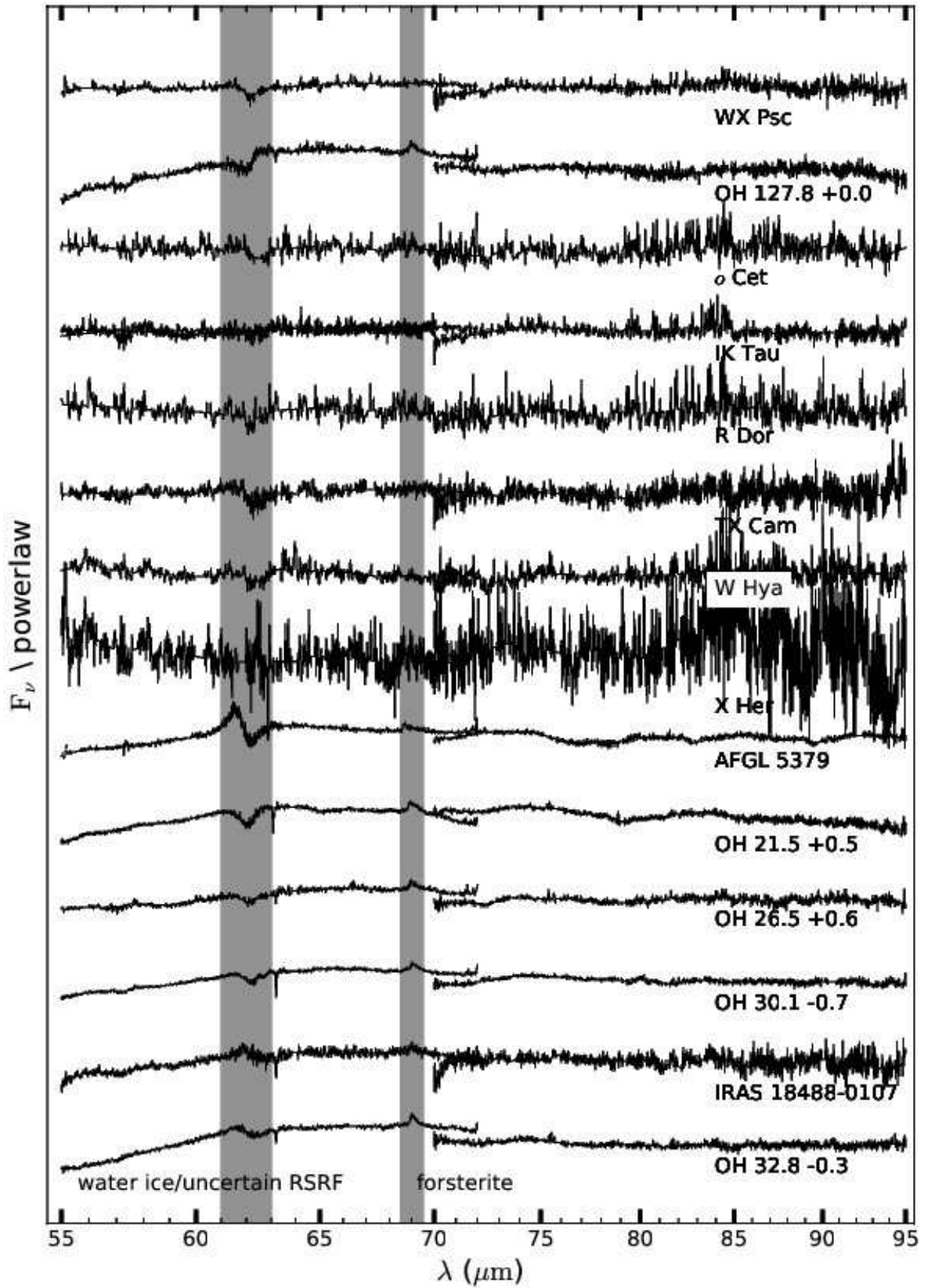}
\caption{Continuum divided spectra of M-type targets. 
The problematic region for the RSRF where the water ice feature is located, and the location of the forsterite feature are indicated.}
\label{Fig:DustOH}
\end{figure*}

\setcounter{figure}{6}
\begin{figure*}
\centering
\includegraphics[width=0.9\hsize]{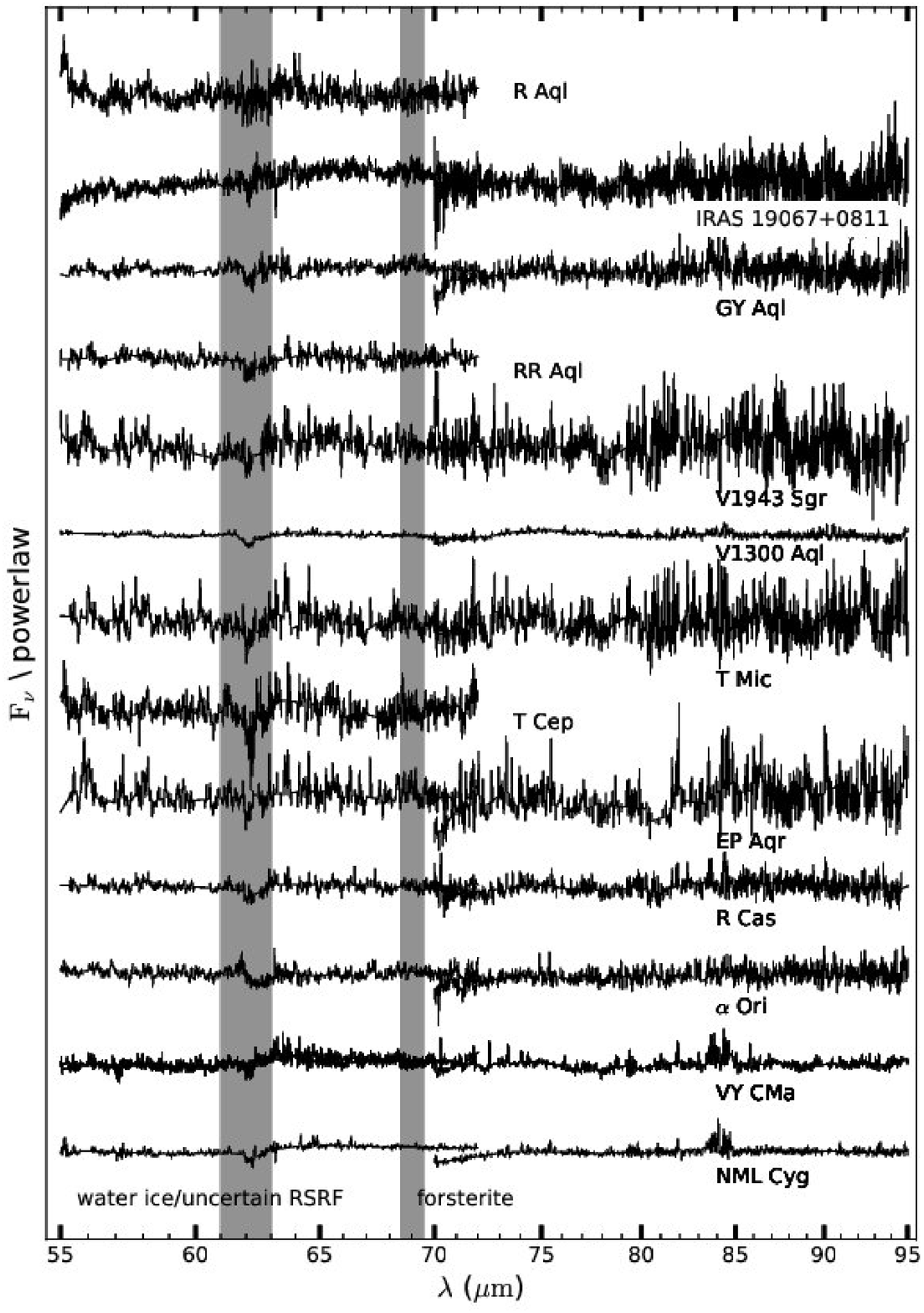}
\caption{Continued.}
%\label{Fig:DustM}
\end{figure*}

\begin{figure*}
\centering
\includegraphics[width=0.9\hsize]{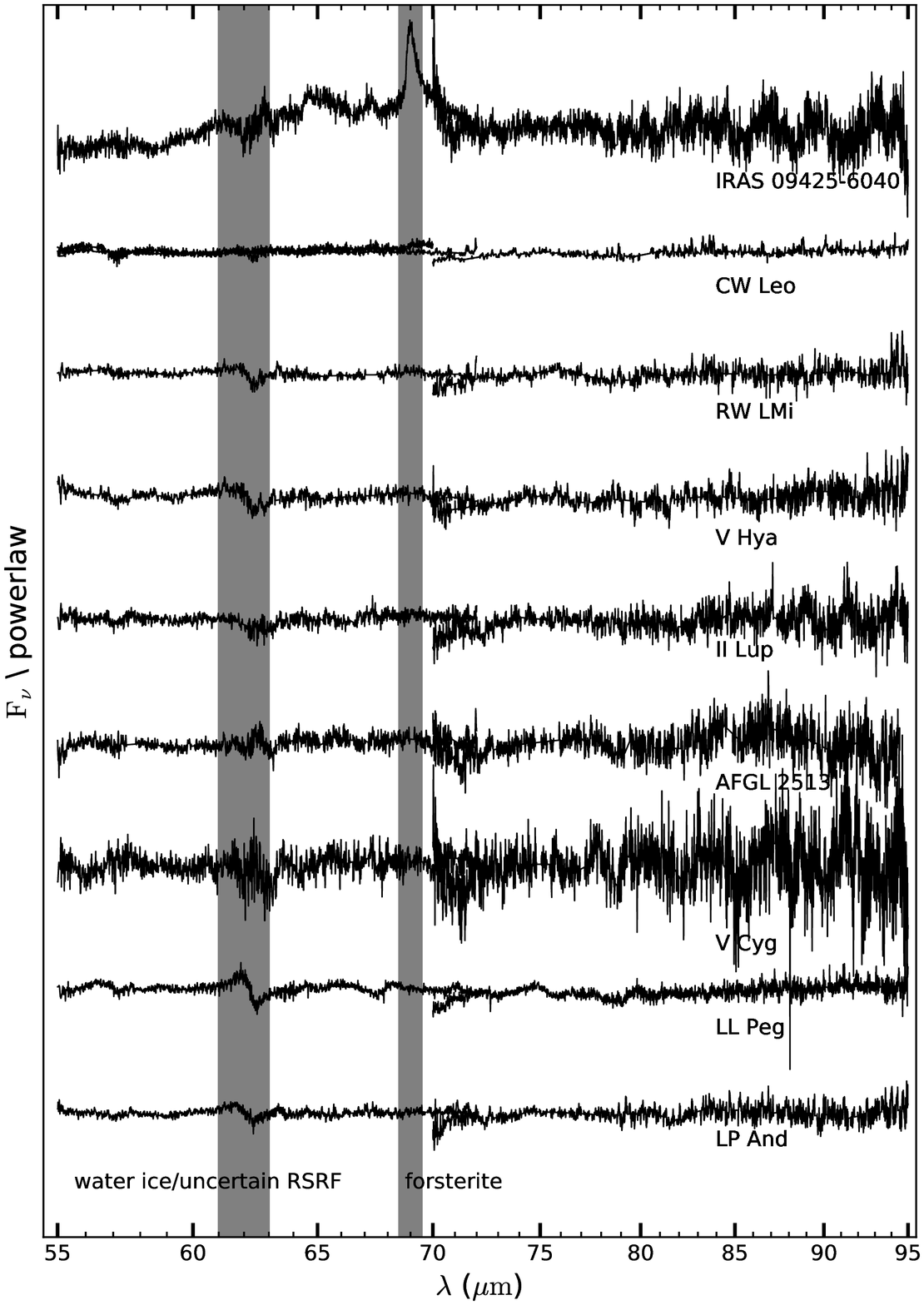}
\caption{As Fig.~\ref{Fig:DustOH} for the C-type targets.}
\label{Fig:DustC}
\end{figure*}

\begin{figure*}
\centering
\includegraphics[width=0.9\hsize]{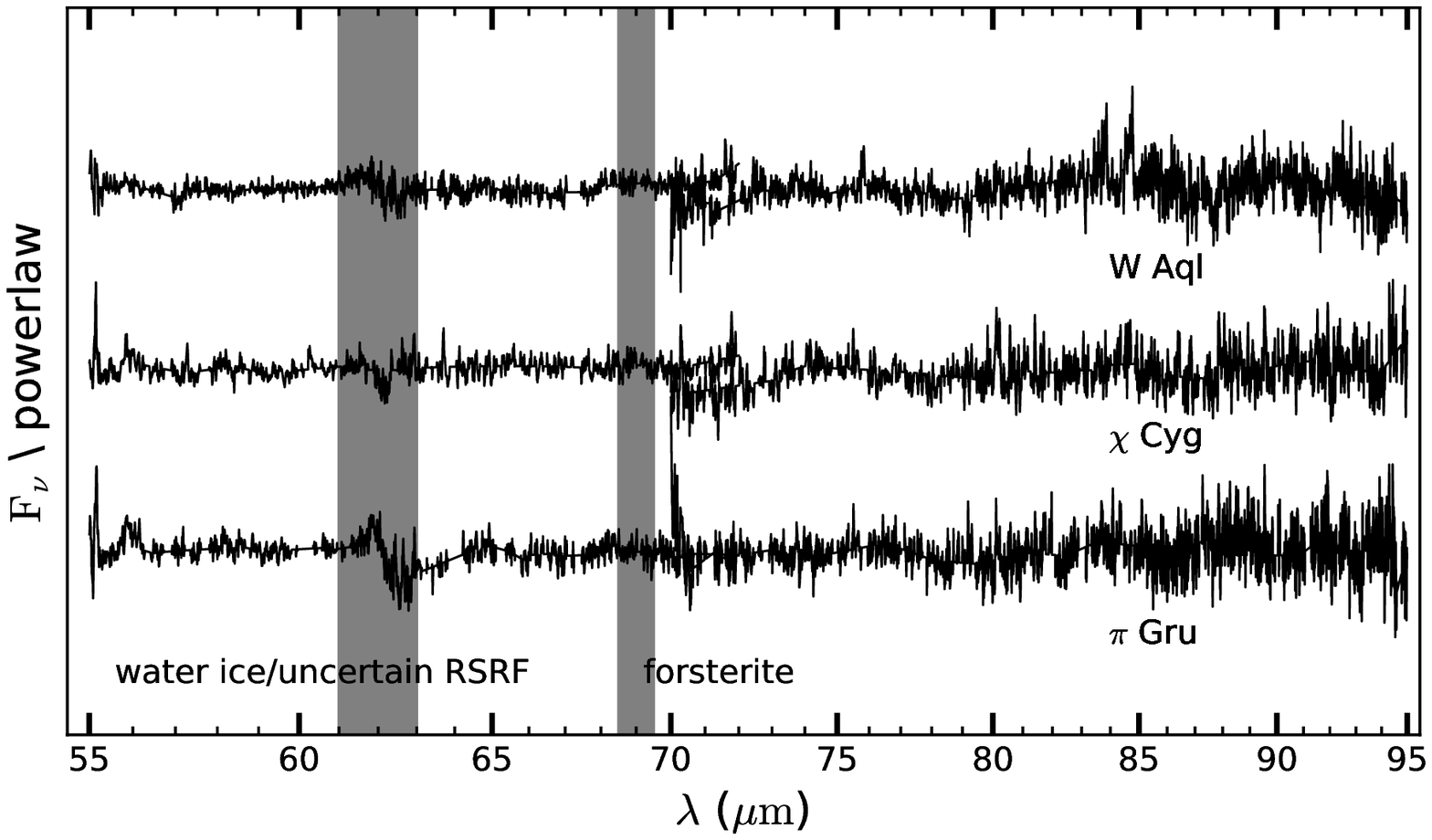}
\caption{As Fig.~\ref{Fig:DustOH} for the S-type targets.}
\label{Fig:DustS}
\end{figure*}

\subsection{Rotational temperatures}
\label{SS-RT}

A two-component model (or a single temperature when not enough data were available) to trace the excitation temperature is
obviously a very crude approximation to the true gas temperature profile in a CSE, which is determined by several heating and 
cooling mechanisms whose relative contributions vary throughout the CSE \citep{Goldreich1976}. It is reassuring to see that for the C-stars 
where we modelled both CO and HCN the temperatures (the single value, or both hot and cool components) agree
within the error bar.
The temperatures are relatively uniform between 80 and 120~K for the cool, and between 450 and 650~K for the hot component.
This justifies a posteriori a maximum temperature of 500~K in the LTE calculations of Sect.~\ref{Sect:Identification}.
The cool component is still relatively warm because the lowest CO transitions (J= 1-0 to 3-2), which that trace the outermost parts of 
the CSE, are outside the SPIRE wavelength range. When no SPIRE spectrum is available, 
the lowest detectable transition is the J= 14-13 transition.

We have compared the rotational temperatures to the gas temperature profiles available in the literature, typically based
on detailed studies of individual objects that solved the thermal balance equation in a self-consistent way
(\citet{Maercker_2016_2} for R Dor, R Cas, TX Cam, and IK Tau; 
\citet{Danilovich2014} for W Aql;
\citet{Schoier_etal_2011} for $\chi$ Cyg; 
\citet{Maercker2008} for W Hya, and IK Tau;
\citet{Decin2006} for VY CMa; 
\citet{Schoier2001} for LP And and R For;  
\citet{Ryde1999} for II Pup; 
\citet{Groenewegen1998} for CW Leo;
\citet{Justtanont1996} for OH 26.5; and
\citet{Groenewegen1994} for OH 32.8).
Such a comparison is qualitative only as it assumes that the gas temperature equals the rotational temperature of the CO molecule, 
which is typically not the case, as there is heating and cooling by other molecules and the excitation temperature of the 
different CO transitions are typically not equal.
Nevertheless,
we find that the hot component traces the gas temperature at $8-10$ to $20-30$ stellar radii, while the cool
component traces the gas temperature in the region  $50$ -- $150$ stellar radii.
The two exceptions are OH~26.5 and OH~32.8.
For OH~26.5 the hot component is typical for the temperature at $\sim 40$ stellar radii and the cool temperature of 90~K is
actually never reached. Due to the superwind nature of the MLR profile adopted in \citet{Justtanont1996} (i.e. a lower MLR in the
outer part of the wind which leads to higher temperatures, due to the lower density and smaller photo dissociation radius of water, 
the major coolant) the temperature never drops below $\sim$200~K.
This model has been refined by \citet{Justtanont2013} but the corresponding temperature profile has not been published.
For OH~32.8 the low temperature of 15~K is only reached at several hundred stellar radii.

%%%%%%%%%%%%%%%%%%%%%%%%%%%%%%%%%%%%%%%%%%%%%%%%%%%%%%%%%%%%%%%%%%%%%%%%%%%%%%%%%%%%%%%%%%%%%%%%%%%%%%%%%%
\section{Summary} \label{sect:summary}

PACS and SPIRE spectra are presented for a sample of 40 AGB stars and red supergiants, reduced according
to the current state of the art reduction strategies.
Molecular lines have been identified and line fluxes measured. The full spectra, and the continuum spectra with 
all identified lines removed are made available to the community.

In addition, we derive rotational diagrams and rotation temperatures for CO. In future works this will be extended to 
line radiative transfer modelling for all detected lines (plus literature data).
Finally, we measure the slope of the dust continua. In future works this will be extended to dust radiative transfer modelling 
of the spectra (and other photometric data) over the entire wavelength region.

%%%%%%%%%%%%%%%%%%%%%%%%%%%%%%%%%%%%%%%%%%%%%%%%%%%%%%%%%%%%%%%%%%%%%%%%%%%%%%%%%%%%%%%%%%%%%%%%%%%%%%%%%%
\begin{acknowledgements}

This research has been funded by the Belgian Science Policy Office under contract BR/143/A2/STARLAB.
TD acknowledges support from the Fund for Scientific Research (FWO), Flanders, Belgium.
\end{acknowledgements}

%%%%%%%%%%%%%%%%%%%%%%%%%%%%%%%%%%%%%%%%%%%%%%%%%%%%%%%%%%%%%%%%%%%%%%%%%%%%%%%%%%%%%%%%%%%%%%%%%%%%%%%%%%
\bibliographystyle{aa}
\bibliography{Lit}

%%%%%%%%%%%%%%%%%%%%%%%%%%%%%%%%%%%%%%%%%%%%%%%%%%%%%%%%%%%%%%%%%%%%%%%%%%%%%%%%%%%%%%%%%%%%%%%%%%%%%%%%%%

\clearpage

\begin{appendix}

\section{Observations} 
\label{Appen:Obs} 

Table~\ref{Table:ObsPreview} presents the basic information of all data sets used in the present paper.
Listed are the target name (also see Table~\ref{Table:sample}), name of the PI (see Sect.~\ref{sect:sample} for 
a description of the programs), \textit{Herschel}s unique Observation Identification number, and a description of the data set.

\begin{table*}	
\caption{Specifications of observations sample targets.} 
\label{Table:ObsPreview}
\setlength{\tabcolsep}{1.0mm}

\centering
\begin{tabular}{lllllclcc}
\hline
Identifier			&		PI		&		Obs. ID			&	Observation &  Remark  \\
\hline
\multicolumn{4}{l}{\textit{AGB-stars}}													\\
\hline							
WX Psc			        &	Groenewegen	&	1342202122			&	PACS B2A, short R1	\\
				&	Groenewegen	&	1342202121			&	PACS B2B, long R1   & mismatch with 70 $\mu$m bolometer flux	\\
				&	Barlow		&	1342246973			&	SPIRE FTS           & small mismatch with PACS spectrum		\\
				&	Groenewegen	&	1342188486/1342188487	&	PACS image 70, 160	\\
R Scl		         	&	Groenewegen	&	1342189545			&	SPIRE FTS   & mismatch with 500 $\mu$m bolometer flux		\\
				&	Groenewegen	&	1342213264/1342213265	&	PACS image 70, 160	\\
				&	Groenewegen	&	1342188657			&	SPIRE image		\\
OH\,127.8\,+0.0	                &	Lombaert        &	1342189956 - 1342189961 &	PACS full spectrum 	\\
				&	Justtanont	&	1342268319				&	SPIRE FTS & alternative background subtraction \\
				&	Groenewegen	&	1342189181/1342189182	&	PACS image 70, 160	\\
				&	Molinari        &	1342226650/1342226651	&	SPIRE image		\\
$o$ Cet			        &	Groenewegen	&	1342213287			&	PACS B2A, short R1	\\
				&	Groenewegen	&	1342213286			&	PACS B2B, long R1	\\
				&	Groenewegen	&	1342189546			&	SPIRE FTS                   &  	\\
				&	Groenewegen	&	1342190335/1342190336	&	PACS image 70, 160	\\
				&	Groenewegen	&	1342189423			&	SPIRE image		\\
IK Tau		        	&	Groenewegen	&	1342203680			&	PACS B2A, short R1	\\
				&	Groenewegen	&	1342203681			&	PACS B3A		 \\
				&	Groenewegen	&	1342203679			&	PACS B2B, long R1	\\
				&	Groenewegen	&	1342192176			&	SPIRE FTS     & mismatch with 250 $\mu$m bolometer flux	\\
                                &                       &                                       &                     & small mismatch with PACS spectrum	\\
				&	Groenewegen	&	1342190343/1342190344	&	PACS image 70, 160	\\
				&	Groenewegen	&	1342248696/1342248697	&	PACS image 100, 160\\
				&	Groenewegen	&	1342191180			&	SPIRE image		\\
R Dor			        &	Groenewegen	&	1342197794			&	PACS B2A, short R1	\\
				&	Groenewegen	&	1342197795			&	PACS B2B, long R1	\\
				&	Barlow		&	1342245114			&	SPIRE FTS		\\
				&	Groenewegen	&	1342197685/1342197686	&	PACS image 70, 160	\\
				&	Groenewegen	&	1342188164			&	SPIRE image		\\
TX Cam			        &	Groenewegen	&	1342225856			&	PACS B2A, short R1 & summed 3$\times$3 spaxel \\
				&	Groenewegen	&	1342225855			&	PACS B2B, long R1  & summed 3$\times$3 spaxel \\
				&	Groenewegen	&	1342251300			&	SPIRE FTS          & alternative background subtraction	\\
                                &                       &                                       &                           &  offset with PACS spectrum  \\
				&	Groenewegen	&	1342217395/1342217396	&	PACS image 70, 160	\\
				&	Royer		&	1342242742/1342242743	&	PACS image 100	\\
IRAS 09425$-$6040        	&	Groenewegen	&	1342225564			&	PACS B2A, short R1	\\
				&	Groenewegen	&	1342225563			&	PACS B2B, long R1	\\
CW Leo			        &	Groenewegen	&	1342186964			&	PACS B2A, B3A, full R1  & extended source \\
				&	Groenewegen	&	1342186965			&	PACS B2B	        & extended source \\
				&	Groenewegen	&	1342197466			&	SPIRE FTS               & extended  source \\
				&	Groenewegen	&	1342186298/1342186299	&	PACS image 70, 160	\\
				&	Groenewegen	&	1342197709/1342197710	&	PACS image 100	\\	
				&	Groenewegen	&	1342207040			&	SPIRE image		\\
RW LMi   			&	Groenewegen	&	1342197800			&	PACS B2A, short R1	\\
				&	Groenewegen	&	1342197799			&	PACS B2B, long R1	\\
				&	Groenewegen	&	1342198264			&	SPIRE FTS		\\
				&	Groenewegen	&	1342210620/1342210621	&	PACS image 70, 160	\\
				&	Groenewegen	&	1342206689			&	SPIRE image		\\
V Hya	         		&	Groenewegen	&	1342197791			&	PACS B2A, short R1	\\
				&	Groenewegen	&	1342197790			&	PACS B2B, long R1	\\
				&	Barlow		&	1342247570			&	SPIRE FTS		\\
				&	Groenewegen	&	1342211997/1342211998	&	PACS image 70, 160	\\
				&	Groenewegen	&	1342188160			&	SPIRE image		\\
W Hya		        	&	Groenewegen	&	1342212604			&	PACS B2A, short R1	\\
				&	Groenewegen	&	1342203453			&	PACS B2B		 \\
				&	Groenewegen	&	1342223808			&	PACS long R1		\\
				&	Groenewegen	&	1342189116			&	SPIRE FTS	  &      \\
				&	Groenewegen	&	1342213848/1342213849	&	PACS image 70, 160	\\
				&	Groenewegen	&	1342189519			&	SPIRE image		\\	
\hline
\end{tabular}
\end{table*}

\setcounter{table}{0}
\begin{table*}	
\caption{Continued.} 
\setlength{\tabcolsep}{1.0mm}
\centering
\begin{tabular}{lllllclcc}
\hline
Identifier			&		PI		&		Obs. ID			&	Observation & Remark	\\
\hline

II Lup			        &	Groenewegen	&	1342215686			&	PACS B2A, short R1	\\
				&	Groenewegen	&	1342215685			&	PACS B2B, long R1  & mismatch with 70, 160 $\mu$m bolometer flux	\\
				&	Barlow		&	1342251281			&	SPIRE FTS          & alternative background subtraction	\\
				&	Groenewegen	&	1342190247/1342190248	&	PACS image 70, 160	\\
X Her			        &	Groenewegen	&	1342197802			&	PACS B2A, short R1	\\
				&	Groenewegen	&	1342202120			&	PACS B2B, long R1	\\
				&	Groenewegen	&	1342188322/1342188323	&	PACS image 70, 160	\\
AFGL 5379		        &	Groenewegen	&	1342228537			&	PACS B2A, short R1 & mispointing, one spaxel	\\
				&	Groenewegen	&	1342228538			&	PACS B2B, long R1  & mispointing, one spaxel	\\
				&	Justtanont	&	1342268287			&	SPIRE FTS  & alternative background subtraction	\\
                                &                       &                                       &                  &  offset with PACS spectrum  \\
				&	Molinari		&	1342204368/1342204369	&	PACS image 70, 160	\\
OH\,21.5\,+0.5		        &	Justtanont		&	1342268778			&	PACS B2A, short R1 & mispointing, one spaxel	\\
				&	Justtanont		&	1342268748			&	PACS B2B, long R1  & mispointing, one spaxel	\\
                                &                               &                                       &                          & mismatch with 70, 160 $\mu$m bolometer flux \\
                                &                               &                                       &                          &  offset with PACS spectrum  \\
				&	Justtanont		&	1342268311			&	SPIRE FTS & alternative background subtraction	\\
				&	Molinari		&	1342218642/1342218643	&	PACS image 70, 160	\\
				&	Molinari		&	1342218642/1342218643	&	SPIRE image		\\
OH\,26.5\,+0.6		        &	Groenewegen	&	1342207777			&	PACS B2A, short R1	\\
				&	Groenewegen	&	1342207776			&	PACS B2B, long R1	\\
				&	Barlow		&	1342243624			&	SPIRE FTS & alternative background subtraction	\\
                                &                       &                                       &                 &  mismatch with 350, 500 $\mu$m bolometer flux \\
				&	Groenewegen	&	1342191817/1342191818	&	PACS image 70, 160	\\
				&	Groenewegen	&	1342218696/1342218697	&	SPIRE image		\\
OH\,30.1\,$-$0.7        	&	Justtanont		&	1342269305			&	PACS B2A, short R1 & c1 spaxel, inhomogeneous background \\
				&	Justtanont		&	1342269304			&	PACS B2B, long R1  & c1 spaxel, inhomogeneous background \\
				&	Justtanont		&	1342268316			&	SPIRE FTS          & alternative background subtraction	 \\
				&	Molinari		&	1342186275/1342186276	&	PACS image 70, 160	\\
				&	Molinari		&	1342186275/1342186276	&	SPIRE image		\\
IRAS 18488$-$0107	        &	Justtanont		&	1342268791			&	PACS B2A, short R1	\\
				&	Justtanont		&	1342268792			&	PACS B2B, long R1  & mismatch with 70,160 $\mu$m bolometer flux	\\
				&	Justtanont		&	1342268317			&	SPIRE FTS          & alternative background subtraction	\\
				&	Molinari		&	1342218692/1342218693	&	PACS image 70, 160	\\
				&	Molinari		&	1342218692/1342218693	&	SPIRE image		\\
OH\,32.8\,$-$0.3	        &	Justtanont		&	1342268793			&	PACS B2A, short R1	\\
				&	Justtanont		&	1342268794			&	PACS B2B, long R1  & mismatch with 70, 160 $\mu$m bolometer flux \\
				&	Justtanont		&	1342268318			&	SPIRE FTS          & alternative background subtraction	\\
				&	Molinari		&	1342218692/1342218693	&	PACS image 70, 160	\\
				&	Molinari		&	1342218692/1342218693	&	SPIRE image		\\
R Aql			        &	Cami			&	1342243900			&	PACS B2A, short R1	\\
				&	Molinari		&	1342207030/1342207031	&	PACS image 70, 160	\\
				&	Molinari		&	1342207030/1342207031	&	SPIRE image		\\
IRAS 19067+0811	                &	Justtanont		&	1342268797			&	PACS B2A, short R1	\\
				&	Justtanont		&	1342268798			&	PACS B2B, long R1	\\
				&	Justtanont		&	1342268308			&	SPIRE FTS          & alternative background subtraction	\\
                                &                               &                                       &                          &  offset with PACS spectrum \\
				&	Molinari		&	1342207030/1342207031	&	PACS image 70, 160	\\
				&	Molinari		&	1342207030/1342207031	&	SPIRE image		\\
W Aql			        &	Groenewegen	&	1342209731			&	PACS B2A, short R1	\\
				&	Groenewegen	&	1342209732			&	PACS B2B, long R1	\\
				&	Groenewegen	&	1342194084/1342194085 	&	PACS image 70, 160	\\
GY Aql			        &	OBSherschel1	&	1342268638			&	PACS B2A, short R1	\\
				&	OBSherschel1	&	1342268449			&	PACS B2B, long R1	\\
$\chi$ Cyg		        &	Groenewegen	&	1342198177			&	PACS B2A, short R1	\\
				&	Groenewegen	&	1342198176			&	PACS B2B, long R1	\\
				&	Groenewegen	&	1342188320/1342188321	&	PACS image 70, 160	\\
RR Aql			        &	OBSherschel1	&	1342269414			&	PACS B2A, short R1	\\
\hline
\end{tabular}
\end{table*}

\setcounter{table}{0}
\begin{table*}	
\caption{Continued.} 
\centering
\begin{tabular}{lllllllcc}
\hline
Identifier			&		PI		&		Obs. ID			&	Observation & Remark	\\
\hline

V1943 Sgr		        &	OBSherschel1	&	1342268730			&	PACS B2A, short R1	\\
				&	OBSherschel1	&	1342268569			&	PACS B2B, long R1	\\
				&	OBSherschel1	&	1342268314			&	SPIRE FTS	  &  \\
				&	Groenewegen	&	1342208468/1342208469	&	PACS image 70, 160	\\
AFGL 2513		        &	OBSherschel1	&	1342270010			&	PACS B2A, short R1	\\
				&	OBSherschel1	&	1342269936			&	PACS B2B, long R1	\\
IRC $-$10 529		        &	OBSherschel1	&	1342269916			&	PACS B2A, short R1	\\
				&	OBSherschel1	&	1342269510			&	PACS B2B, long R1   & mismatch with 70 $\mu$m bolometer flux	\\
				&	OBSherschel1	&	1342216902			&	SPIRE FTS		\\
				&	Groenewegen	&	1342196034/1342196035	&	PACS image 70, 160	\\	
T Mic			&	OBSherschel1	&	1342268729			&	PACS B2A, short R1	\\
				&	OBSherschel1	&	1342268788			&	PACS B2B, long R1	\\
				&	Groenewegen	&	1342193036/1342193037	&	PACS image 70, 160	\\
V Cyg			&	Groenewegen	&	1342208939			&	PACS B2A, short R1	\\
				&	Groenewegen	&	1342208940			&	PACS B2B, long R1	\\
				&	Groenewegen	&	1342188462/1342188463	&	PACS image 70, 160	\\
T Cep			&	Cami			&	1342246557			&	PACS B2A, short R1	\\
				&	Andr\'{e}		&	1342188652/1342188653	&	PACS image 70, 160	\\
				&	Andr\'{e}		&	1342188652/1342188653	&	PACS image 70, 160	\\
EP Aqr			&	OBSherschel1	&	1342270639			&	PACS B2A, short R1	\\
				&	OBSherschel1	&	1342270684			&	PACS B2B, long R1	\\
				&	Groenewegen	&	1342195460/1342195461	&	PACS image 70, 160	\\
				&	Cox			&	1342257155/1342257156	&	PACS image 100, 160\\
$\pi$ Gru			&	Groenewegen	&	1342210397			&	PACS B2A, short R1	\\
				&	Groenewegen	&	1342210398			&	PACS B2B, long R1	\\
				&	Groenewegen	&	1342196799/1342196800	&	PACS image 70, 160	\\
				&	Groenewegen	&	1342193791			&	SPIRE image		\\
LL Peg			        &	Groenewegen	&	1342199417			&	PACS B2A, short R1	\\
				&	Groenewegen	&	1342199418			&	PACS B2B, long R1	\\
				&	Groenewegen	&	1342189126			&	SPIRE FTS         & small mismatch with PACS spectrum	\\
				&	Groenewegen	&	1342188378/1342188379	&	PACS image 70, 160	\\
				&	Royer		&	1342237362/1342237363	&	PACS image 100, 160\\
				&	Groenewegen	&	1342188178			&	SPIRE image		\\
LP And			        &	Groenewegen	&	1342212512			&	PACS B2A, short R1	\\
				&	Groenewegen	&	1342212513			&	PACS B2B, long R1	\\
				&	Barlow		&	1342246288			&	SPIRE FTS		\\
				&	Groenewegen	&	1342188490/1342188491	&	PACS image 70, 160	\\			
R Cas			        &	Groenewegen	&	1342212576			&	PACS B2A, short R1	\\
				&	Groenewegen	&	1342212577			&	PACS B2B, long R1	\\
				&	Barlow		&	1342246981			&	SPIRE FTS	     &	\\
				&	Groenewegen	&	1342222422/1342222423	&	PACS image 70, 160	\\
				&	Royer		&	1342237158/1342237159	&	PACS image 100, 160 \\
				&	Groenewegen	&	1342188578			&	SPIRE image		\\
\hline							
\multicolumn{4}{l}{\textit{Red Super Giants}}											\\
\hline						
$\alpha$ Ori		        &	Groenewegen	&	1342218757			&	PACS B2A, short R1	\\
				&	Groenewegen	&	1342218756			&	PACS B2B, long R1	\\
				&	Groenewegen	&	1342193663			&	SPIRE FTS          & mismatch with 250 $\mu$m bolometer flux \\
                                &                       &                                       &                          & mismatch with PACS spectrum \\ 
				&	Groenewegen	&	1342204435/1342204436	&	PACS image 70, 160	\\
				&	Royer		&	1342242656/1342242657	&	PACS image 100, 160\\
				&	Groenewegen	&	1342192099			&	SPIRE image		\\
VY CMa			        & 	Groenewegen	&	1342186653			&	PACS B2A, B3A, full R1\\
				&	Groenewegen	&	1342186654			&	PACS B2B 		\\
				&	Groenewegen	&	1342192834			&	SPIRE FTS		\\
				&	Groenewegen	&	1342194070/1342194071	&	PACS image 70, 160	\\
NML Cyg			        &	Groenewegen	&	1342198175			&	PACS B2A, short R1	\\
				&	Groenewegen	&	1342198174			&	PACS B2B, long R1	\\
				&	Barlow		&	1342243592			&	SPIRE FTS		\\
				&	Groenewegen	&	1342195485/1342195486	&	PACS image 70, 160	\\
\hline

\end{tabular}
%\tablebib{ (1)~\citet{Kamath2014}; (2)~\citet{Kamath2015}; (3)~\citet{Gielen2009}; }
%\tablefoot{
%\tablefoottext{a}{This star was studied by \citet{Boyer_2011} who classified it (erroneously) as an extreme-AGB star.}
%}
\end{table*}

\clearpage

\section{SPIRE background subtraction} 
\label{Appen:Background}

The correction of the spectral shape of the SPIRE spectra that suffer from high background contamination was done by 
the HIPE \textit{SpectrometerBackgroundSubtraction} script.  
This method is illustrated below for OH~30.1\,$-$0.7. 

\begin{figure}
\centering

%\begin{minipage}{0.48\textwidth}
\resizebox{\hsize}{!}{\includegraphics[angle=-0]{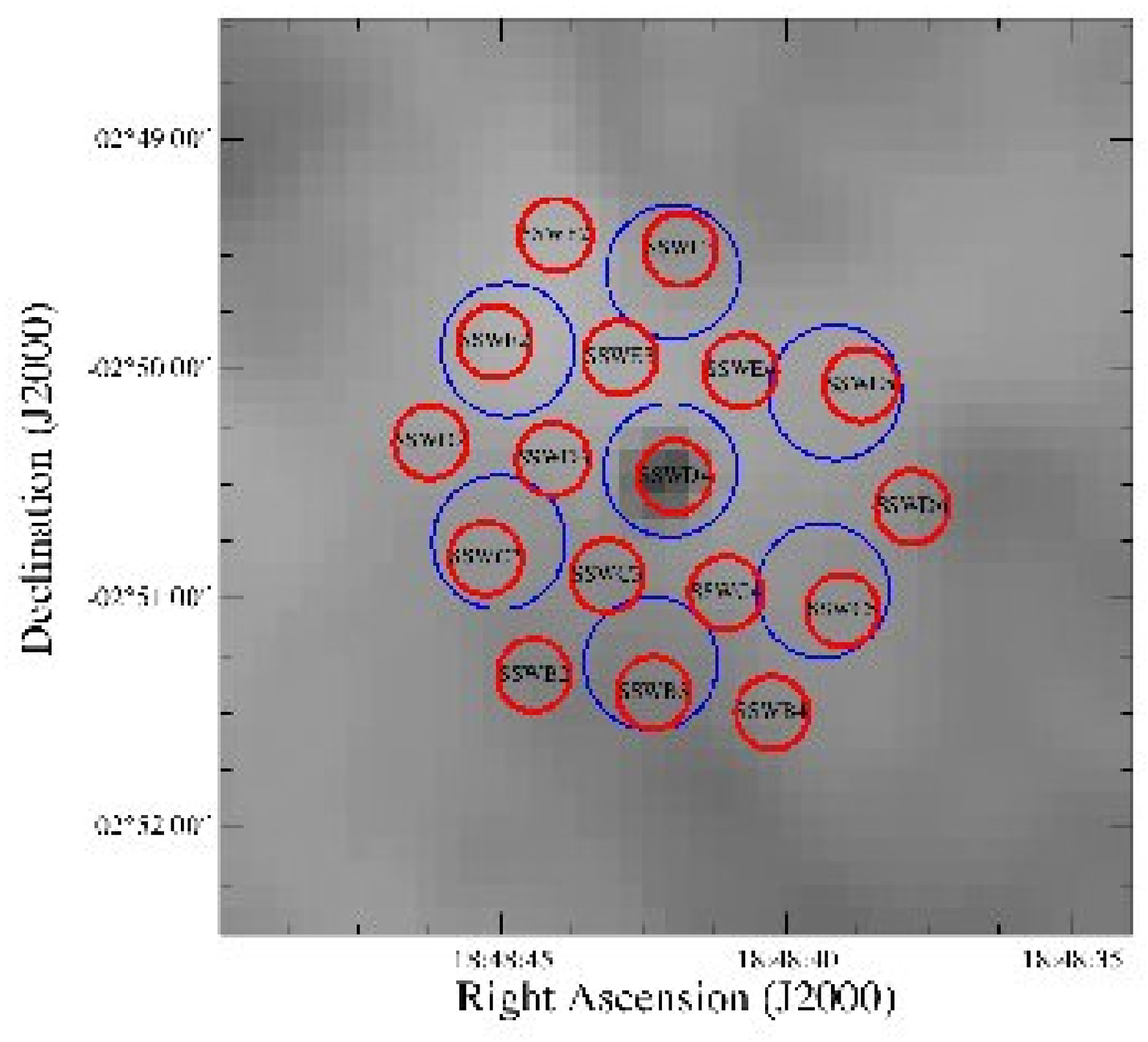}} 
%\end{minipage}

%\begin{minipage}{0.48\textwidth}
\resizebox{\hsize}{!}{\includegraphics[angle=-0]{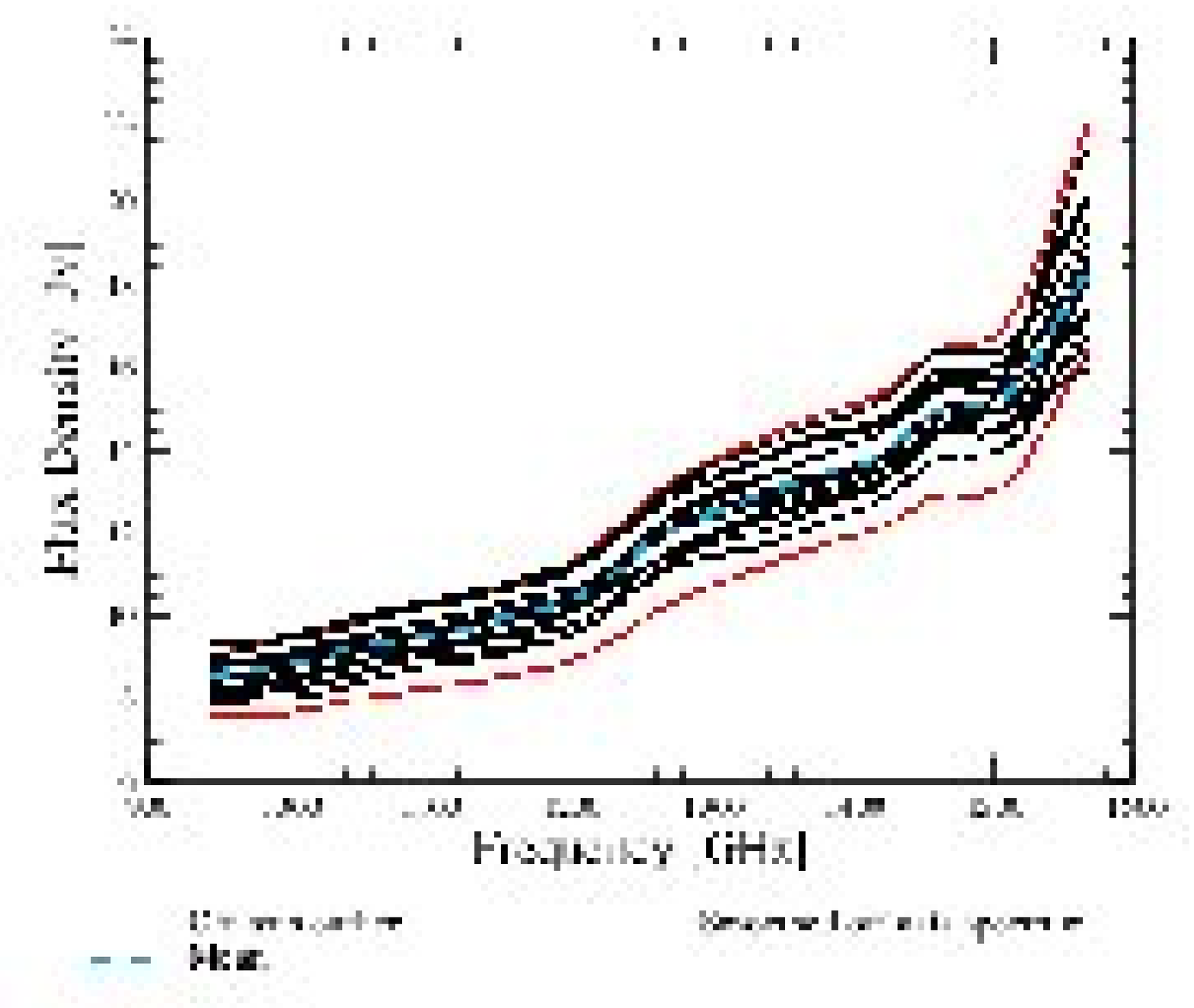}} 
%\end{minipage}

\caption{Top panel: Footprint of the SPIRE spectrometer detectors on the sky for OH~30.1\,$-$0.7. 
The blue and red circles represent the beams of the SLW and SSW arrays, respectively (the SSW arrays are identified individually). 
Two SSW detectors are not considered for background subtraction, and are not represented, which explains 
the apparent asymmetry of the SSW detector array.
Lower panel: Smoothed off-axis detector intensities and mean off-axis intensities for the SSW array. 
The black and red curves represent the smoothed spectra measured in the off-axis detectors, i.e. off-source. 
The spectra marked in red at the edge of the distribution are not used for background subtraction. 
The others (in black) are averaged, and the average (dotted blue curve) is used for background subtraction.
}
\label{Fig:Background}
\end{figure}

\begin{figure}
\centering
\resizebox{\hsize}{!}{\includegraphics{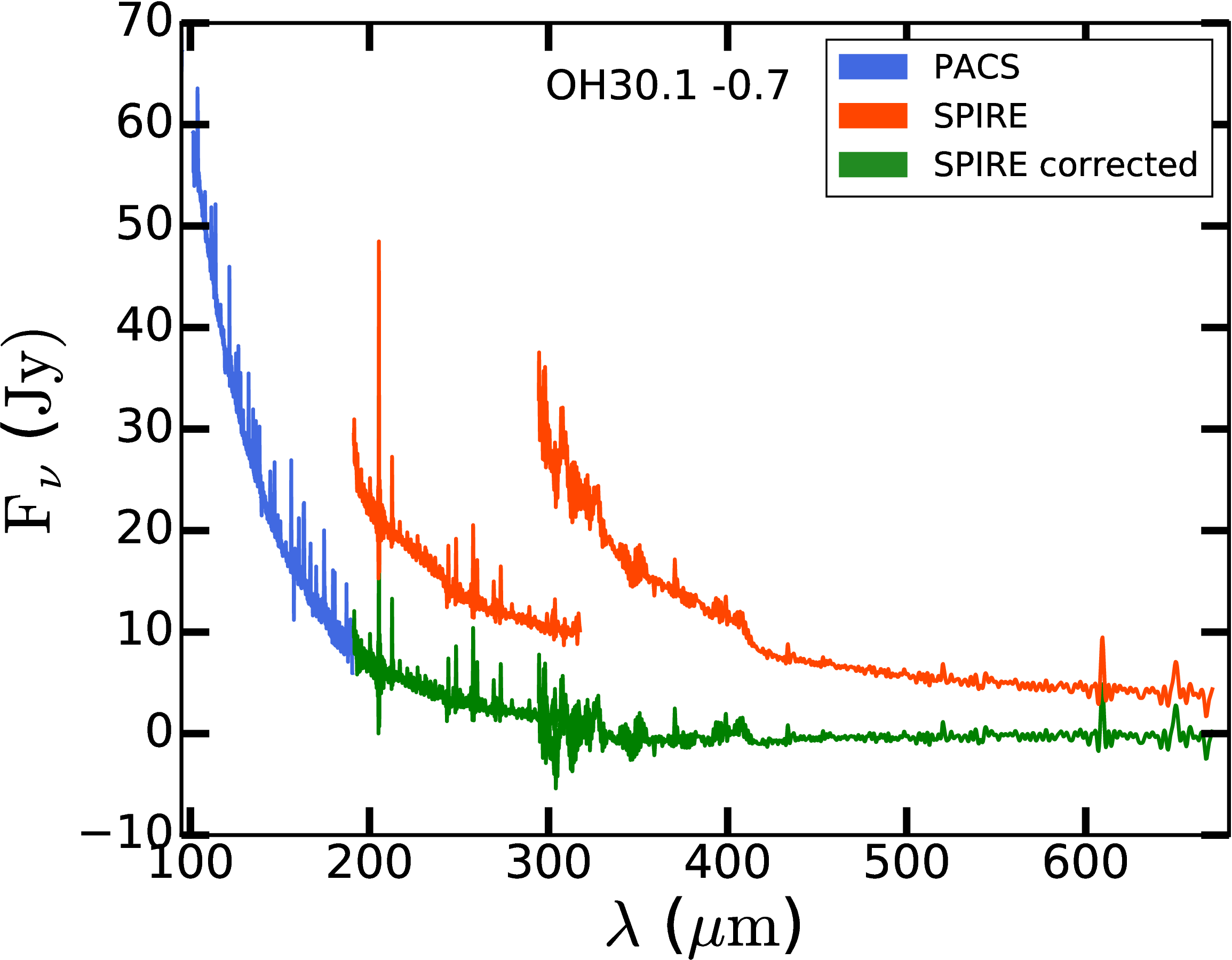}}
\caption{Effect of the background correction on the SPIRE spectrum of OH~30.1\,$-$0.7.}
\label{Fig:Background1}
\end{figure}

The top panel in Fig.~\ref{Fig:Background} shows the footprint of the SPIRE spectrometer on the SPIRE image 
of OH~30.1\,$-$0.7 at 250 \mic. 
The off-axis detectors that surround the on-source central detector were used to measure the sky background. This is done by smoothing 
the off-axis detectors with a wide kernel to extract low-frequency background information from the data. The smoothed spectra of the 
different detectors were visually inspected and the most outlying shapes were rejected, as illustrated in the lower panel in Fig.~\ref{Fig:Background}. 
The off-axis detectors that were used for OH~30.1\,$-$0.7 and the other targets are listed in Table~\ref{Table:Bolometers}. 
The large-scale shape that arises from the averaging of the remaining off-axis spectra is then subsequently subtracted from 
the on-source spectrum. 
Figure~\ref{Fig:Background1} shows that the corrected spectrum closely matches the long wavelength end of the PACS spectrum and that 
the discontinuity between the SSW and SLW bands has disappeared. 

Subtracting the bright off-axis spectrum from a bright background emission may introduce an extra 
source of uncertainty to the absolute and relative flux calibration of the SPIRE spectra and the analysis of both the corrected continuum 
and the SPIRE spectral lines must be performed with care.

\begin{table}
\caption{SPIRE bolometers used for the background subtraction of the spectra of the background contaminated sources.}
\label{Table:Bolometers}				
\centering

\begin{tabular}{lll}
\hline															
Name	&	SSW 	&	SLW \\
\hline		
OH\,127.8\,+0.0	        &	B2, B3, B4	&	B2, B3	\\		
			&	C2, C5		&	C2, C4	\\		
			&	D2, D6		&	D2, D3	\\	
			&	E2, E5		&			\\		
			&	F2, F3		&			\\
TX Cam     		&	B2, B3, B4	&	B2, B3	\\		
			&	C2, C5		&	C2, C4	\\		
			&	D2, D6		&	D2, D3	\\	
			&	E2, E5		&			\\		
			&	F2, F3		&			\\
II Lup	        	&	B2, B3, B4	&	B2, B3	\\		
			&	C2, C5		&	C2, C4	\\		
			&	D2, D6		&	D3	\\	
			&	E2, E5		&			\\		
			&	F2, F3		&			\\
AFGL 5379	        &	B2, B3, B4	&	B2, B3	\\		
			&	C2, C5		&	C2, C4	\\		
			&	D2, D6		&	D2, D3	\\	
			&	E2, E5		&			\\		
			&	F2, F3		&			\\
OH21.5\,+0.5	        &	B2, B3, B4	&	B2, B3	\\		
			&	C2, C5		&	C2, C4	\\		
			&	D2, D6		&	D2		\\	
			&	E2, E5		&			\\		
			&	F2, F3		&			\\
OH26.6\,+0.6	        &	B2, B3, B4	&	B2, B3	\\		
			&	C2, C5		&	C2, C4	\\		
			&	D2, D6		&	D2, D3	\\	
			&	E2, E5		&			\\		
			&	F2, F3		&			\\								
OH30.1\,$-$0.7	        &	B2, B3, B4	&	B2, B3	\\		
			&	C2, C5		&	C2		\\		
			&	D2, D6		&	D3		\\	
			&	E2, E5		&			\\		
			&	F5			&			\\
IRAS 18488$-$0107	&	B2, B3, B4	&	B2, B3	\\		
			&	C2, C5		&	C2, C4	\\		
			&	D2, D6		&	D2, D3	\\	
			&	E2			&			\\		
			&	F2			&			\\	
OH32.8\,$-$0.3	        &	B3, B4		&	B2, B3	\\		
			&	C5			&	C2		\\		
			&	D2, D6		&	D2		\\	
			&	E2, E5		&			\\		
			&	F2, F3		&			\\	
IRAS 19067+0811	        &	B2, B3, B4	&	B2, B3	\\		
			&	C2, C5		&	C2, C4	\\		
			&	D2, D6		&	D2, D3	\\	
			&	E2, E5		&			\\		
			&	F2, F3		&			\\ 
\hline			
\end{tabular}				
\end{table}

\clearpage

\section{Population-temperature diagrams}
\label{Appen:RotDiag}

This Appendix shows the rotation diagrams for all sources, following the methodology outlined in Sect.~\ref{S-RD}.
Rotational temperatures are listed in Table~\ref{Table:pop-tempdiagram}.
Red and yellow points are excluded from the fitting (see Sect.~\ref{S-RD}).

\begin{figure*}
 \label{Fig:RotDiagramsCO}

 \begin{subfigure}{0.49\textwidth}
 \centering
 \includegraphics[width = \textwidth]{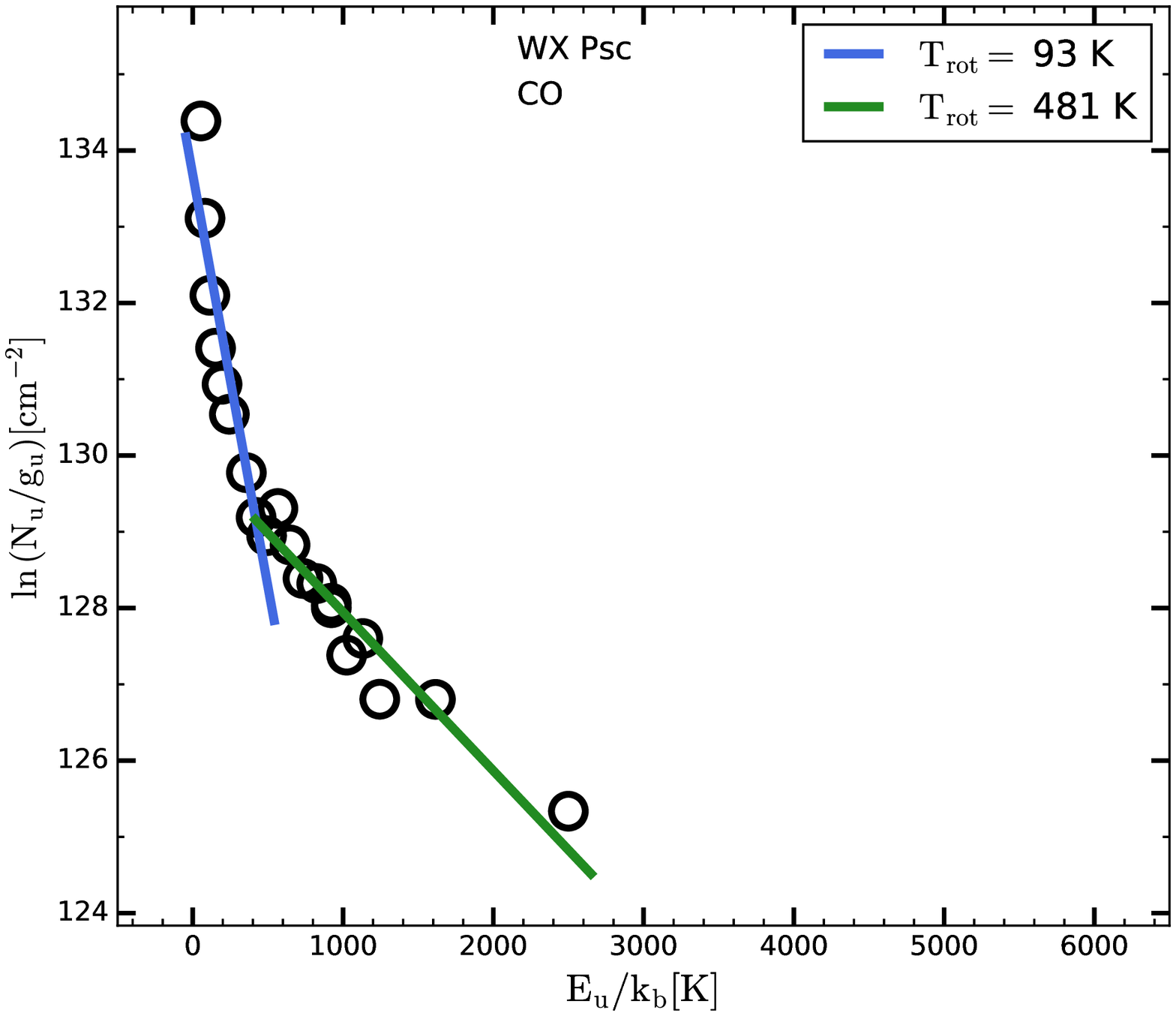}
 \end{subfigure}
\hfill
 \begin{subfigure}{0.49\textwidth}
 \centering
 \includegraphics[width = \textwidth]{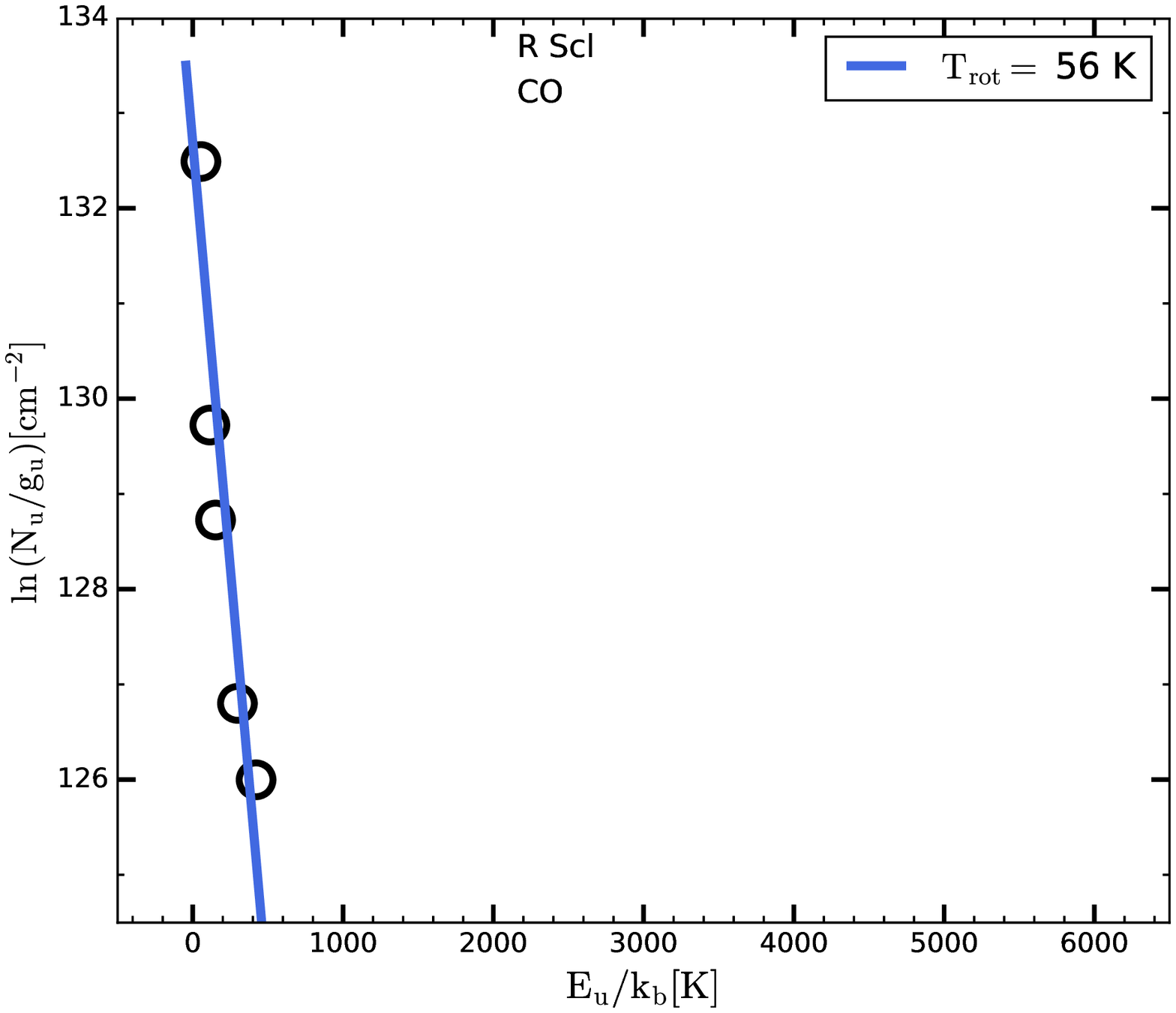}
 \end{subfigure}
  \begin{subfigure}{0.49\textwidth}
 \centering
 \includegraphics[width = \textwidth]{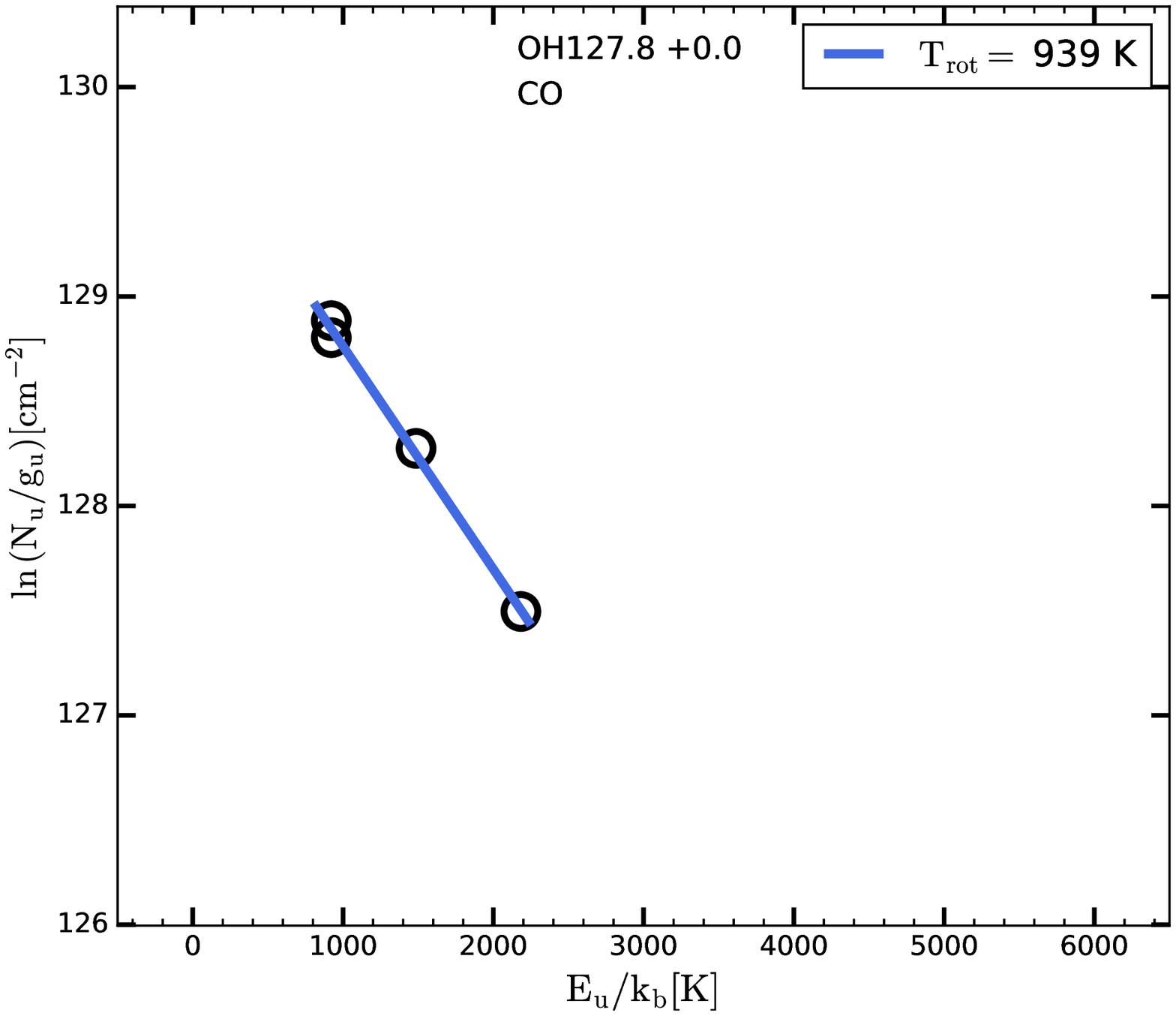}
 \end{subfigure}
\hfill
 \begin{subfigure}{0.49\textwidth}
 \centering
 \includegraphics[width = \textwidth]{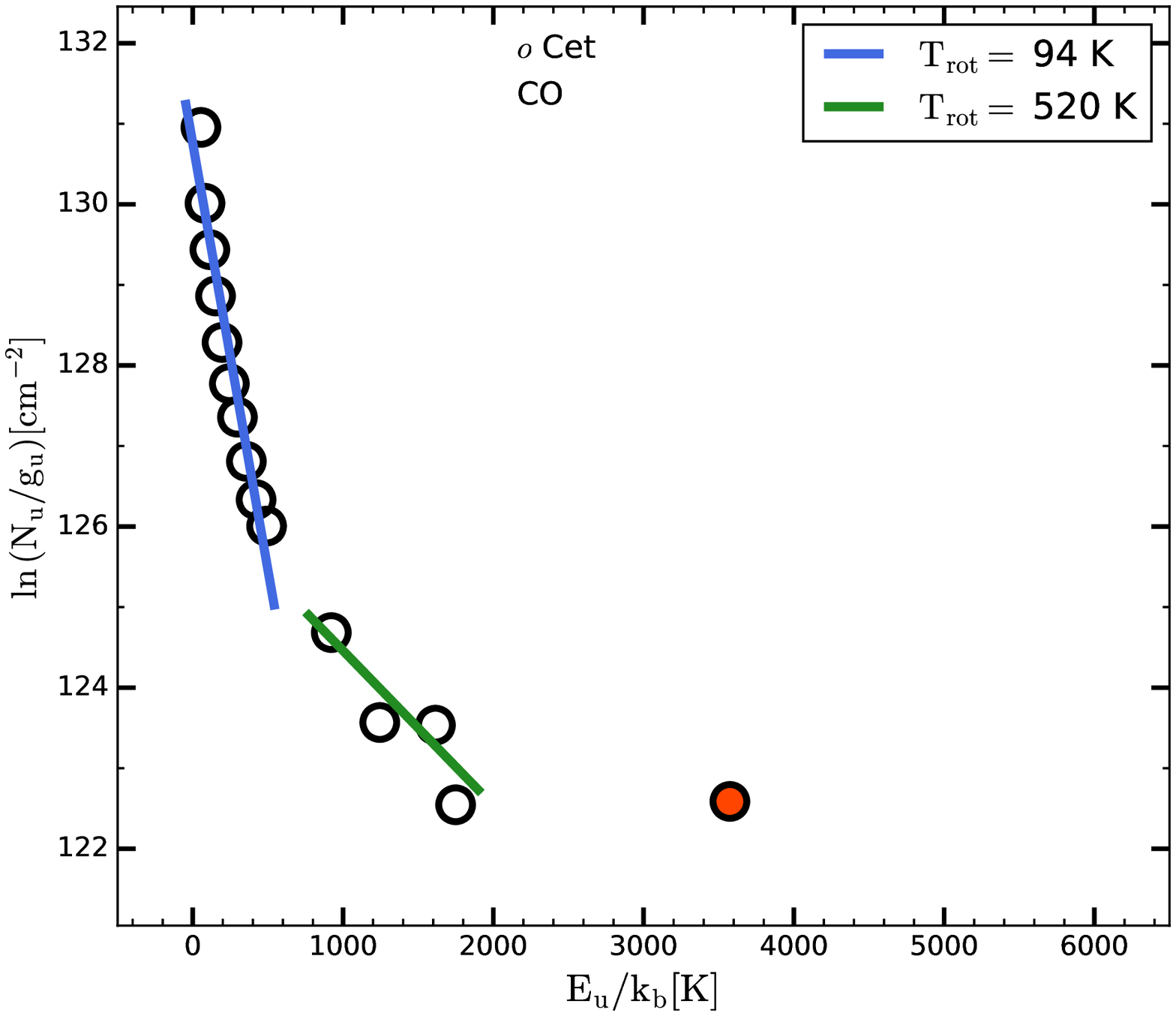} 
 \end{subfigure}
 \begin{subfigure}{0.49\textwidth}
 \centering
 \includegraphics[width = \textwidth]{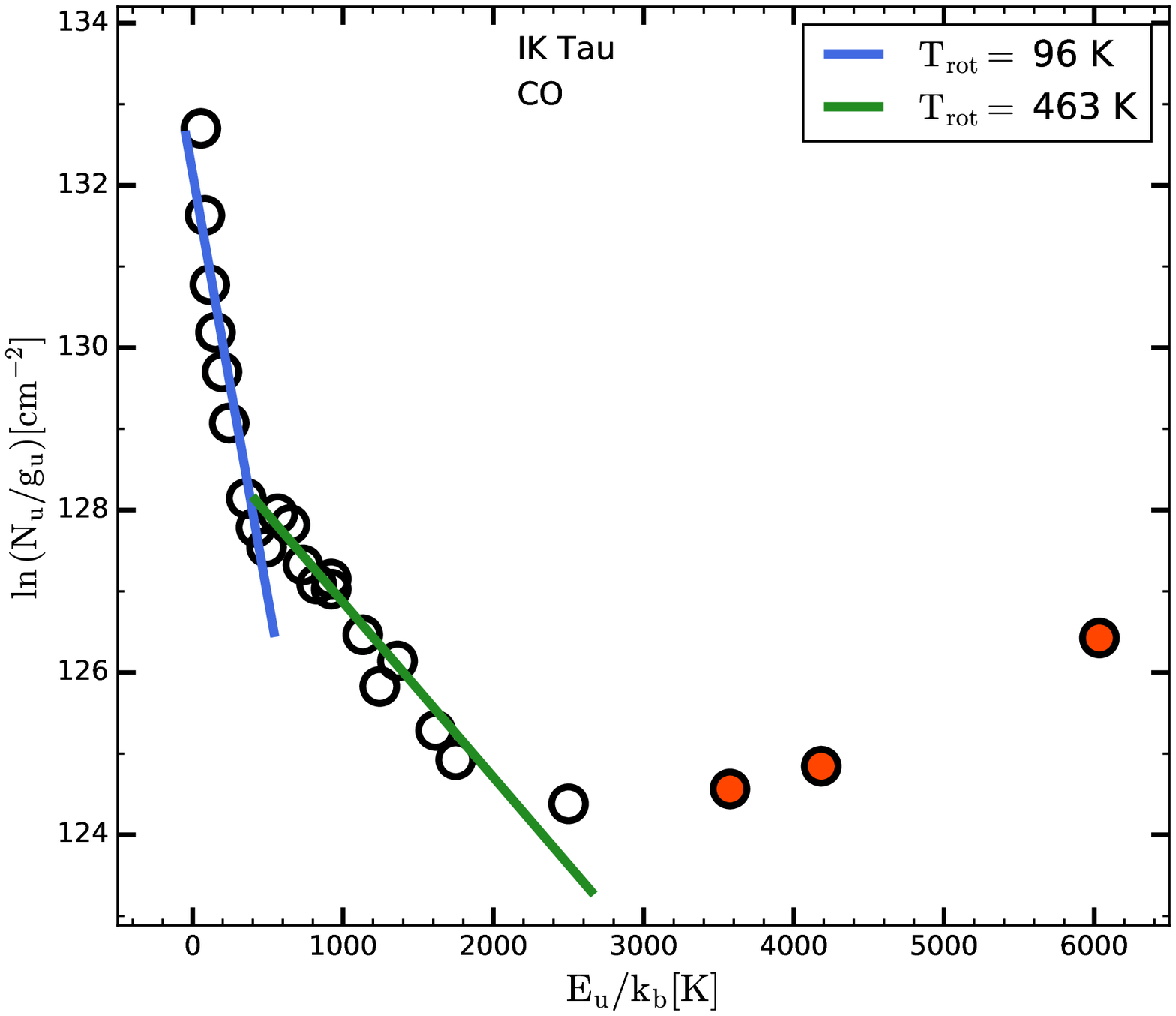}    %IK Tau
 \end{subfigure}
\hfill
 \begin{subfigure}{0.49\textwidth}
 \centering
 \includegraphics[width = \textwidth]{Rotdiagram_RDOR_COv=0.eps}
 \end{subfigure}
  \caption{CO population-temperature diagrams. Yellow and red points are excluded from the fit, see Sect.~\ref{S-RD}}.
\end{figure*}

\begin{figure*}
\ContinuedFloat
  \begin{subfigure}{0.49\textwidth}
 \centering
 \includegraphics[width = \textwidth]{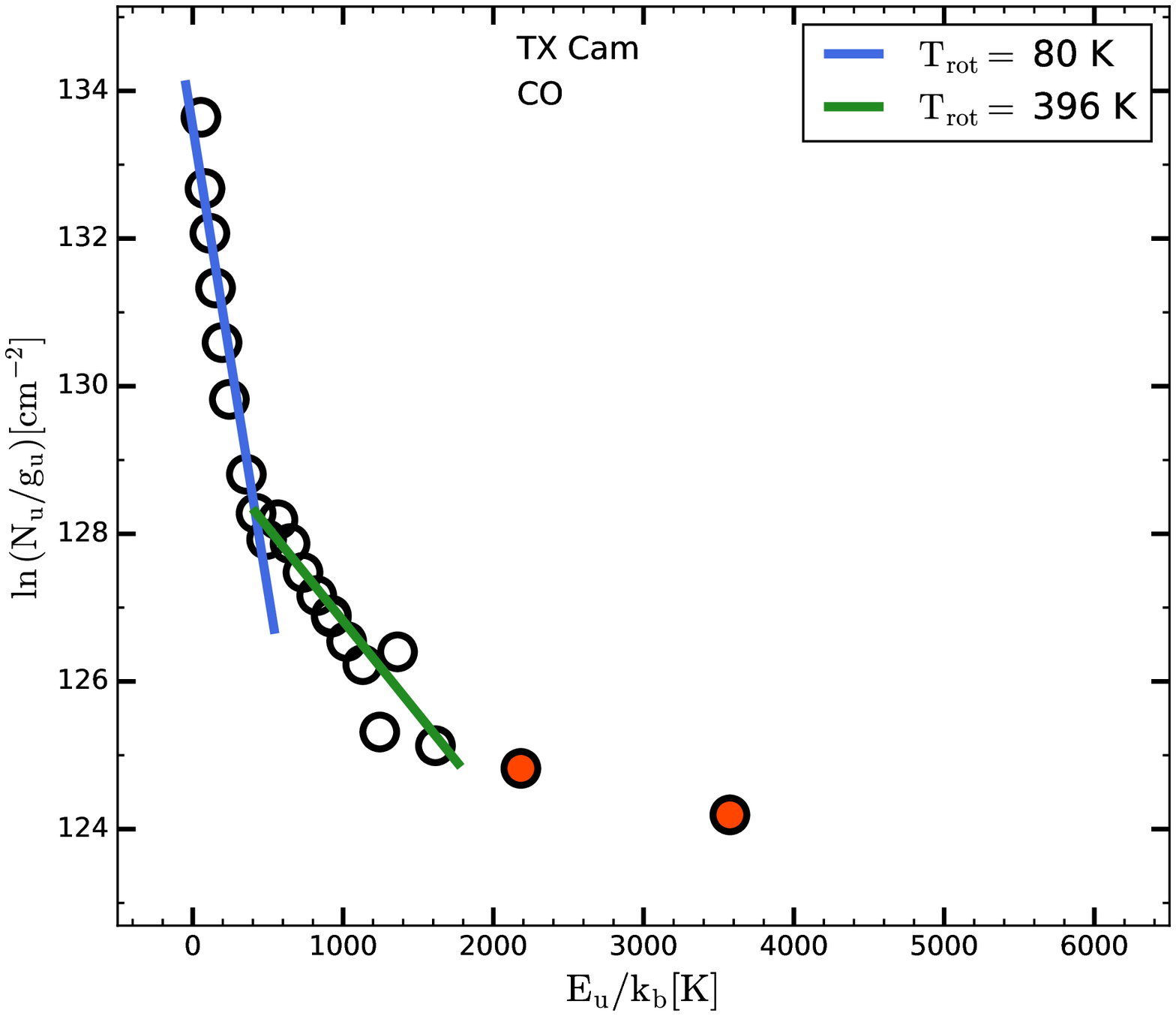} 
 \end{subfigure}
\hfill
 \begin{subfigure}{0.49\textwidth}
 \centering
 \includegraphics[width = \textwidth]{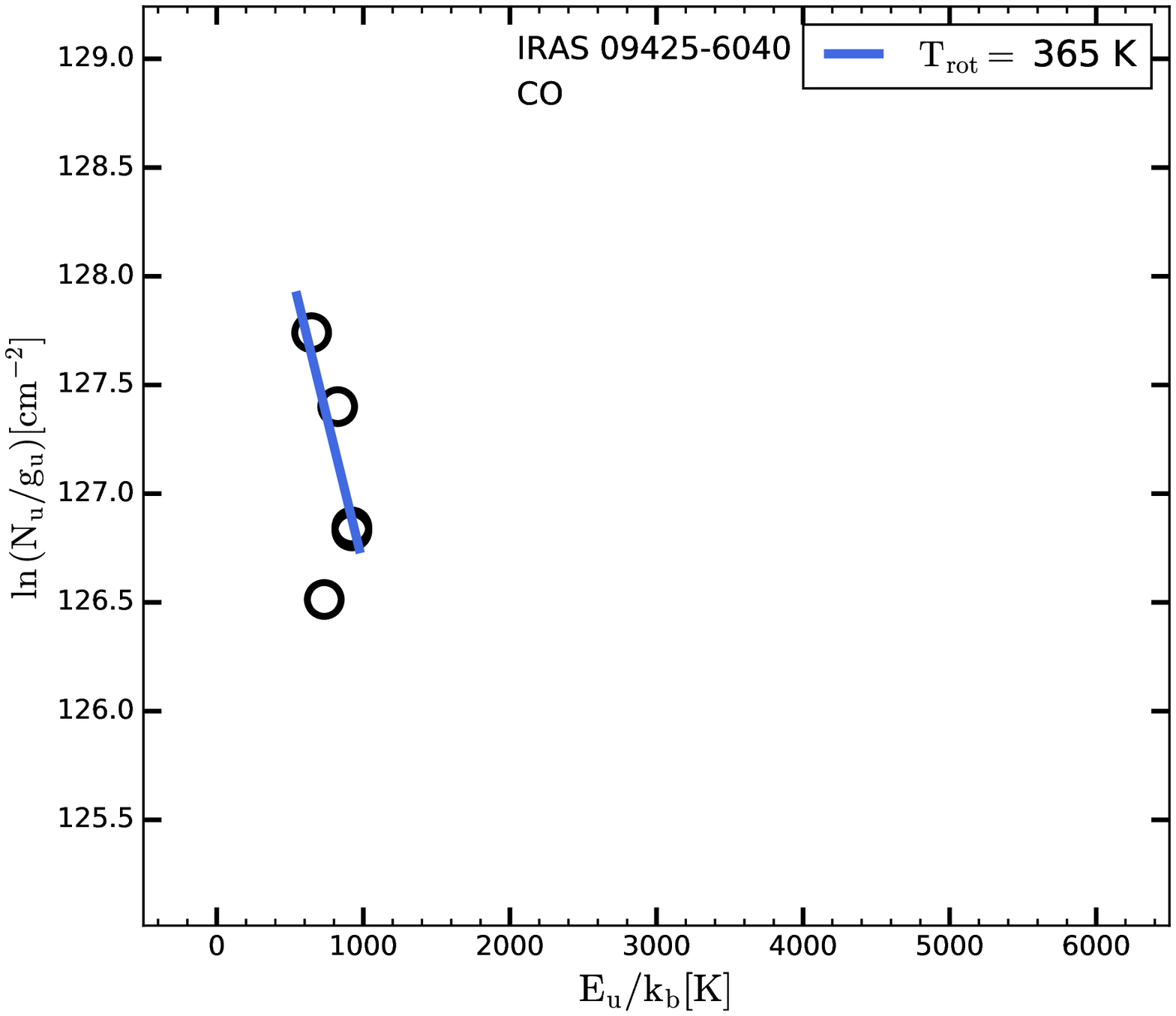} 
 \end{subfigure}
  \begin{subfigure}{0.49\textwidth}
 \centering
 \includegraphics[width = \textwidth]{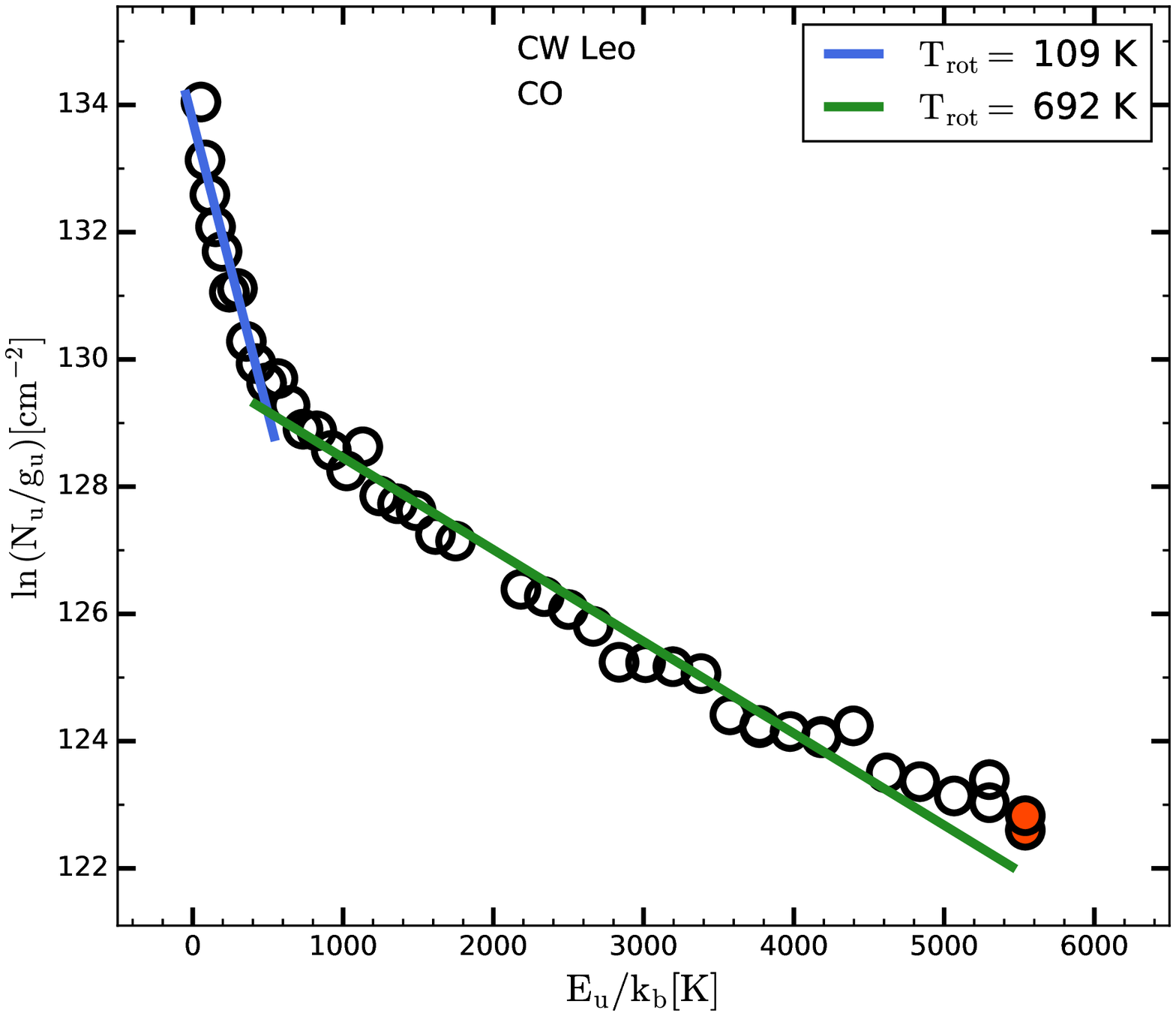}
 \end{subfigure}
\hfill
  \begin{subfigure}{0.49\textwidth}
 \centering
 \includegraphics[width = \textwidth]{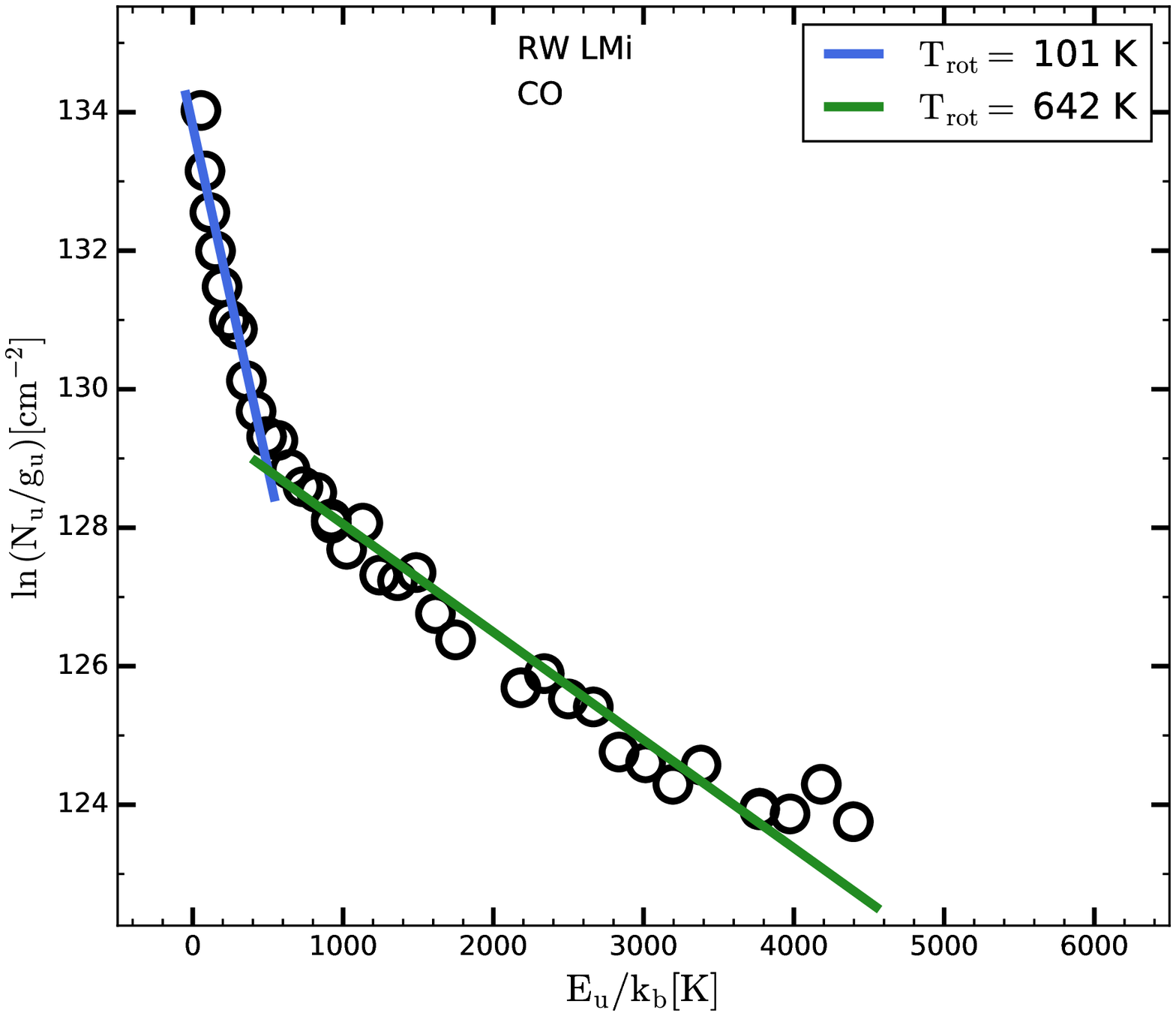}          % RW LMi
 \end{subfigure}
   \begin{subfigure}{0.49\textwidth}
 \centering
 \includegraphics[width = \textwidth]{Rotdiagram_VHYA_COv=0.eps}
 \end{subfigure}
\hfill
 \begin{subfigure}{0.49\textwidth}
 \centering
 \includegraphics[width = \textwidth]{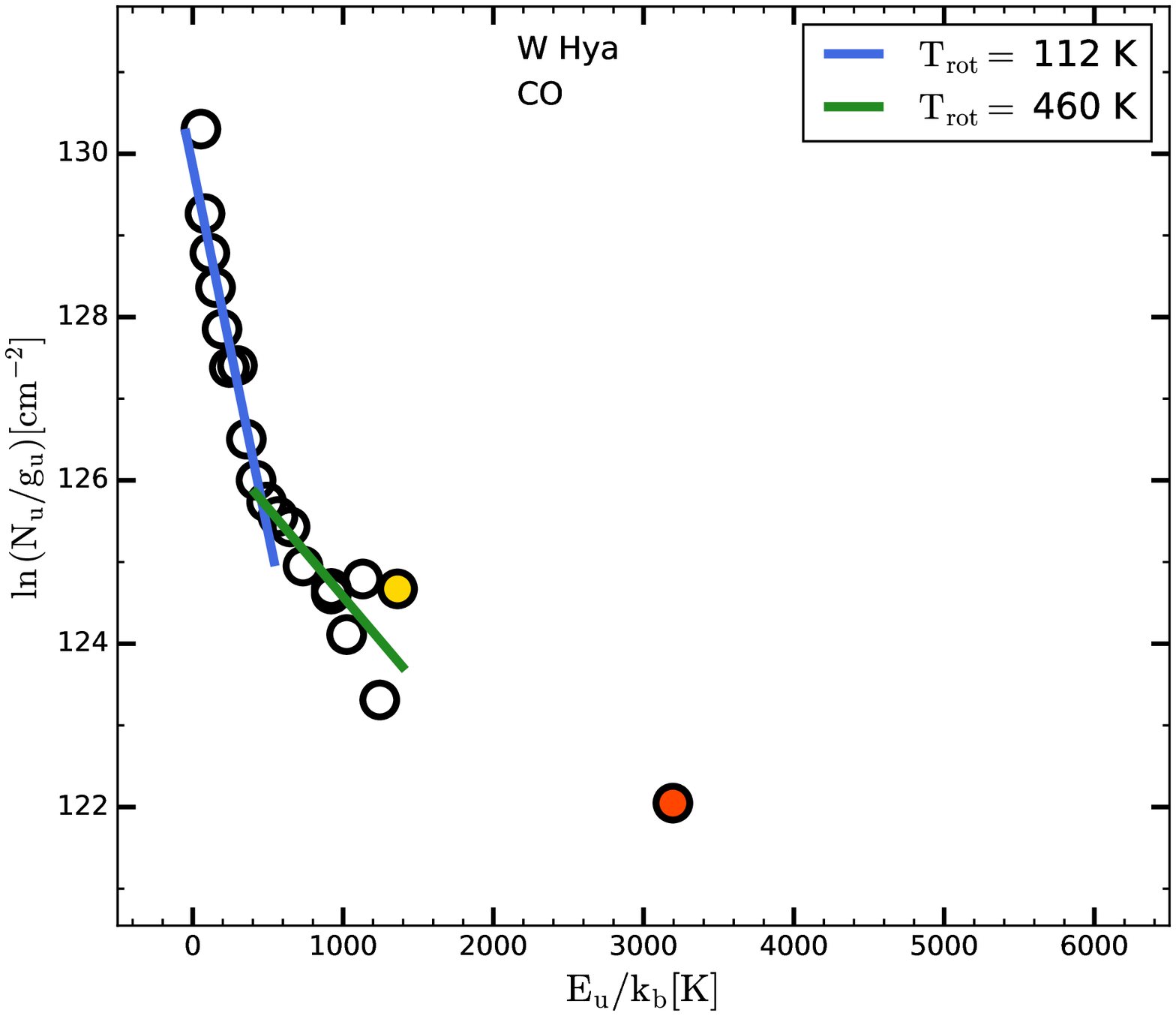} 
 \end{subfigure} 
  \caption{Continued.}
\end{figure*}

\begin{figure*}
\ContinuedFloat
  \begin{subfigure}{0.49\textwidth}
 \centering
 \includegraphics[width = \textwidth]{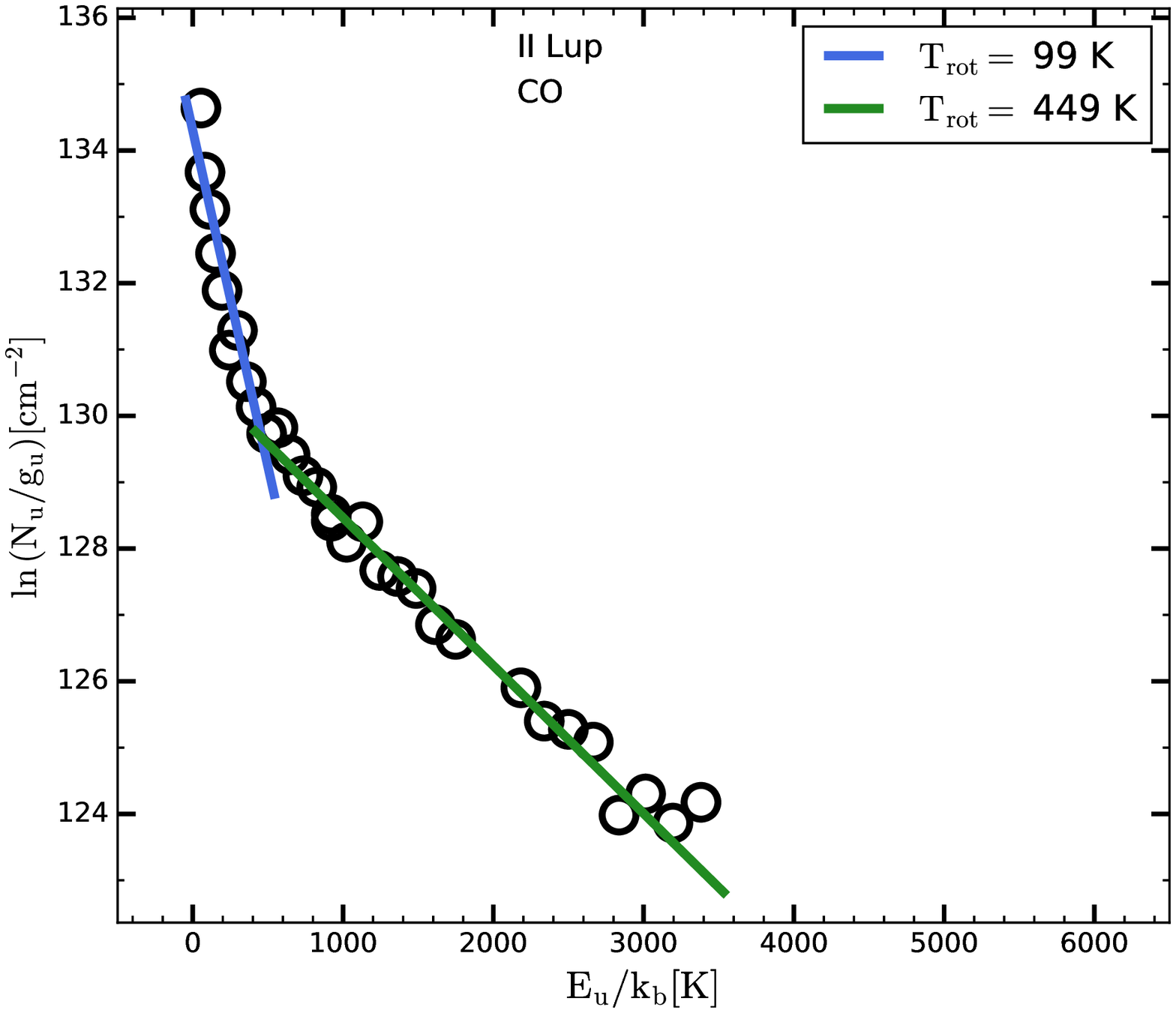} % II Lup   
 \end{subfigure}
\hfill
 \begin{subfigure}{0.49\textwidth}
 \centering
 \includegraphics[width = \textwidth]{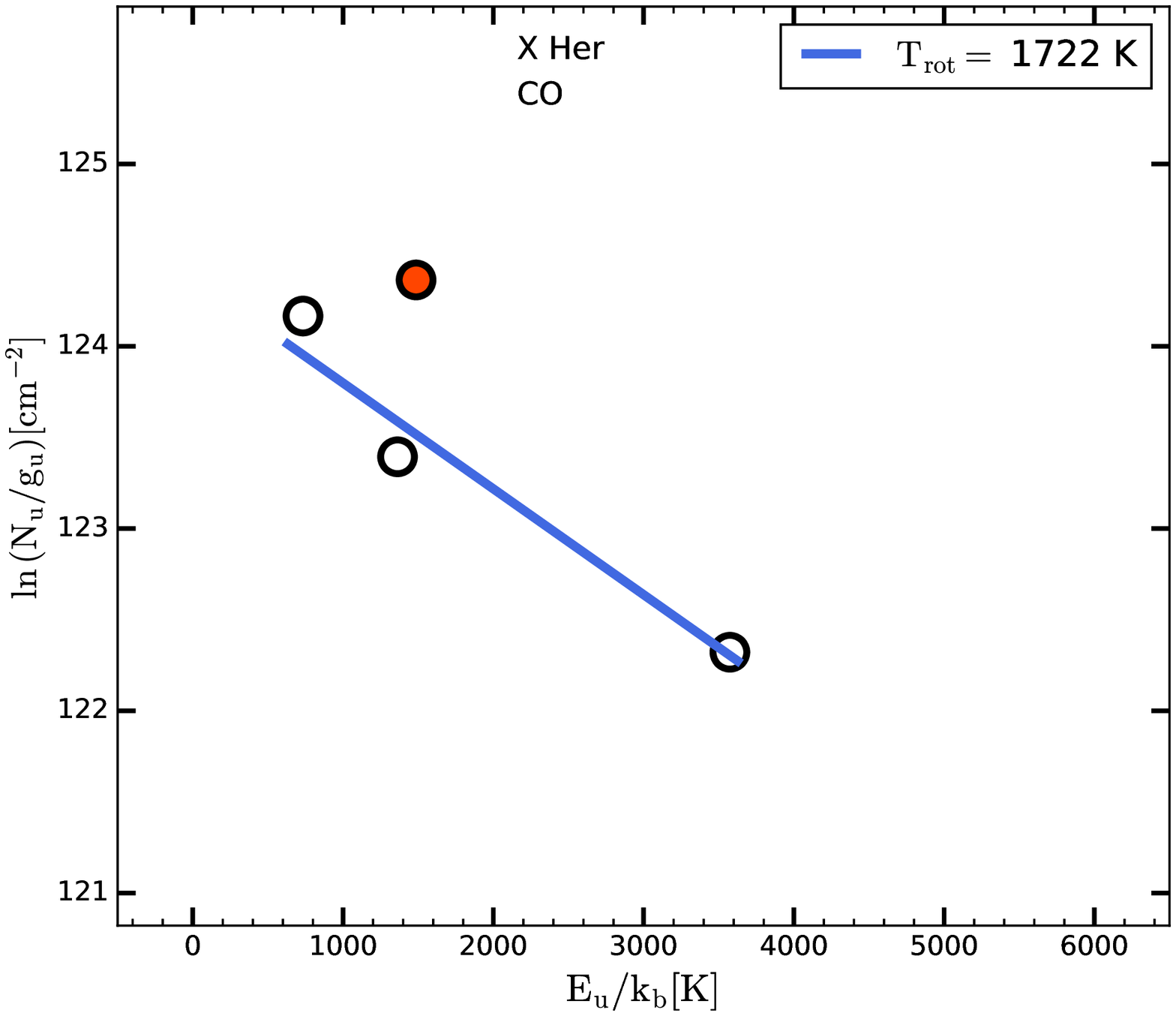}
 \end{subfigure}
 \begin{subfigure}{0.49\textwidth}
 \centering
 \includegraphics[width = \textwidth]{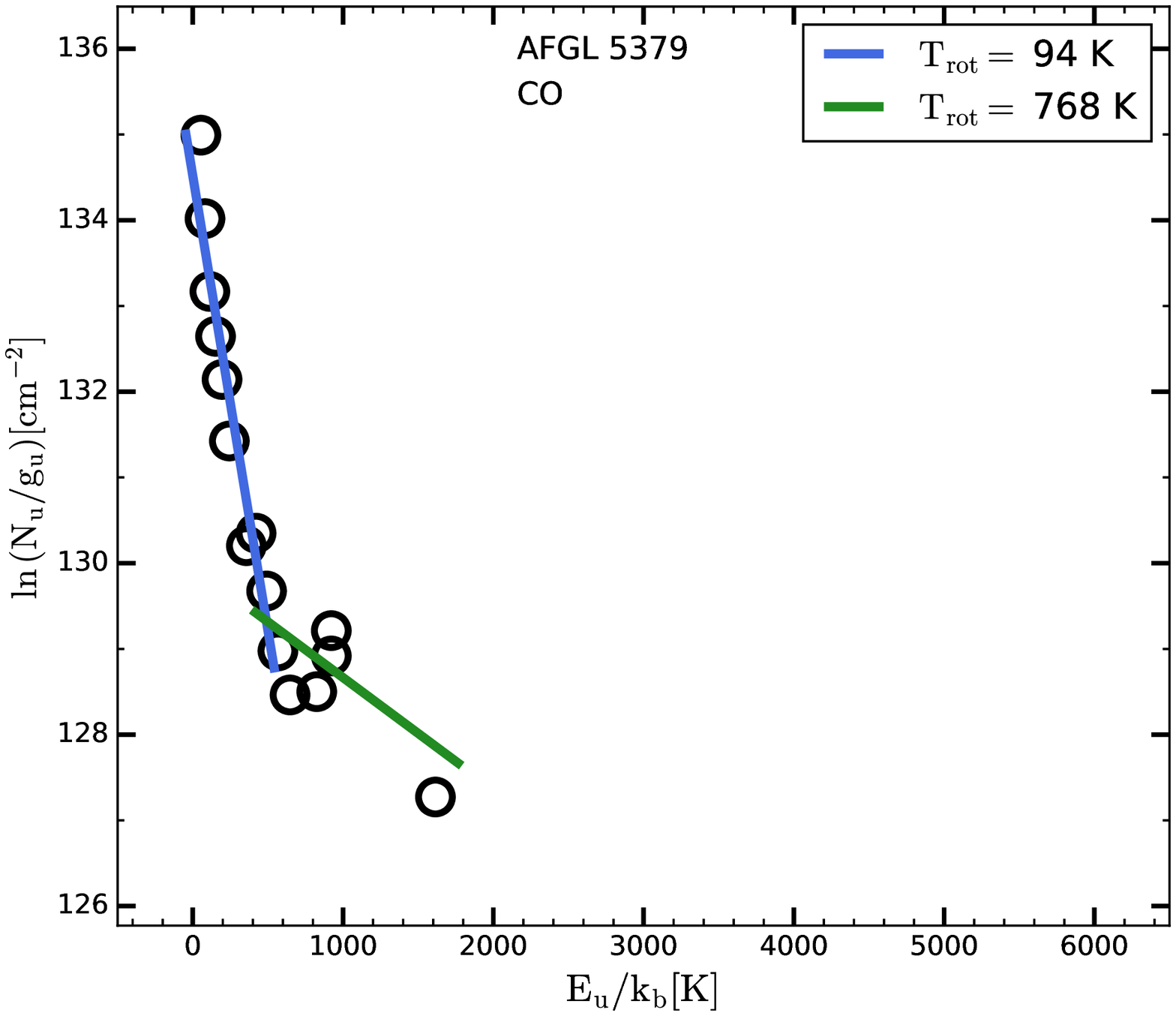}
 \end{subfigure}
\hfill
 \begin{subfigure}{0.49\textwidth}
 \centering
 \includegraphics[width = \textwidth]{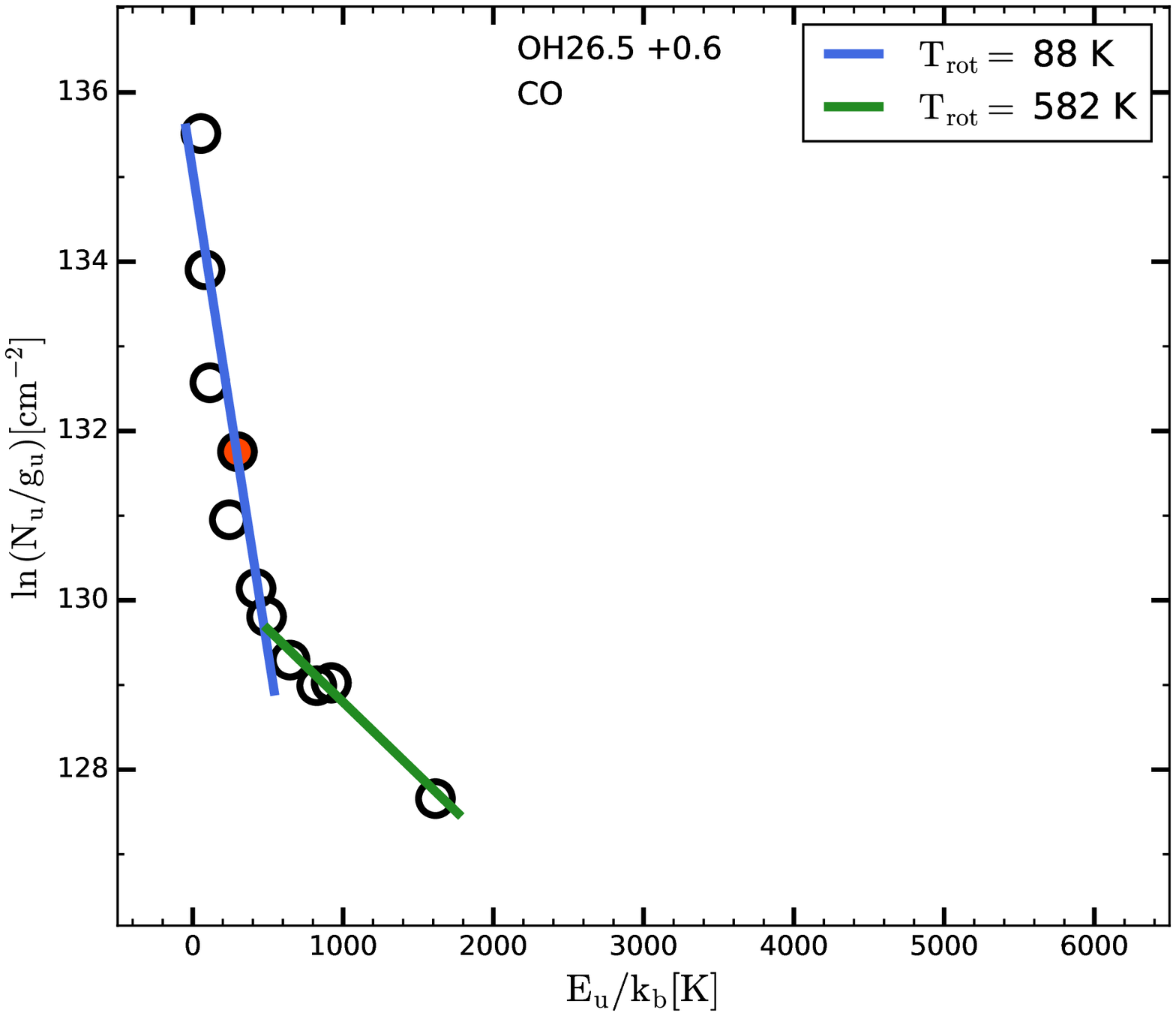} 
 \end{subfigure}
    \begin{subfigure}{0.49\textwidth}
 \centering
 \includegraphics[width = \textwidth]{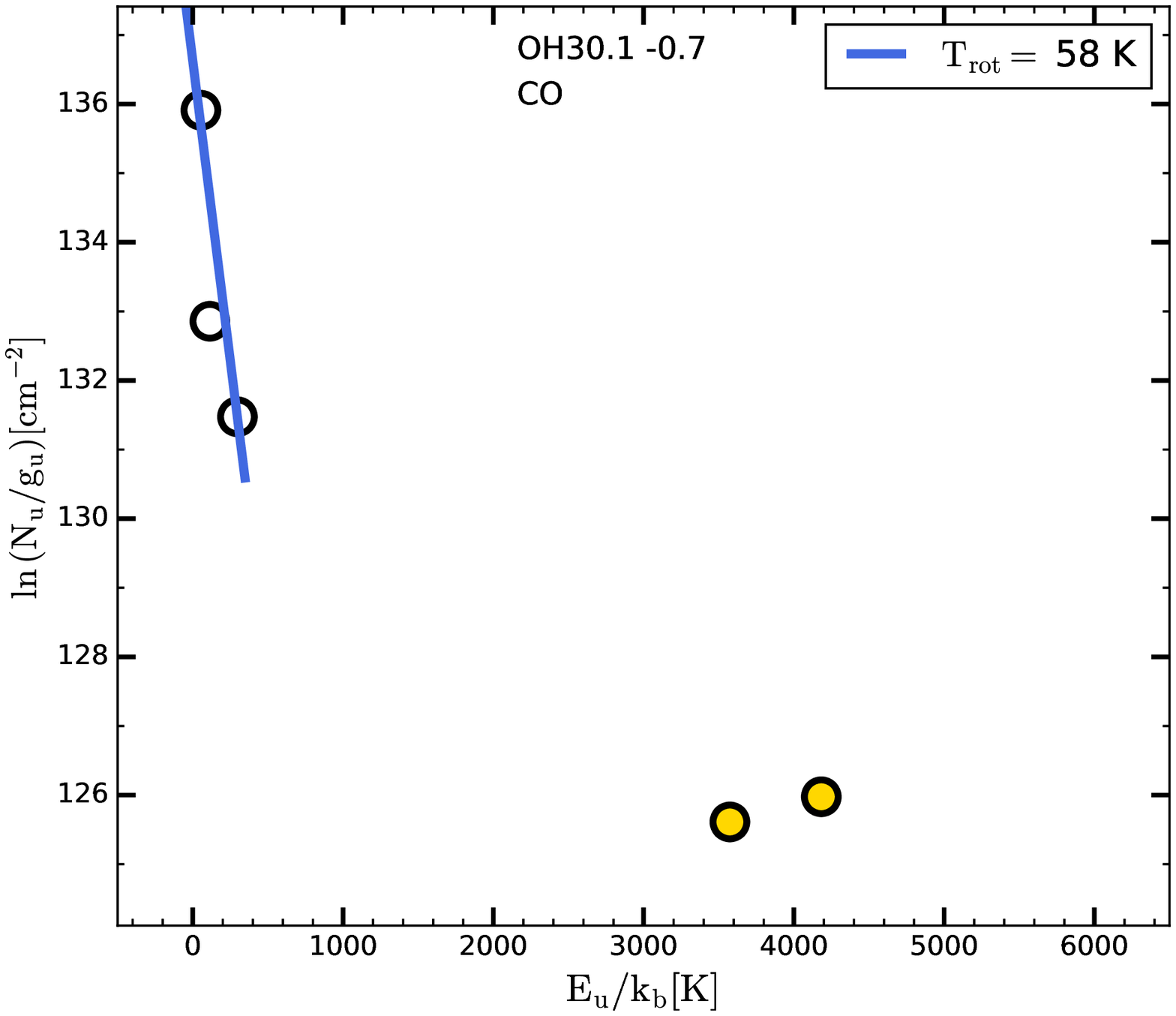}
 \end{subfigure}
 \hfill 
 \begin{subfigure}{0.49\textwidth}
 \centering
 \includegraphics[width = \textwidth]{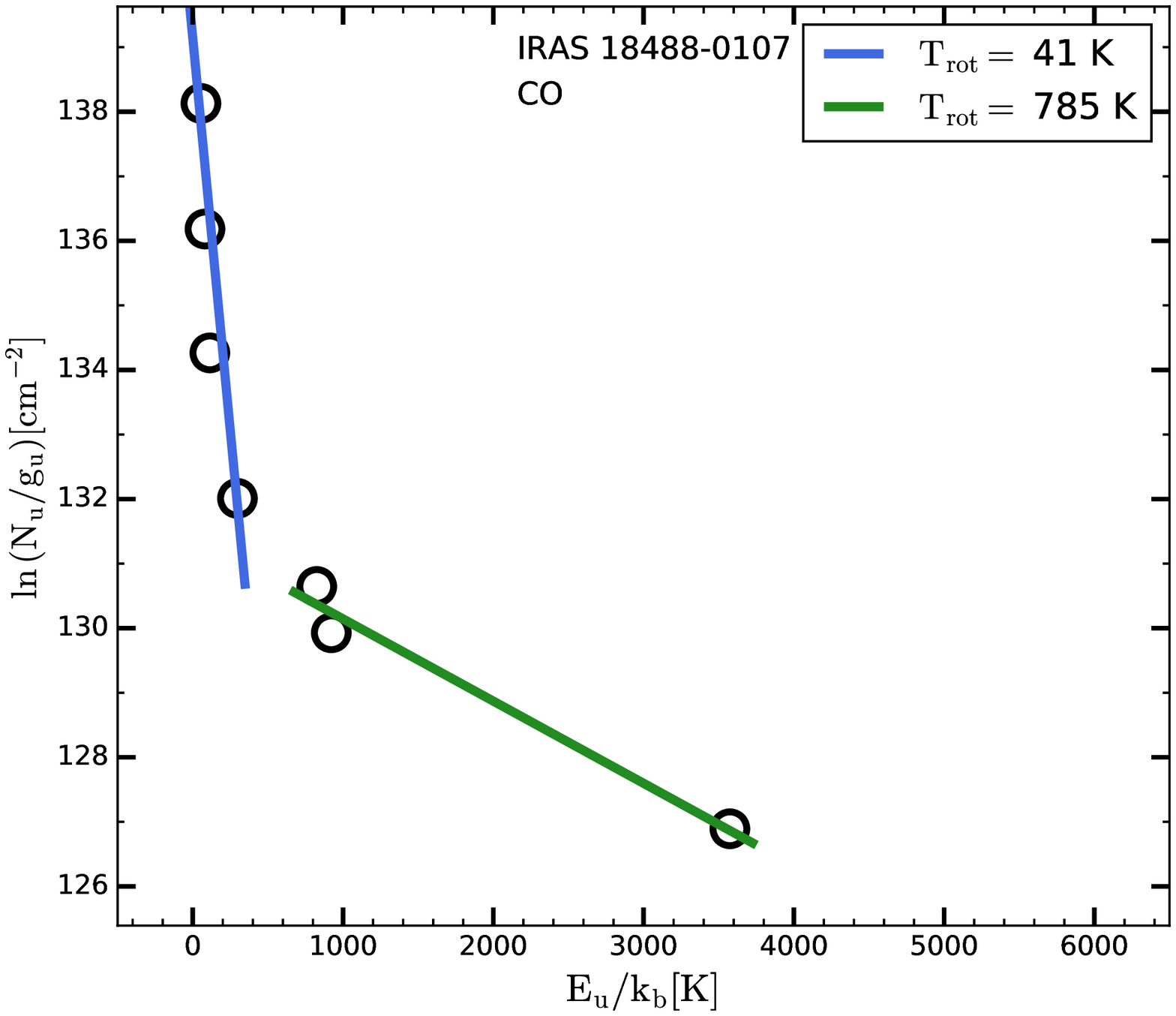} 
 \end{subfigure}
  \caption{Continued.}
\end{figure*}

\begin{figure*}
\ContinuedFloat  
  \begin{subfigure}{0.49\textwidth}
 \centering
 \includegraphics[width = \textwidth]{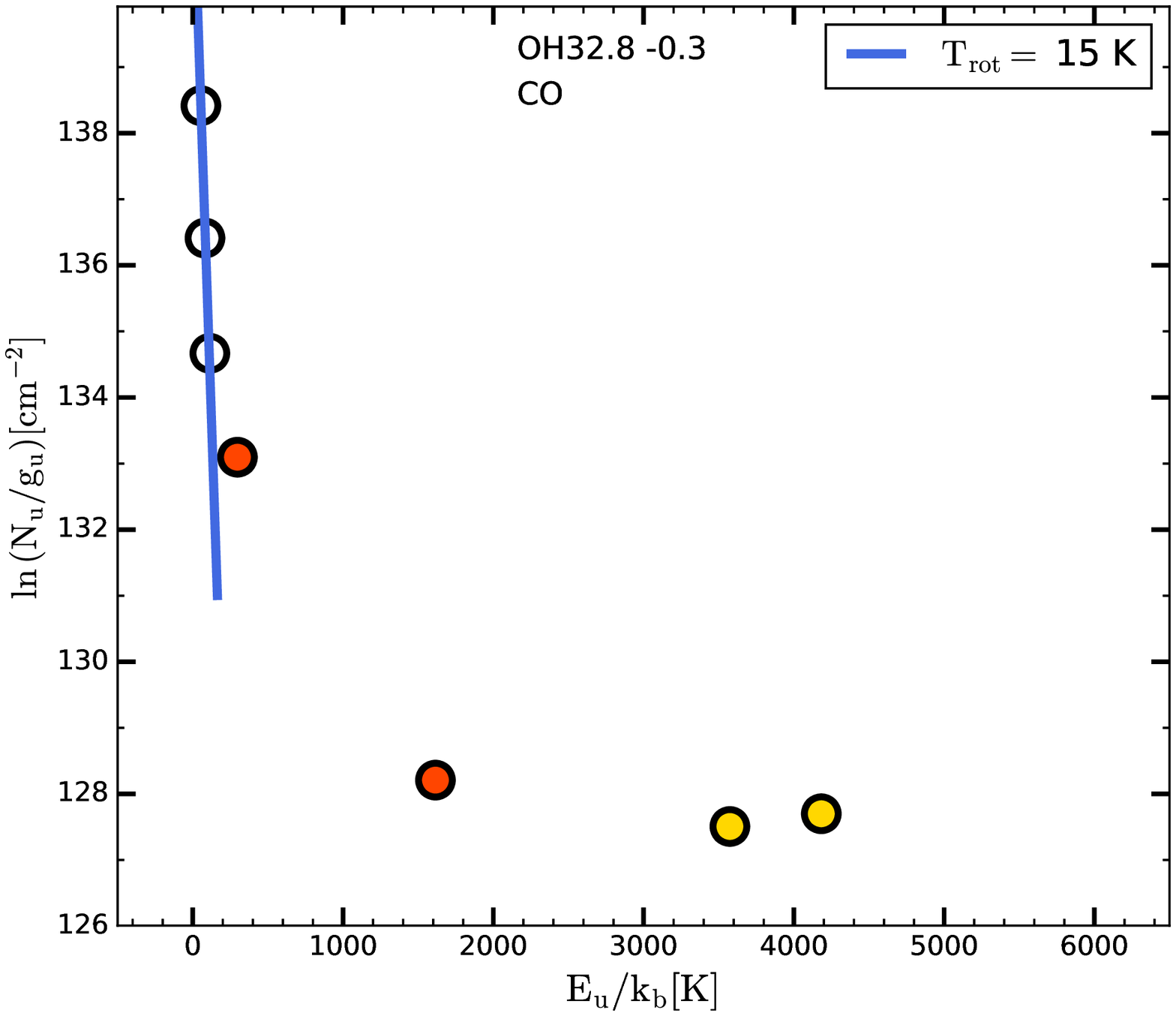} 
 \end{subfigure}
 \hfill  
 \begin{subfigure}{0.49\textwidth}
 \centering
 \includegraphics[width = \textwidth]{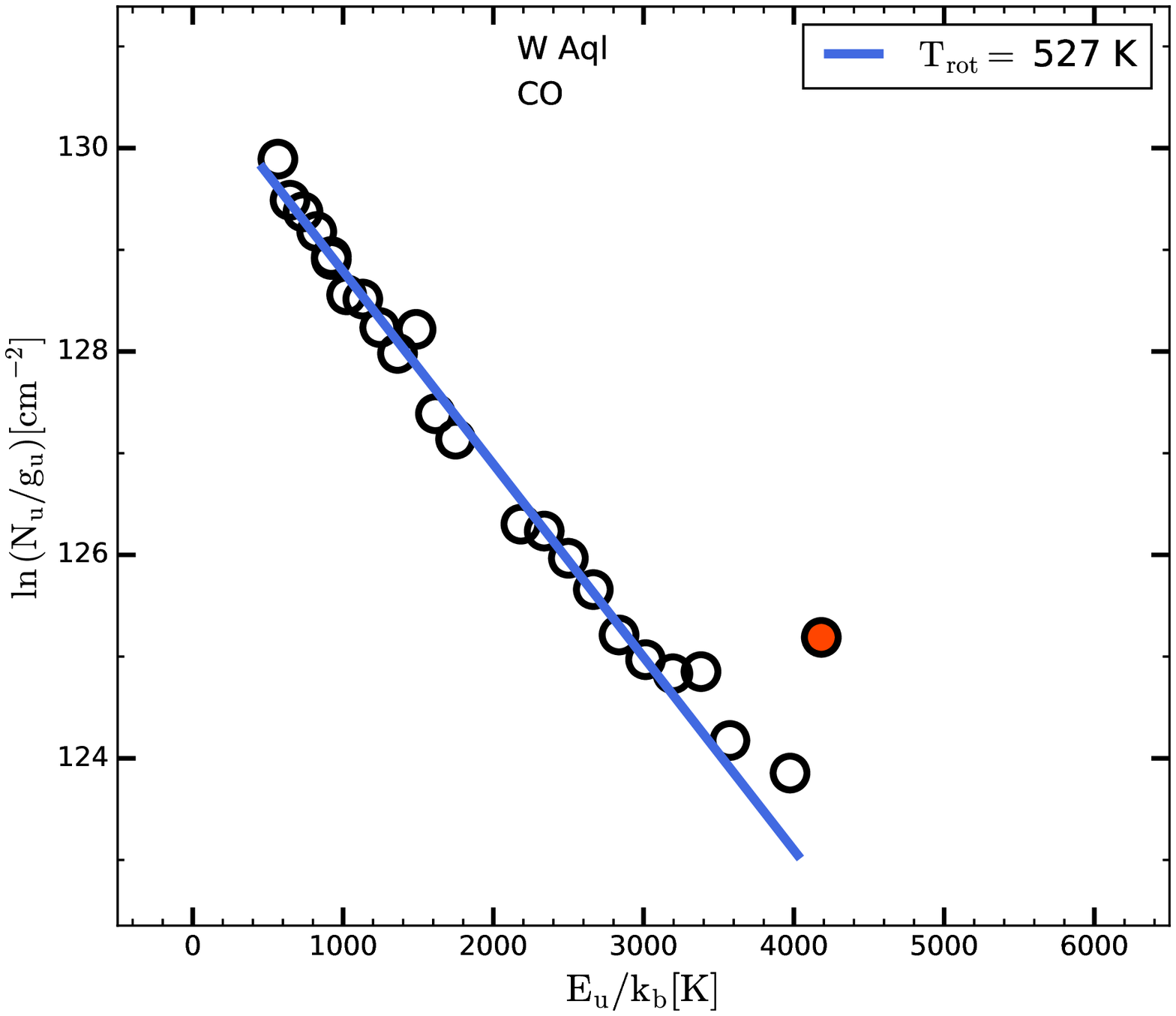} 
 \end{subfigure}
  \begin{subfigure}{0.49\textwidth}
 \centering
 \includegraphics[width = \textwidth]{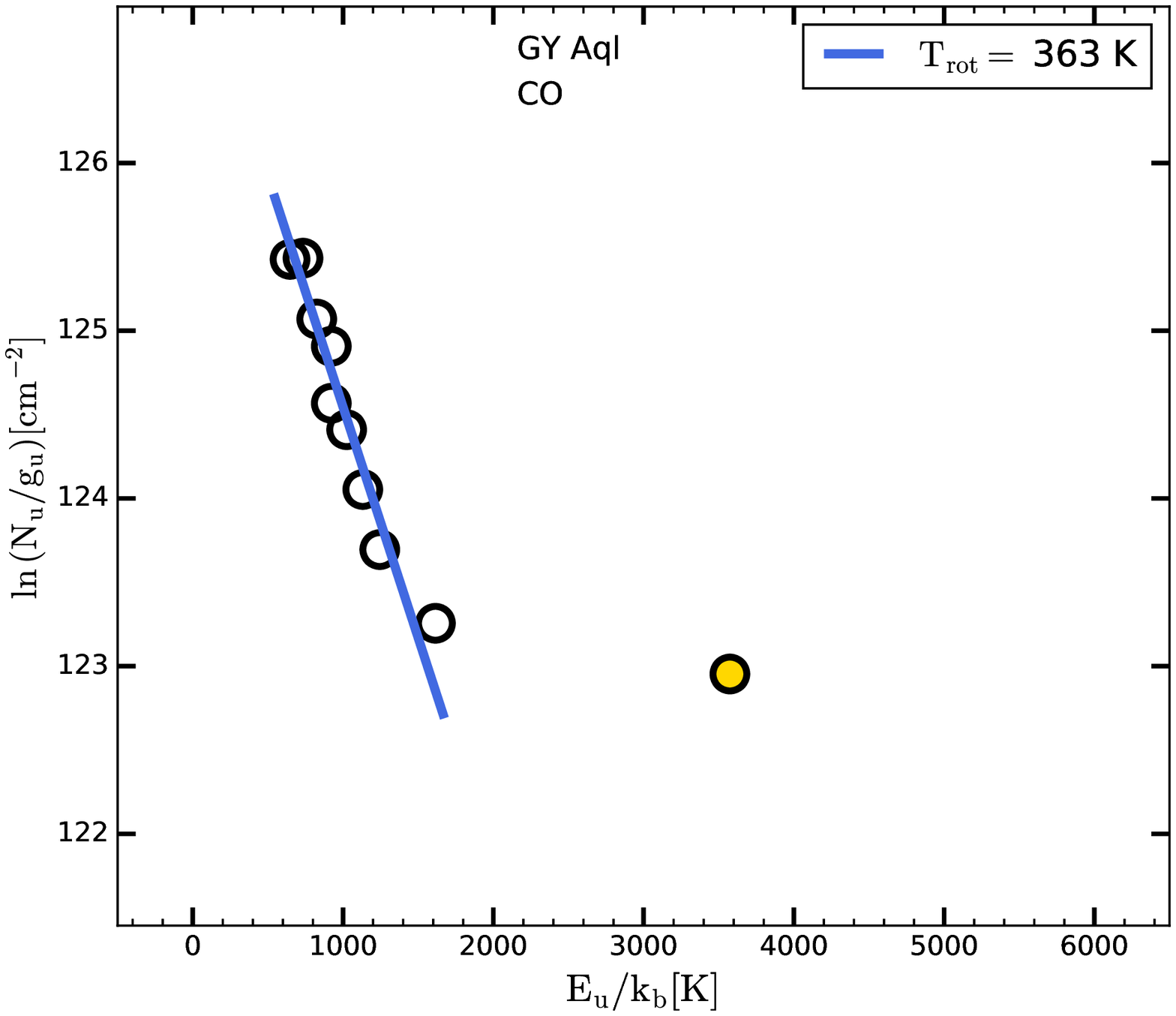} %GY Aql       
 \end{subfigure}
 \hfill   
  \begin{subfigure}{0.49\textwidth}
 \centering
 \includegraphics[width = \textwidth]{Rotdiagram_KHICYG_COv=0.eps}  
 \end{subfigure}
 \begin{subfigure}{0.49\textwidth}
 \centering
 \includegraphics[width = \textwidth]{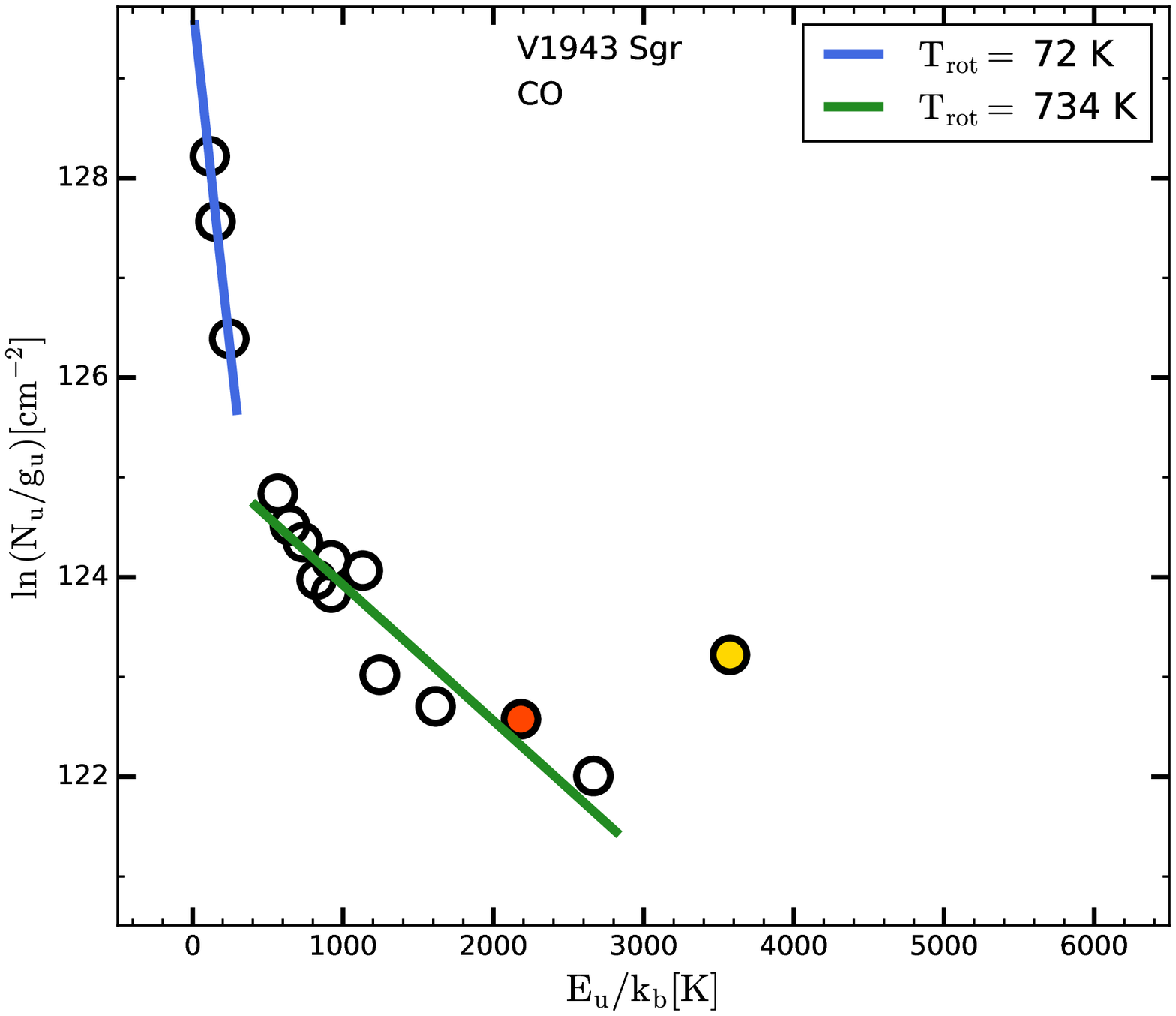} %V1943  Sgr
 \end{subfigure}
\hfill
 \begin{subfigure}{0.49\textwidth}
 \centering
 \includegraphics[width = \textwidth]{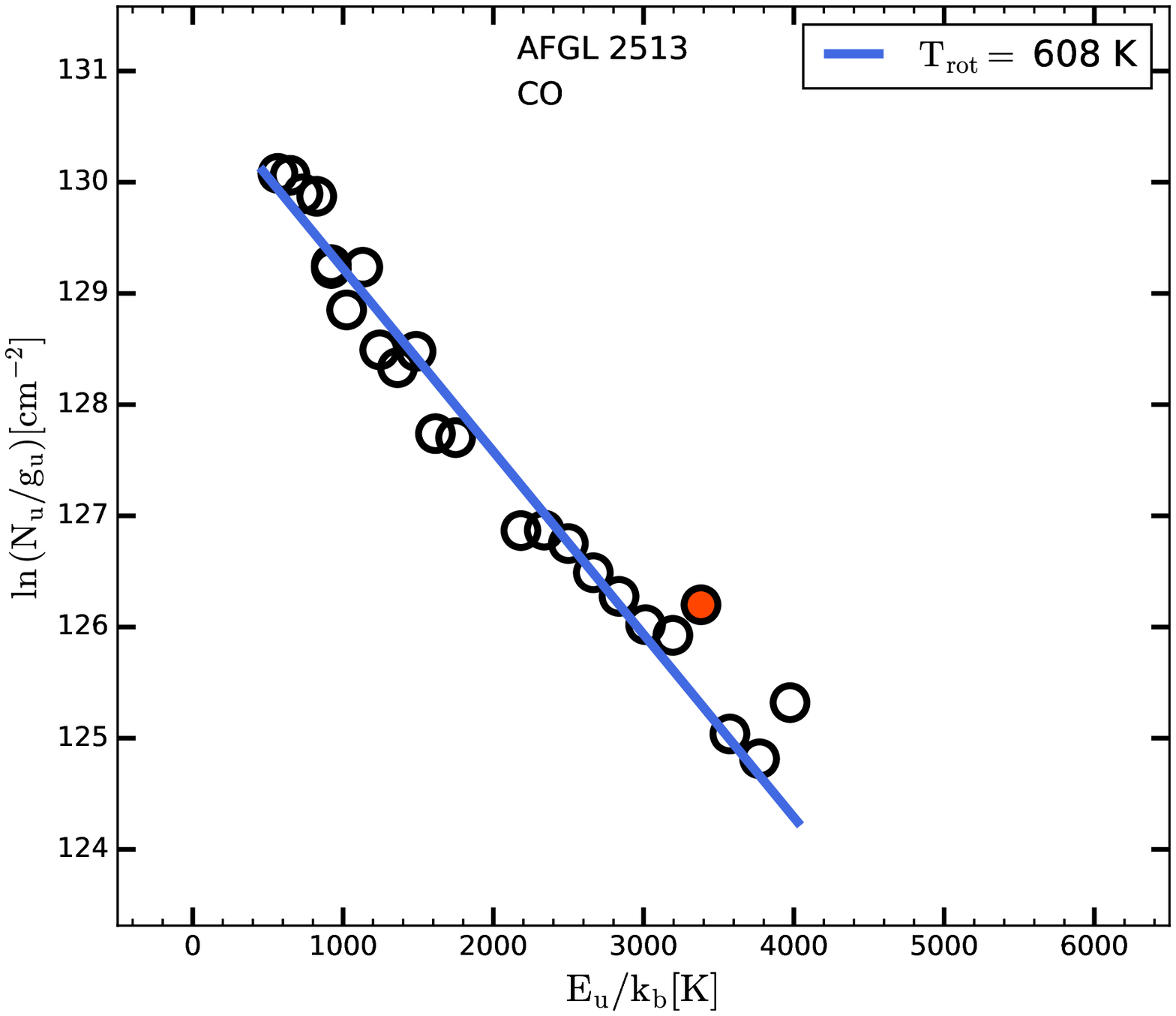}
 \end{subfigure}
  \caption{Continued.}
\end{figure*}

\begin{figure*}
\ContinuedFloat  
 \begin{subfigure}{0.49\textwidth}
 \centering
 \includegraphics[width = \textwidth]{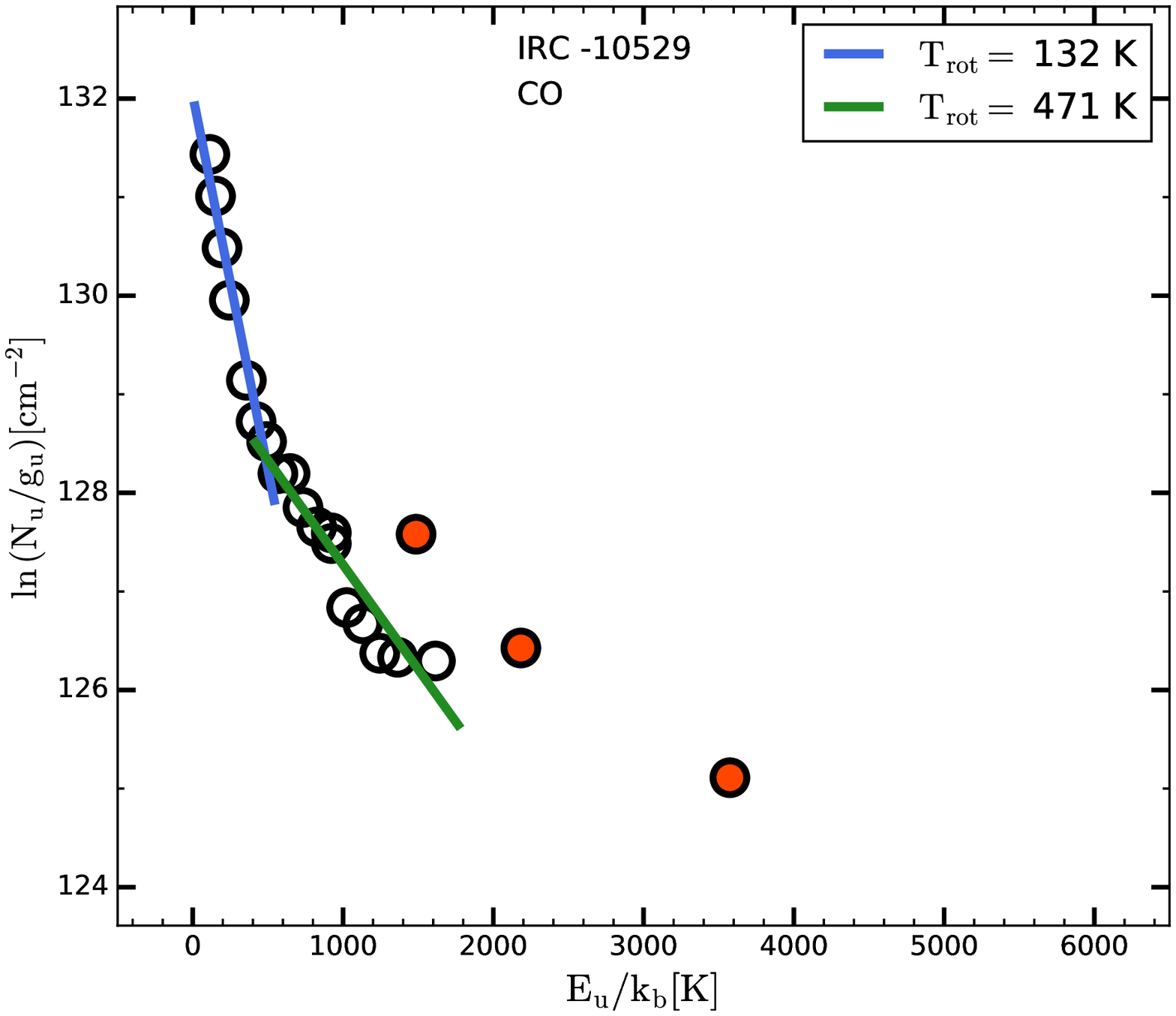} % irc -10 529 
 \end{subfigure}
\hfill
 \begin{subfigure}{0.49\textwidth}
 \centering
 \includegraphics[width = \textwidth]{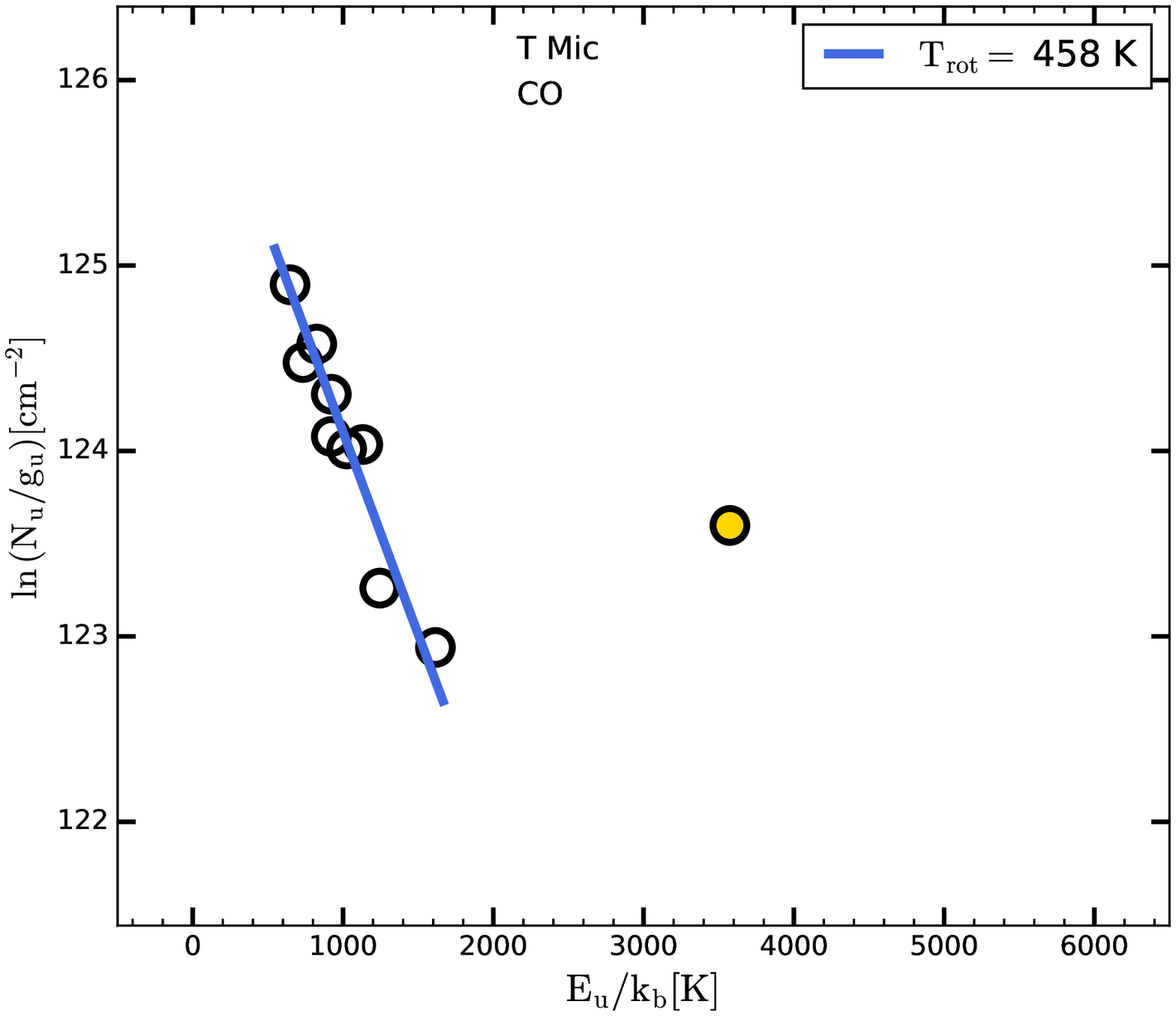}  %T Mic   
 \end{subfigure}
    \begin{subfigure}{0.49\textwidth}
 \centering
 \includegraphics[width = \textwidth]{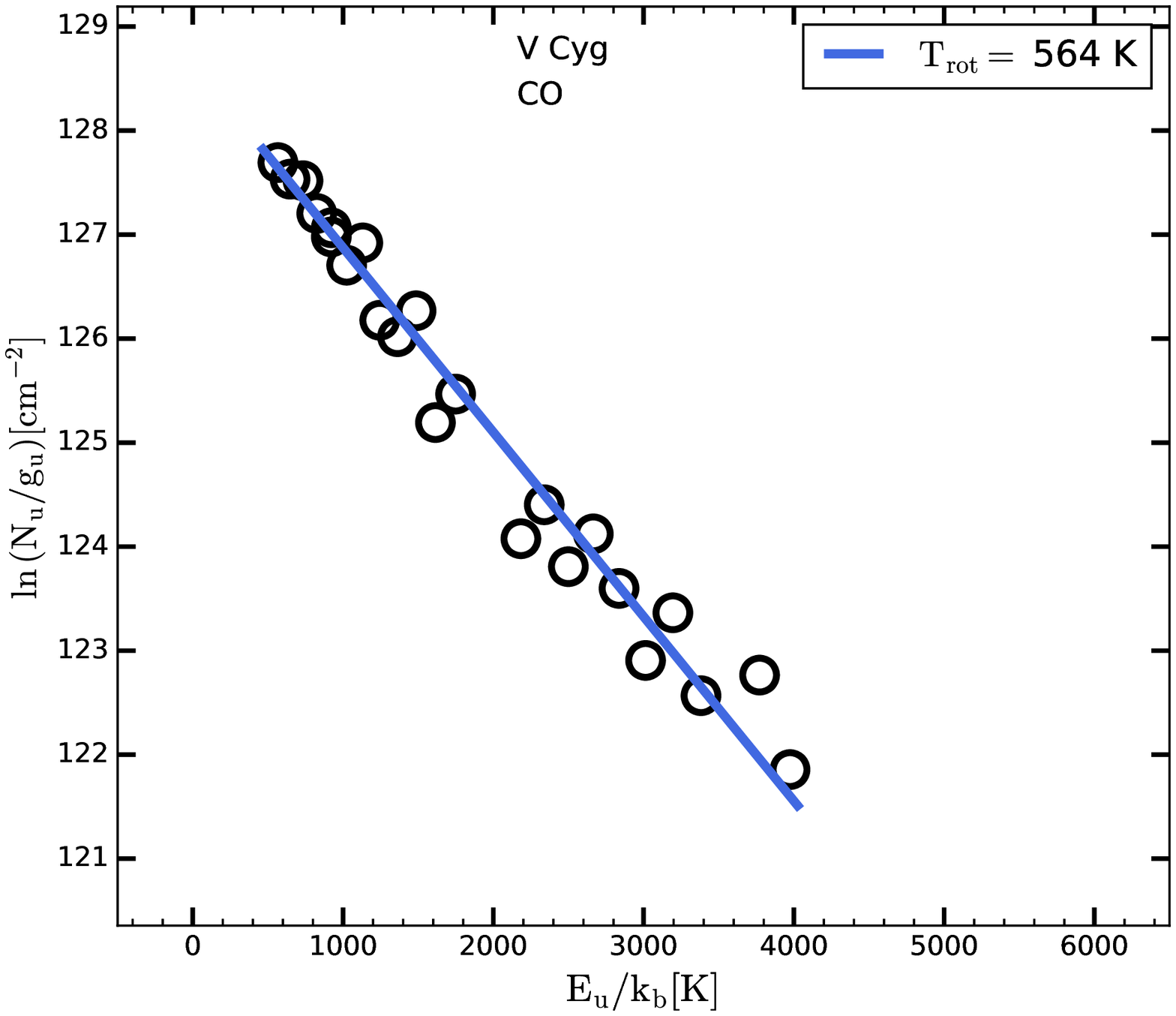} 
 \end{subfigure}
\hfill
 \begin{subfigure}{0.49\textwidth}
 \centering
 \includegraphics[width = \textwidth]{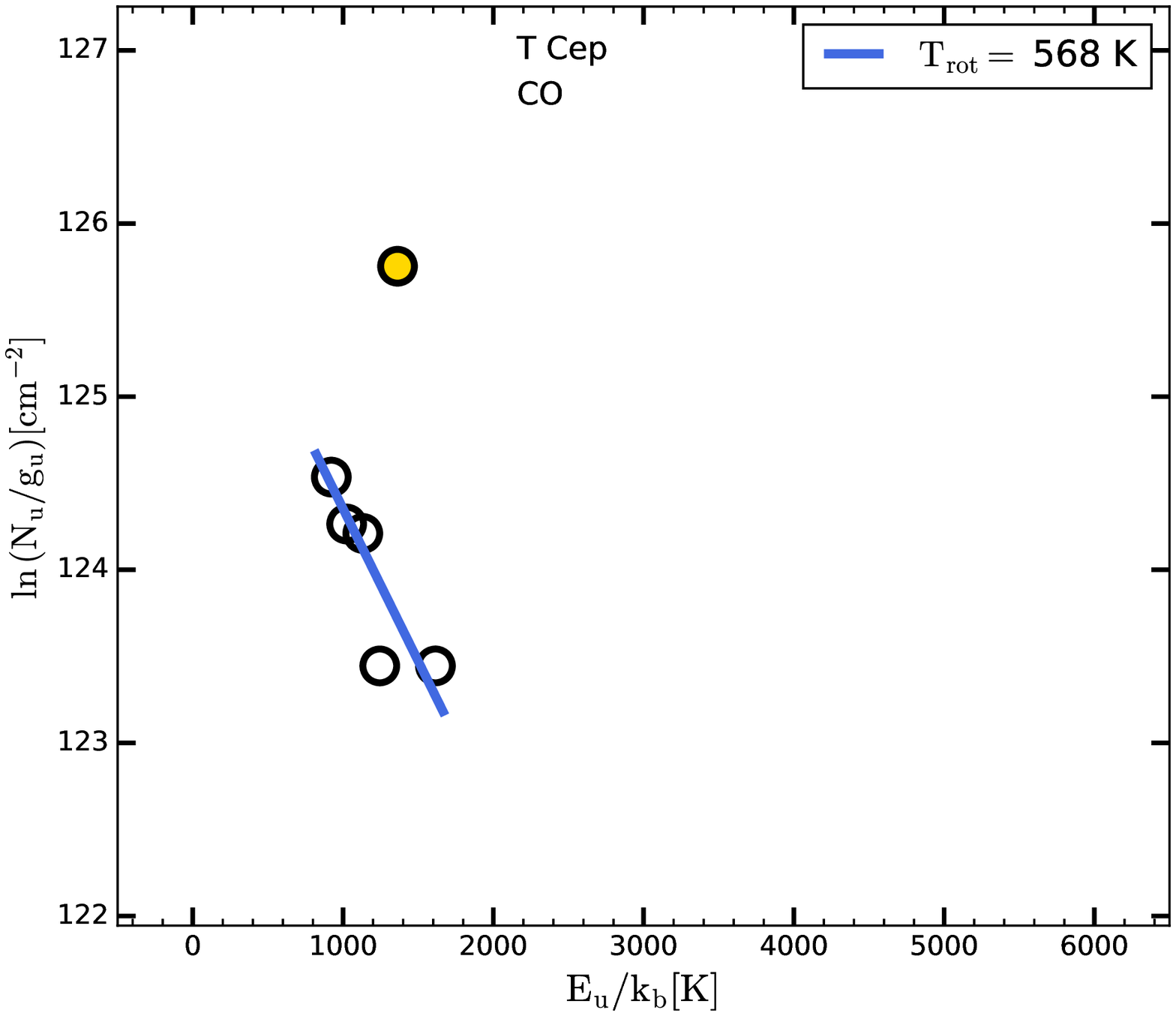} 
 \end{subfigure}
   \begin{subfigure}{0.49\textwidth}
 \centering
 \includegraphics[width = \textwidth]{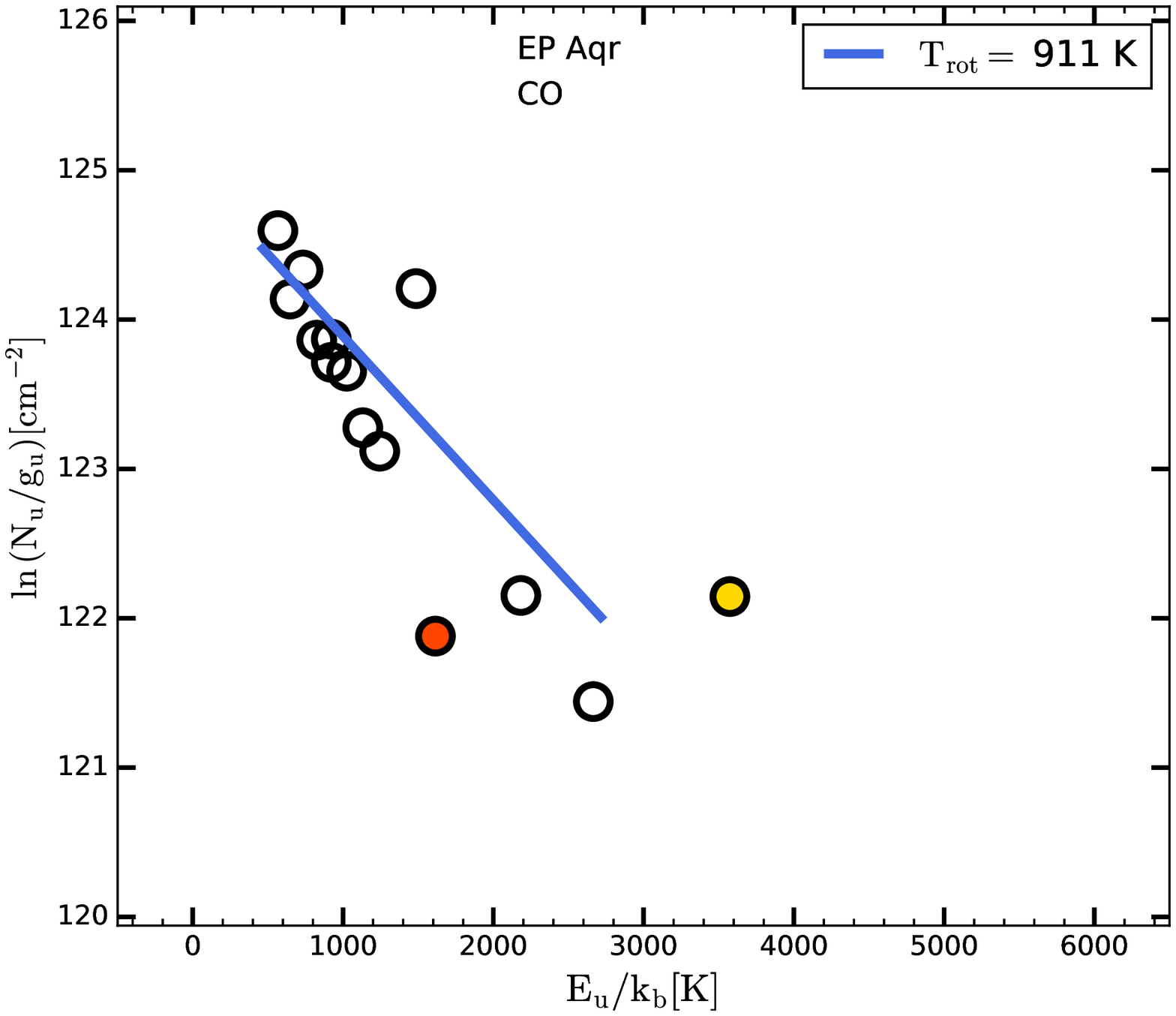} 
 \end{subfigure}
\hfill
 \begin{subfigure}{0.49\textwidth}
 \centering
 \includegraphics[width = \textwidth]{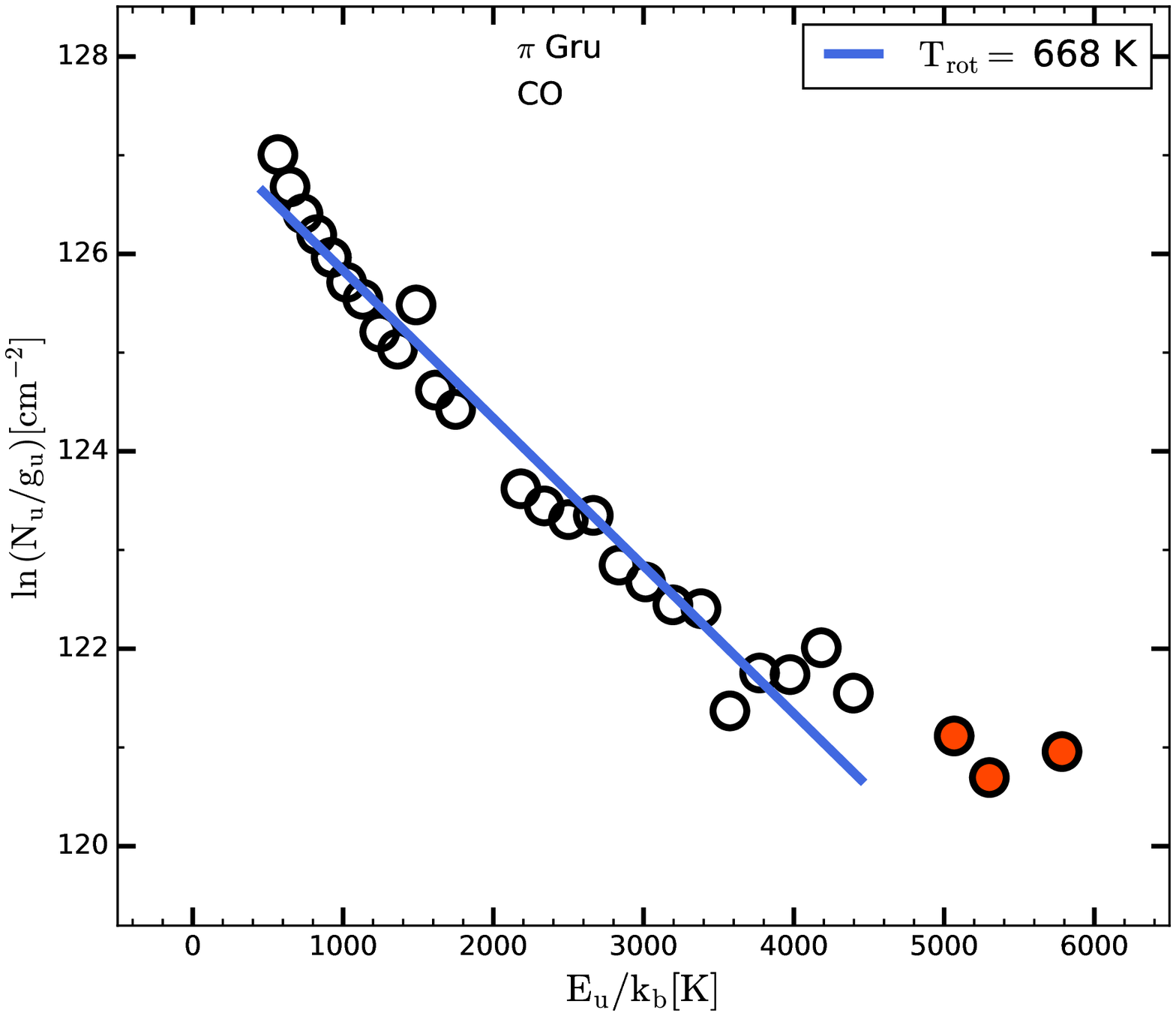}
 \end{subfigure}
   \caption{Continued.}
\end{figure*}

\begin{figure*}
\ContinuedFloat   
   \begin{subfigure}{0.49\textwidth}
 \centering
 \includegraphics[width = \textwidth]{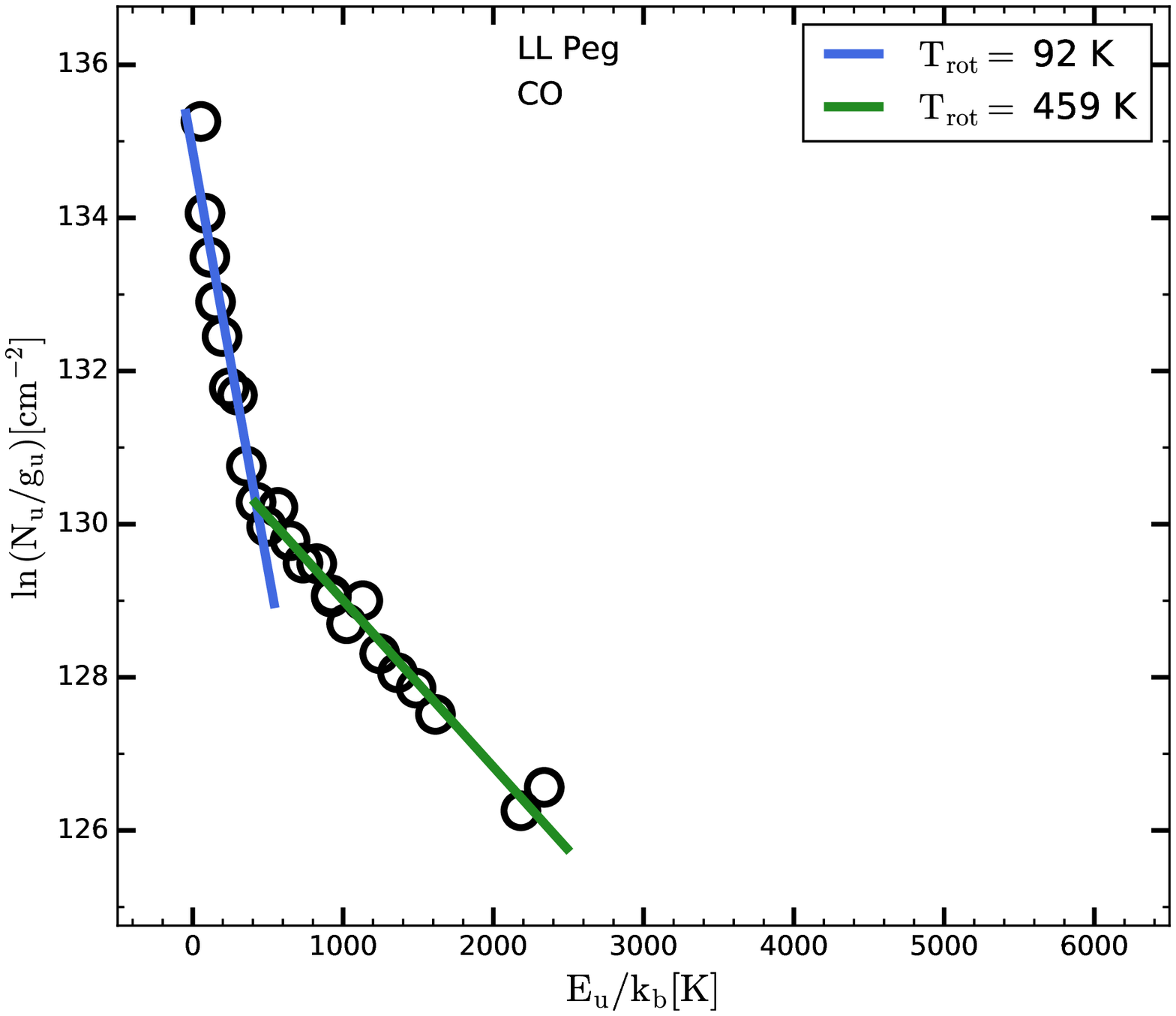} % LL Peg
 \end{subfigure}
\hfill
   \begin{subfigure}{0.49\textwidth}
 \centering
 \includegraphics[width = \textwidth]{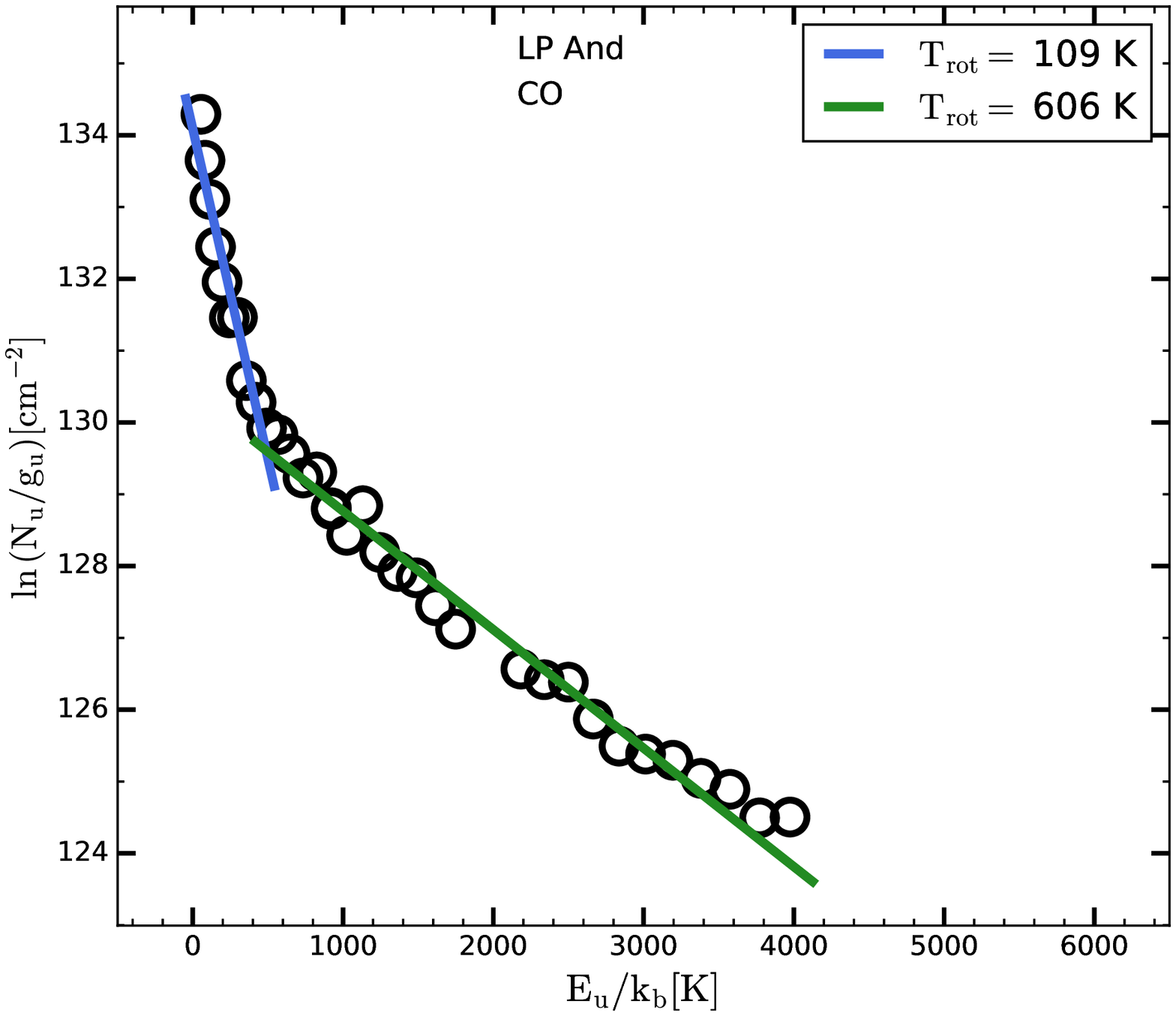}  % LP And
 \end{subfigure}
 \begin{subfigure}{0.49\textwidth}
 \centering
 \includegraphics[width = \textwidth]{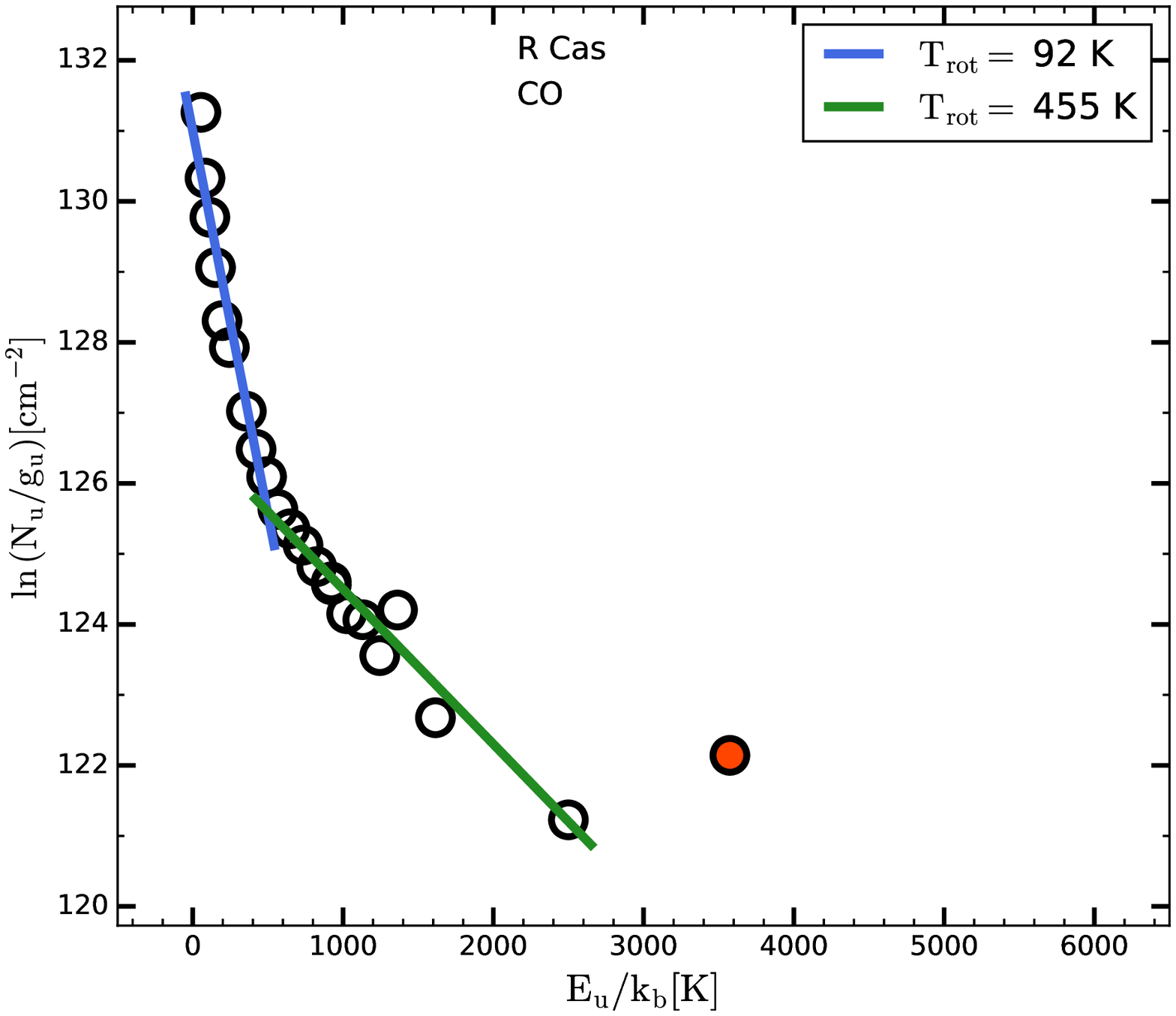}
 \end{subfigure}
 \hfill
  \begin{subfigure}{0.49\textwidth}
 \centering
 \includegraphics[width = \textwidth]{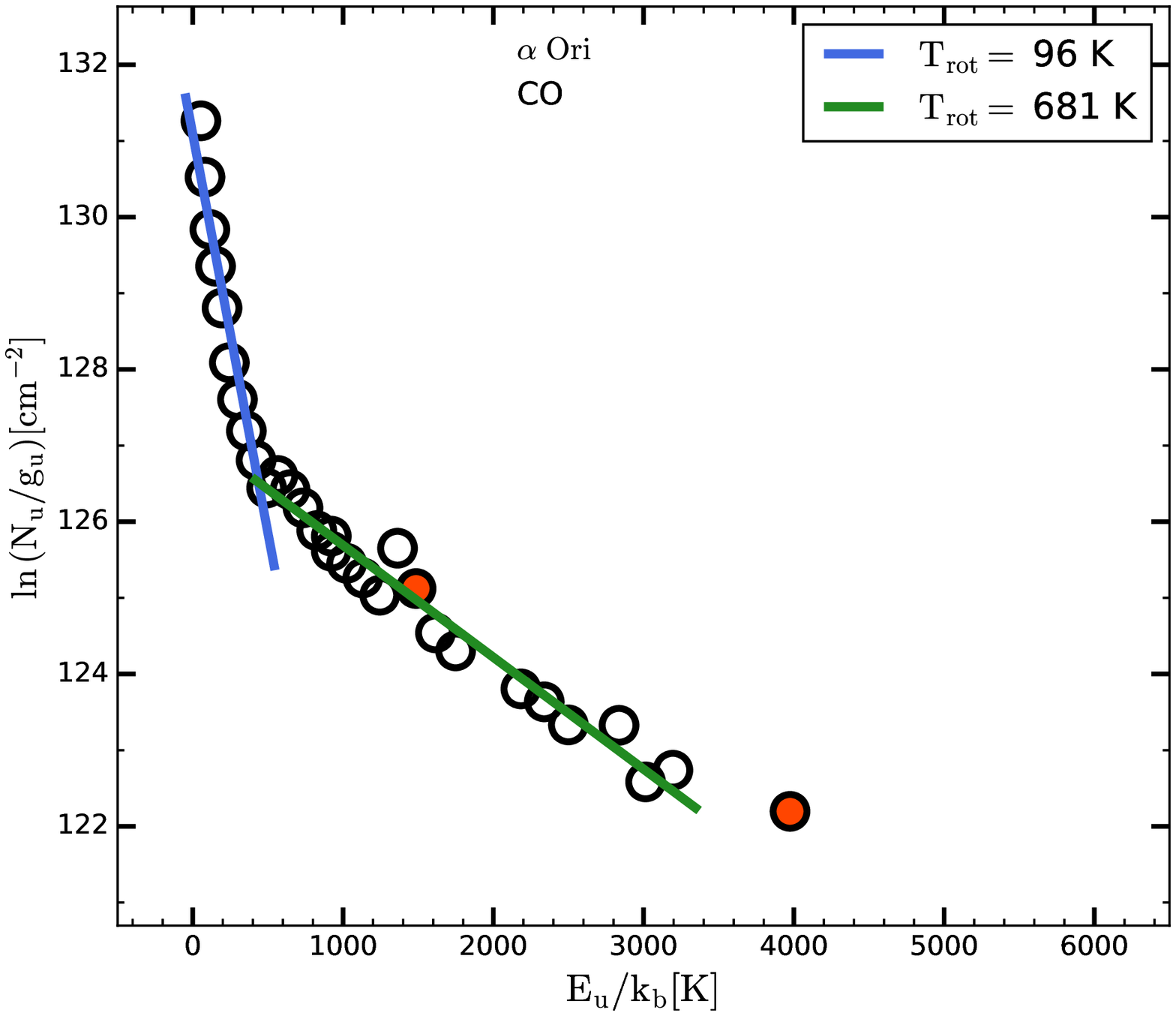}
 \end{subfigure}
  \begin{subfigure}{0.49\textwidth}
 \centering
 \includegraphics[width = \textwidth]{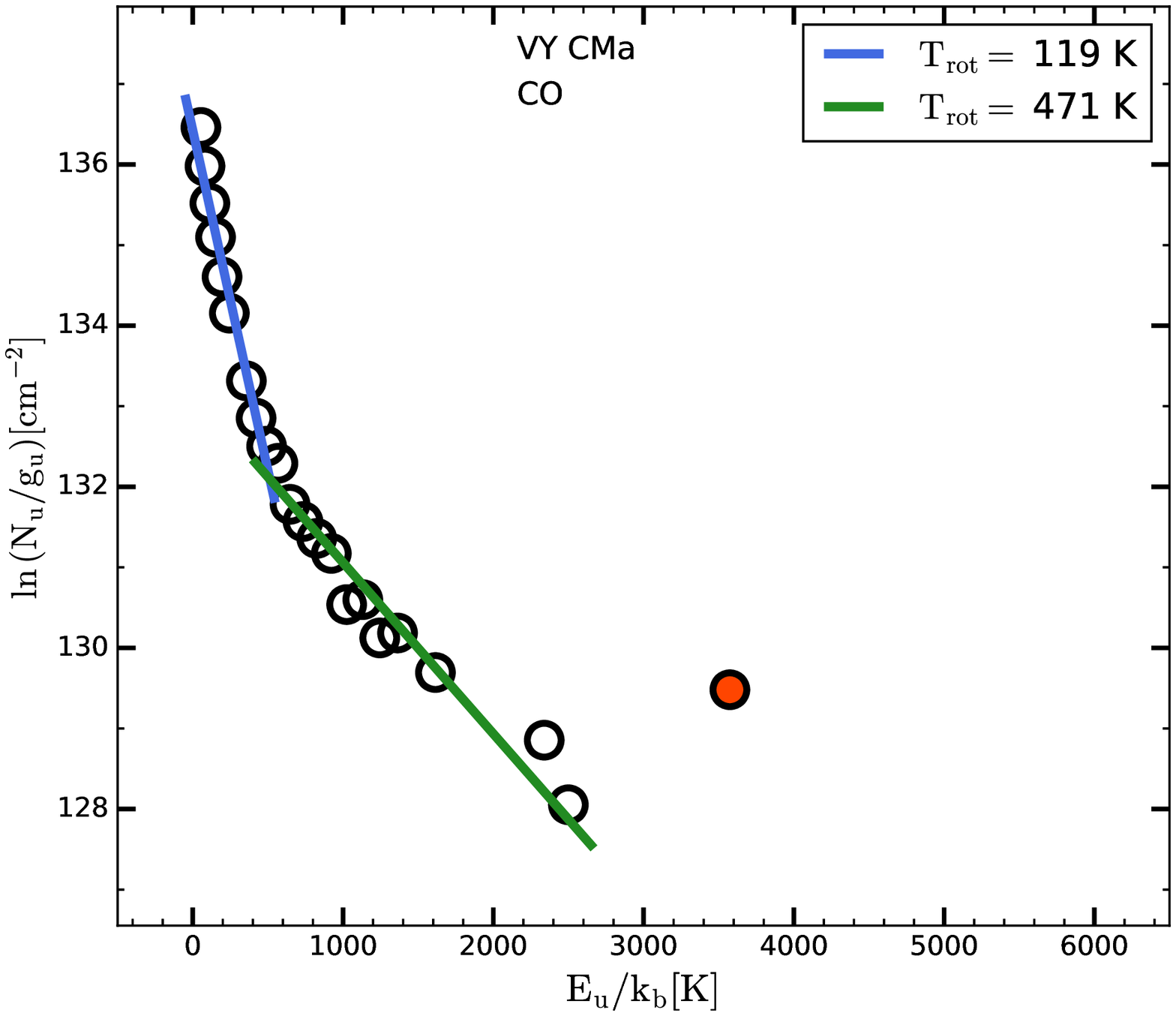}
 \end{subfigure}
   \begin{subfigure}{0.49\textwidth}
 \centering
 \includegraphics[width = \textwidth]{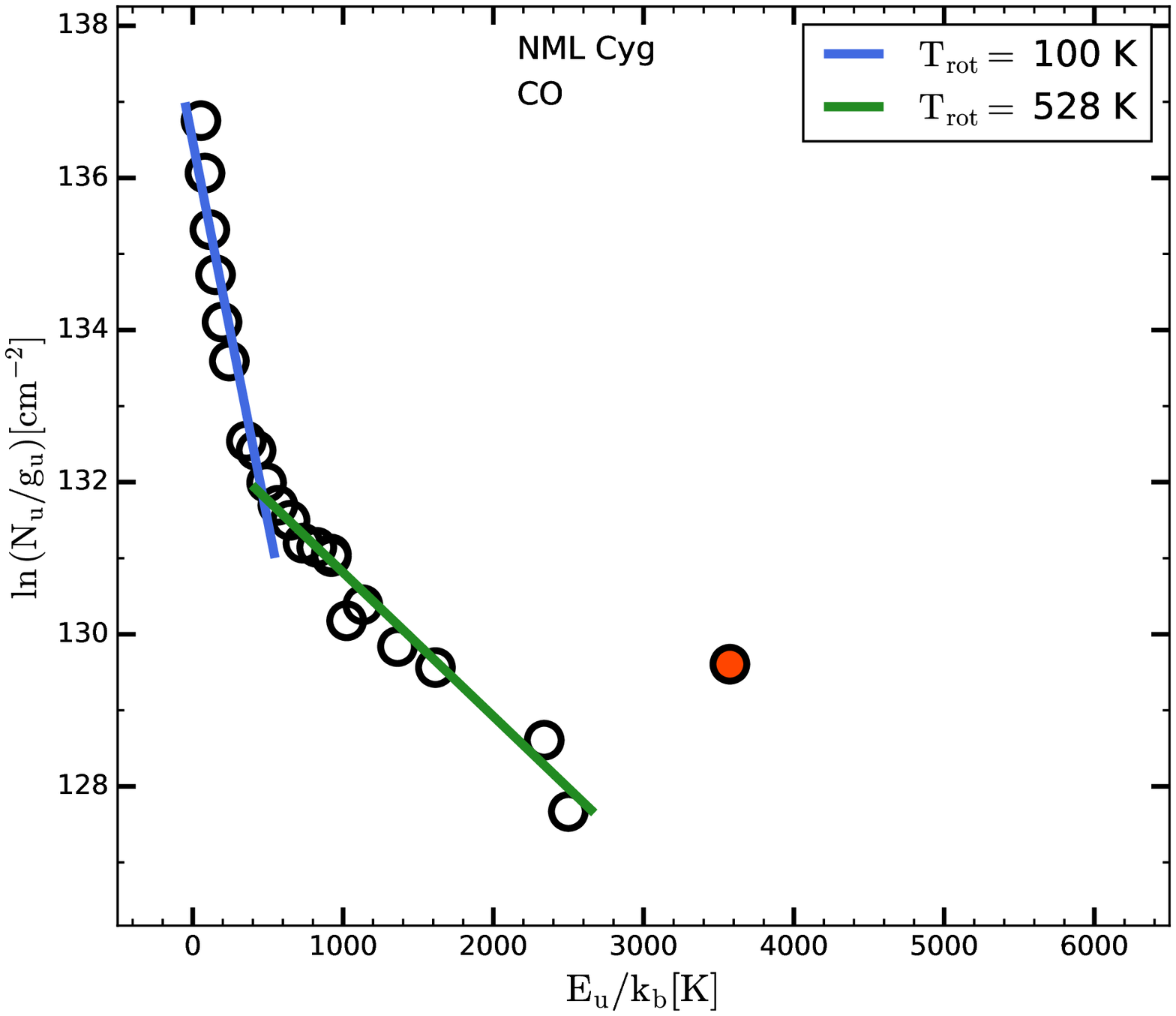}
 \end{subfigure}
  \caption{Continued.}
\end{figure*}

%%%%%%%%%%%%%%%%%%%%%%%%%%%%%%%%%%%%%%%%%%%%%%%%%%%%%%%%%%%%%%%%%%%%%%%%%%%%%%%%%

\begin{figure*}
 \label{Fig:RotDiagramsC}

 \begin{subfigure}{0.49\textwidth}
 \centering
 \includegraphics[width = \textwidth]{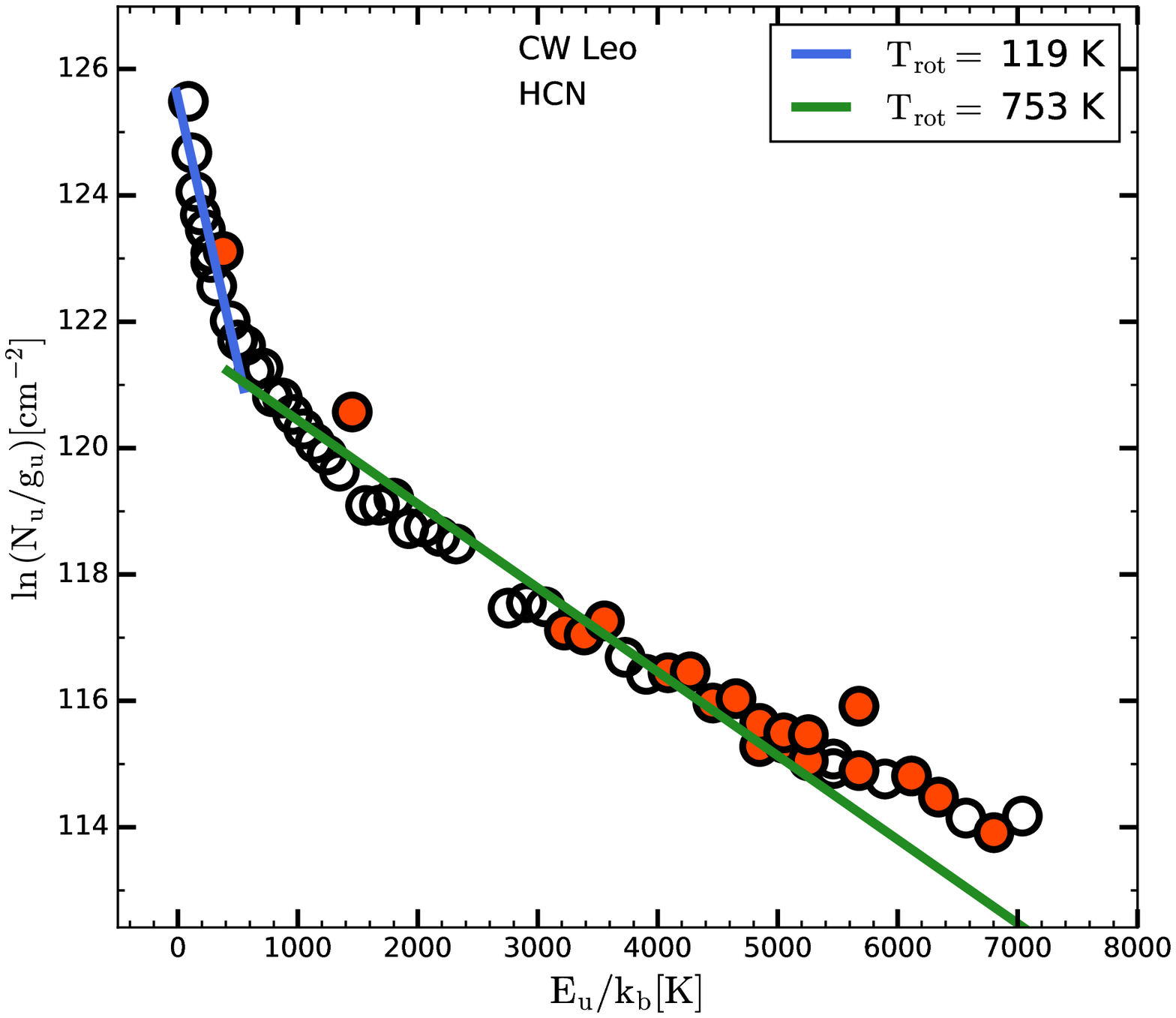}
 \end{subfigure}
  \hfill
   \begin{subfigure}{0.49\textwidth}
 \centering
 \includegraphics[width = \textwidth]{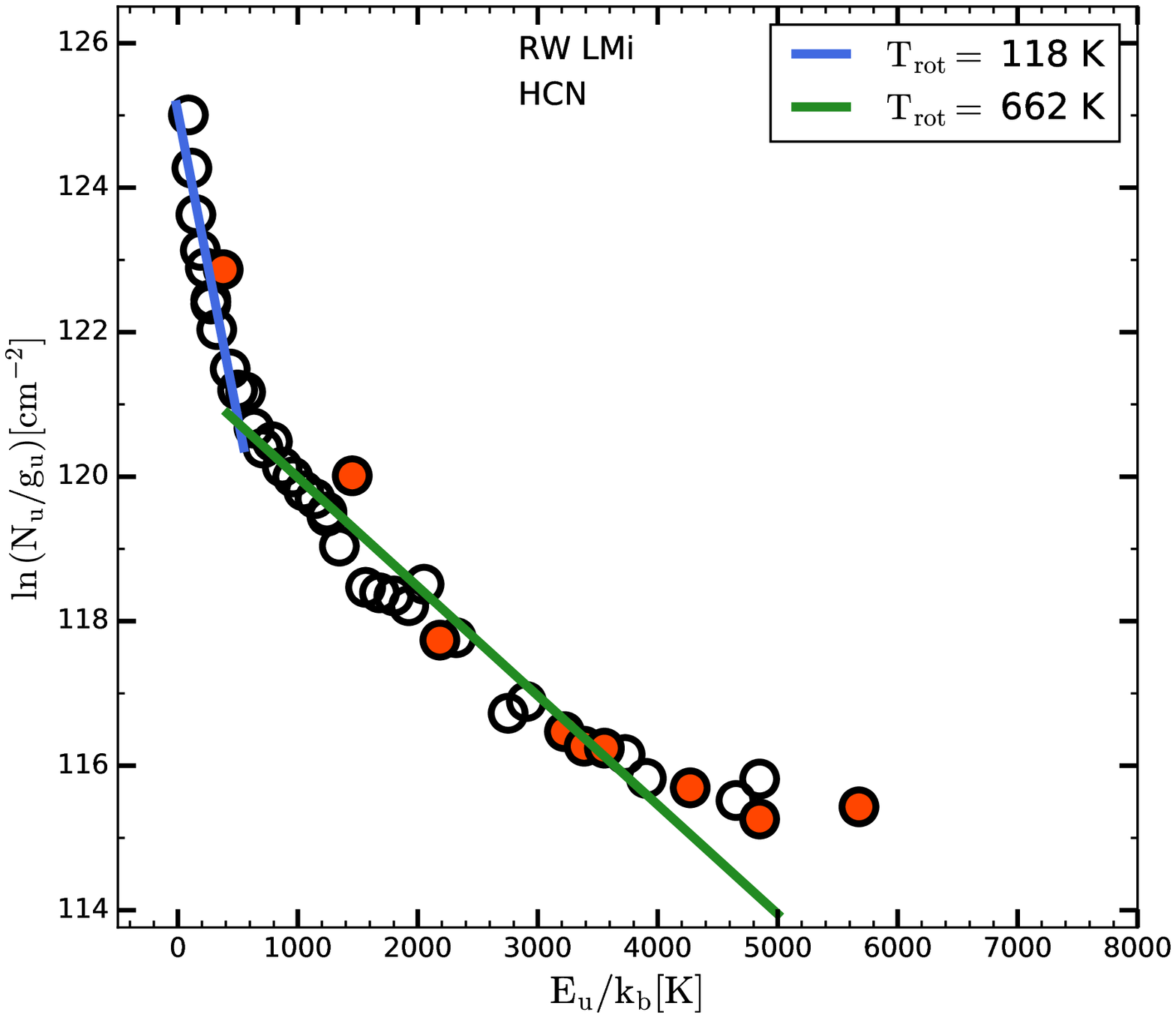}
 \end{subfigure}
  \begin{subfigure}{0.49\textwidth}
 \centering
 \includegraphics[width = \textwidth]{Rotdiagram_VHYA_HCNv=0.eps}
 \end{subfigure}
 \hfill
 \begin{subfigure}{0.49\textwidth}
 \centering
 \includegraphics[width = \textwidth]{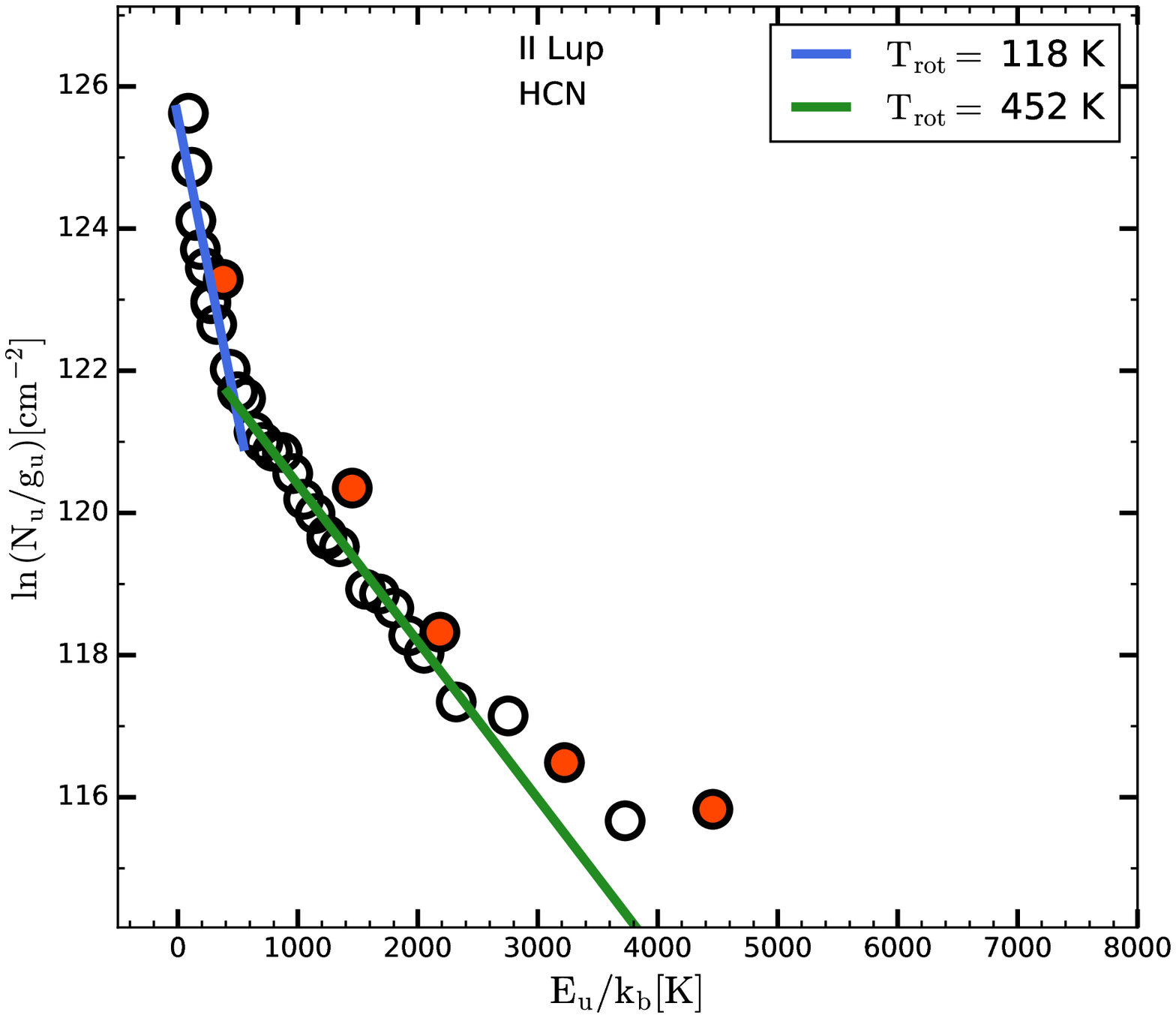}
 \end{subfigure}
  \begin{subfigure}{0.49\textwidth}
 \centering
 \includegraphics[width = \textwidth]{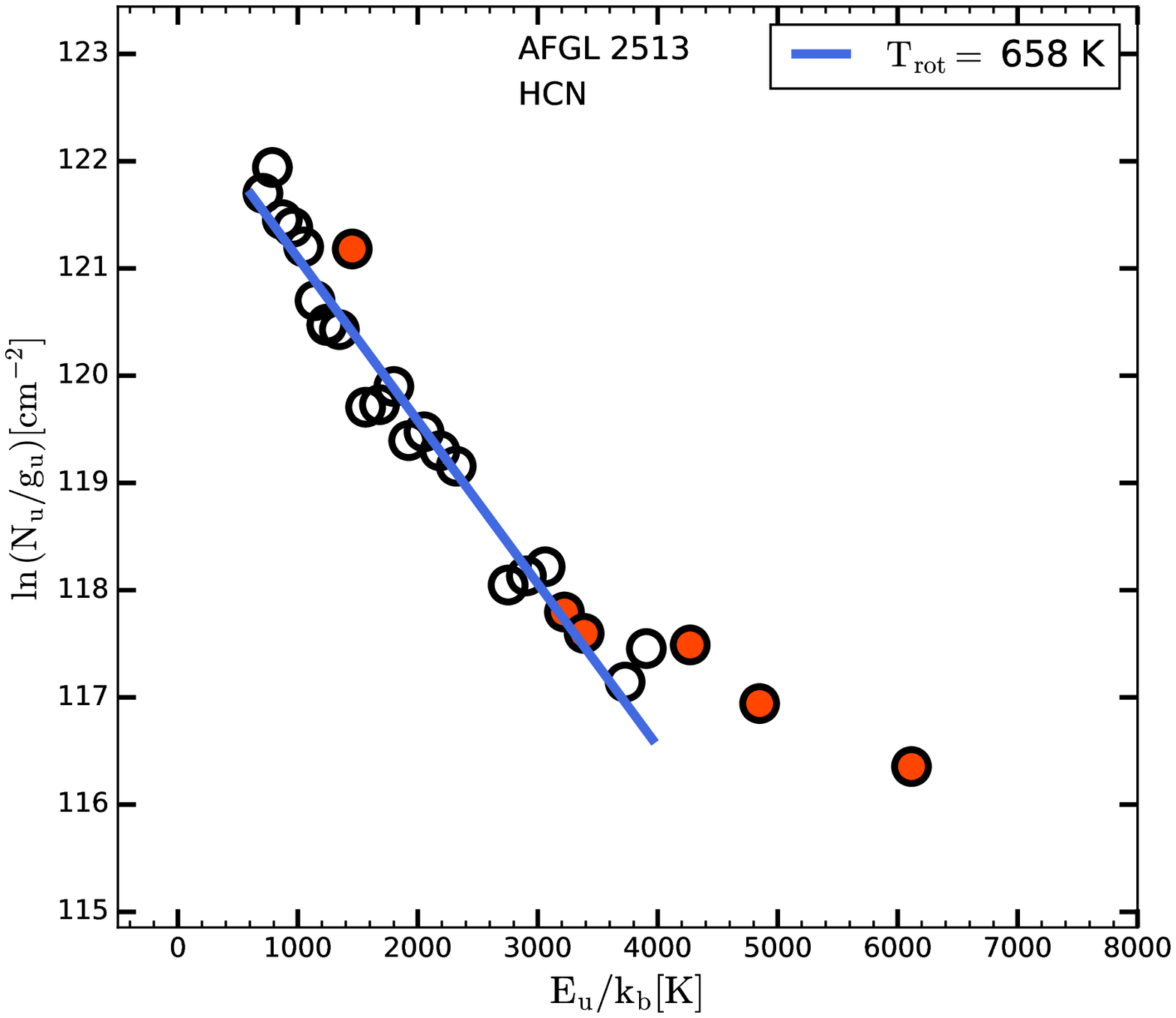}
 \end{subfigure}
  \hfill
  \begin{subfigure}{0.49\textwidth}
 \centering
 \includegraphics[width = \textwidth]{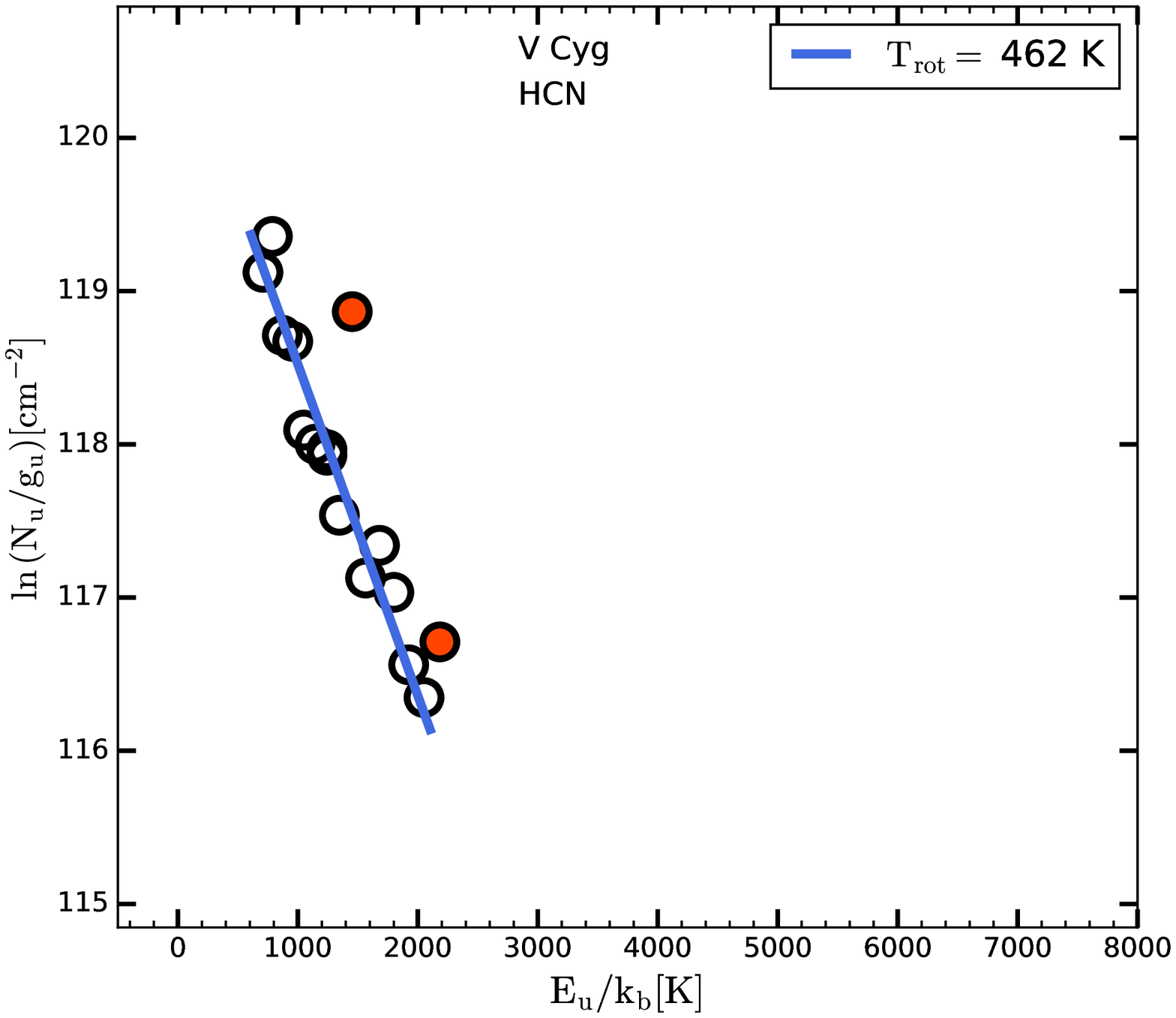}
 \end{subfigure}
 \caption{HCN population-temperature diagrams for C-type stars.}
\end{figure*}

\begin{figure*}
\ContinuedFloat  
 \begin{subfigure}{0.49\textwidth}
 \centering
 \includegraphics[width = \textwidth]{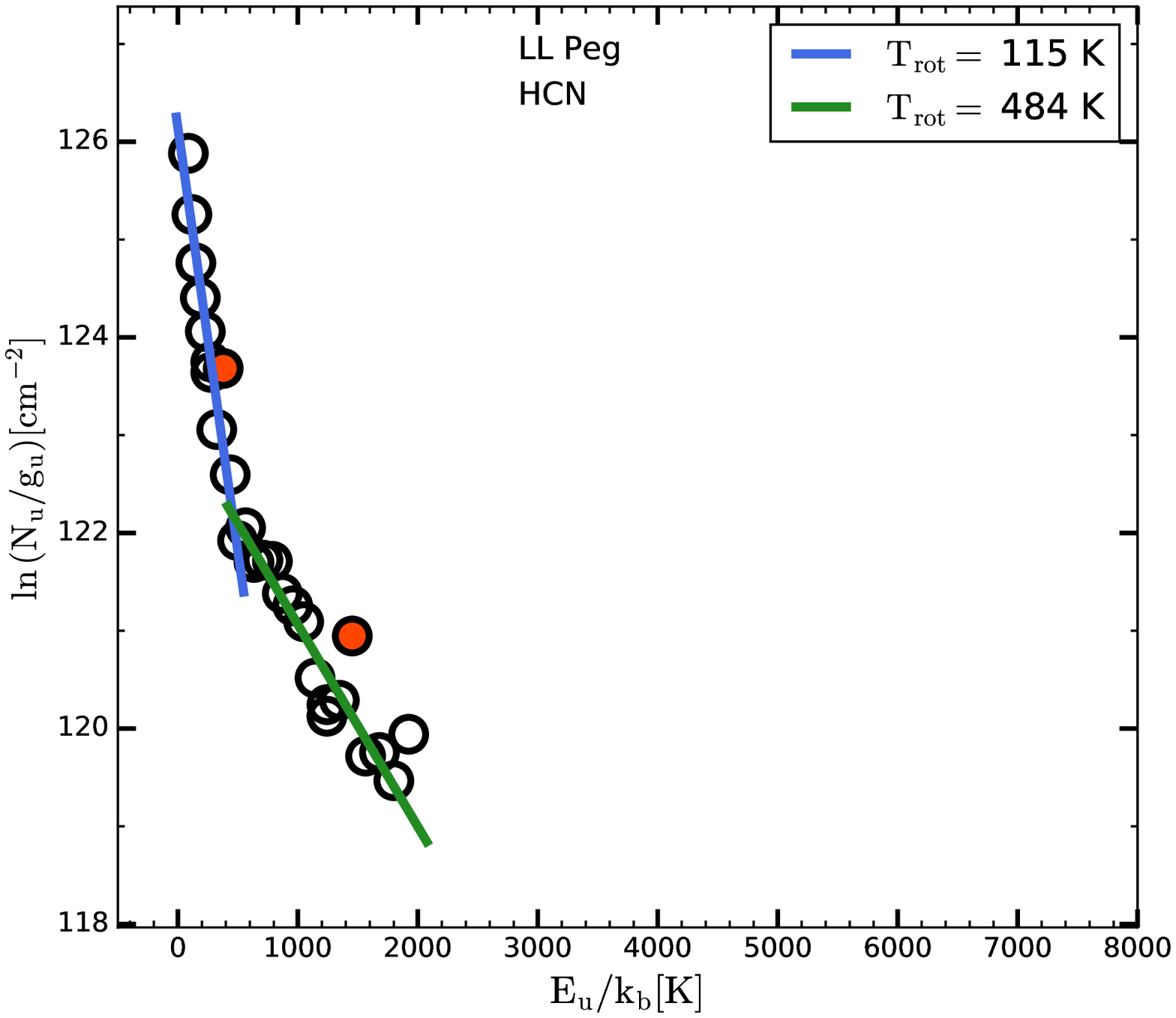}
 \end{subfigure}
 \hfill
 \begin{subfigure}{0.49\textwidth}
 \centering
 \includegraphics[width = \textwidth]{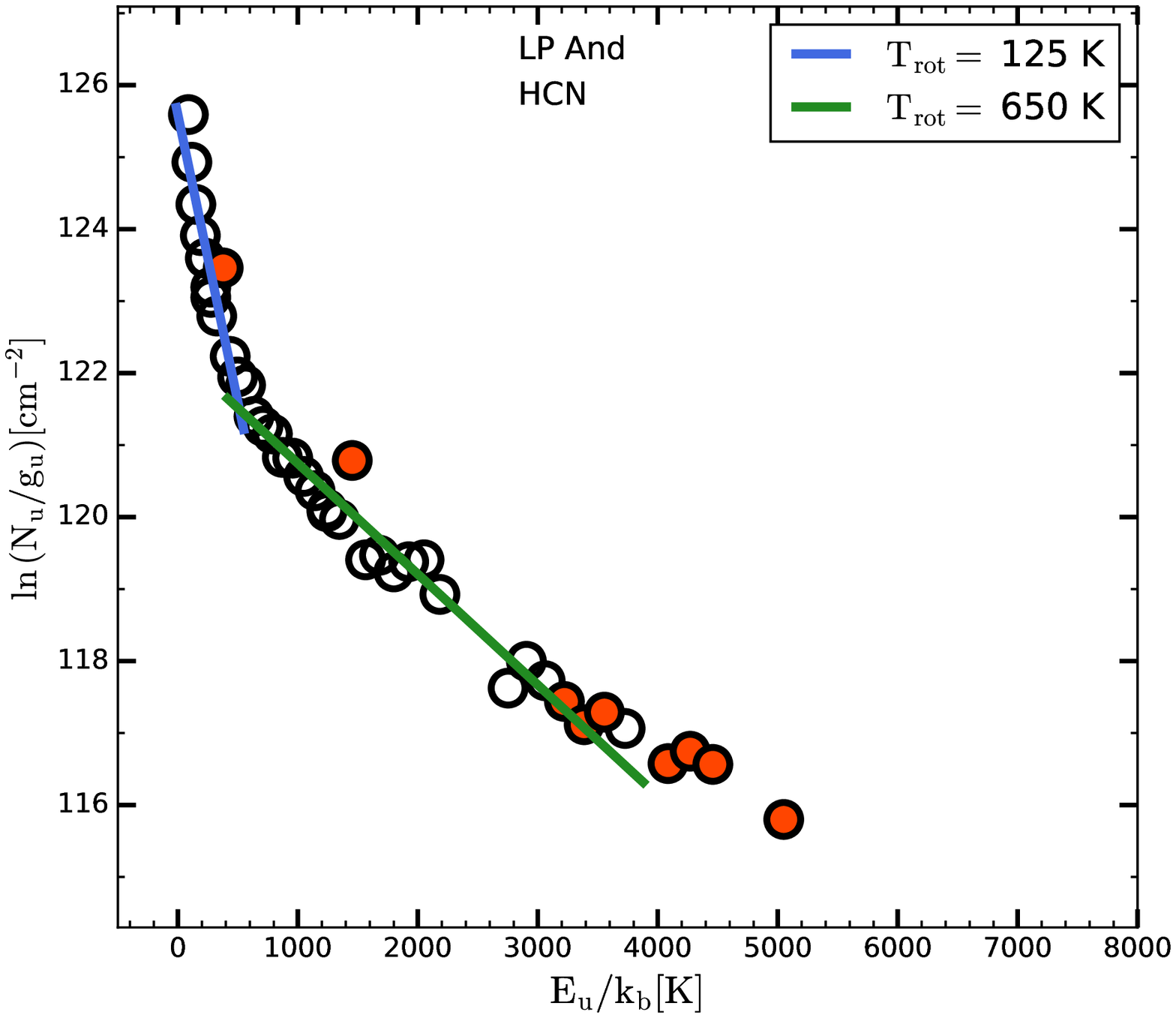}
 \end{subfigure}
 \caption{Continued.}
% \label{Fig:RotDiagramsC}
\end{figure*}

\clearpage

\section{Power law fitting}
\label{Appen:Powerlaw}

This Appendix shows the continuum spectra fitted with a segmented power law, following the methodology outlined in Sect.~\ref{S-PL}.
The indices of the power law are listed in Table~\ref{Table:powerlawresults}.

\begin{figure*}
 \label{Fig:AppenPowerlaw}

 \begin{subfigure}{0.49\textwidth}
 \centering
 \includegraphics[width = \textwidth]{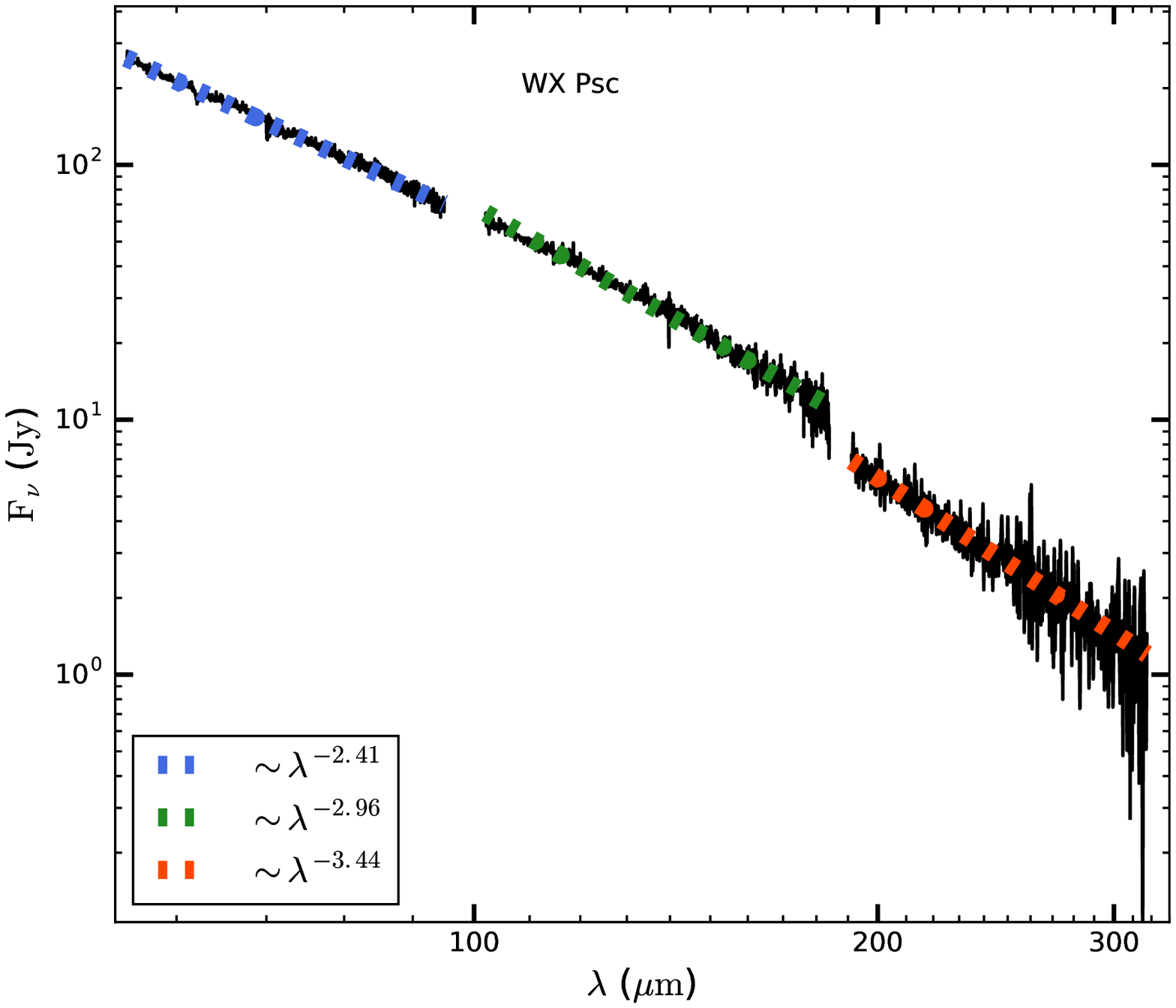}
 \end{subfigure}
\hfill
 \begin{subfigure}{0.49\textwidth}
 \centering
 \includegraphics[width = \textwidth]{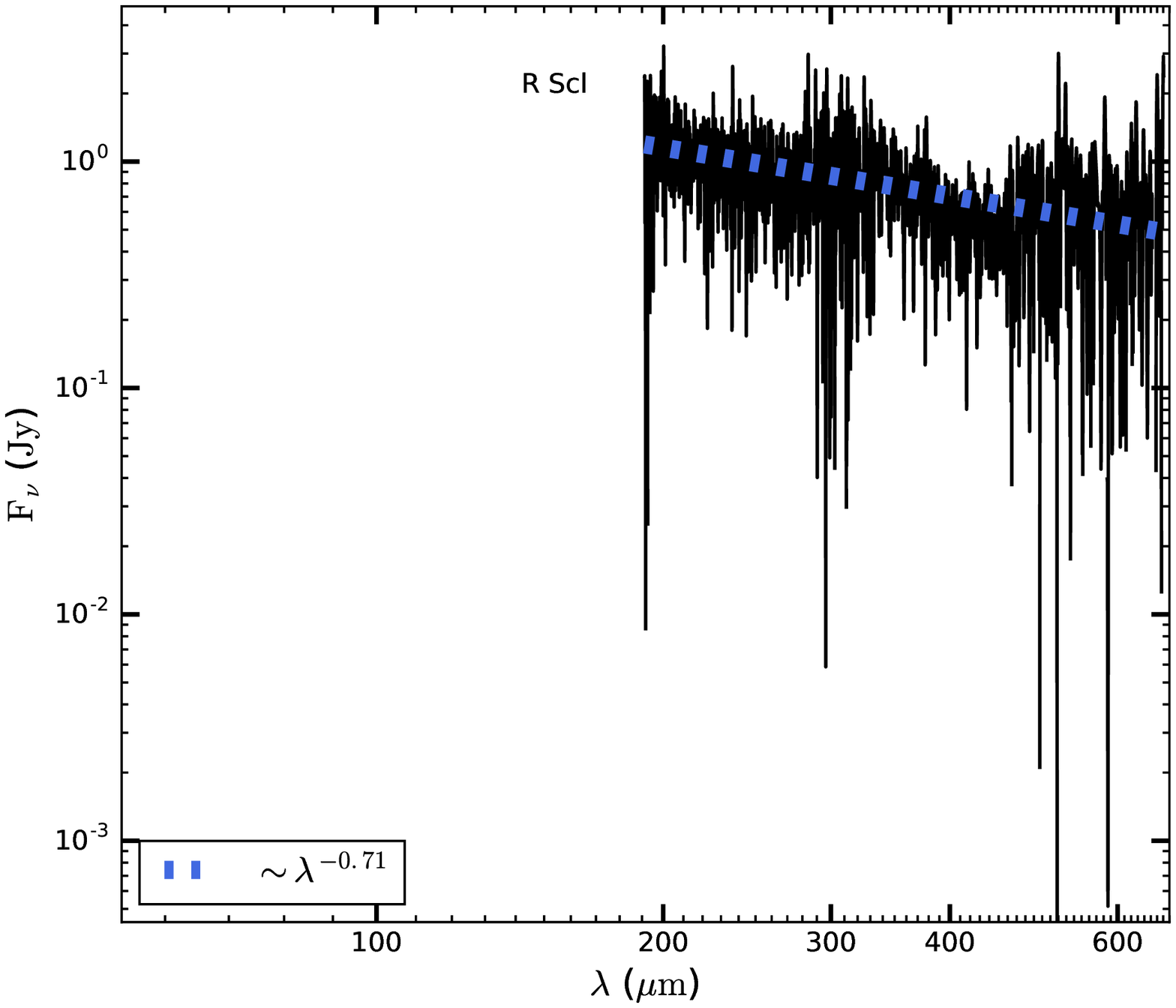}
 \end{subfigure}
  \begin{subfigure}{0.49\textwidth}
 \centering
 \includegraphics[width = \textwidth]{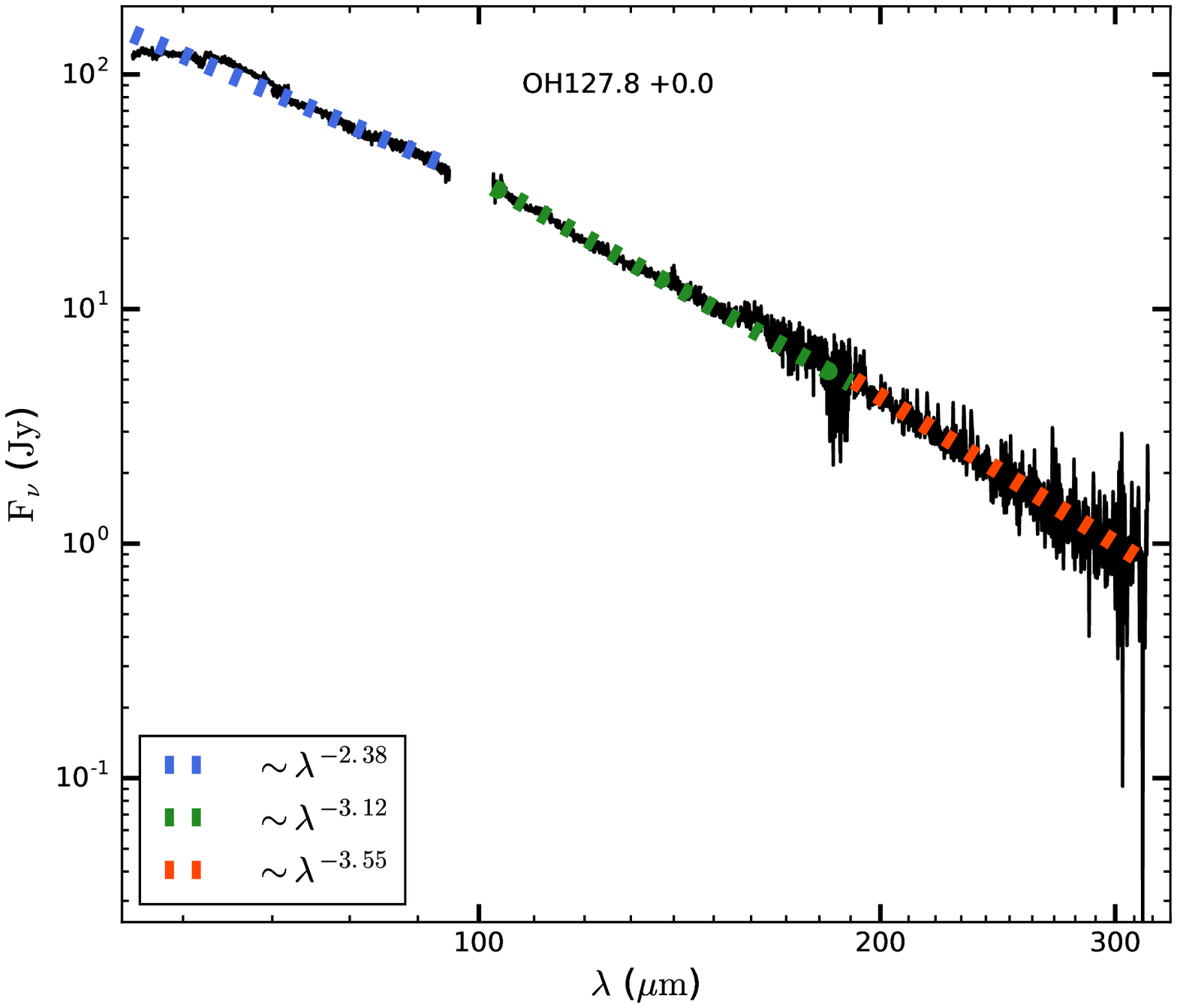}
 \end{subfigure}
\hfill
 \begin{subfigure}{0.49\textwidth}
 \centering
 \includegraphics[width = \textwidth]{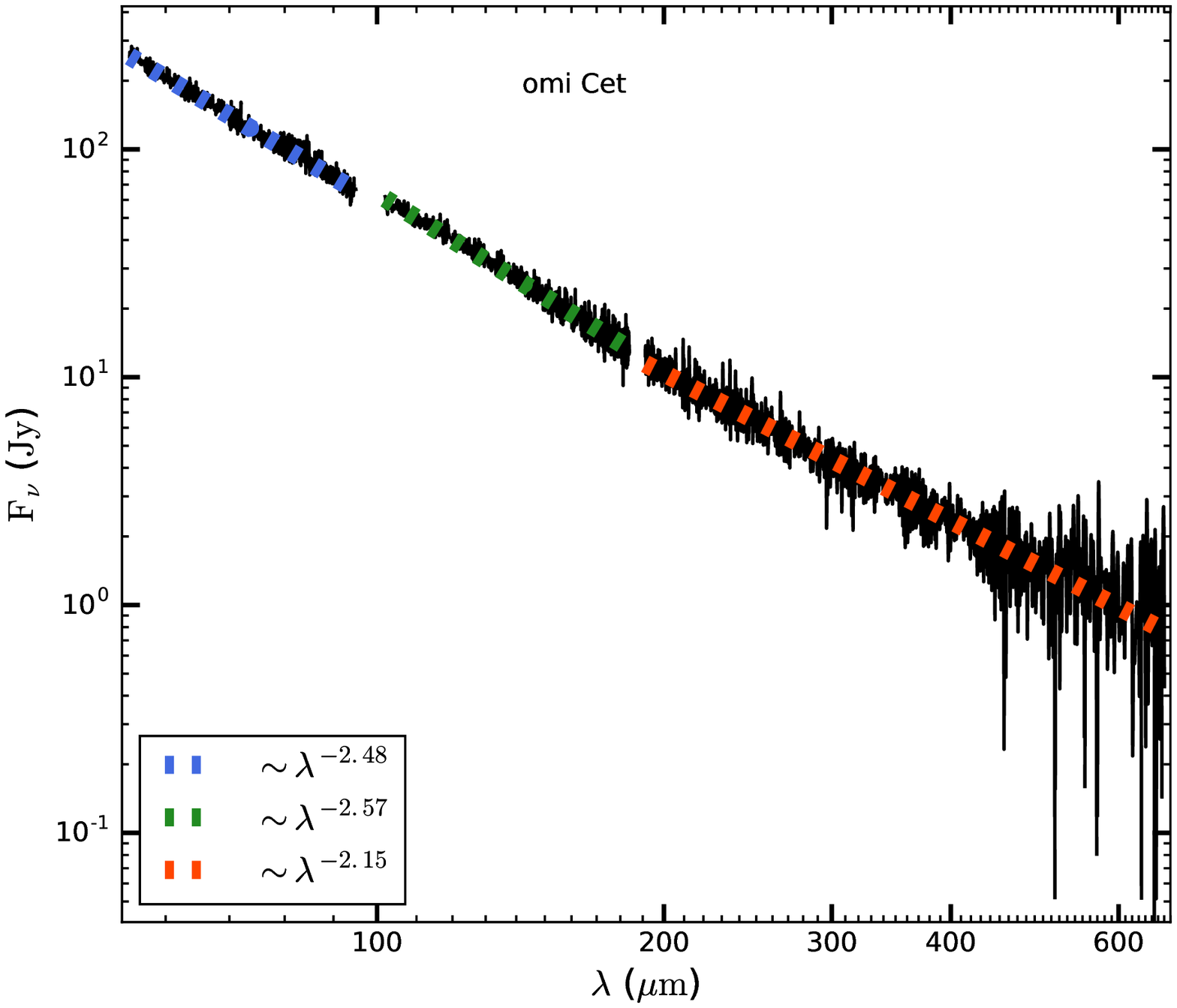} 
 \end{subfigure}
 \begin{subfigure}{0.49\textwidth}
 \centering
 \includegraphics[width = \textwidth]{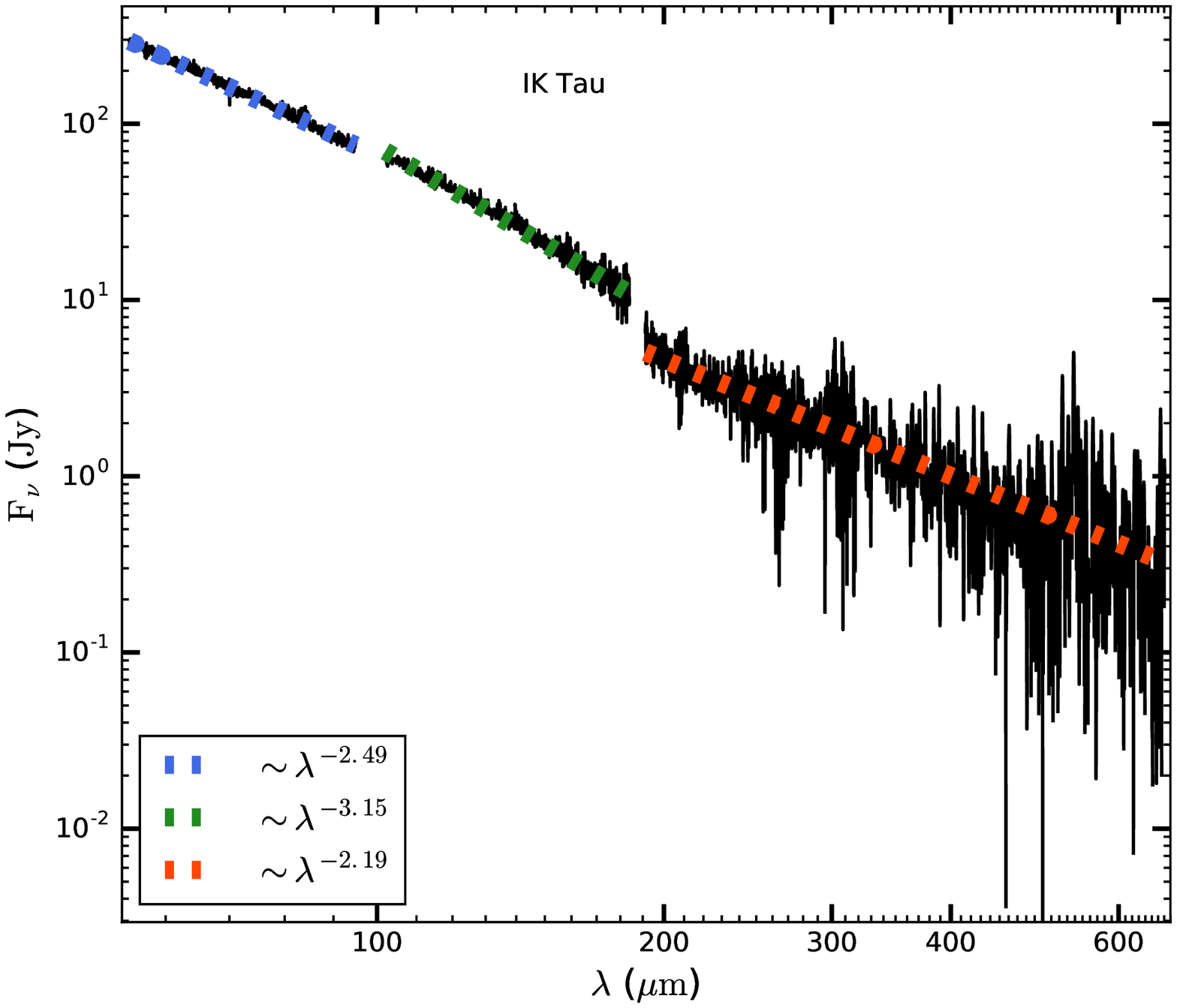}    %IK Tau
 \end{subfigure}
\hfill
 \begin{subfigure}{0.49\textwidth}
 \centering
 \includegraphics[width = \textwidth]{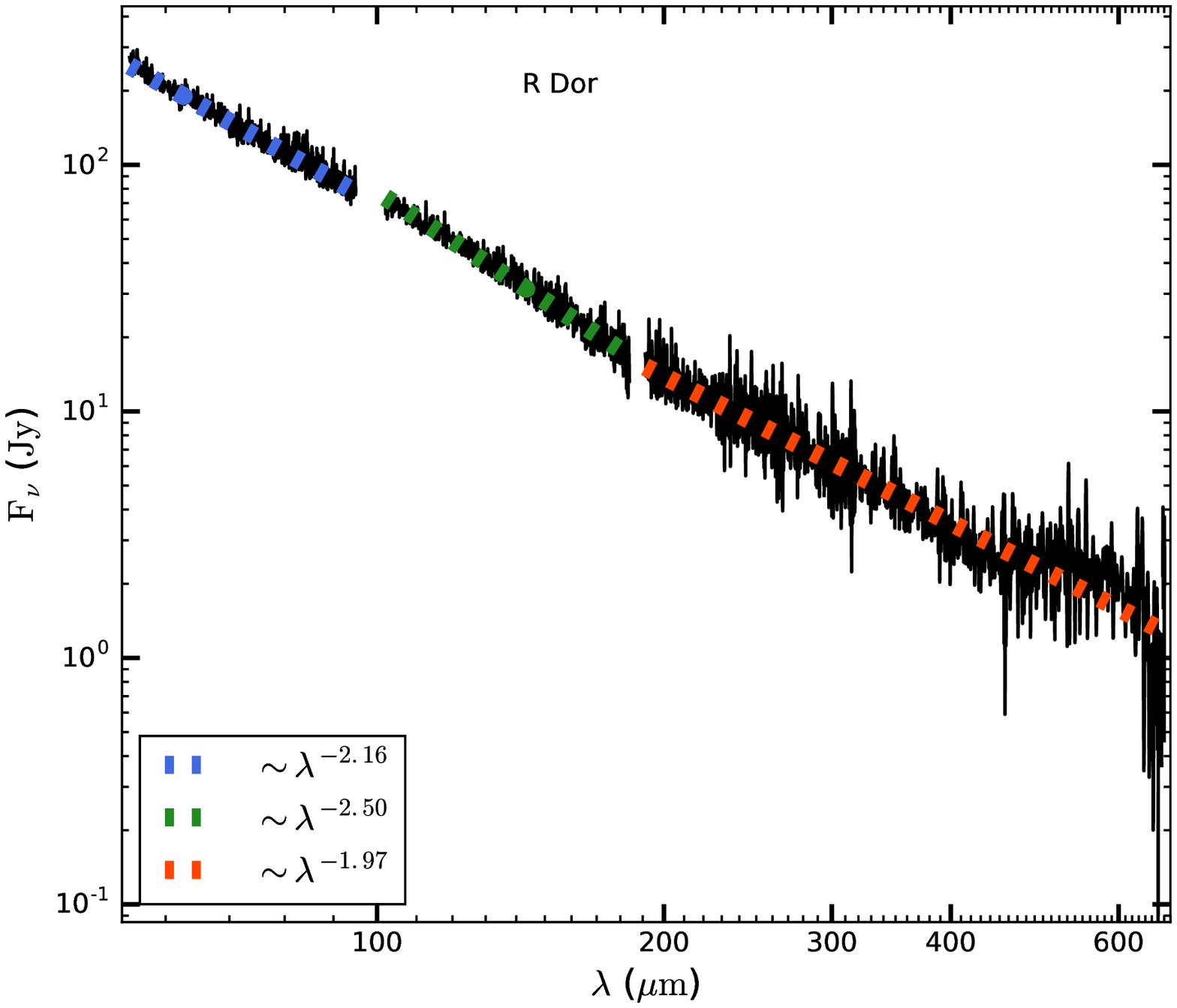}
 \end{subfigure}
  \caption{Power law fits to three segments of a full PACS and SPIRE line-clipped continuum spectrum.}
\end{figure*}

\begin{figure*}
\ContinuedFloat
  \begin{subfigure}{0.49\textwidth}
 \centering
 \includegraphics[width = \textwidth]{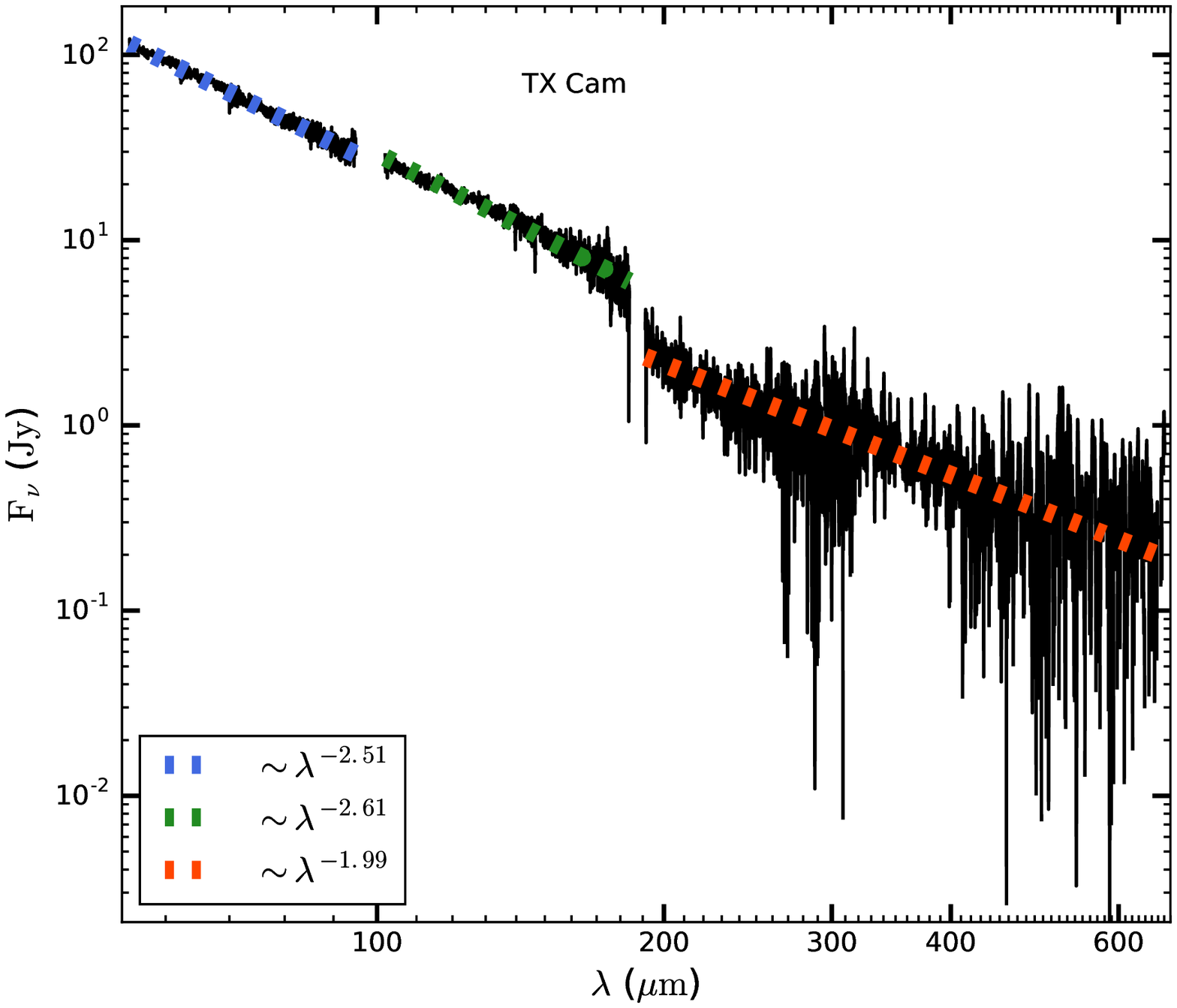} 
 \end{subfigure}
\hfill
 \begin{subfigure}{0.49\textwidth}
 \centering
 \includegraphics[width = \textwidth]{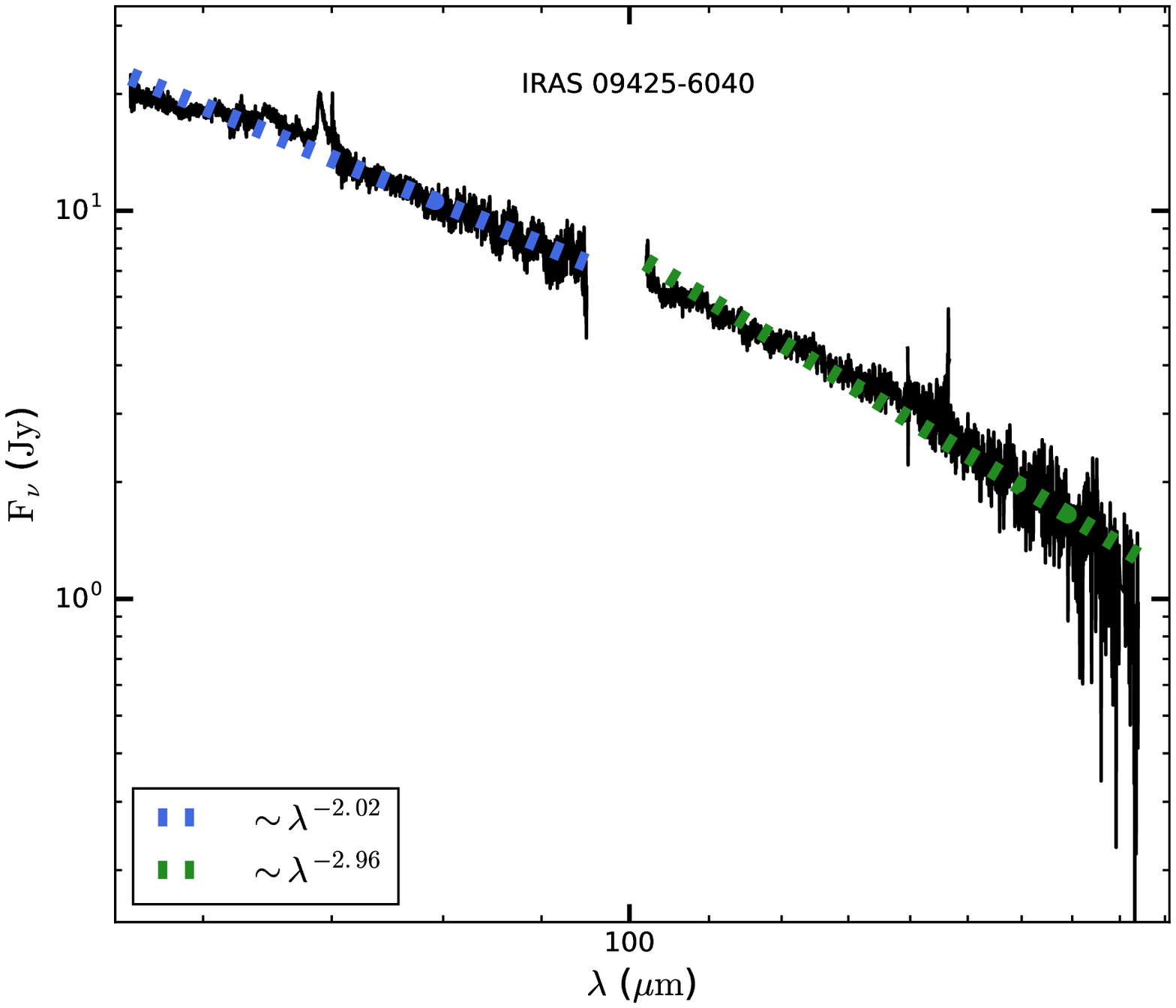} 
 \end{subfigure}
  \begin{subfigure}{0.49\textwidth}
 \centering
 \includegraphics[width = \textwidth]{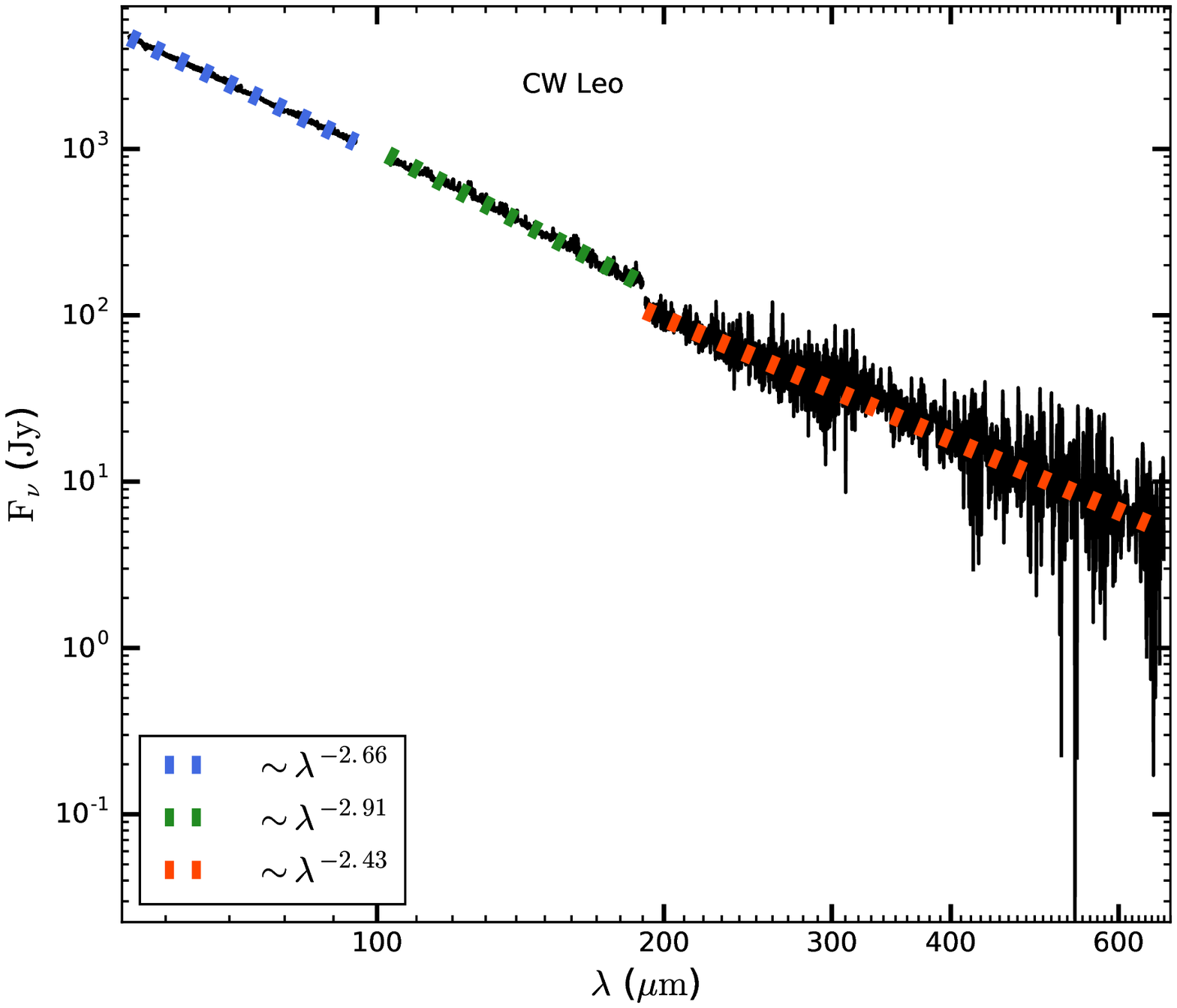}
 \end{subfigure}
\hfill
  \begin{subfigure}{0.49\textwidth}
 \centering
 \includegraphics[width = \textwidth]{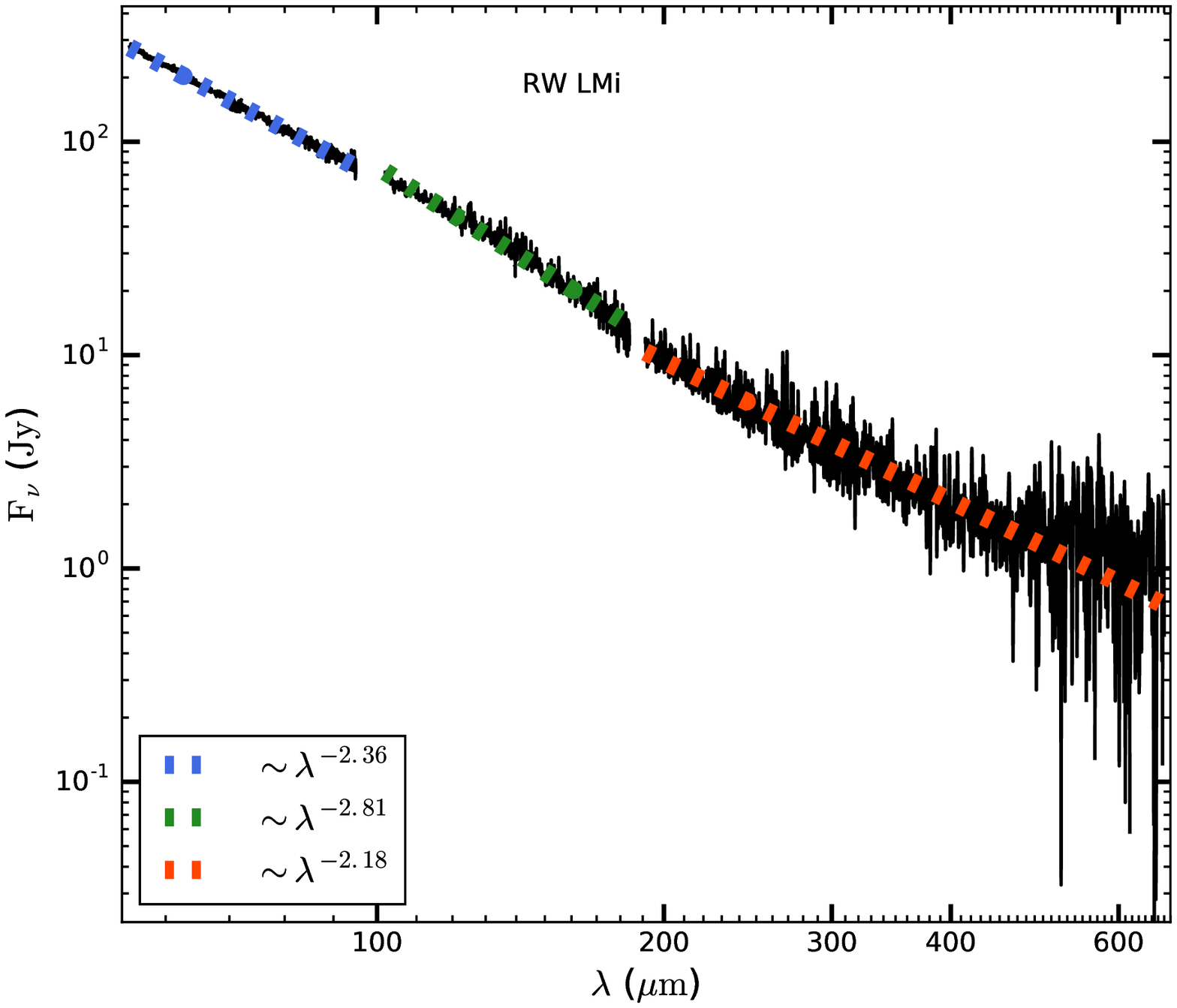}          % RW LMi
 \end{subfigure}
   \begin{subfigure}{0.49\textwidth}
 \centering
 \includegraphics[width = \textwidth]{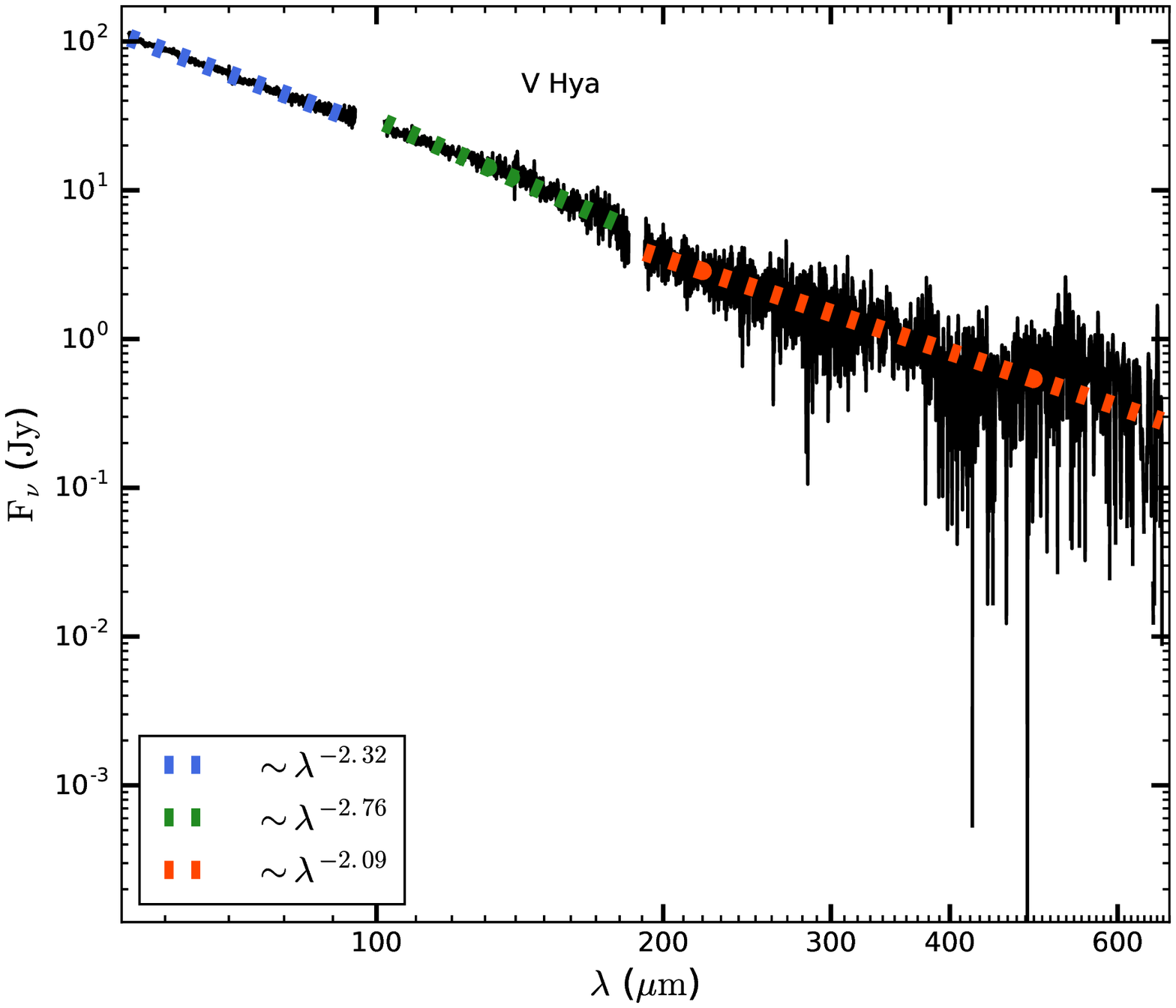}
 \end{subfigure}
\hfill
 \begin{subfigure}{0.49\textwidth}
 \centering
 \includegraphics[width = \textwidth]{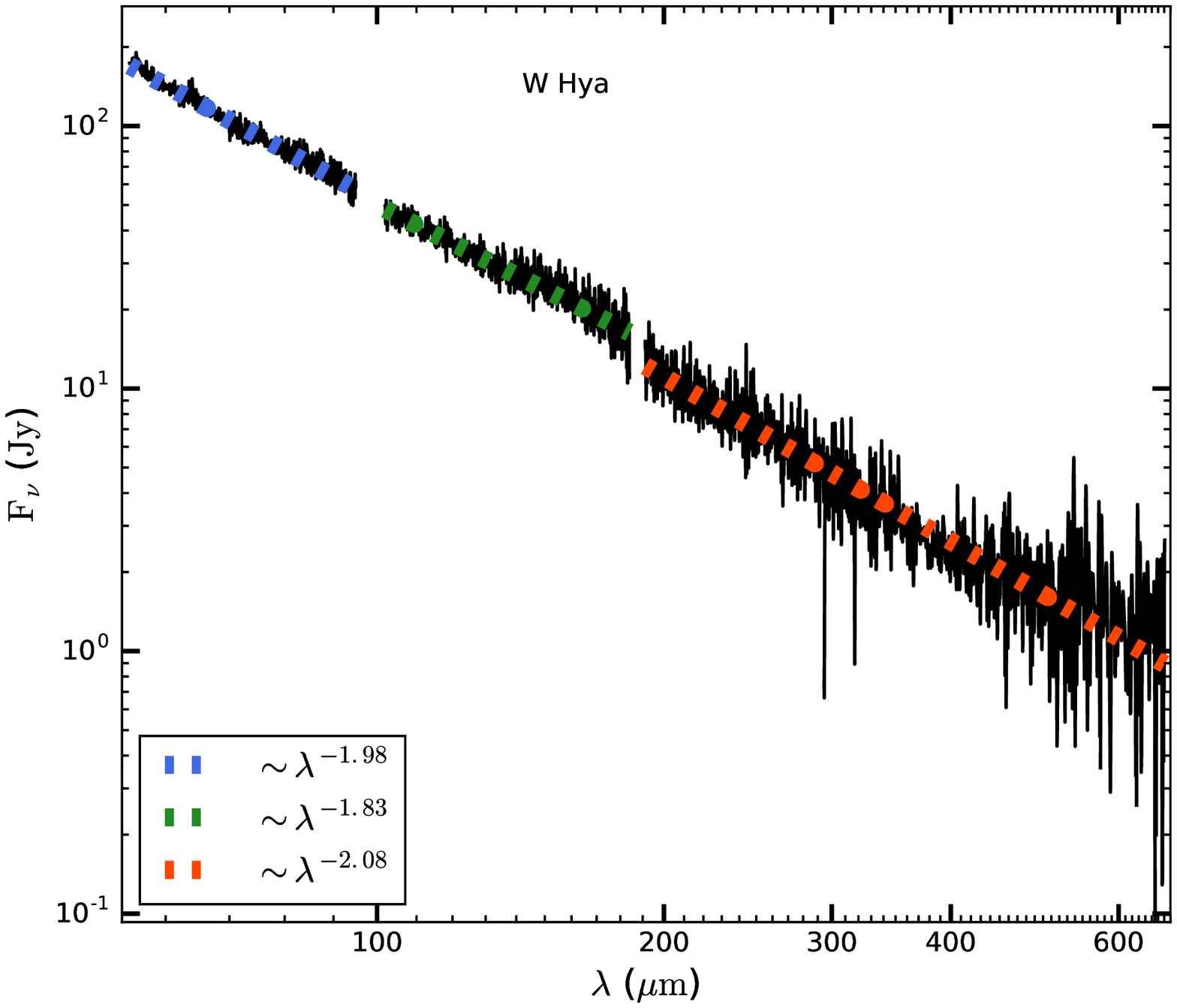} 
 \end{subfigure} 
  \caption{Continued.}
\end{figure*}

\begin{figure*}
\ContinuedFloat
  \begin{subfigure}{0.49\textwidth}
 \centering
 \includegraphics[width = \textwidth]{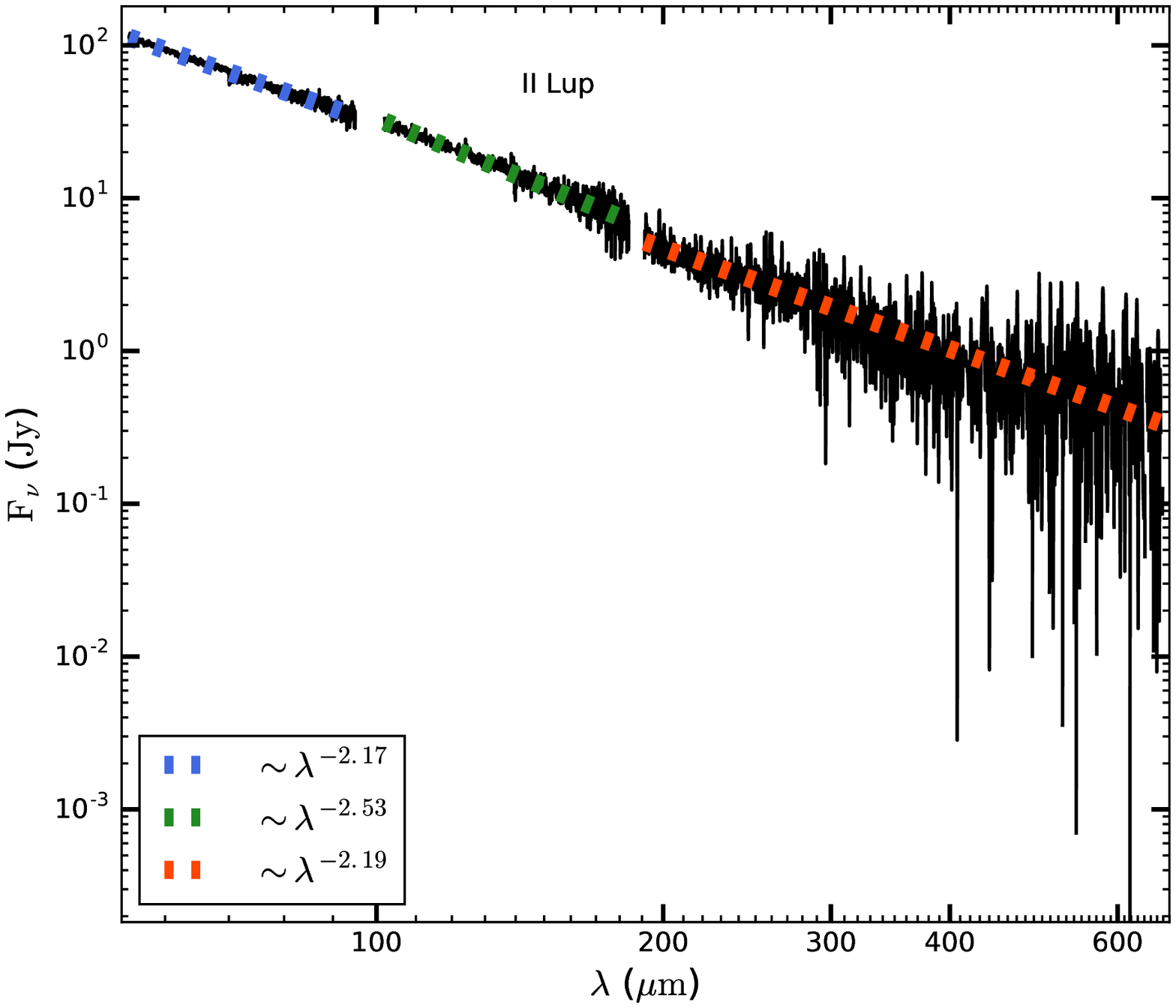} % II Lup   
 \end{subfigure}
\hfill
 \begin{subfigure}{0.49\textwidth}
 \centering
 \includegraphics[width = \textwidth]{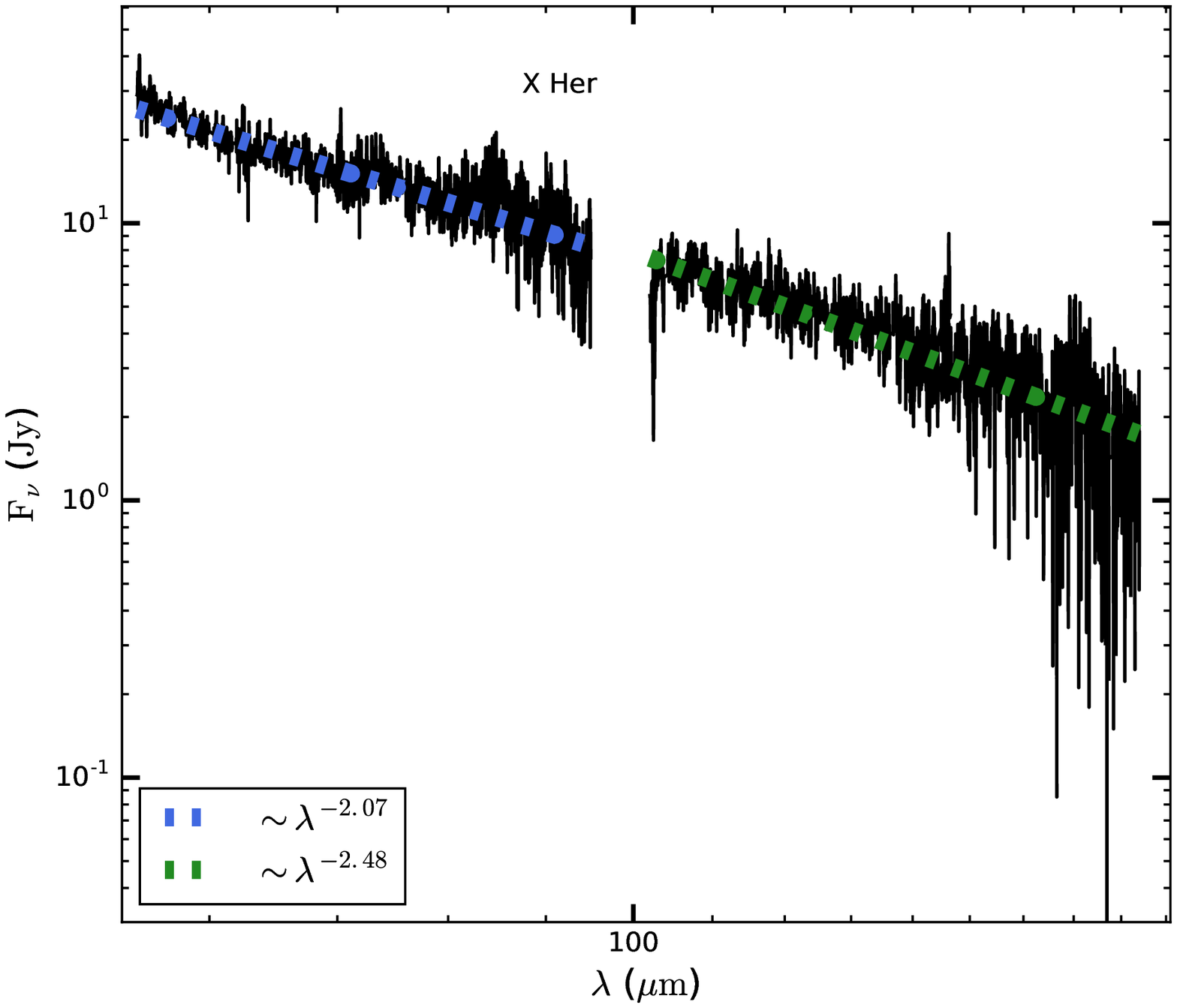}
 \end{subfigure}
 \begin{subfigure}{0.49\textwidth}
 \centering
 \includegraphics[width = \textwidth]{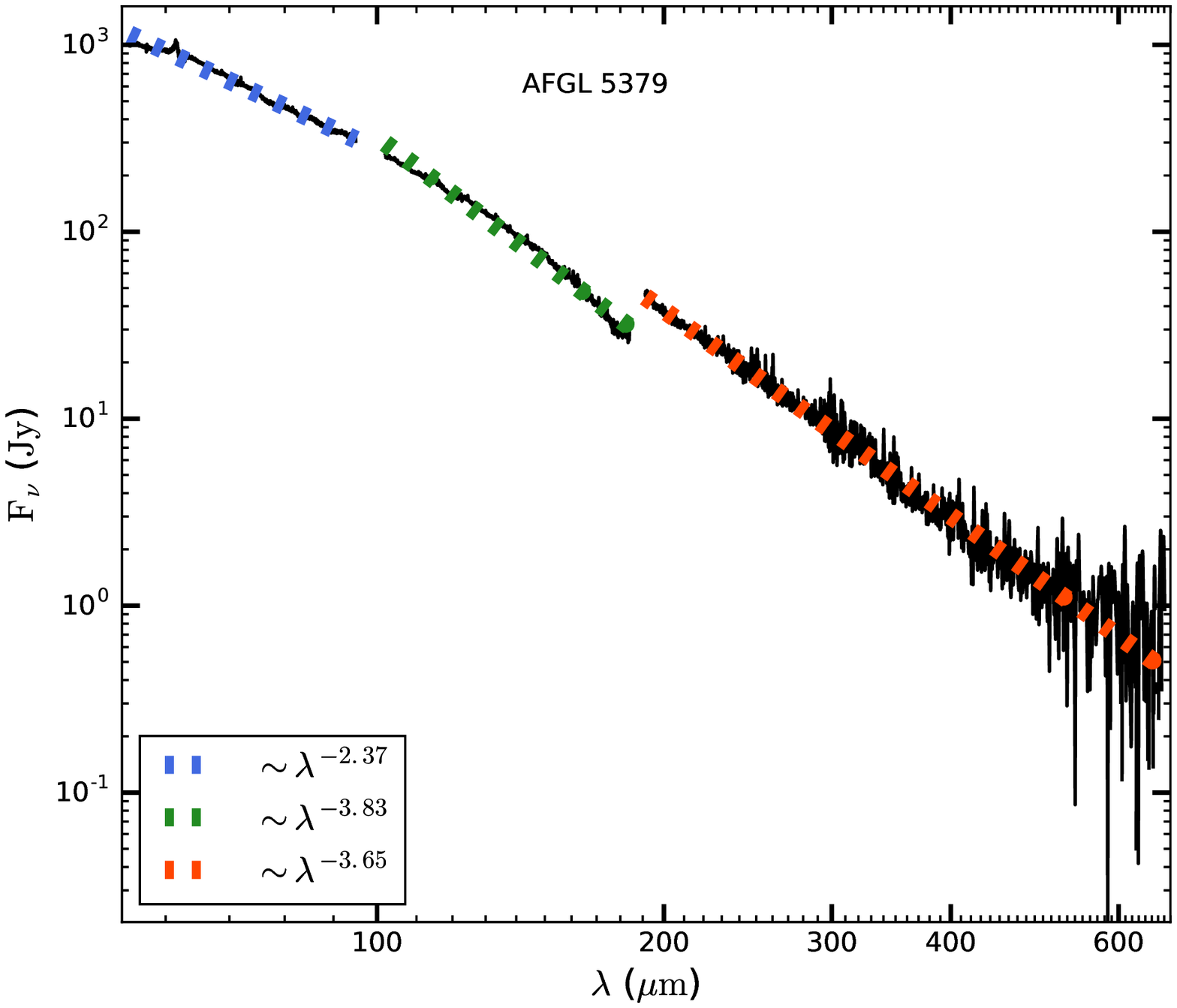}
 \end{subfigure}
\hfill
\begin{subfigure}{0.49\textwidth}
 \centering
 \includegraphics[width = \textwidth]{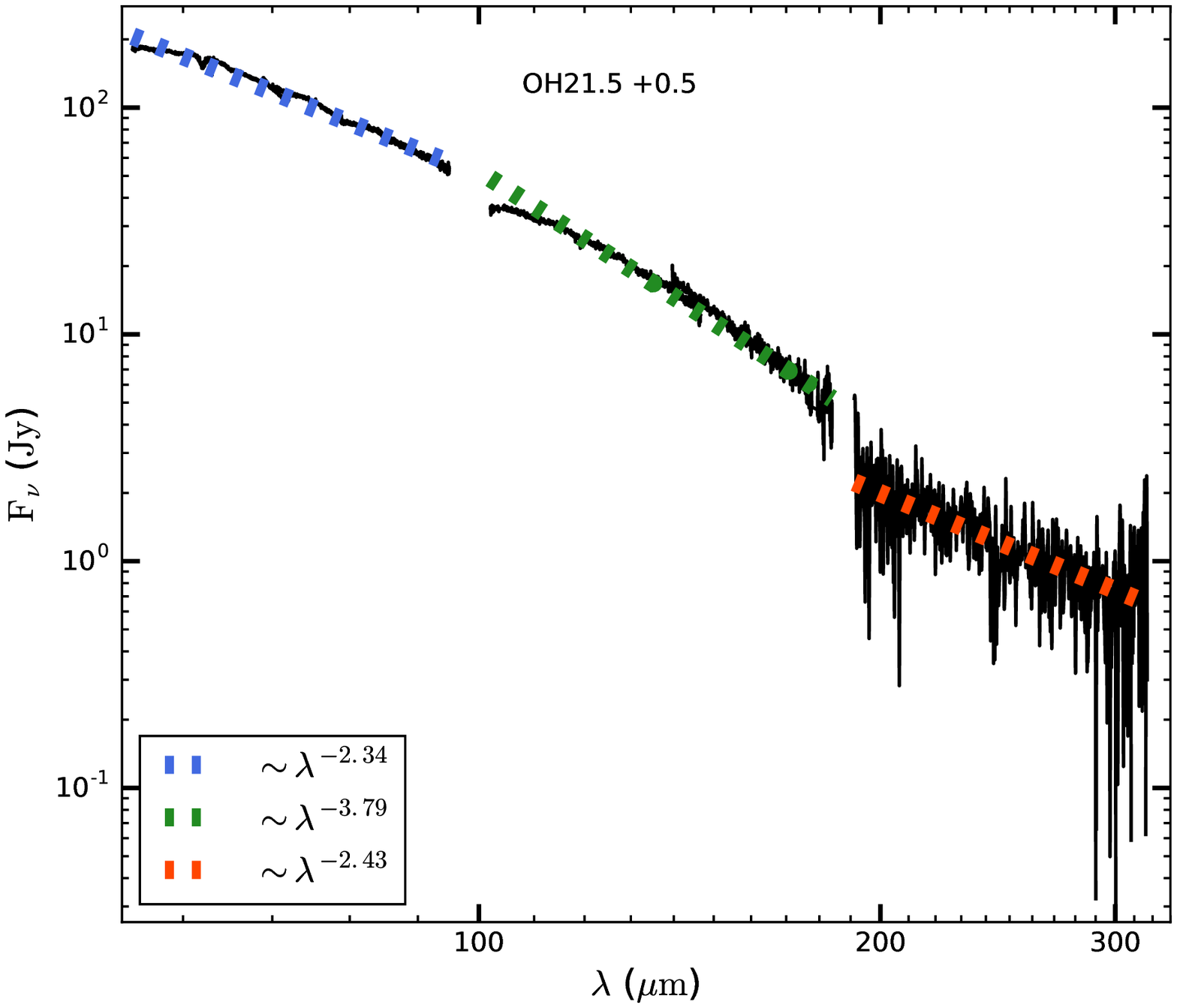} 
 \end{subfigure}
   \begin{subfigure}{0.49\textwidth}
 \centering
 \includegraphics[width = \textwidth]{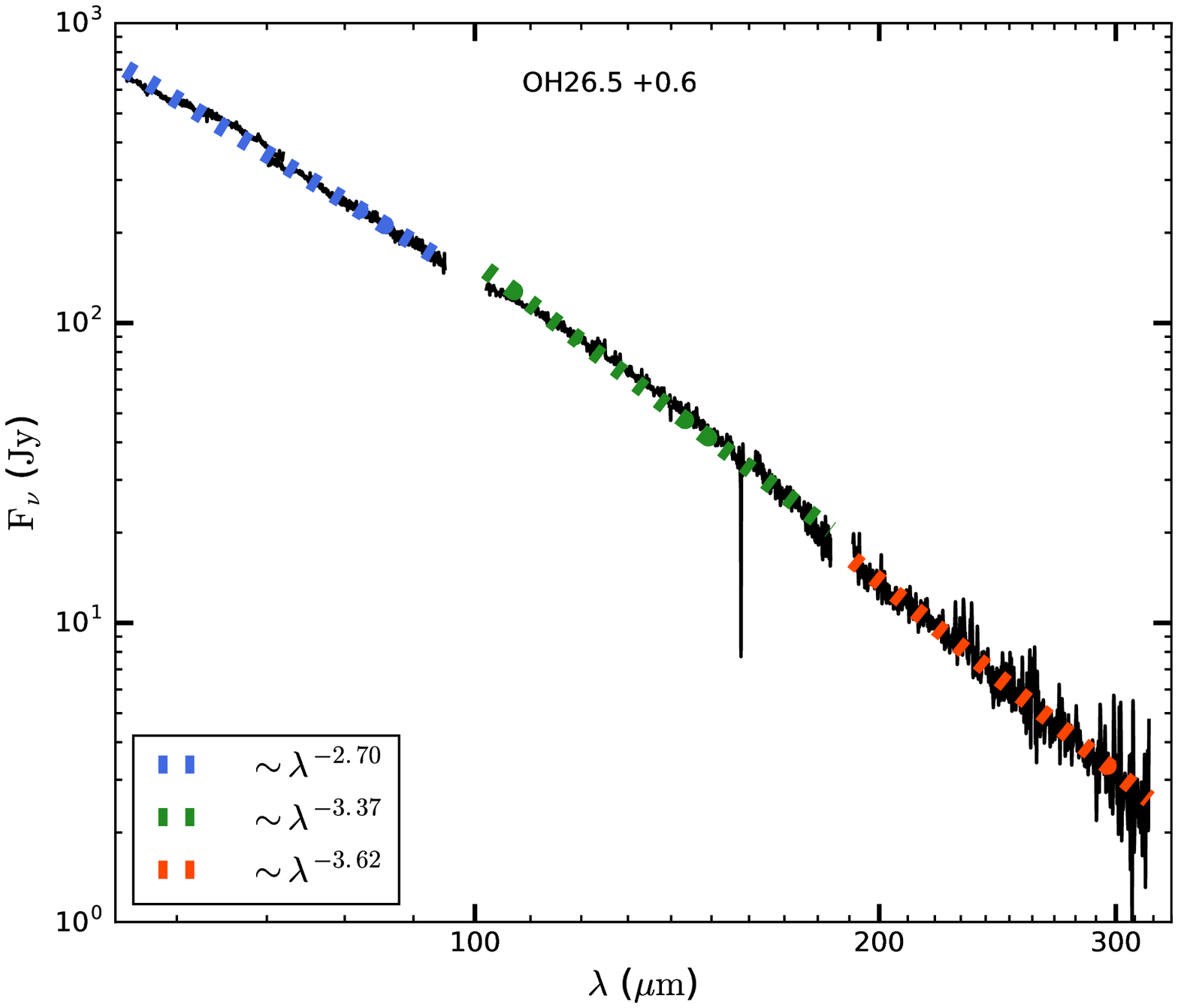} 
 \end{subfigure}
 \hfill 
  \begin{subfigure}{0.49\textwidth}
 \centering
 \includegraphics[width = \textwidth]{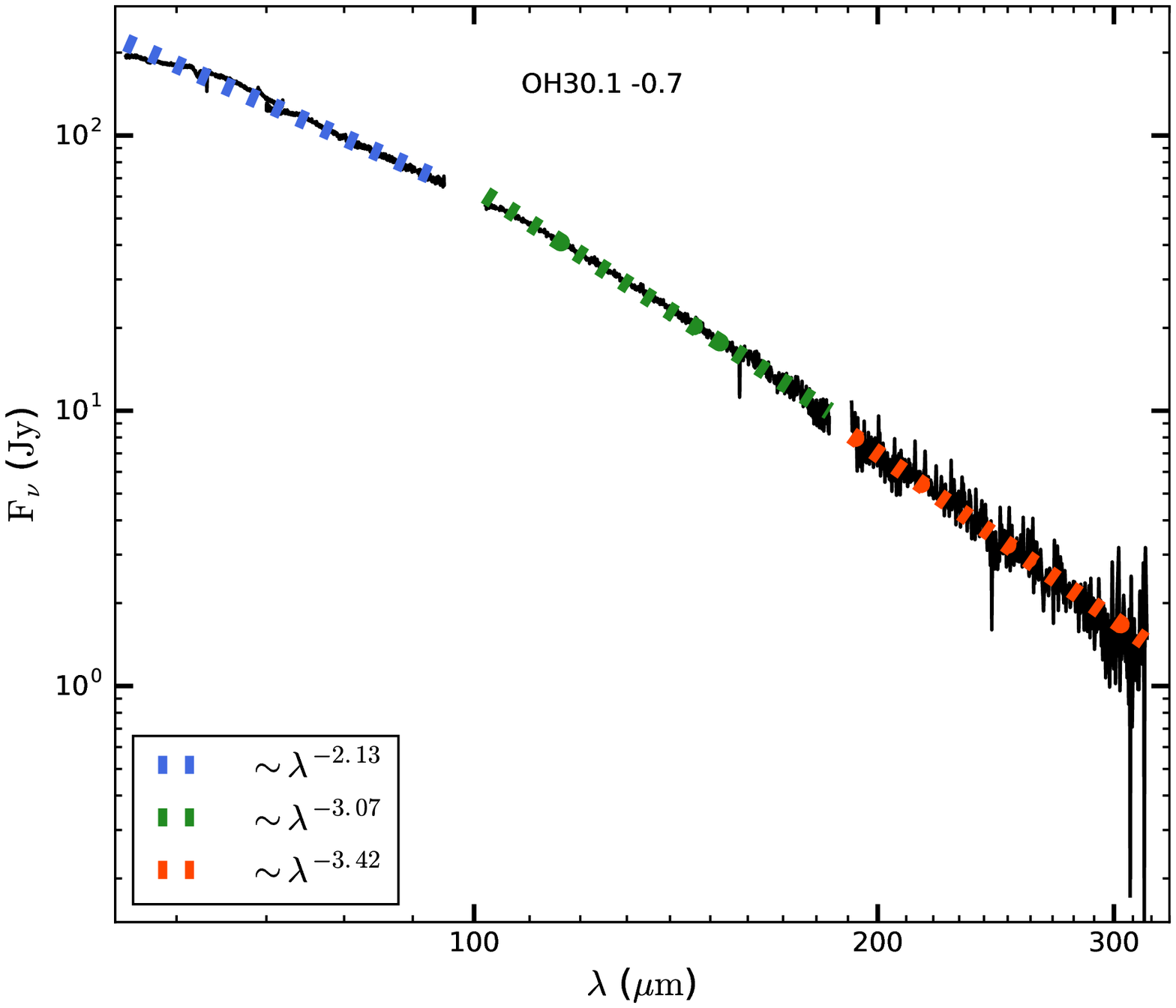}
 \end{subfigure}
 \caption{Continued.}
\end{figure*}

\begin{figure*}
\ContinuedFloat  
 \begin{subfigure}{0.49\textwidth}
 \centering
 \includegraphics[width = \textwidth]{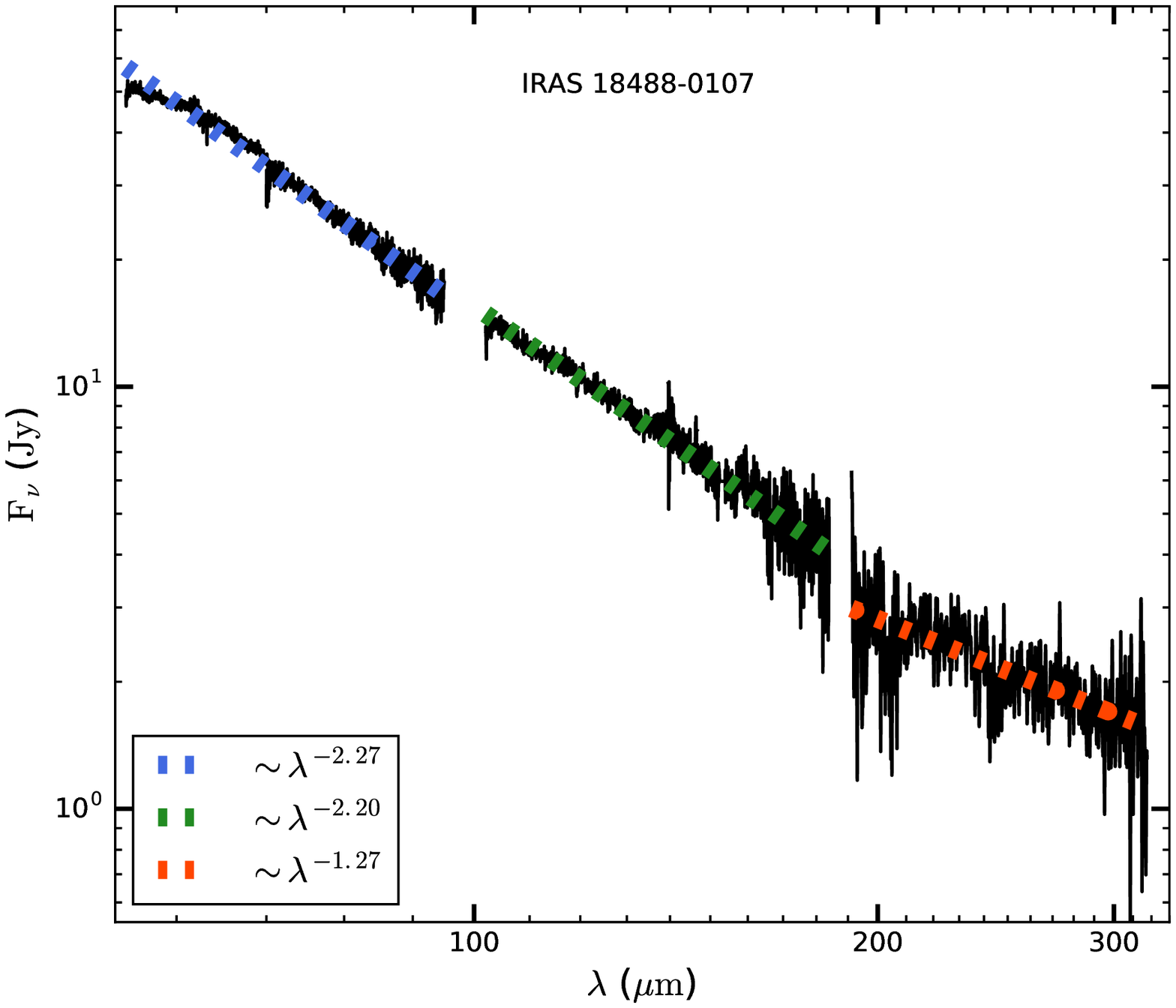} 
 \end{subfigure}
 \hfill 
     \begin{subfigure}{0.49\textwidth}
 \centering
 \includegraphics[width = \textwidth]{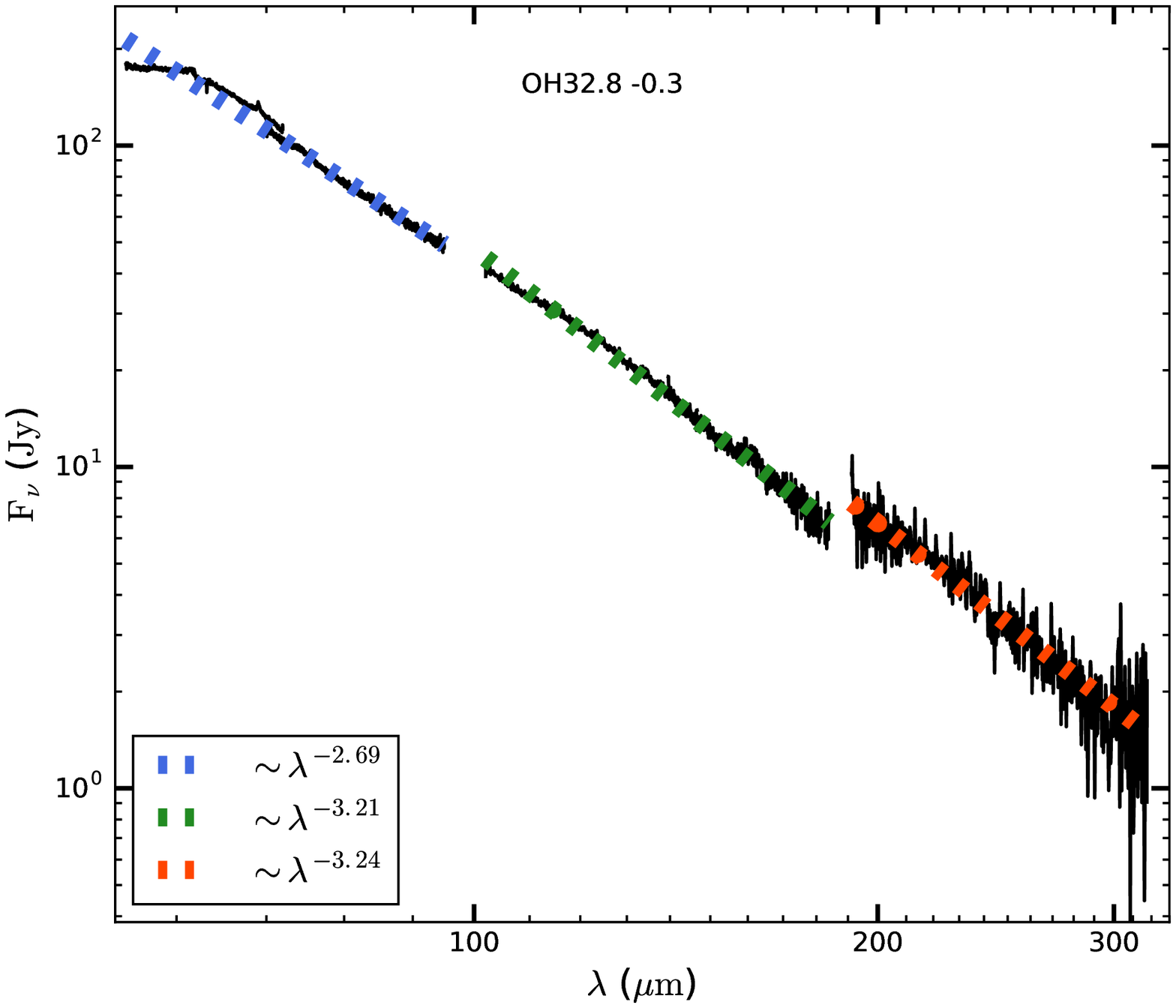} 
 \end{subfigure}
  \begin{subfigure}{0.49\textwidth}
 \centering
 \includegraphics[width = \textwidth]{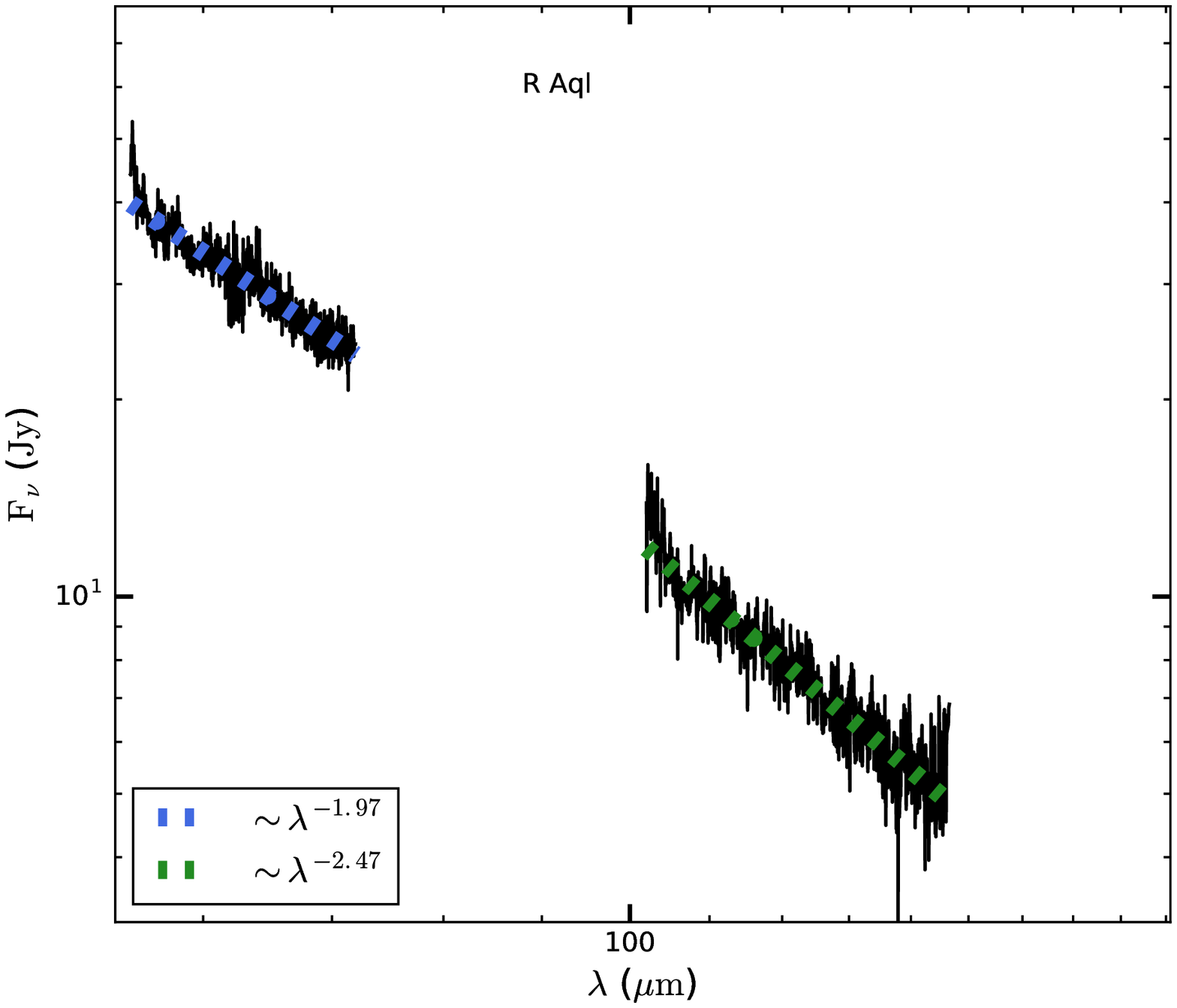} 
 \end{subfigure}
 \hfill 
     \begin{subfigure}{0.49\textwidth}
 \centering
 \includegraphics[width = \textwidth]{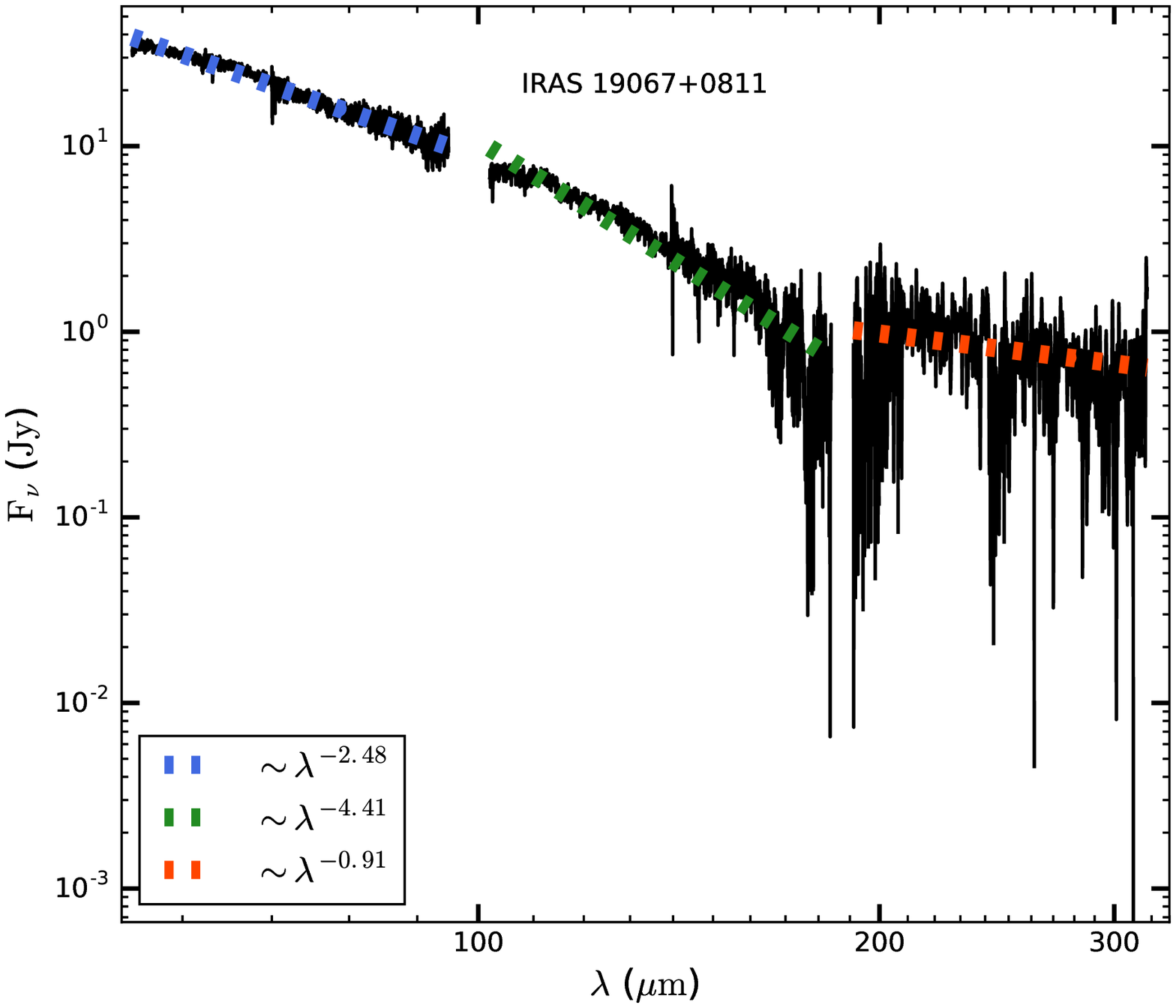} 
 \end{subfigure}
 \begin{subfigure}{0.49\textwidth}
 \centering
 \includegraphics[width = \textwidth]{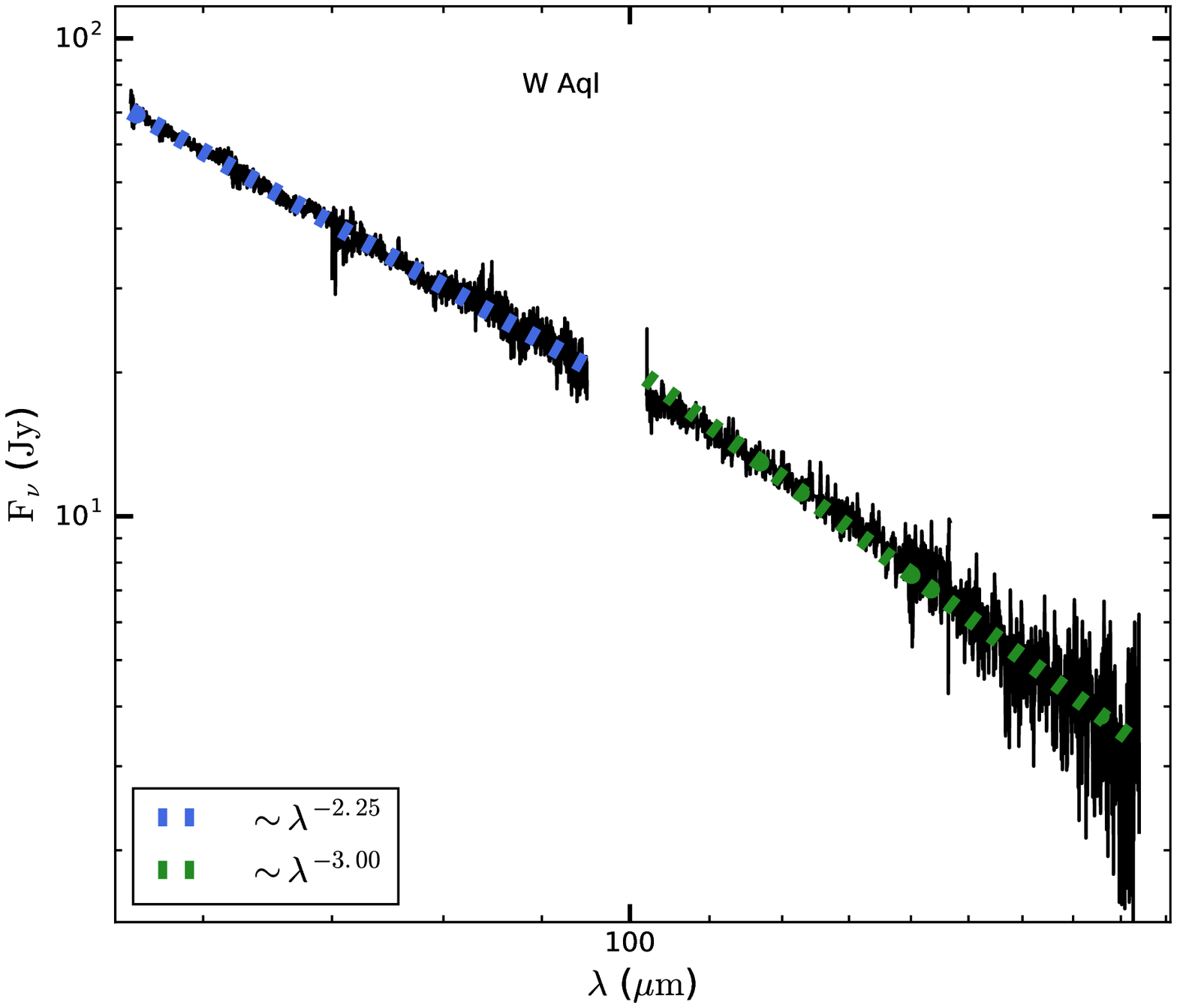} 
 \end{subfigure}
 \hfill  
  \begin{subfigure}{0.49\textwidth}
 \centering
 \includegraphics[width = \textwidth]{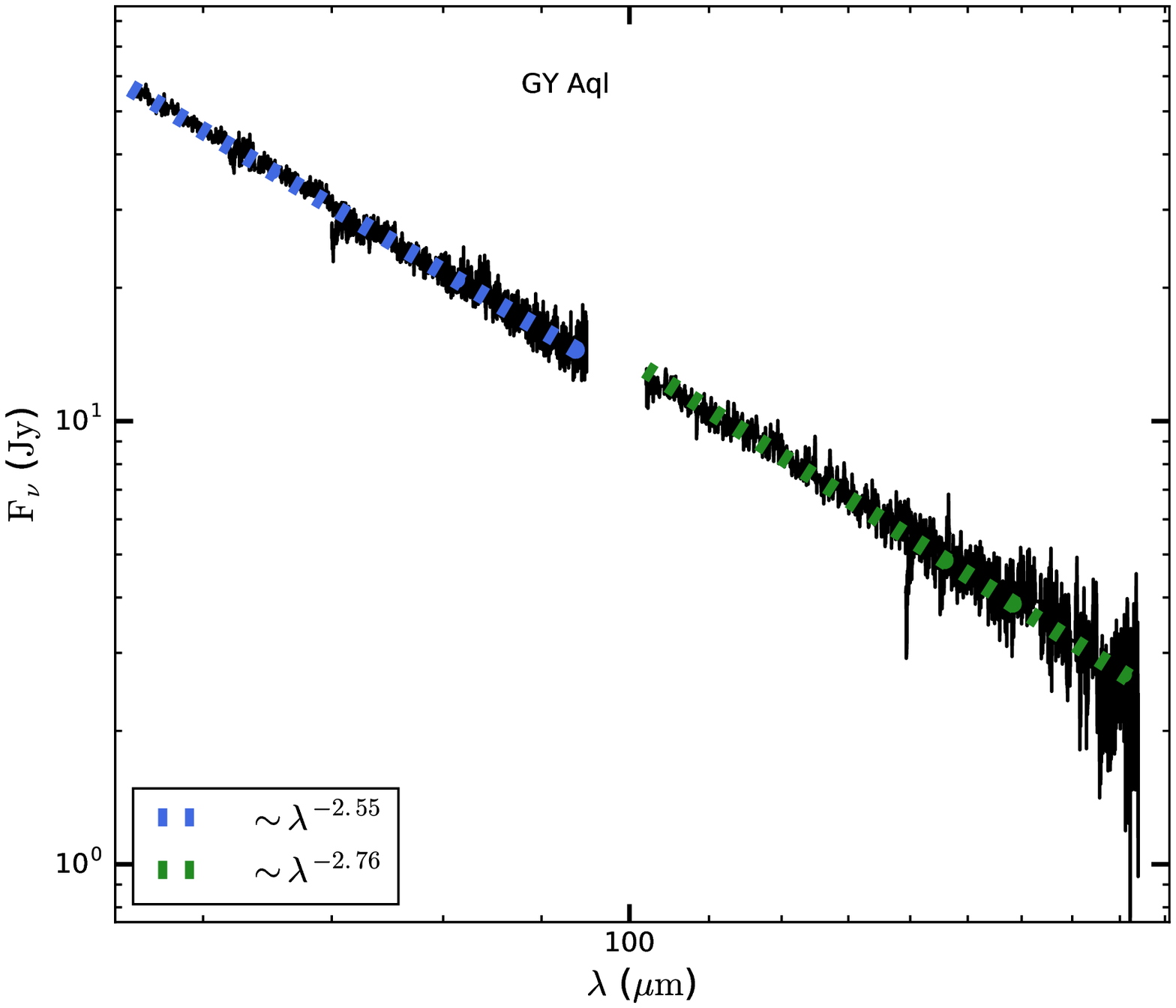} %GY Aql       
 \end{subfigure}
  \caption{Continued.}
\end{figure*}

\begin{figure*}
\ContinuedFloat  
  \begin{subfigure}{0.49\textwidth}
 \centering
 \includegraphics[width = \textwidth]{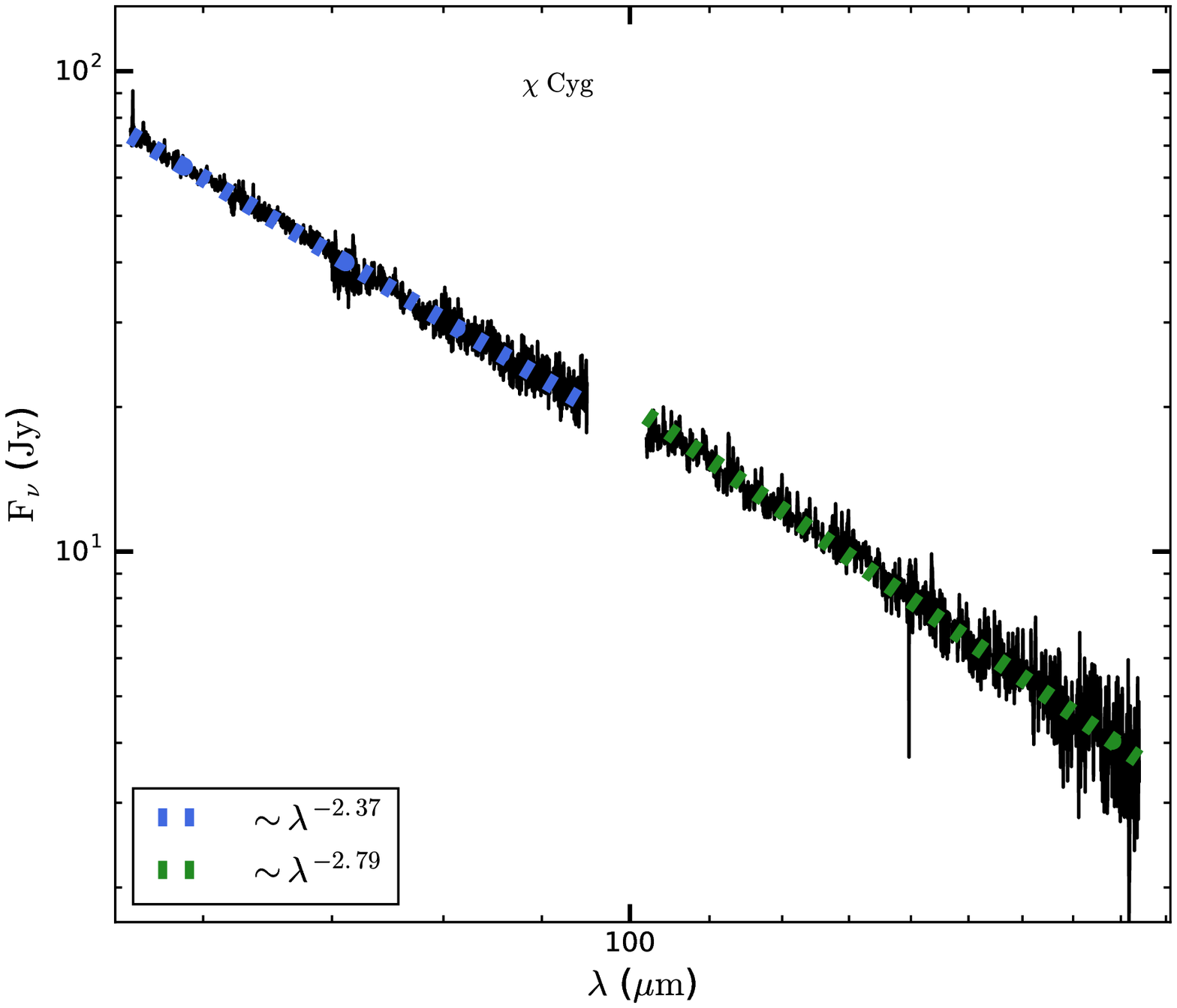}  
 \end{subfigure}
 \hfill 
   \begin{subfigure}{0.49\textwidth}
 \centering
 \includegraphics[width = \textwidth]{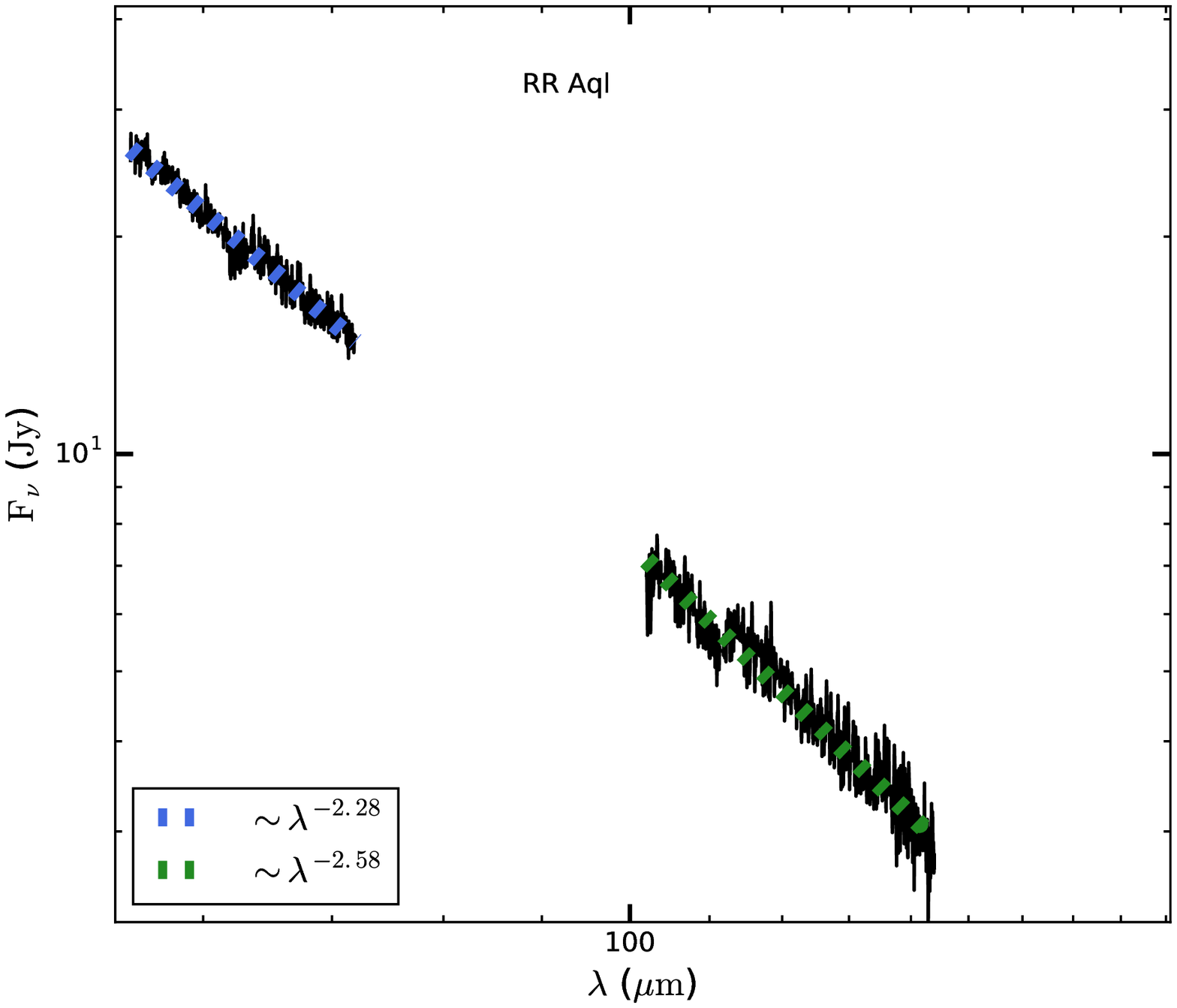}  
 \end{subfigure}
 \begin{subfigure}{0.49\textwidth}
 \centering
 \includegraphics[width = \textwidth]{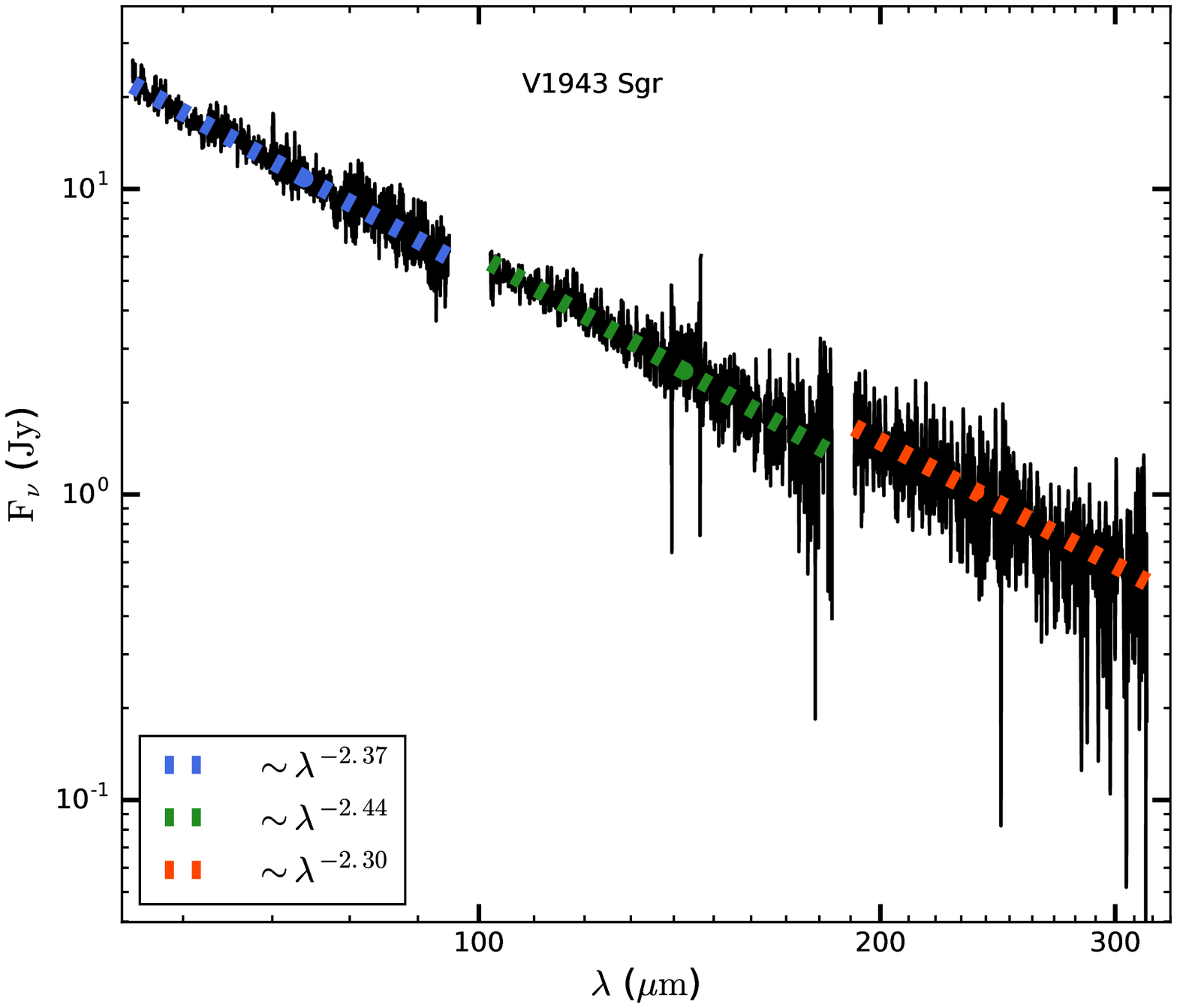} %V1943  Sgr
 \end{subfigure}
 \hfill 
 \begin{subfigure}{0.49\textwidth}
 \centering
 \includegraphics[width = \textwidth]{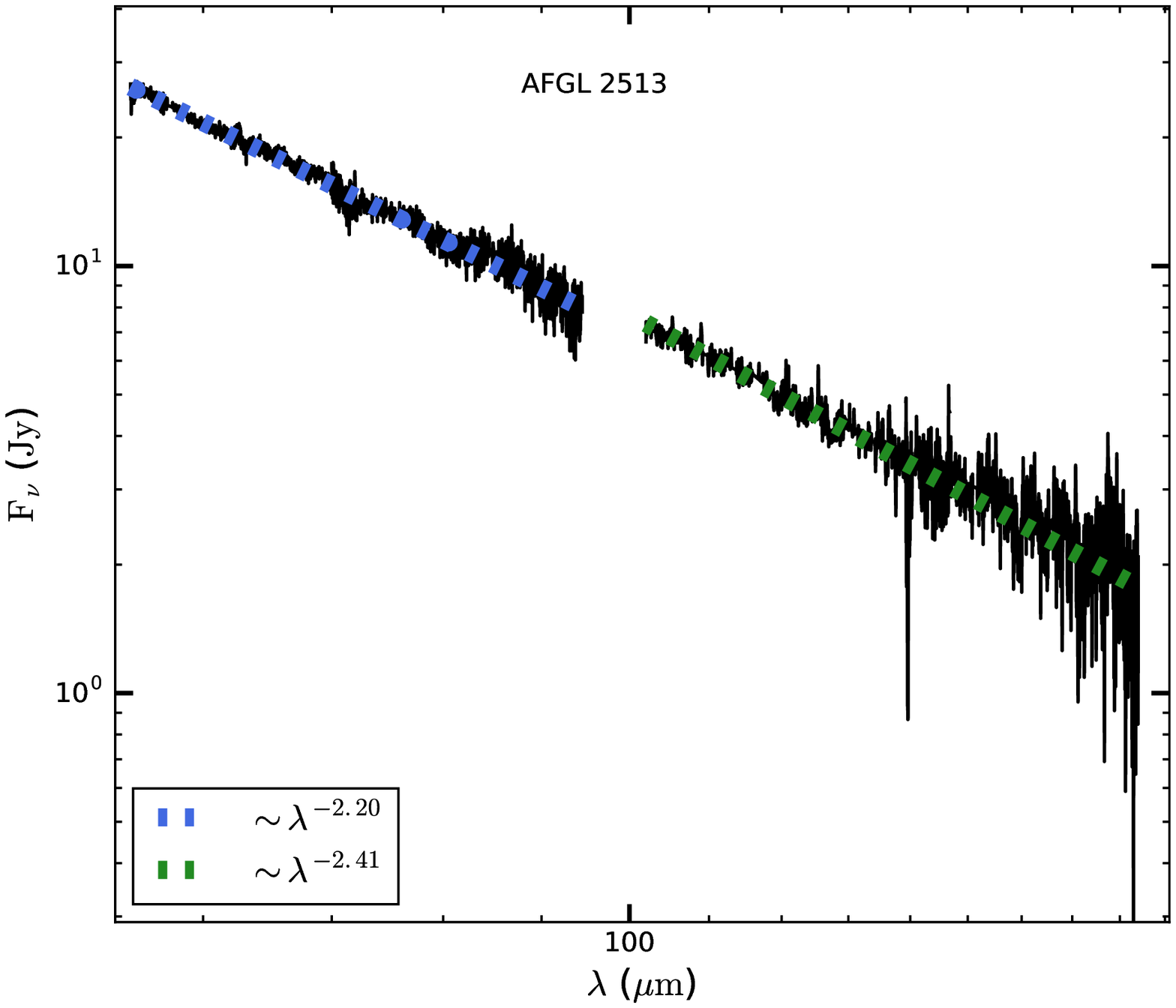}
 \end{subfigure}
 \begin{subfigure}{0.49\textwidth}
 \centering
 \includegraphics[width = \textwidth]{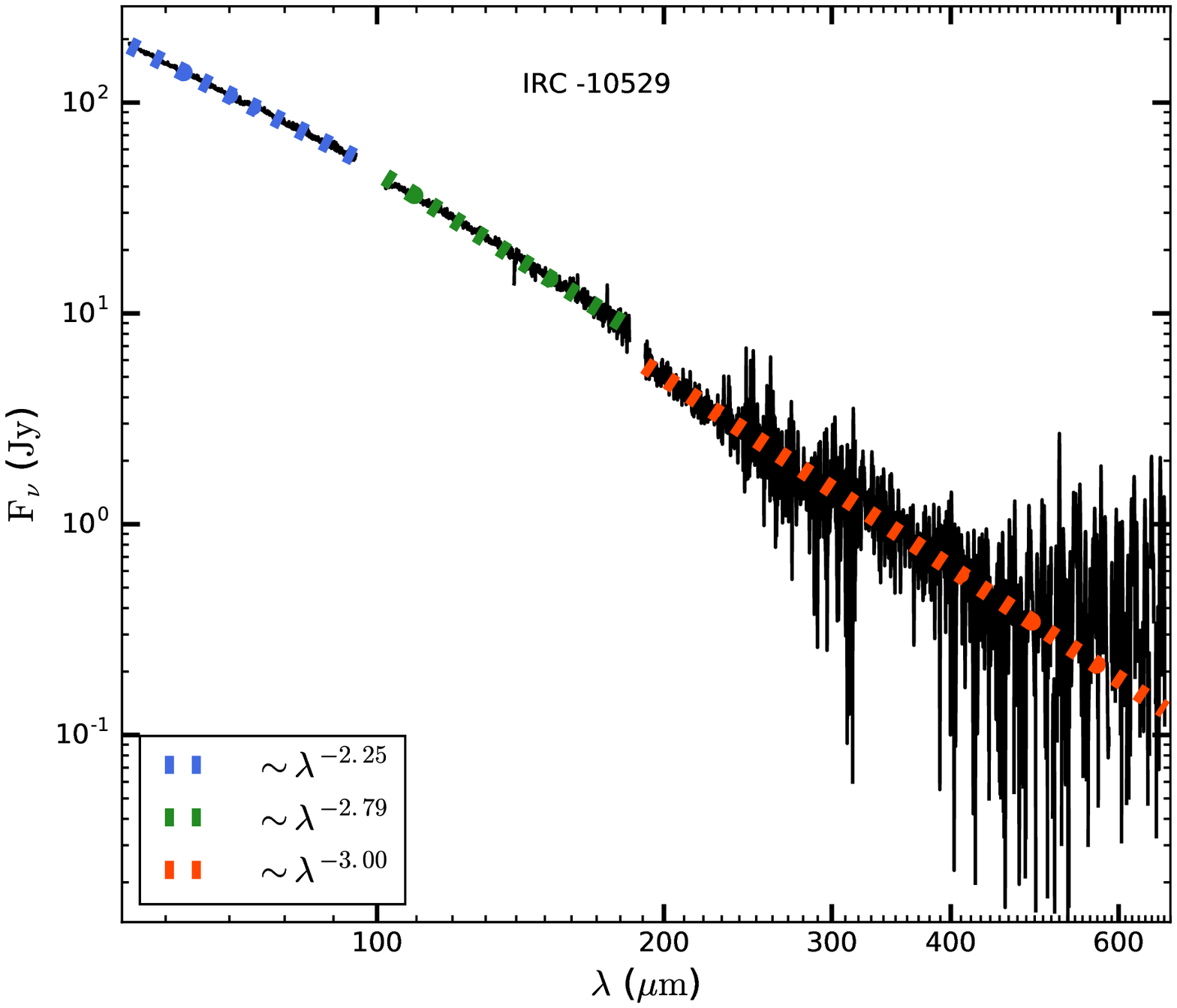} % irc -10 529 
 \end{subfigure}
\hfill
 \begin{subfigure}{0.49\textwidth}
 \centering
 \includegraphics[width = \textwidth]{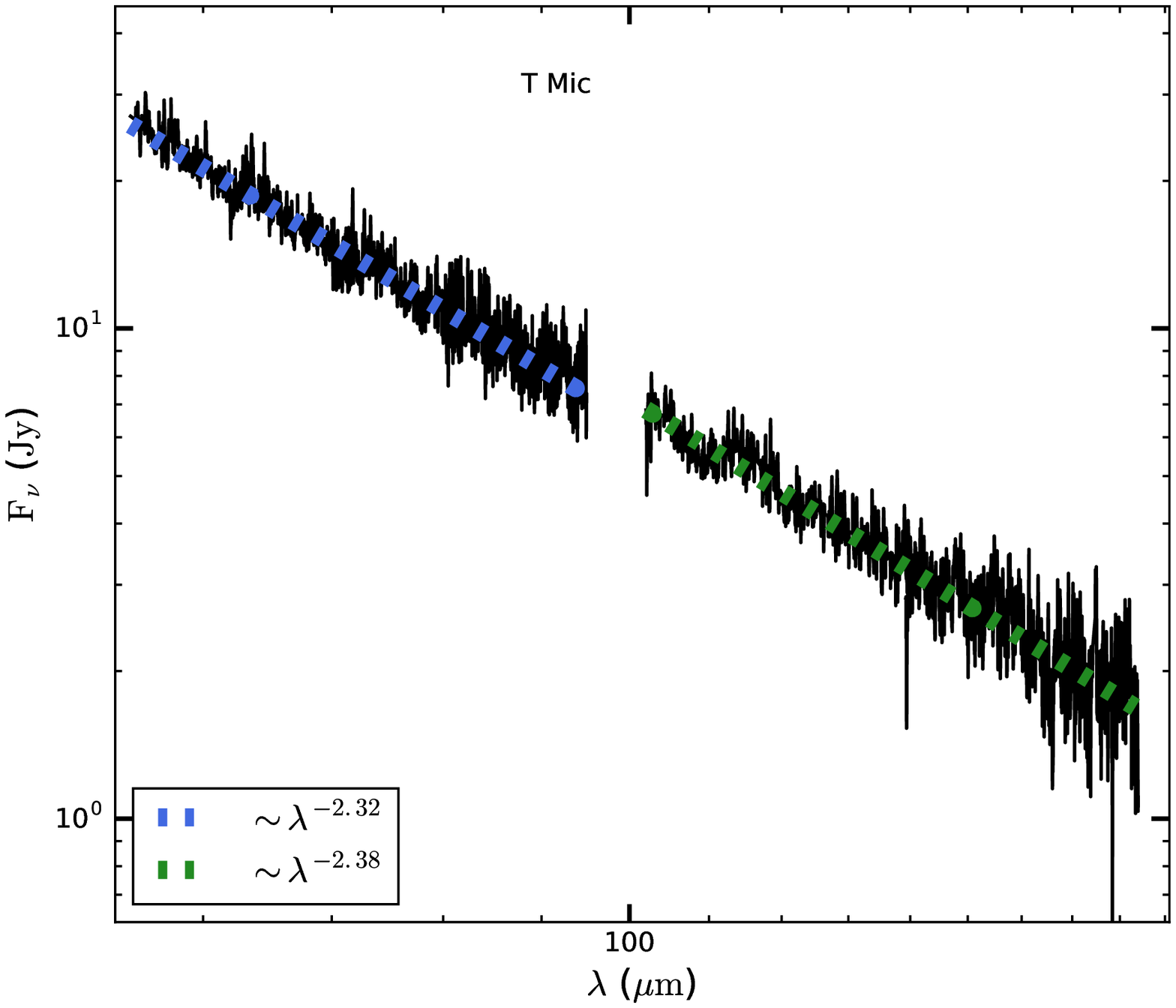}  %T Mic   
 \end{subfigure}
   \caption{Continued.}
\end{figure*}

\begin{figure*}
\ContinuedFloat   
  \begin{subfigure}{0.49\textwidth}
 \centering
 \includegraphics[width = \textwidth]{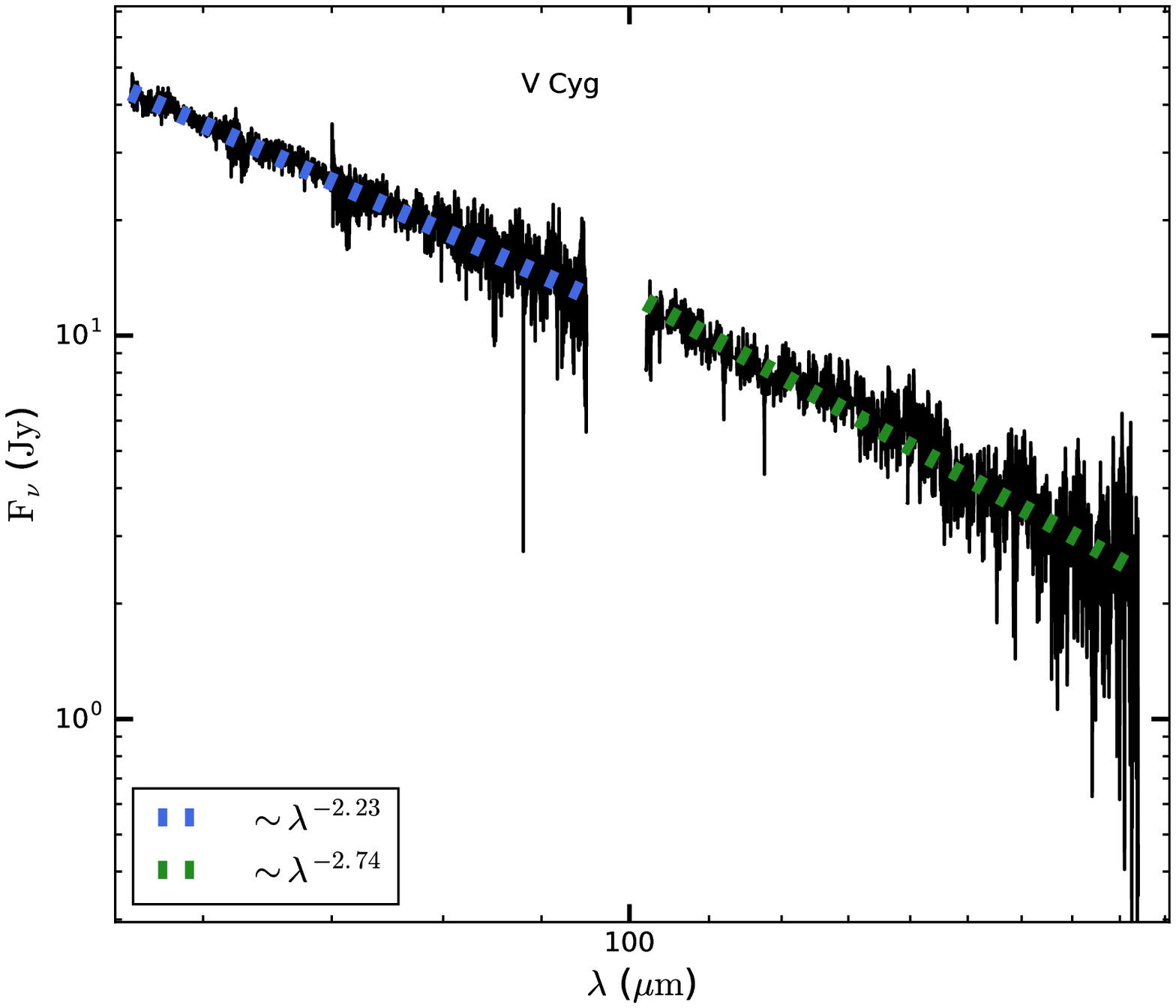} 
 \end{subfigure}
\hfill
 \begin{subfigure}{0.49\textwidth}
 \centering
 \includegraphics[width = \textwidth]{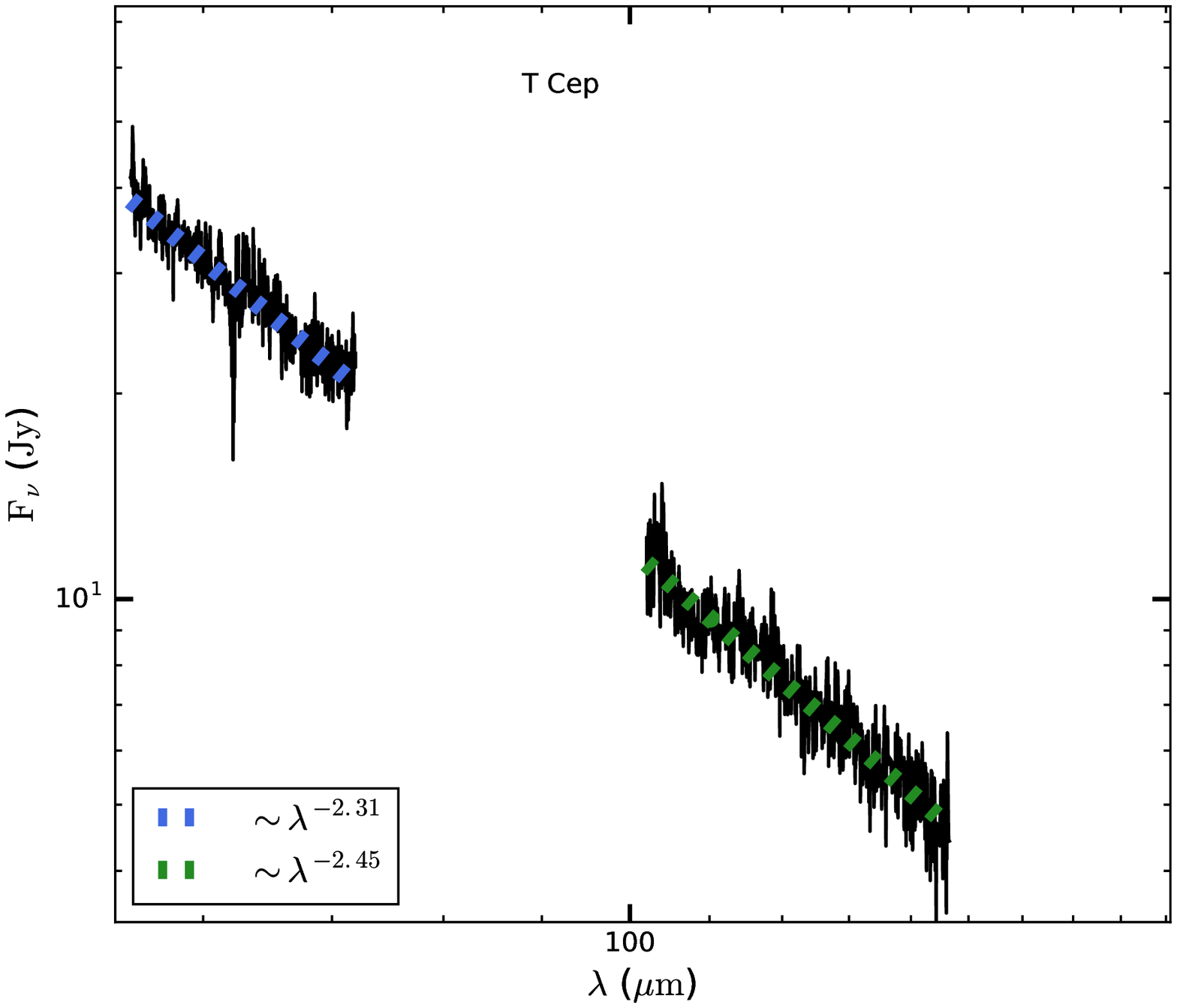} 
 \end{subfigure}
   \begin{subfigure}{0.49\textwidth}
 \centering
 \includegraphics[width = \textwidth]{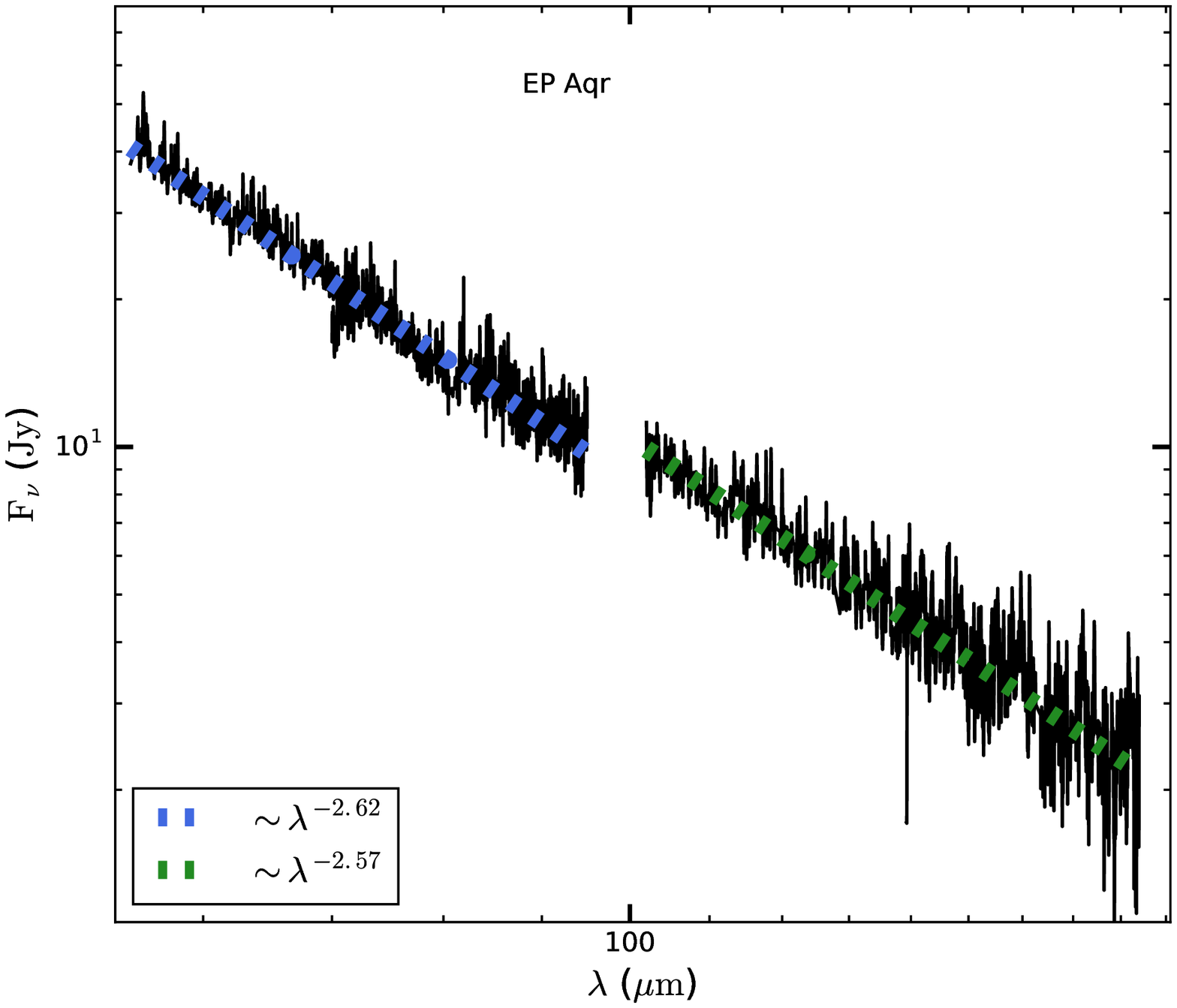} 
 \end{subfigure}
\hfill
 \begin{subfigure}{0.49\textwidth}
 \centering
 \includegraphics[width = \textwidth]{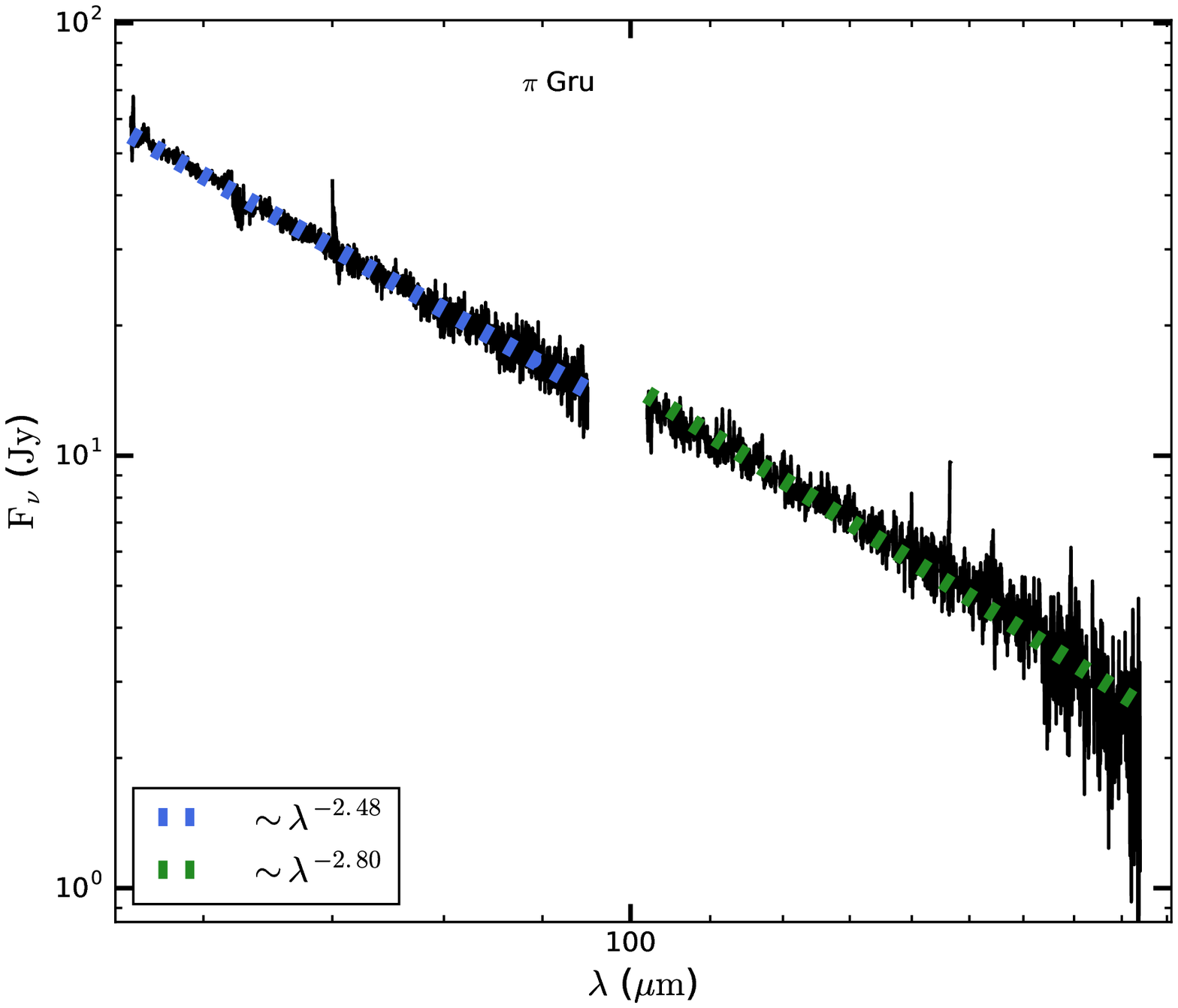}
 \end{subfigure}
   \begin{subfigure}{0.49\textwidth}
 \centering
 \includegraphics[width = \textwidth]{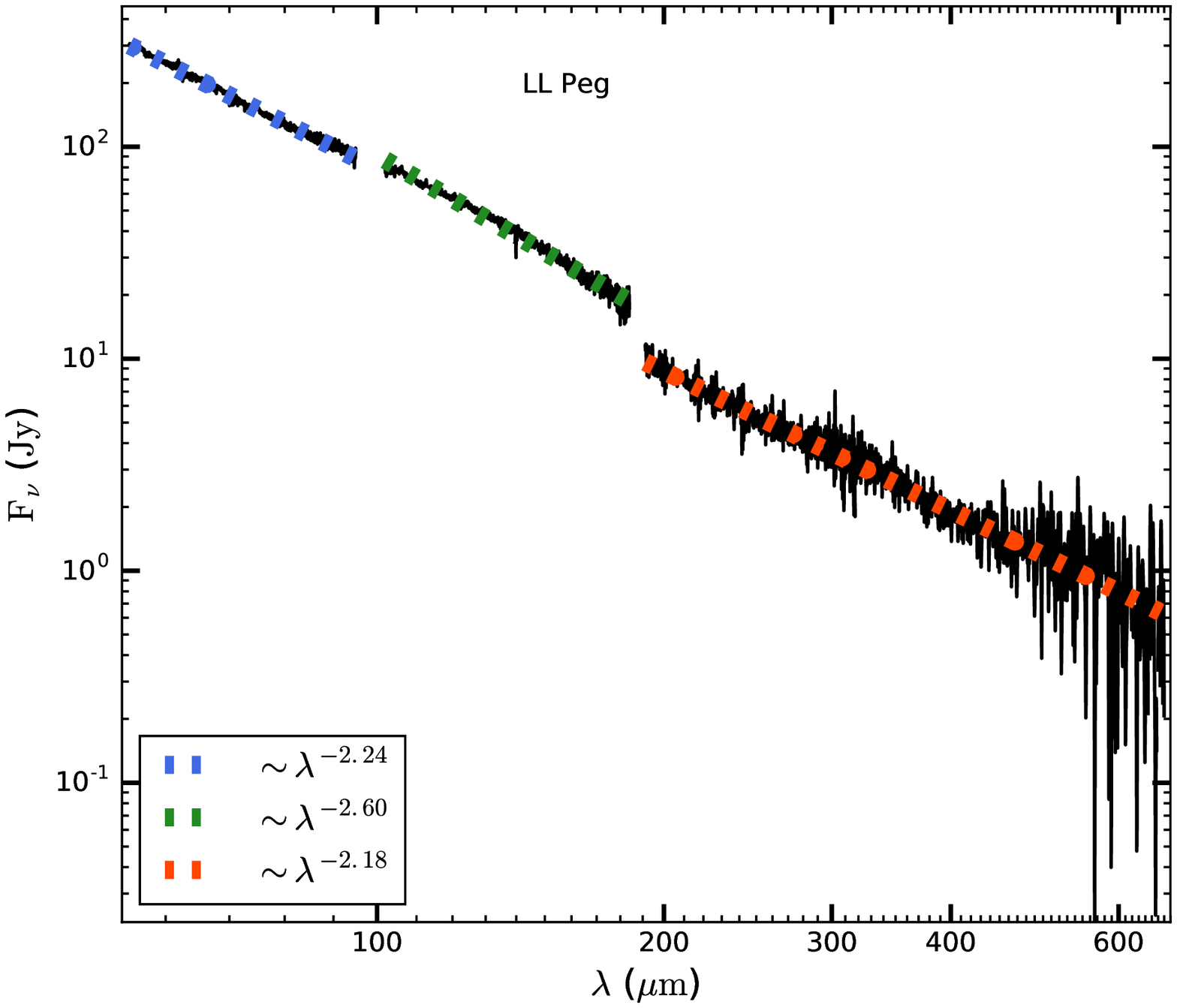} % LL Peg
 \end{subfigure}
\hfill
   \begin{subfigure}{0.49\textwidth}
 \centering
 \includegraphics[width = \textwidth]{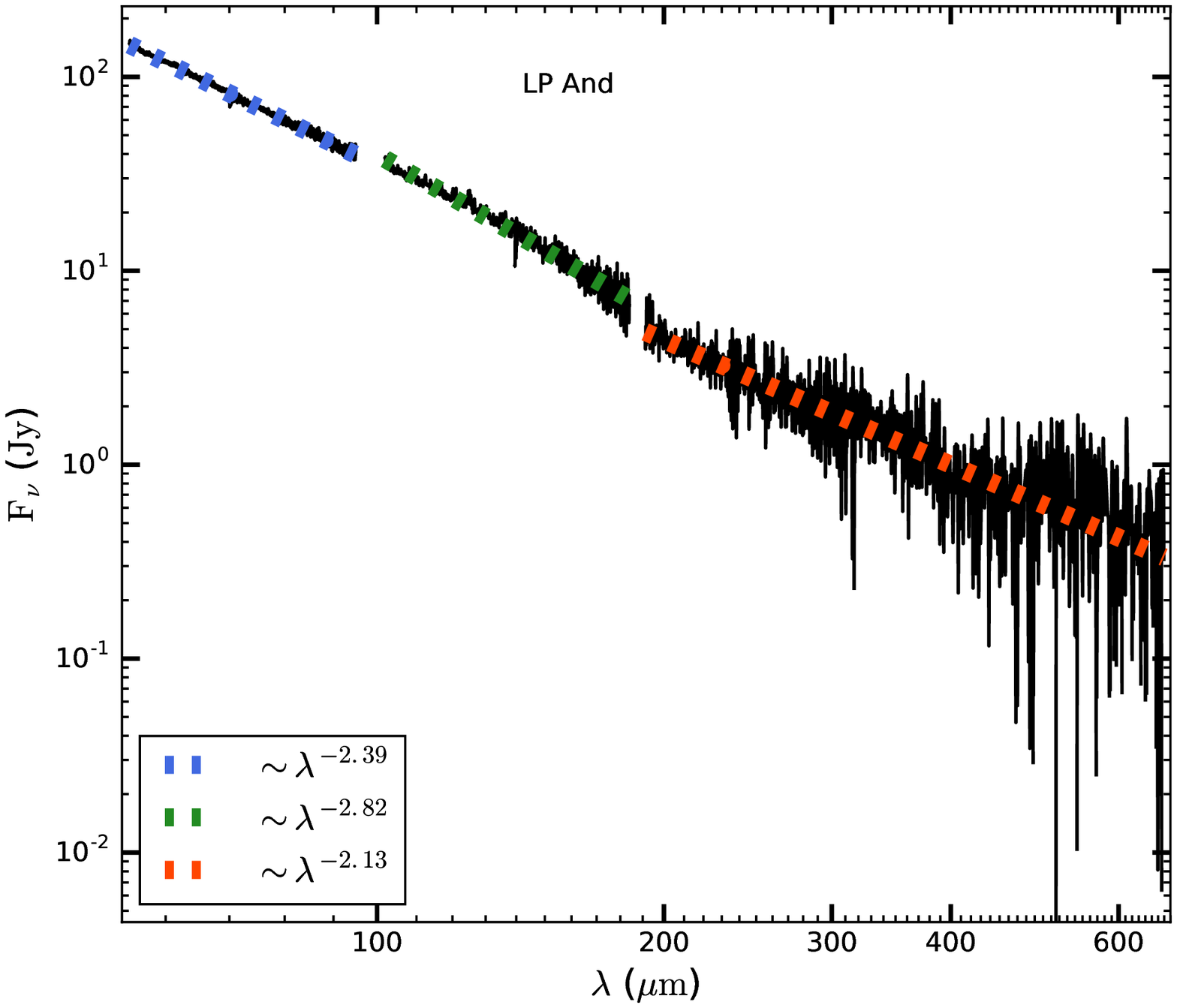}  % LP And
 \end{subfigure}
   \caption{Continued.}
\end{figure*}
 
 \begin{figure*}
\ContinuedFloat  
 \begin{subfigure}{0.49\textwidth}
 \centering
 \includegraphics[width = \textwidth]{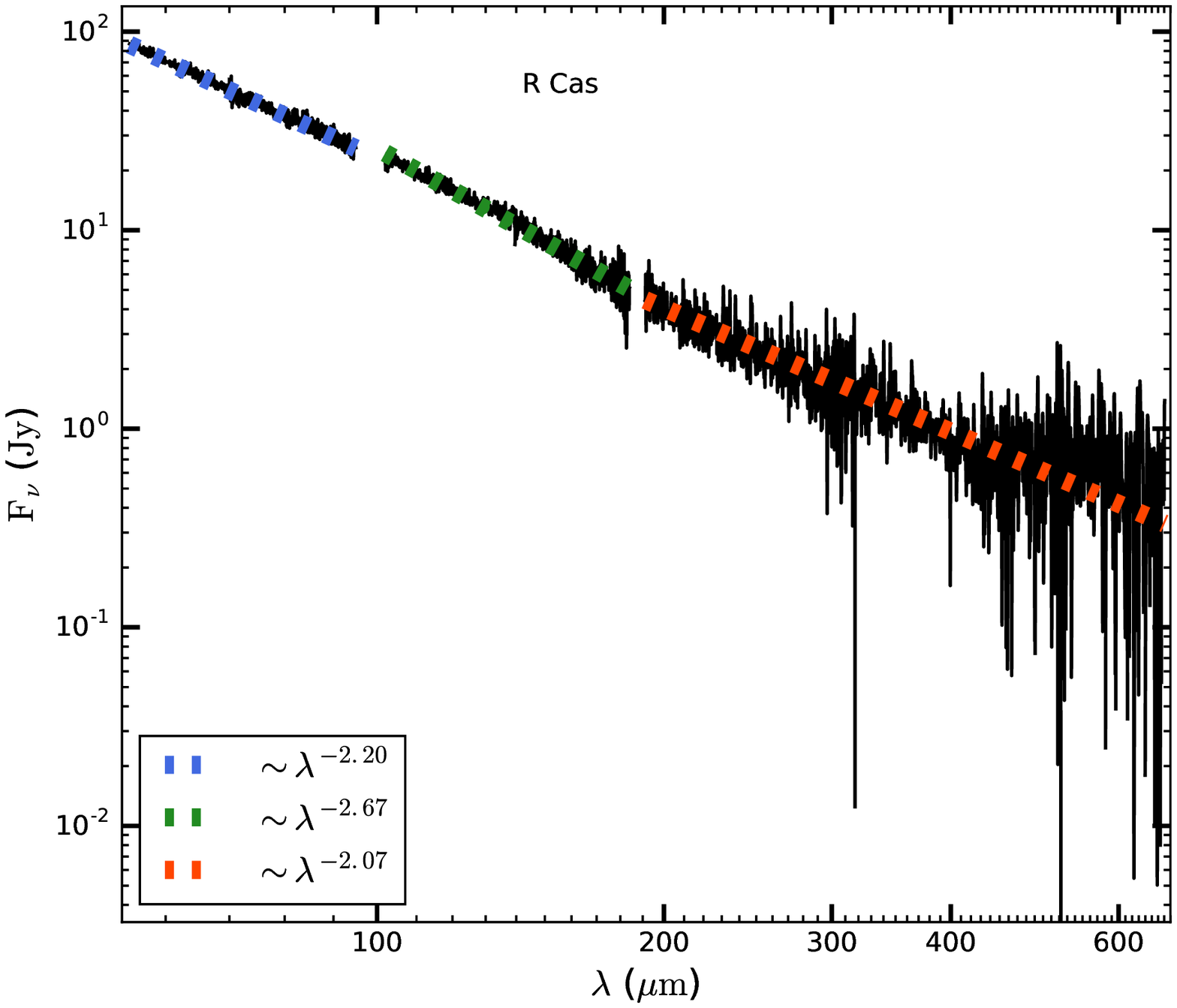}
 \end{subfigure}
 \hfill
  \begin{subfigure}{0.49\textwidth}
 \centering
 \includegraphics[width = \textwidth]{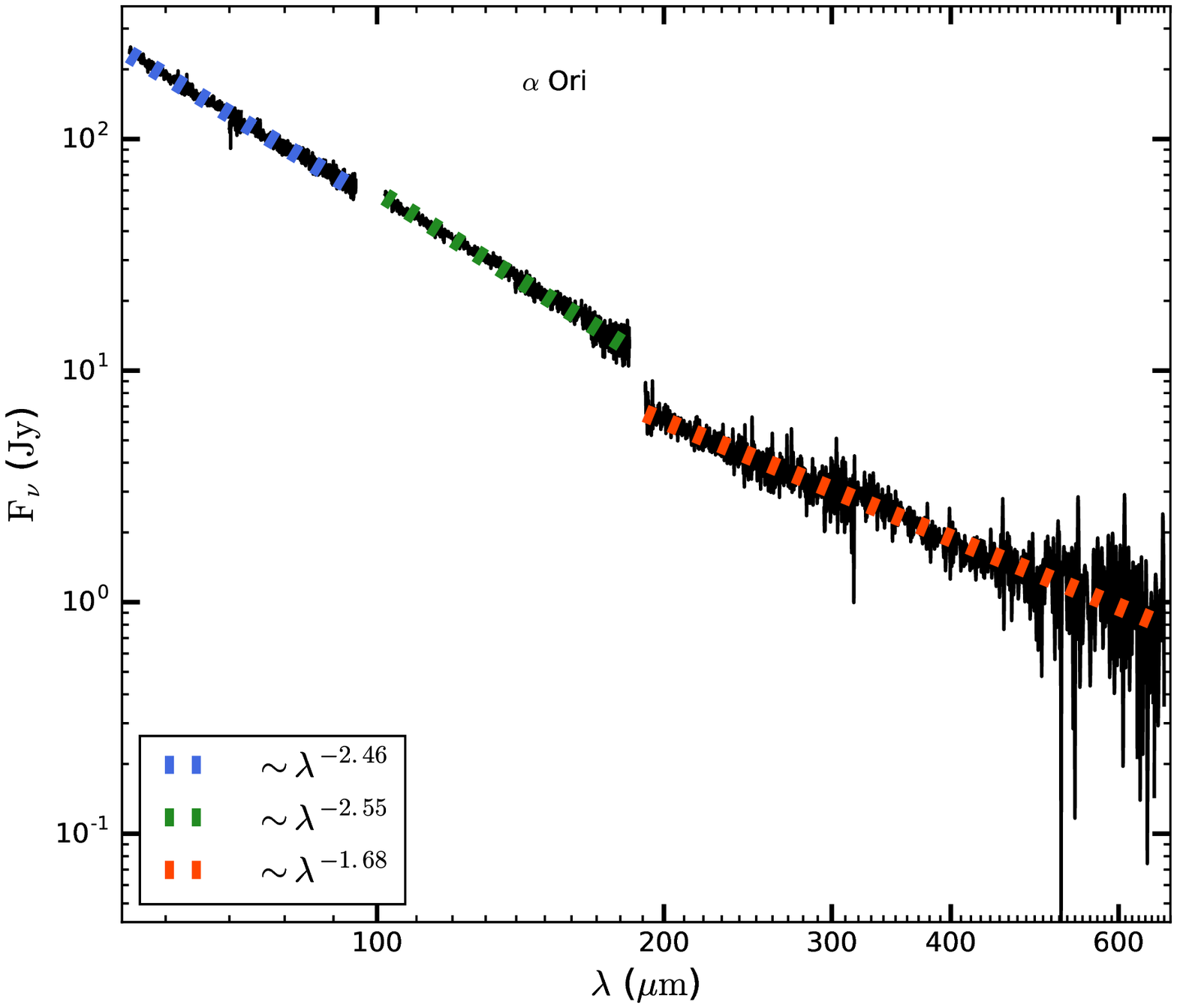}
 \end{subfigure}
  \begin{subfigure}{0.49\textwidth}
 \centering
 \includegraphics[width = \textwidth]{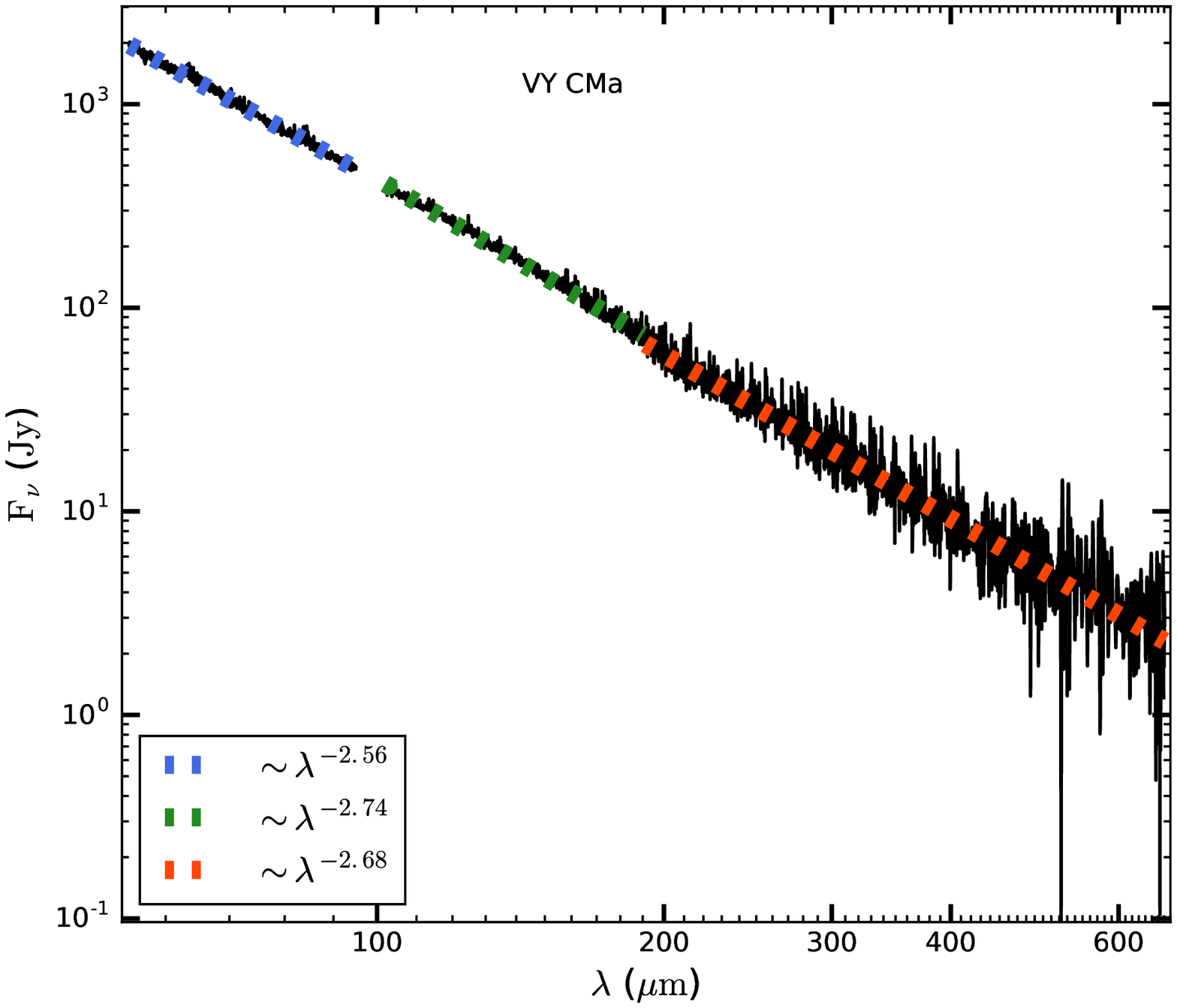}
 \end{subfigure}
 \hfill
   \begin{subfigure}{0.49\textwidth}
 \centering
 \includegraphics[width = \textwidth]{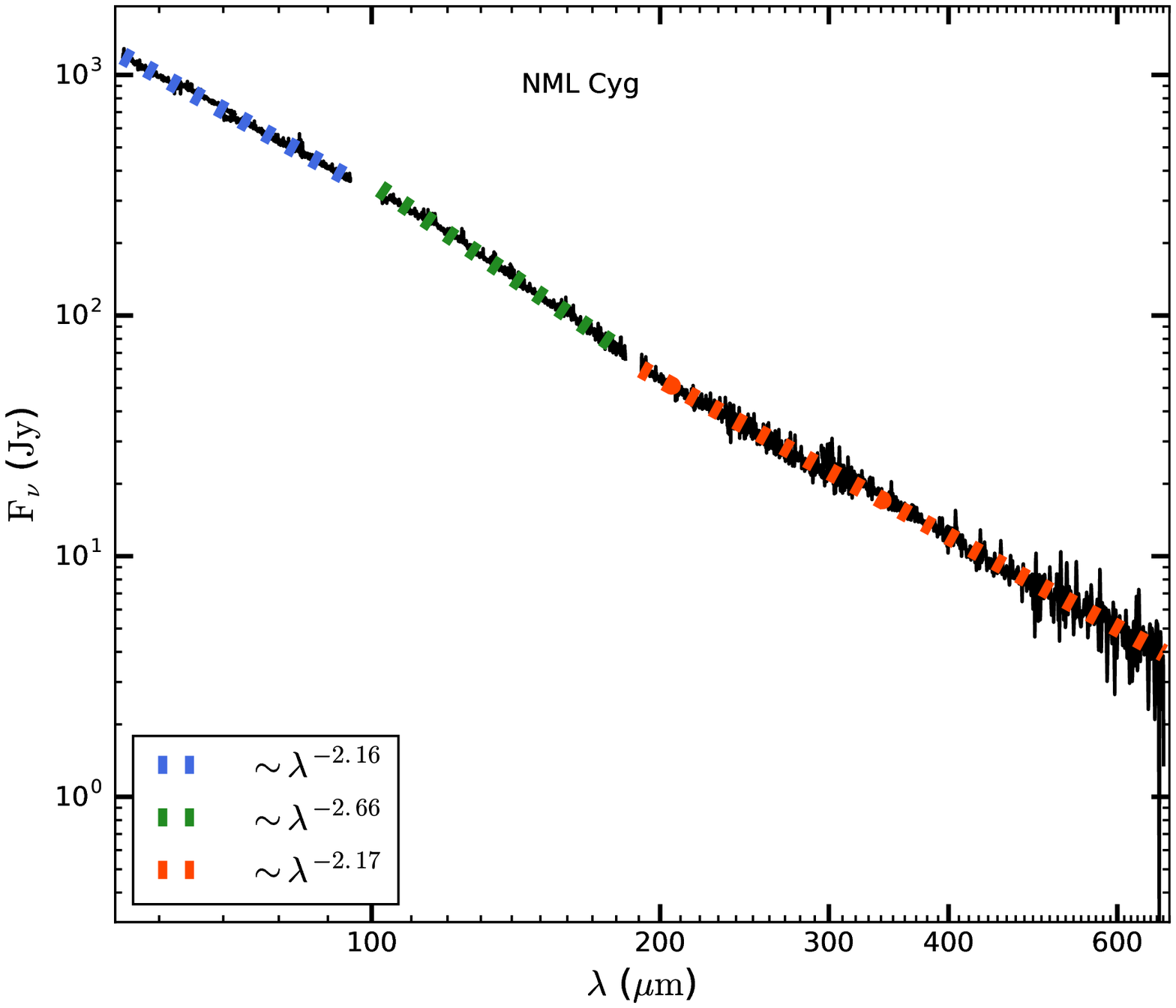}
 \end{subfigure}
  \caption{Continued.}
\end{figure*}

\clearpage

\section{Line list and identification}
\label{Appen:lines}

This Appendix gives the detected lines and the derived line fluxes following the methodology outlined 
in Sect.~\ref{LineMeasurement}, and the possible line identifications as outlined in Sect.~\ref{Sect:Identification}.

Table~\ref{Table:LineF} lists all the information.
Only a portion of this table is shown here to demonstrate its form and content.
The full table with all measured lines and possible line identifications for all targets is available at the CDS.

The first six columns are quantities that are derived from the spectra and the line fitting: 
the spectral band, observed wavelength and frequency, integrated line flux and error, and
the relative FWHM.
The reported uncertainty in the line flux includes only the fitting uncertainty.
The relative FWHM represents the measured width of the lines relative to the theoretical instrumental spectral resolution 
at the corresponding wavelength. High values can indicate possible broadening due to the blending of multiple molecular transitions.
The next seven columns are related to the line identification: the molecular species and transition, the laboratory wavelength and
frequency, and the contribution of that laboratory line to the observed line for 
temperatures of 75, 300, and 500~K assuming LTE and optically thin line emission (see Sect.~\ref{Sect:Identification} for details).

\begin{sidewaystable*}
\caption{Integrated line strengths and possible identifications for selected spectral lines in WX Psc. }
\label{Table:LineF} 
%\centering

\begin{tabular}{lllllllllllll}
\hline
band & $\mathrm{\lambda_{obs}}$ & $\mathrm{\nu_{obs}}$ & $F_{\rm int}$      &  $\sigma(F_{\rm int})$  & rel. fwhm &  Species       
 & Transition  &  $\mathrm{\lambda_0}$ & $\mathrm{\nu_0}$ & Con. 75K        & Con. 300K        &  Con. 500K  \\
    & $\mathrm{(\mu m)}$     &        (GHz)        & $\mathrm{({W}{m^-2})}$ & $\mathrm{({W}{m^-2})}$ &           &      
         &             &$\mathrm{(\mu m)}$    &         (GHz)    & $\mathrm{(\%)}$  & $\mathrm{(\%)}$ & $\mathrm{(\%)}$ \\
\hline

\multicolumn{13}{l}{ PACS } \\
%\multicolumn{13}{l}{\multirow{2}{*}{PACS}} \\
%\multicolumn{13}{l}{ } \\
\hline
B2A&55.82&5370.5073&2.60e-15&2.88e-16&3.142&$\mathrm{H_{2}O}$&$9_{ 7, 3} \rightarrow 9_{ 6, 4}$&55.76&5376.4121&$\mathrm{< 0.01}$&3.05&5.65\\
&   &   &   &   &   &$\mathrm{H_{2}O}$&$11_{ 7, 5} \rightarrow 11_{ 6, 6}$&55.80&5372.7418&$\mathrm{< 0.01}$&0.41&2.00\\
&   &   &   &   &   &$\mathrm{H_{2}O}$&$9_{ 7, 2} \rightarrow 9_{ 6, 3}$&55.80&5372.4280&$\mathrm{< 0.01}$&9.13&16.91\\
&   &   &   &   &   &$\mathrm{NH_{3}}$&$J_{K,i} = 9_{ 8, 1} \rightarrow 8_{ 8, 0}$&55.81&5371.4285&99.28&1.29&$\mathrm{< 0.01}$\\
&   &   &   &   &   &$\mathrm{H_{2}O}$&$10_{ 2, 9} \rightarrow 10_{ 1,10}$&55.84&5368.8073&0.48&61.28&42.10\\
&   &   &   &   &   &$\mathrm{H_{2}O}$&$8_{ 7, 2} \rightarrow 8_{ 6, 3}$&55.84&5368.6418&$\mathrm{< 0.01}$&18.27&22.19\\
&   &   &   &   &   &$\mathrm{H_{2}O}$&$8_{ 7, 1} \rightarrow 8_{ 6, 2}$&55.85&5367.7484&$\mathrm{< 0.01}$&6.09&7.39\\
&   &   &   &   &   &$\mathrm{H_{2}O}$&$12_{ 7, 6} \rightarrow 12_{ 6, 7}$&55.87&5366.2658&$\mathrm{< 0.01}$&0.35&3.02\\
B2A&56.01&5352.9588&1.01e-15&2.04e-16&1.661&$\mathrm{NH_{3}}$&$J_{K,i} = 9_{ 7, 1} \rightarrow 8_{ 7, 0}$&55.97&5355.9080&99.66&3.39&$\mathrm{< 0.01}$\\
&   &   &   &   &   &$\mathrm{H_{2}O}$&$7_{ 7, 1} \rightarrow 7_{ 6, 2}$&55.98&5355.0246&0.02&14.91&16.56\\
&   &   &   &   &   &$\mathrm{H_{2}O}$&$7_{ 7, 0} \rightarrow 7_{ 6, 1}$&55.98&5354.8838&0.07&44.72&49.66\\
&   &   &   &   &   &$\mathrm{H_{2}O}$&$10_{ 1, 9} \rightarrow 10_{ 0,10}$&56.03&5350.8728&0.25&36.98&33.78\\
B2A&56.80&5277.8504&1.58e-15&2.02e-16&1.738&$\mathrm{H_{2}O}$&$9_{ 1, 9} \rightarrow 8_{ 0, 8}$&56.77&5280.7345&24.95&24.99&25.04\\
&   &   &   &   &   &$\mathrm{H_{2}O}$&$9_{ 0, 9} \rightarrow 8_{ 1, 8}$&56.82&5276.5179&74.73&74.81&74.96\\
B2A&57.65&5200.4867&2.25e-15&1.92e-16&1.271&$\mathrm{H_{2}O}$&$4_{ 2, 2} \rightarrow 3_{ 1, 3}$&57.64&5201.4306&100.00&100.00&99.92\\

R1 short&108.09&2773.4422&1.02e-15&4.09e-17&0.954&$\mathrm{H_{2}O}$&$2_{ 2, 1} \rightarrow 1_{ 1, 0}$&108.07&2773.9766&100.00&100.00&100.00\\
R1 short&108.75&2756.7630&1.97e-16&4.16e-17&1.009&$\mathrm{SO_{2}}$&$43_{20,24} \rightarrow 42_{19,23}$&108.72&2757.3522&$\mathrm{< 0.01}$&0.42&1.13\\
&   &   &   &   &   &$\mathrm{H_{2}\element[][17]{O}}$&$2_{ 2, 1} \rightarrow 1_{ 1, 0}$&108.74&2756.8424&49.98&37.93&12.82\\
&   &   &   &   &   &$\mathrm{H_{2}\element[][18]{O}}$&$2_{ 2, 1} \rightarrow 1_{ 1, 0}$&108.74&2756.8424&50.02&37.93&12.88\\
&   &   &   &   &   &CO&24 $\rightarrow$ 23&108.76&2756.3876&$\mathrm{< 0.01}$&23.66&72.84\\
R1 long&144.79&2070.5615&1.73e-16&1.25e-17&0.852&SiO, v =  0        &48 $\rightarrow$ 47&144.75&2071.1379&$\mathrm{< 0.01}$&9.99&44.62\\
&   &   &   &   &   &$\mathrm{SO_{2}}$&$32_{15,17} \rightarrow 31_{14,18}$&144.78&2070.7199&0.67&1.95&0.94\\
&   &   &   &   &   &$\mathrm{SO_{2}}$&$27_{16,12} \rightarrow 26_{15,11}$&144.78&2070.6996&1.56&2.41&1.06\\
&   &   &   &   &   &$\mathrm{H_{2}S}$&$3_{ 2, 1} \rightarrow 2_{ 1, 2}$&144.78&2070.6804&65.37&0.03&$\mathrm{< 0.01}$\\
&   &   &   &   &   &CO&18 $\rightarrow$ 17&144.78&2070.6160&32.30&85.57&53.27\\
\hline
\multicolumn{13}{l}{ SPIRE } \\
%\multicolumn{13}{l}{\multirow{2}{*}{SPIRE}} \\
%\multicolumn{13}{l}{ } \\
\hline
SSW&191.33&1566.9171&3.31e-17&3.49e-18&1.073&$\mathrm{SO_{2}}$&$48_{ 5,43} \rightarrow 48_{ 2,46}$&191.27&1567.3843&0.14&1.37&1.15\\
&   &   &   &   &   &$\mathrm{SO_{2}}$&$84_{ 5,79} \rightarrow 83_{ 6,78}$&191.30&1567.1178&$\mathrm{< 0.01}$&0.17&2.61\\
&   &   &   &   &   &$\mathrm{SO_{2}}$&$85_{ 5,81} \rightarrow 84_{ 4,80}$&191.33&1566.8877&$\mathrm{< 0.01}$&0.18&2.93\\
&   &   &   &   &   &\element[][13]{C}S&34 $\rightarrow$ 33&191.34&1566.8118&0.10&10.81&16.02\\
&   &   &   &   &   &$\mathrm{SO_{2}}$&$44_{ 5,39} \rightarrow 43_{ 4,40}$&191.34&1566.7757&17.05&29.04&19.38\\
&   &   &   &   &   &$\mathrm{SO_{2}}$, $\nu_{\mathrm{2}}$ = 1&$25_{11,15} \rightarrow 24_{10,14}$&191.35&1566.7451&0.53&33.71&36.42\\
&   &   &   &   &   &$\mathrm{SO_{2}}$&$40_{ 4,36} \rightarrow 39_{ 3,37}$&191.37&1566.5652&82.18&23.97&12.65\\
&   &   &   &   &   &$\mathrm{SO_{2}}$&$86_{ 3,83} \rightarrow 85_{ 4,82}$&191.38&1566.4728&$\mathrm{< 0.01}$&0.20&3.23\\
&   &   &   &   &   &$\mathrm{SO_{2}}$&$87_{ 3,85} \rightarrow 86_{ 2,84}$&191.40&1566.2905&$\mathrm{< 0.01}$&0.21&3.52\\
SSW&194.47&1541.6022&8.11e-17&3.19e-18&0.996&$\mathrm{H_{2}O}$&$6_{ 3, 3} \rightarrow 5_{ 4, 2}$&194.42&1541.9670&15.33&82.84&85.96\\
&   &   &   &   &   &\element[][13]{C}O&14 $\rightarrow$ 13&194.55&1540.9883&84.67&16.77&13.35\\
SSW&216.95&1381.8594&7.12e-17&3.04e-18&0.909&CO&12 $\rightarrow$ 11&216.93&1381.9951&100.00&99.82&99.56\\
SSW&226.80&1321.8572&1.22e-16&3.24e-18&1.038&$\mathrm{H_{2}O}$&$6_{ 2, 5} \rightarrow 5_{ 3, 2}$&226.76&1322.0648&43.26&94.77&95.81\\
&   &   &   &   &   &\element[][13]{C}O&12 $\rightarrow$ 11&226.90&1321.2655&56.64&5.22&4.12\\
SLW&294.89&1016.6164&3.29e-17&2.70e-18&1.044&$\mathrm{SO_{2}}$&$47_{ 5,43} \rightarrow 46_{ 4,42}$&294.92&1016.5113&0.71&46.00&53.66\\
&   &   &   &   &   &$\mathrm{SO_{2}}$&$32_{17,15} \rightarrow 33_{16,18}$&294.98&1016.3064&0.03&4.63&6.03\\
&   &   &   &   &   &$\mathrm{SO_{2}}$&$27_{16,12} \rightarrow 28_{15,13}$&295.01&1016.1947&0.38&6.46&6.30\\
&   &   &   &   &   &\element[][13]{C}S&22 $\rightarrow$ 21&295.09&1015.9294&65.54&40.94&33.00\\
&   &   &   &   &   &$\mathrm{SO_{2}}$&$28_{ 3,25} \rightarrow 28_{ 0,28}$&295.10&1015.8995&33.34&1.98&0.91\\
\hline
\end{tabular}
\end{sidewaystable*}

\section{Atomic lines}
\label{Appen-Atomic}

In this Appendix the presence of atomic lines in the PACS and SPIRE spectra is discussed. 
There is relatively little known about atomic lines around AGB stars.
The C{\sc i} (1-0) line at 609~$\mu$m has been detected in a few AGB stars using heterodyne techniques.
The first detection of C{\sc i} in an AGB star was made by \citet{Keene93} around CW Leo, and was confirmed by \citet{vdVeen98}.
The modelling of these observations showed that the C{\sc i} is located in a shell (or shells) located at $\sim 14$\arcsec\ (or larger).
The same applies to the detection in R Scl \citep{Olofsson15} where the C{\sc i} is located in a shell at about  18\arcsec.
In both cases the origin of the atomic carbon is believed to be due to the photo dissociation of CO.
\citet{Saberi18} detected emission around $o$ Ceti and confirmed the tentative detection in V Hya by \citet{Knapp2000}.
The emission in $o$ Cet is also believed to be in a shell located at $\sim 36$\arcsec,  while \citet{Knapp2000} suggest shock dissociation 
of circumstellar CO (close to the star) for V Hya.
Strong emission has been seen in $\alpha$ Ori \citep{Huggins94,vdVeen98} believed to be related to the chromosphere in this supergiant.
Non-detections have been reported for the following stars in our sample which have spectra: IK Tau and VY CMa \citep{vdVeen98}.

Figure~\ref{Fig:Atomic} shows the region around the C{\sc i} (1-0) line at 609~$\mu$m and the (2-1) line at 370~$\mu$m for the stars discussed above.
The spectrum used is the one that remains after removing all identified molecular lines, as described in Sect.~\ref{Sect:Identification} 
and Appendix~\ref{Appen:lines}. No formal fitting has been attempted and the noise level can be judged from the spectral region shown.

The stars where previous analysis has shown that the atomic carbon is located in a shell (CW Leo, R Scl and $o$ Cet) are not detected.
The surface brightness of the emission must be faint, and partly outside the central $ 3\times 3$ spaxels.
Emission is detected in the sources where previous analysis suggested that the C{\sc i} is located close to the central star (V Hya and $\alpha$ Ori)

The next two sources have been reported as non-detections from heterodyne observations and this is confirmed here (NML Tau and VY CMa).
The remaining sources, OH 127.8, AFGL 5379, CIT 6, and AFGL 3068, appear to show one or both C{\sc i} lines, and are targets suggested
for follow-up observations at higher spectral resolution.

The O{\sc i} lines at 63.2 and 145.5~$\mu$m were also inspected but no obvious emission lines appear present.
Finally, the C{\sc ii} 157.7~$\mu$m lines were plotted but no convincing emission lines were found, except some background subtraction problems 
in OH\,30.1\,$-$0.7 and OH\,26.5\,+0.6.

\begin{figure*}

\begin{minipage}{0.49\textwidth}
\resizebox{\hsize}{!}{\includegraphics[angle=0]{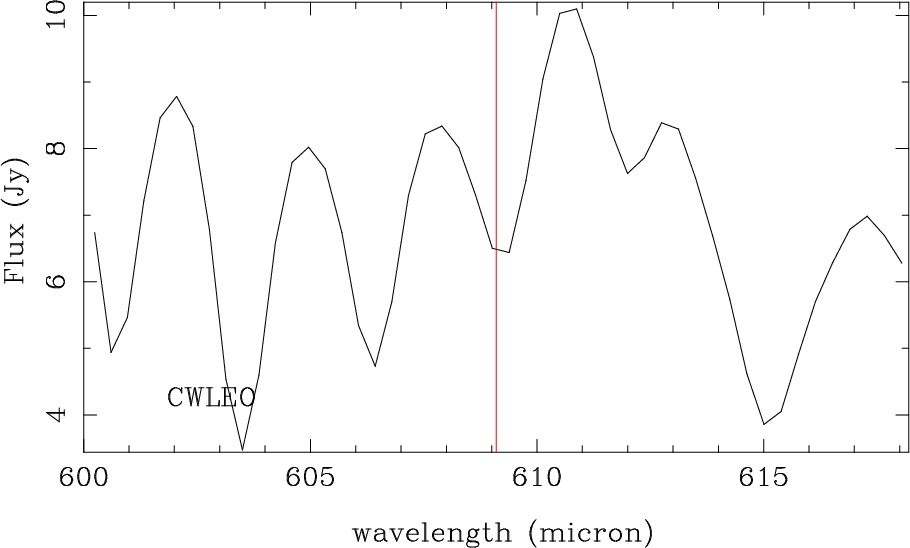}} 
\end{minipage}
\begin{minipage}{0.49\textwidth}
\resizebox{\hsize}{!}{\includegraphics[angle=0]{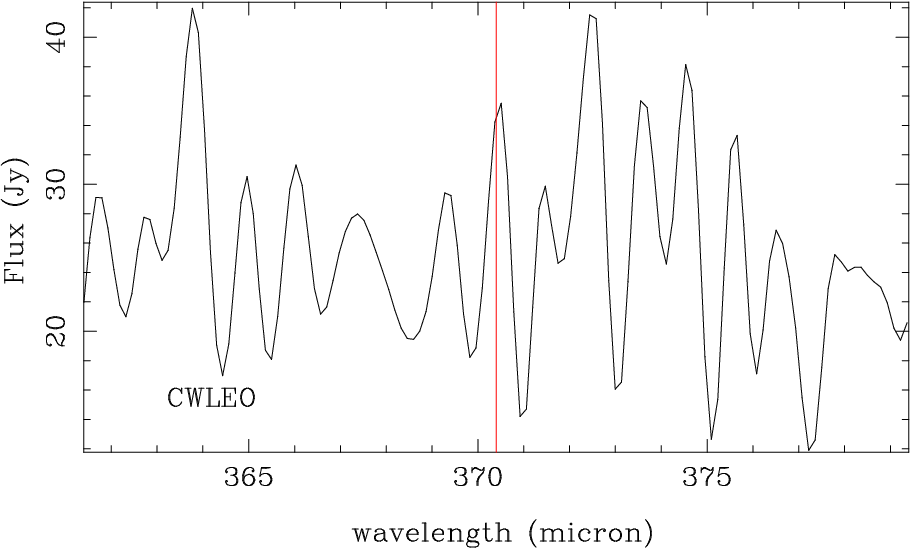}} 
\end{minipage}

\begin{minipage}{0.49\textwidth}
\resizebox{\hsize}{!}{\includegraphics[angle=0]{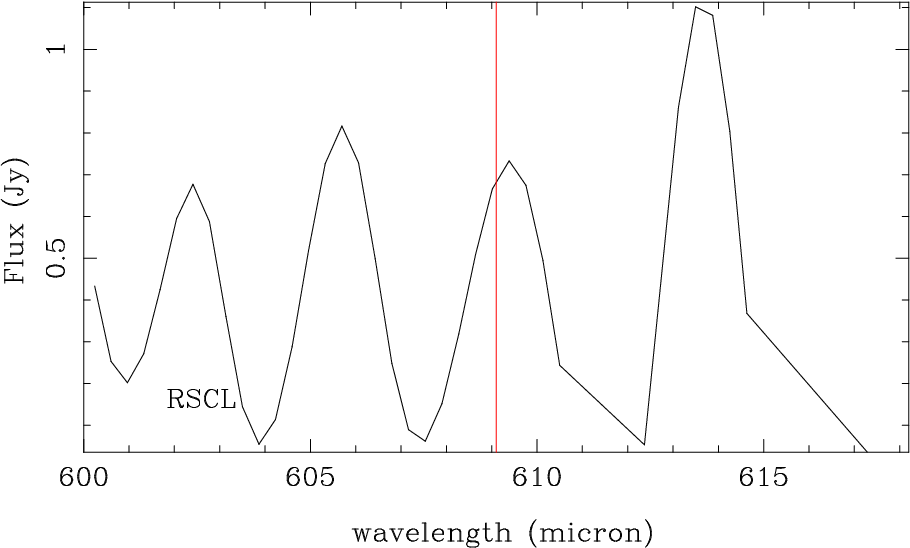}} 
\end{minipage}
\begin{minipage}{0.49\textwidth}
\resizebox{\hsize}{!}{\includegraphics[angle=0]{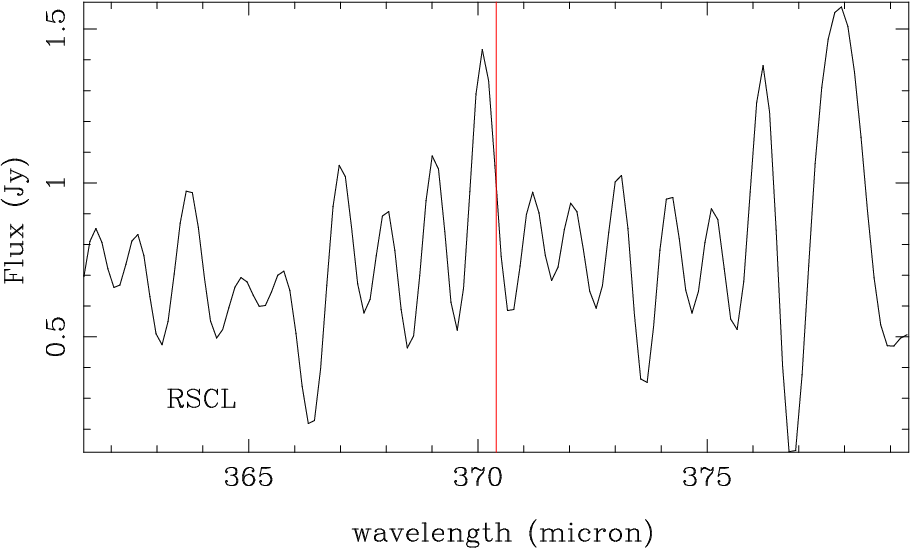}} 
\end{minipage}

\begin{minipage}{0.49\textwidth}
\resizebox{\hsize}{!}{\includegraphics[angle=0]{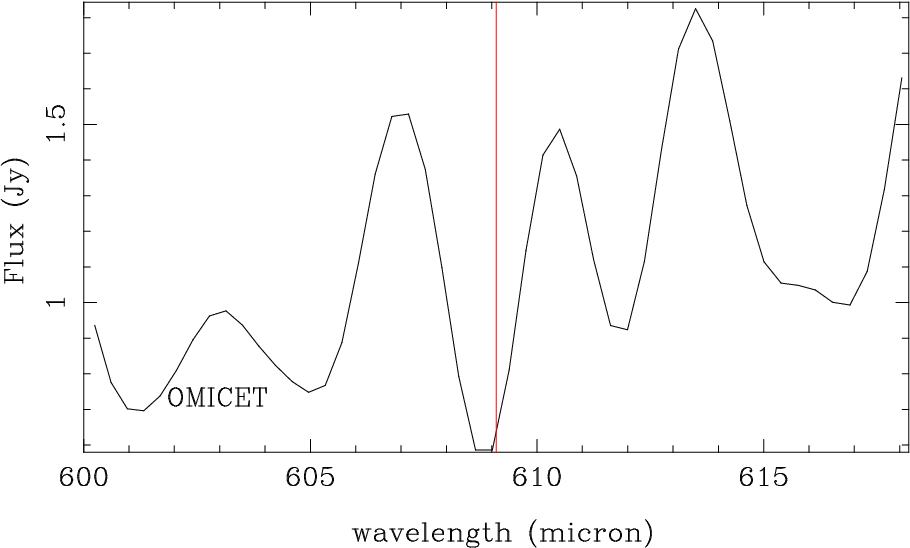}} 
\end{minipage}
\begin{minipage}{0.49\textwidth}
\resizebox{\hsize}{!}{\includegraphics[angle=0]{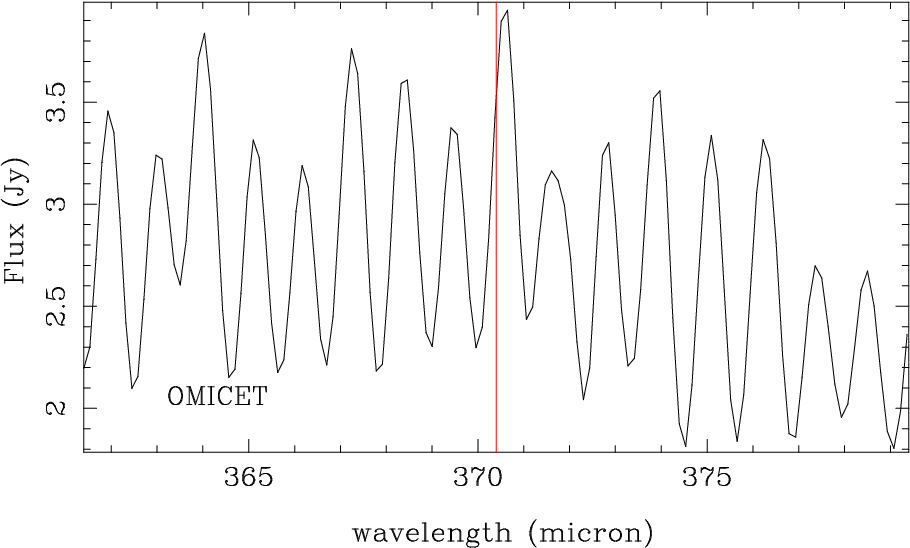}} 
\end{minipage}

\begin{minipage}{0.49\textwidth}
\resizebox{\hsize}{!}{\includegraphics[angle=0]{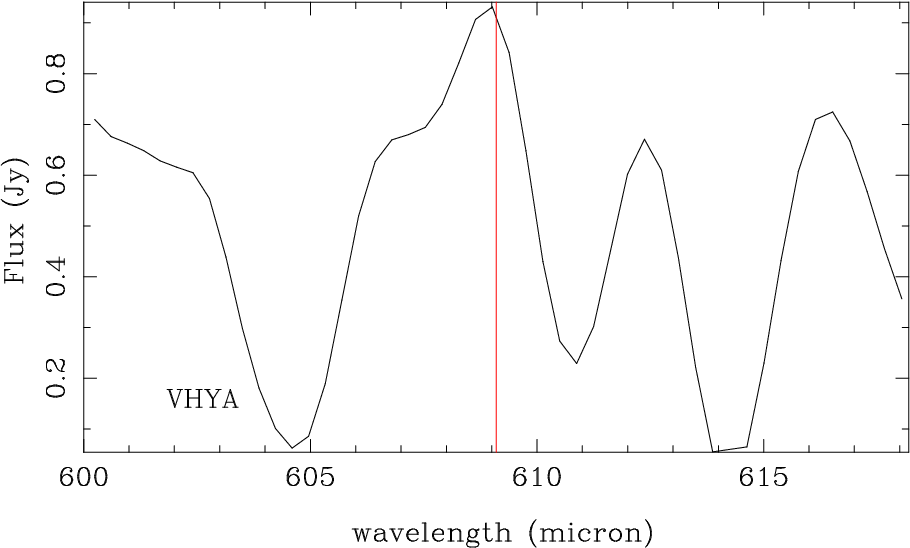}} 
\end{minipage}
\begin{minipage}{0.49\textwidth}
\resizebox{\hsize}{!}{\includegraphics[angle=0]{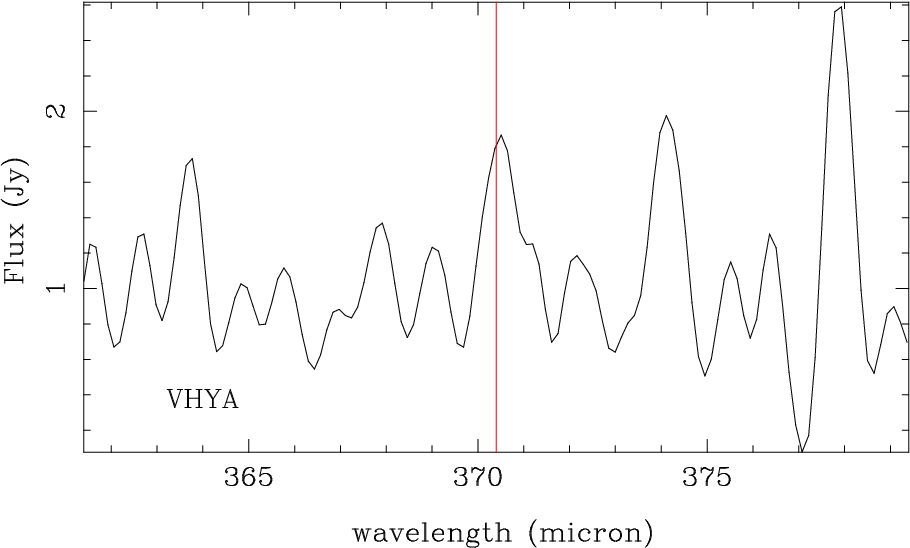}} 
\end{minipage}
   \caption{C{\sc i} lines at 609 (left) and 370 $\mu$m (right). See text for comments on individual sources.}

\label{Fig:Atomic}
\end{figure*}

\setcounter{figure}{0}
\begin{figure*}

\begin{minipage}{0.49\textwidth}
\resizebox{\hsize}{!}{\includegraphics[angle=0]{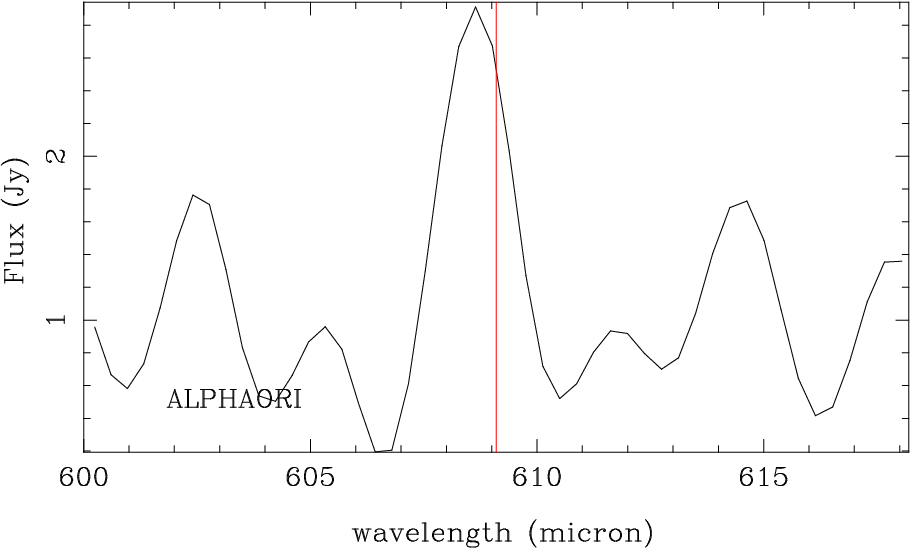}} 
\end{minipage}
\begin{minipage}{0.49\textwidth}
\resizebox{\hsize}{!}{\includegraphics[angle=0]{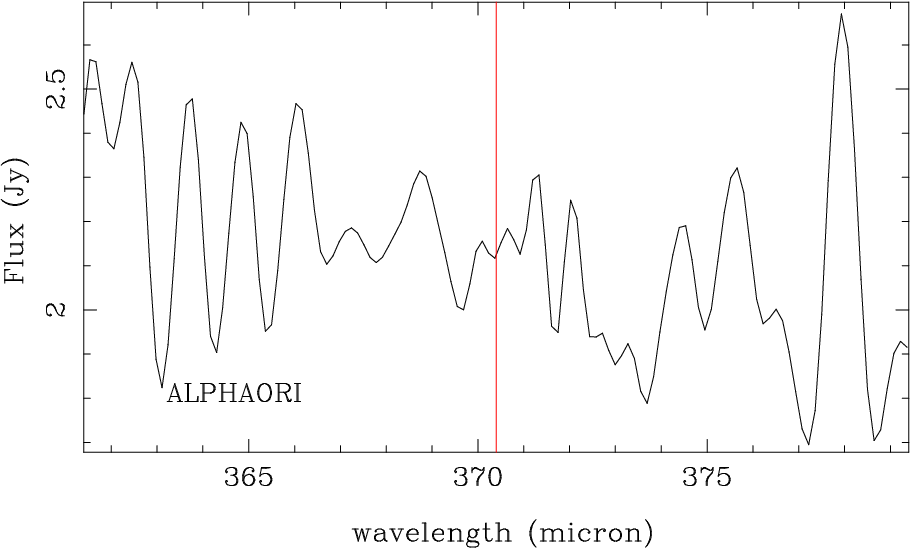}} 
\end{minipage}

\begin{minipage}{0.49\textwidth}
\resizebox{\hsize}{!}{\includegraphics[angle=0]{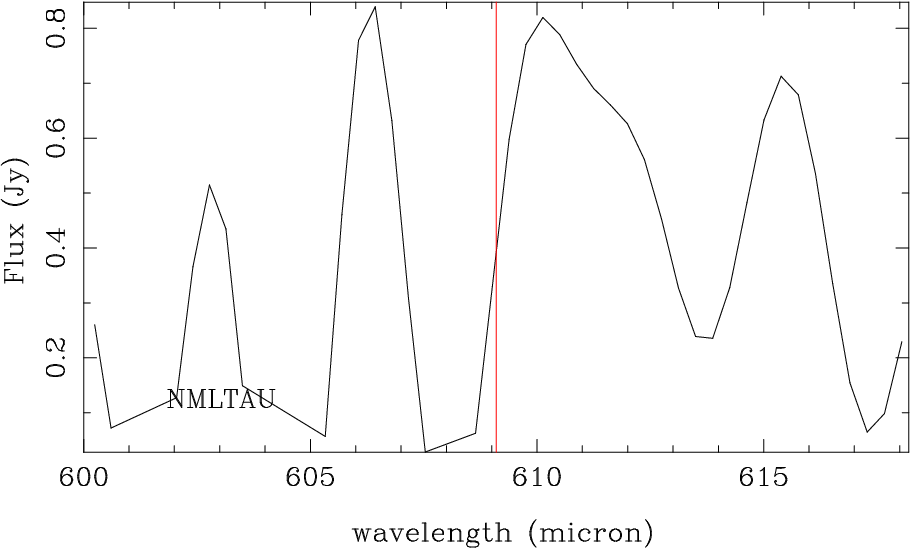}} 
\end{minipage}
\begin{minipage}{0.49\textwidth}
\resizebox{\hsize}{!}{\includegraphics[angle=0]{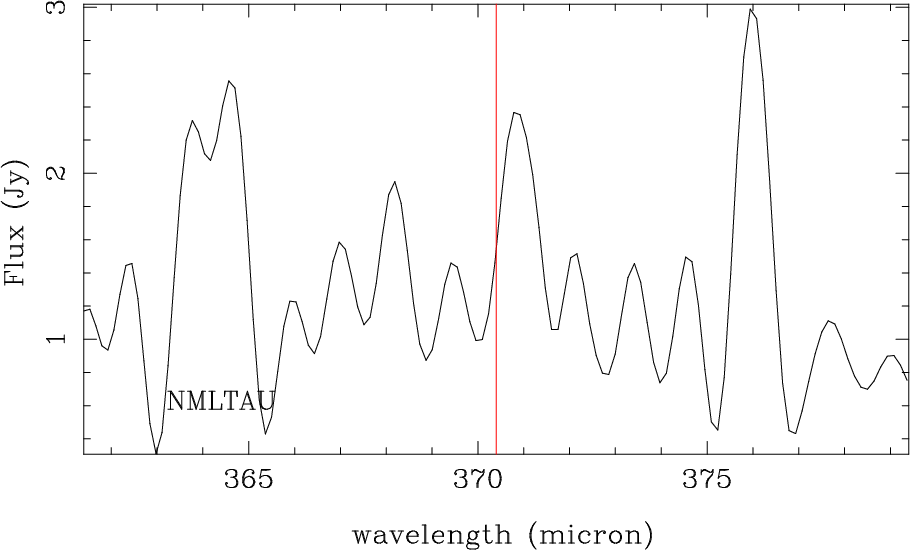}} 
\end{minipage}

\begin{minipage}{0.49\textwidth}
\resizebox{\hsize}{!}{\includegraphics[angle=0]{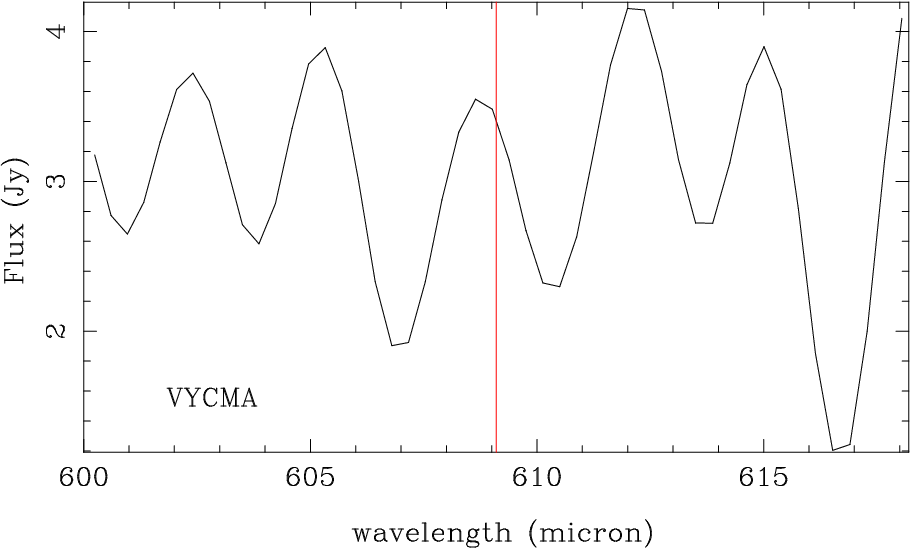}} 
\end{minipage}
\begin{minipage}{0.49\textwidth}
\resizebox{\hsize}{!}{\includegraphics[angle=0]{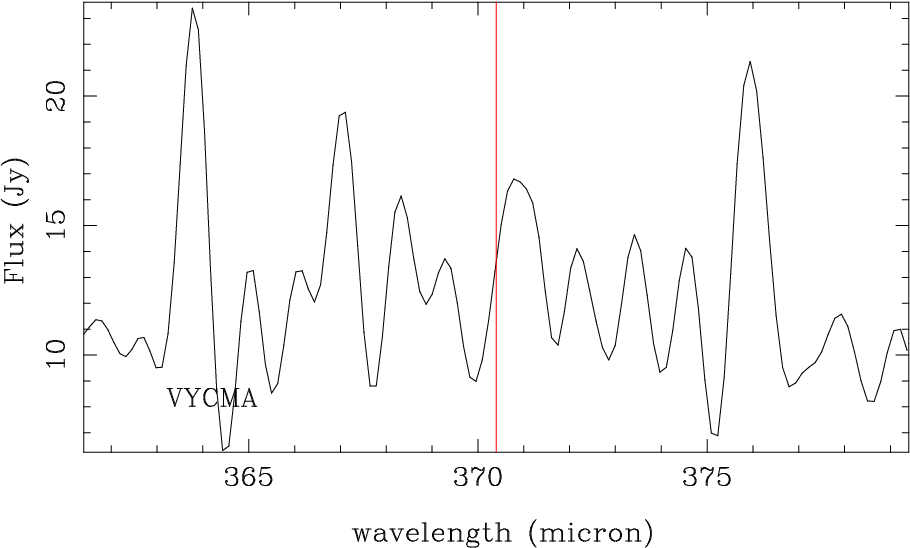}} 
\end{minipage}

\begin{minipage}{0.49\textwidth}
\resizebox{\hsize}{!}{\includegraphics[angle=0]{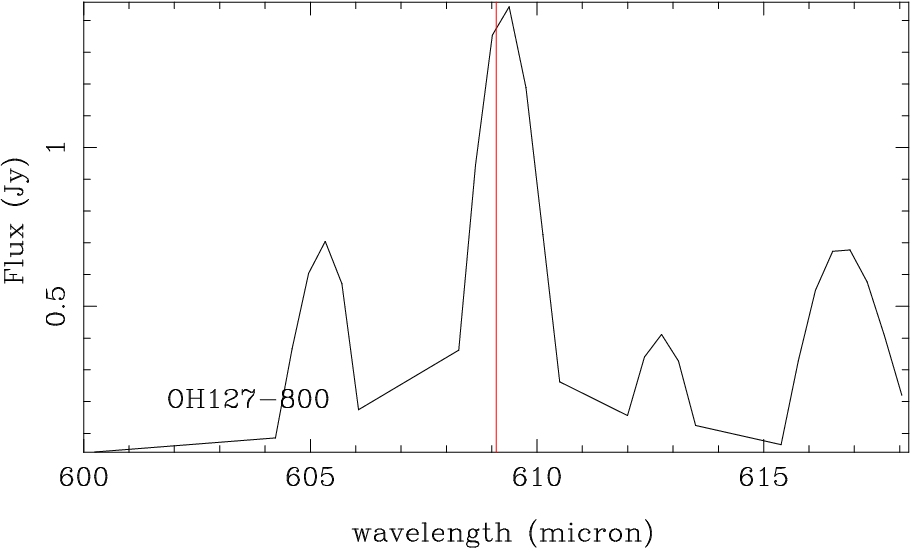}} 
\end{minipage}
\begin{minipage}{0.49\textwidth}
\resizebox{\hsize}{!}{\includegraphics[angle=0]{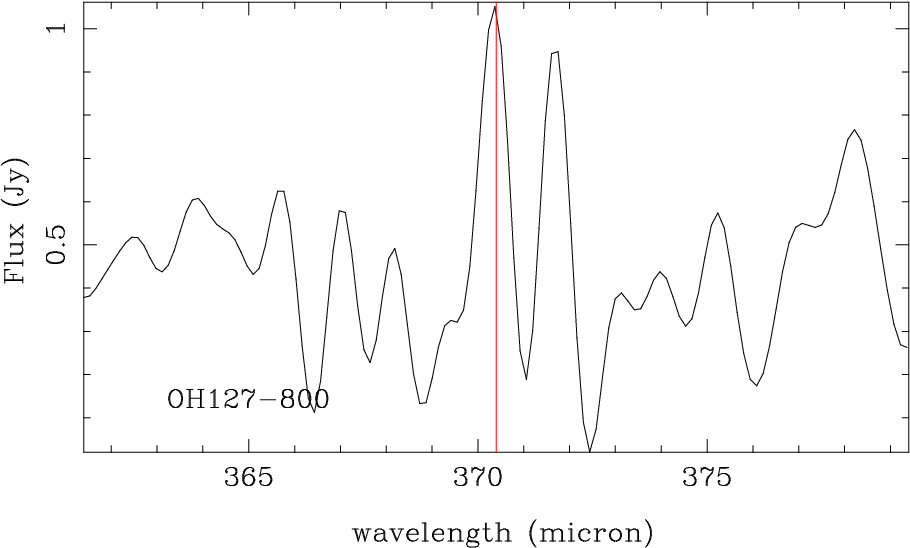}} 
\end{minipage}

   \caption{Continued.}

\end{figure*}

\setcounter{figure}{0}
\begin{figure*}

\begin{minipage}{0.49\textwidth}
\resizebox{\hsize}{!}{\includegraphics[angle=0]{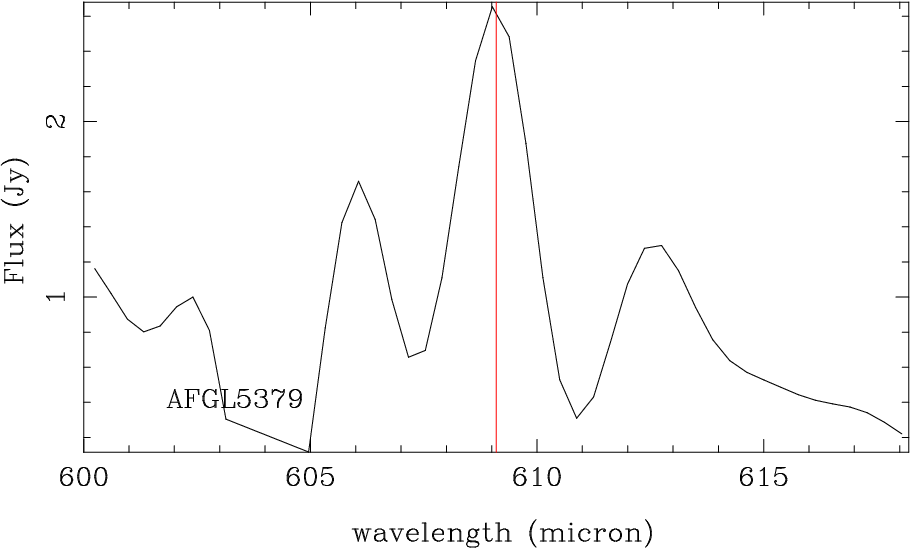}} 
\end{minipage}
\begin{minipage}{0.49\textwidth}
\resizebox{\hsize}{!}{\includegraphics[angle=0]{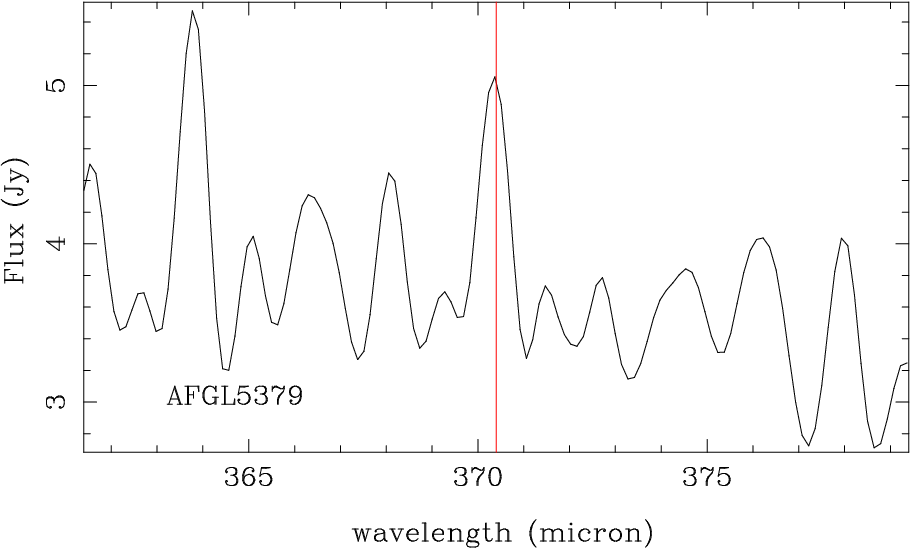}} 
\end{minipage}

\begin{minipage}{0.49\textwidth}
\resizebox{\hsize}{!}{\includegraphics[angle=0]{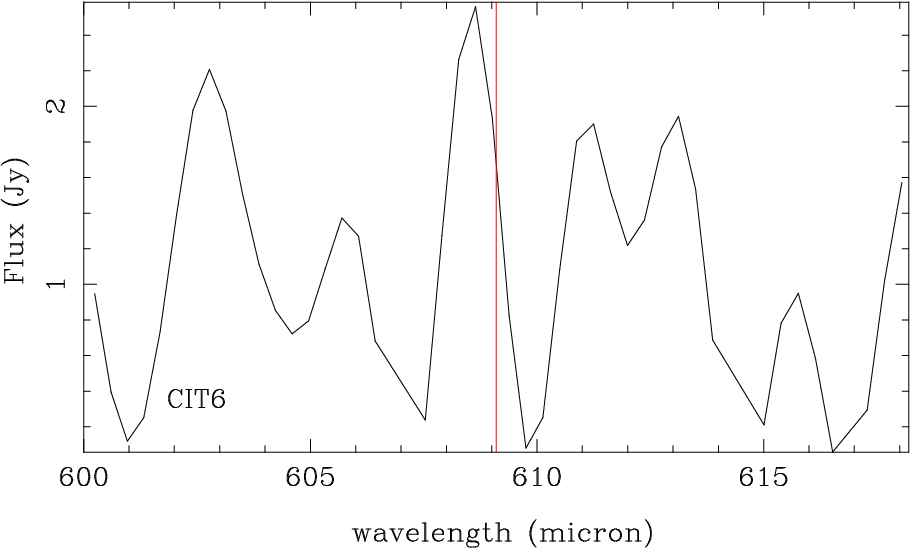}} 
\end{minipage}
\begin{minipage}{0.49\textwidth}
\resizebox{\hsize}{!}{\includegraphics[angle=0]{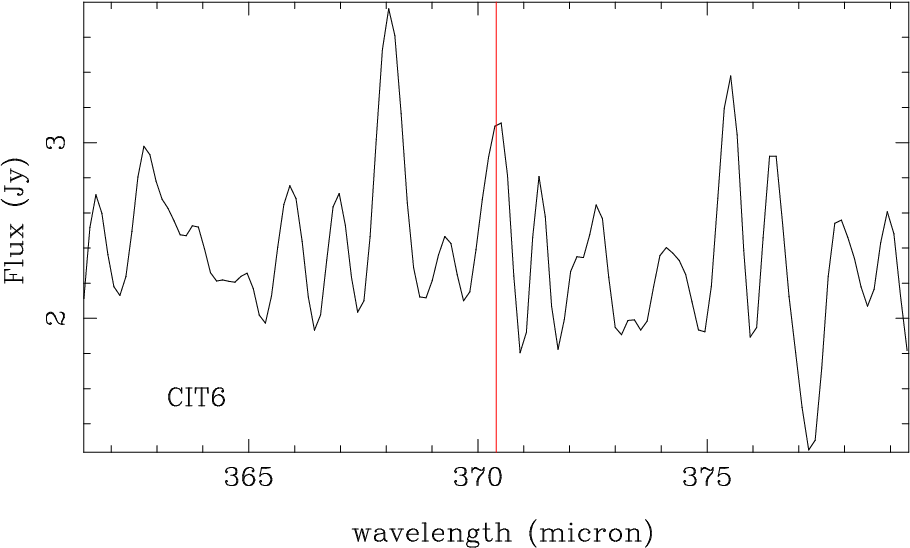}} 
\end{minipage}

\begin{minipage}{0.49\textwidth}
\resizebox{\hsize}{!}{\includegraphics[angle=0]{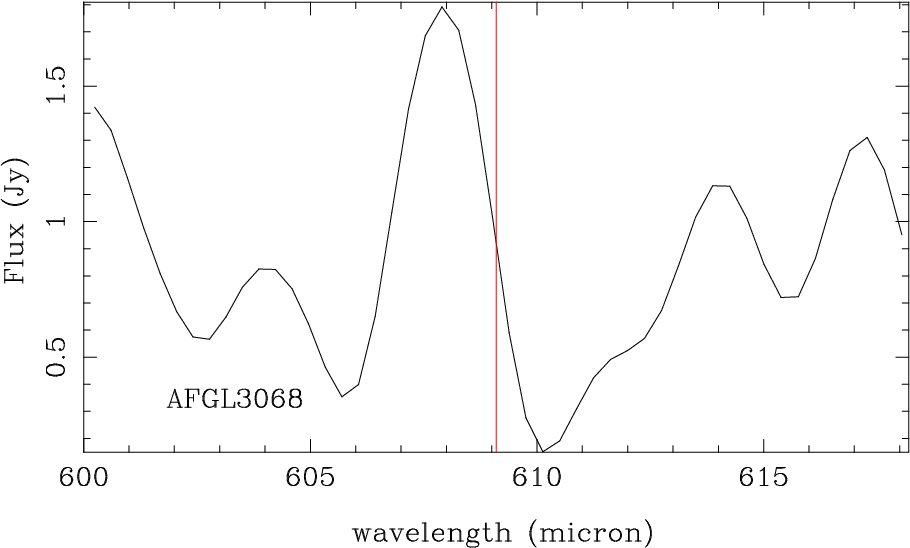}} 
\end{minipage}
\begin{minipage}{0.49\textwidth}
\resizebox{\hsize}{!}{\includegraphics[angle=0]{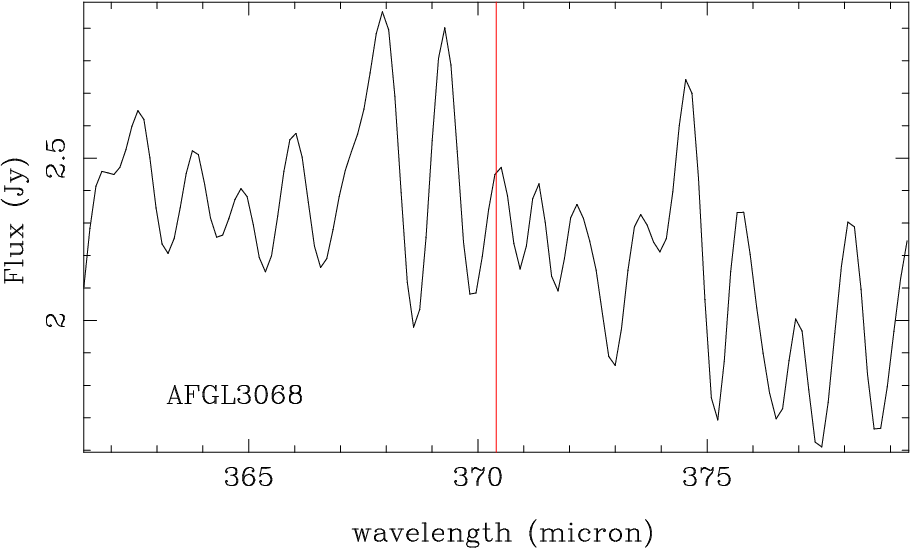}} 
\end{minipage}

    \caption{Continued.}
\end{figure*}

\end{appendix}

\end{document}